**Implementation of a practical Markov chain Monte Carlo sampling algorithm in PyBioNetFit**

Jacob Neumann[1], Yen Ting Lin[2], Abhishek Mallela[4], Ely F. Miller[1], Joshua Colvin[1], Abell T. Duprat[1], Ye Chen[1], William S. Hlavacek[3*] and Richard G. Posner[1*]

[1]Department of Biological Sciences, Northern Arizona University, Flagstaff, Arizona 86011, USA
[2]Computer, Computational, and Statistical Sciences Division and [3]Theoretical Division, Los Alamos National Laboratory, Los Alamos, New Mexico 87545, USA
[4]Department of Mathematics University of California, Davis, California 95616, USA
[*]Corresponding authors: W.S.H. (wish@lanl.gov), R.G.P. (richard.posner@nau.edu)

**Abstract**
Bayesian inference in biological modeling commonly relies on Markov chain Monte Carlo (MCMC) sampling of a multidimensional and non-Gaussian posterior distribution that is not analytically tractable. Here, we present the implementation of a practical MCMC method in the open-source software package PyBioNetFit (PyBNF), which is designed to support parameterization of mathematical models for biological systems. The new MCMC method, am, incorporates an adaptive move proposal distribution. For warm starts, sampling can be initiated at a specified location in parameter space and with a multivariate Gaussian proposal distribution defined initially by a specified covariance matrix. Multiple chains can be generated in parallel using a computer cluster. We demonstrate that am can be used to successfully solve real-world Bayesian inference problems, including forecasting of new Coronavirus Disease 2019 case detection with Bayesian quantification of forecast uncertainty. PyBNF version 1.1.9, the first stable release with am, is available at PyPI and can be installed using the pip package-management system on platforms that have a working installation of Python 3. PyBNF relies on libRoadRunner and BioNetGen for simulations (e.g., numerical integration of ordinary differential equations defined in SBML or BNGL files) and Dask.Distributed for task scheduling on Linux computer clusters.

**PyBioNetFit documentation:** https://pybnf.readthedocs.io/en/latest/
**PyBioNetFit source code:** https://github.com/lanl/pybnf

**Introduction**

Given a model structure, a dataset, and a prior distribution for model parameter values, Bayesian inference (Gelman et al., 2014) is a dataset-dependent transformation from the prior distribution, which encodes pre-existing knowledge (external to the dataset of interest) about parameter values, to a posterior distribution, which probabilistically quantifies new data-informed parameter estimates. The posterior is typically characterized through Markov chain Monte Carlo (MCMC) sampling (Andrieu et al., 2003). The MCMC sampling problems that arise in biological modeling applications of Bayesian inference are often challenging, because the posterior is typically far from Gaussian and multidimensional. A major benefit of Bayesian inference is the ability to quantify uncertainties in parameter estimates (in terms of the parameter posterior), and also in predictions, which is accomplished by performing an array of simulations based on different samples from the parameter posterior. Thus, practical MCMC sampling methods are important for assessing the reliability of predictions obtained from data-driven model parameterizations.

In recent work (Lin et al., 2021), we used an adaptive MCMC sampling method described by Andrieu and Thoms (2008) to make daily 7-d ahead forecasts of newly detected Coronavirus Disease 2019 (COVID-19) cases with Bayesian quantification of forecast uncertainties. We implemented this method in PyBioNetFit (PyBNF), a general-purpose software package designed to support parameterization of biological models. We also added various other features, such as support for a negative binomial likelihood function $NB(r, p)$ wherein the hyperparameter $r$ may be specified before inference or jointly inferred with model parameters. We evaluated new PyBNF features by solving an array of test problems and comparing the results against independently generated results.

**Methods and Implementation**

PyBNF's new MCMC sampler, am, is based on the pseudocode labeled "Algorithm 4" in Andrieu and Thoms (2008). See also Supplementary Methods. The am method is adaptive, meaning that the covariance matrix of the multivariate Gaussian proposal distribution (i.e., the distribution used to stochastically generate move proposals), is learned/optimized on-the-fly during sampling. Both the starting point for sampling and the algorithmic parameters of the proposal distribution are initialized on the basis of user-supplied inputs. Warm starts and continuation are supported. Multiple chains may be generated in an embarrassingly parallel fashion on a computer cluster and combined to decrease the wall-clock time required to complete an inference job. Usage of the am method is fully explained in the online PyBNF documentation.

Inference job setup files for test problems are included in the latest PyBNF distribution. The files are also provided as Supplementary Files S1–S4, which are ZIP archives. As described in Mitra et al. (2019), inference job setup requires a configuration file (marked by a CONF filename extension), a model-definition file compatible with either BioNetGen (Harris et al., 2015) or libRoadRunner (Somogyi et al., 2015) (marked by a BNGL or XML filename extension), and one or more files containing data in a tabular format (marked by an EXP filename extension).

**Results**

To assess the correctness and practicality of the am method, we used it to solve a series of increasingly challenging test problems. We evaluated am relative to mh, a sampler implemented in PyBNF version 1.01 (Mitra et al., 2019) that uses a fixed move proposal distribution, and against PyBNF-independent problem-specific solutions.

We started with a linear regression problem involving synthetic data (Supplementary File S1, Supplementary Methods). For this problem, an analytical expression for the posterior exists. PyBNF's am method was able to reconstruct the analytical posterior nearly exactly (Supplementary Fig. S1), whereas mh failed with the same computational budget.

We next considered a non-linear regression problem: inferring the values of four parameters in a two-phase exponential decay model for viral dynamics under therapy (Ho et al., 1995; Perelson et al., 1996; Supplementary File S2, Supplementary Methods). Results generated using PyBNF's am method were found to be consistent with results generated using problem-specific code (Supplementary Figs. S2–S4).

Using a previously established benchmark problem (Harmon et al., 2017; Mitra et al., 2019; Supplementary File S3, Supplementary Methods), we evaluated the efficiency of am relative to mh. Results obtained using the two methods are consistent (Supplementary Fig. S5). For each of the 16 adjustable parameters in the benchmark problem, we calculated the effective sample size (ESS) for an equal-length chain of parameter values obtained using either am or mh (Supplementary Methods, Supplementary Table S1). ESS for am was typically larger than for mh, indicating greater efficiency. The ratio $\mathrm{ESS_{am}/ESS_{mh}}$ ranged from 3.03 to 27.7, with a mean (median) of 11.5 (10.5).

Finally, we used PyBNF's am method to reproduce inferences performed in the epidemiological forecasting study of Lin et al. (2021) (Supplementary File S4, Supplementary Methods). As illustrated in Figure 1, marginal posteriors found using the problem-specific code of Lin et al. (2021) and am are indistinguishable. Supplementary Figs. S6–S20 show the full results of our comparison of am against the code of Lin et al. (2021).

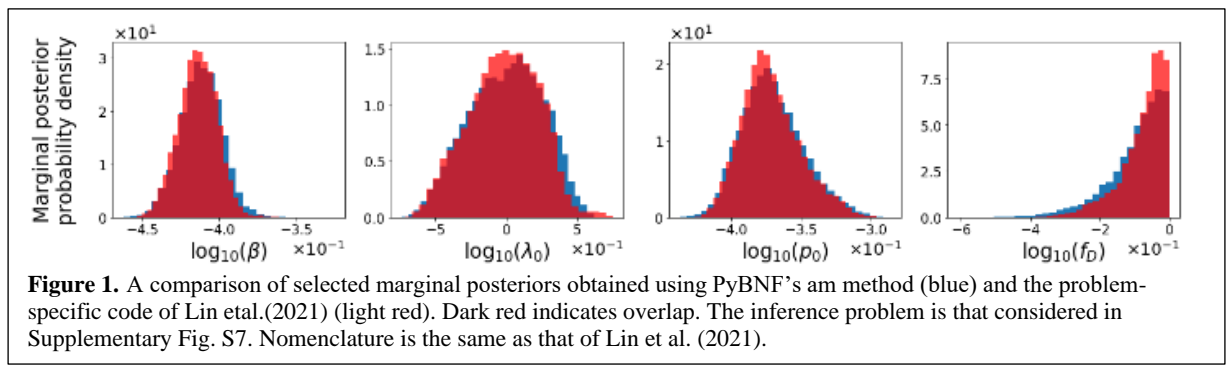

**Figure 1.** A comparison of selected marginal posteriors obtained using PyBNF's am method (blue) and the problem-specific code of Lin etal.(2021) (light red). Dark red indicates overlap. The inference problem is that considered in Supplementary Fig. S7. Nomenclature is the same as that of Lin et al. (2021).

**Conclusions**

We find that the new am method available in PyBNF version 1.1.9 is significantly more efficient than the mh method available in earlier versions of PyBNF. We also find that the adaptive sampler is practical, in that it can be used to solve real-world inference problems, including challenging inference problems that arose in a recent Coronavirus Disease 2019 (COVID-19) forecasting effort (Lin et al., 2021) (Figure 1, Supplementary Figs. S6–S20).

**Funding**

This work was supported by NIH/NIGMS grant R01GM111510 and the LDRD program at Los Alamos National Laboratory.

**SUPPLEMENTARY DATA**

**Supplementary Information:**

1. Cover sheet and table of contents – p. 1
2. References cited in Supplementary Information – p. 2
3. Supplementary Methods – included in this PDF file (pp. 3–7)
4. Supplementary Table S1 – included in this PDF file (p. 8) [Effective sample sizes for $a_m$ and $m_h$ for Test Problem 3]
5. Supplementary Fig. S1 – included in this PDF file (pp. 9–13) [Results from Test Problem 1]
6. Supplementary Fig. S2–S4 – included in this PDF file (pp. 14–28) [Results from Test Problems 2A–2C]
7. Supplementary Fig. S5 – included in this PDF file (pp. 29–33) [Results from Test Problem 3]
8. Supplementary Figs. S6–S20 – included in this PDF file (pp. 34–108) [Results from Test Problems 4A–4O]
9. Supplementary File S1 (in ZIP file) [CONF, BNGL, and EXP files for Test Problem 1] – these files are also available at https://github.com/lanl/PyBNF/tree/master/examples/LinearRegression_aMCMC
10. Supplementary File S2 (in ZIP file) [CONF, BNGL, and EXP files for Test Problems 2A–2C] – these files are also available at https://github.com/lanl/PyBNF/tree/master/examples/HIVdynamics_aMCMC
    10.1. The subdirectory pt303 contains files for Patient 303 viral decay dynamics
    10.2. The subdirectory pt403 contains files for Patient 403 viral decay dynamics
    10.3. The subdirectory pt409 contains files for Patient 409 viral decay dynamics
11. Supplementary File S3 (in ZIP file) [CONF, BNGL, and EXP files for Test Problem 3] – these files are also available at https://github.com/lanl/PyBNF/tree/master/examples/Degranulation_aMCMC
12. Supplementary File S4 (in ZIP file) [CONF, BNGL, and EXP files for Test Problems 4A–4O] – these files are also available at https://github.com/lanl/PyBNF/tree/master/examples/COVID19forecasting_aMCMC
    12.1. The subdirectory BigApple contains files for the New York City MSA epidemic
    12.2. The subdirectory LaLaLand contains files for the Los Angeles MSA epidemic
    12.3. The subdirectory WindyCity contains files for the Chicago MSA epidemic
    12.4. The subdirectory BigD contains files for the Dallas MSA epidemic
    12.5. The subdirectory HTown contains files for the Houston MSA epidemic
    12.6. The subdirectory DC contains files for the Washington DC MSA epidemic
    12.7. The subdirectory MagicCity contains files for the Miami MSA epidemic
    12.8. The subdirectory BrotherlyLove contains files for the Philadelphia MSA epidemic
    12.9. The subdirectory Hotlanta contains files for the Atlanta MSA epidemic
    12.10. The subdirectory Valley_of_the_Sun contains files for the Phoenix MSA epidemic
    12.11. The subdirectory Beantown contains files for the Boston MSA epidemic
    12.12. The subdirectory Frisco contains files for the San Francisco MSA epidemic
    12.13. The subdirectory InlandEmpire contains files for the Riverside MSA epidemic
    12.14. The subdirectory MotorCity contains files for the Detroit MSA epidemic
    12.15. The subdirectory EmeraldCity contains files for the Seattle MSA epidemic

## Supplementary Methods

This report is about version 1.1.9 of the PyBioNetFit (PyBNF) software package (Mitra et al., 2019) (https://github.com/lanl/PyBNF/). Features introduced in version 1.1.9 are summarized in the release history: https://github.com/lanl/PyBNF/releases

Version 1.1.9 can be installed using the following command:

`pip install pybnf==1.1.9`

Additional information about installation is available in the online documentation: https://pybnf.readthedocs.io/en/latest/

The documentation has been updated to cover usage of new features. Tutorial videos are available: https://www.youtube.com/channel/UCJfiDUiRlUR6-MZAt1k__Pg

The most significant change in 1.1.9 is addition of the am method, an adaptive Markov chain Monte Carlo (aMCMC) sampler. Below, this sampler is explained, as are the test problems that we used to evaluate the sampler. We also discuss calculation of effective sample sizes, trace plots, and pairs plots, which are important diagnostic tools for assessing sampling efficiency, and posterior predictive checking, which is important for assessing model fit to data.

### Methodological background and overview of PyBNF's am method

**MCMC sampling with PyBNF.** We will be concerned with MCMC sampling (Andrieu et al., 2003) as used to solve Bayesian inference problems wherein inference is conditioned on a specified model with $N$ parameters $\theta \in \mathbb{R}^N$ taken to have uncertain values. PyBNF, through its interfaces to BioNetGen (Faeder et al., 2009) and libRoadRunner (Somogyi et al., 2015), is compatible with models specified in BNGL and SBML formats (Faeder et al., 2009; Keating et al., 2020). Given a dataset $D$ (providing information about $\theta$), the goal of Bayesian inference is to find the posterior probability density function $p(\theta|D)$, which is often called the parameter posterior or simply the posterior. Problematically, $p(\theta|D)$ is not analytically tractable except in special cases. MCMC methods generate samples of the parameter posterior by leveraging Bayes' formula. This formula states that $p(\theta|D) = p(D|\theta)\, p(\theta)/Z$, where $p(D|\theta)$ is the probability density of $D$ given $\theta$, $p(\theta)$ is the prior (which is specified so as to capture available information about $\theta$ that is external to $D$), and $Z$ is a normalization constant. If no information about $\theta$ is available except for $D$, it is common to specify a flat or uniform prior, which may be proper (i.e., constant over a specified range for each parameter and 0 elsewhere) or improper (taken to be constant everywhere). PyBNF supports specification of proper uniform priors, as well as normal priors. PyBNF supports various types of model-constrained likelihood functions, i.e., specific functional forms for the likelihood $\mathcal{L}(\theta|D) = p(D|\theta)$. Currently, the best supported are a normal likelihood function $\mathcal{N}(\mu, \sigma^2)$, wherein the hyperparameter $\sigma$ may be fixed before inference or jointly inferred with model parameters, and a negative binomial likelihood function $\text{NB}(p, r)$, wherein the hyperparameter $r$ may be fixed before inference or jointly inferred with model parameters. The normalization constant $Z \coloneqq p(D) = \int p(D|\theta)\, p(\theta)\, d\theta$ is often referred to as the Bayes evidence or marginalized likelihood. Calculation of $Z$ is usually difficult (even computationally) as it entails high-dimensional integration over the entire parameter space. This difficulty is bypassed with MCMC sampling, which relies on calculating ratios of posterior densities (in the Metropolis–Hastings sampling criterion, which is discussed below). These ratios do not depend on $Z$. MCMC sampling can be viewed as a special kind of random movement of a walker or sampler within $N$-dimensional parameter space. This movement, a discrete-time Markov stochastic process, is designed such that the long-time stationary distribution of the process converges to the posterior $p(D|\theta)$. Thus, if the stochastic process is performed for a sufficiently long time, the distribution of the samples collected along the process trajectory through parameter space will provide a reasonable approximation of $p(D|\theta)$. MCMC sampling methods akin to the well-known Metropolis–Hastings sampler, like the aMCMC sampler implemented in PyBNF, involve many iterations of three key steps. The first step is generation of a random proposal parameter set $\theta'$. The proposal $\theta'$ may depend on $\theta_t$, the location of the sampler in parameter space at time $t$ (i.e., the initial parameter set at $t = 0$ or the last accepted parameter set at time $t > 0$). The proposal is usually generated from a proposal kernel, which we will denote as $K(\theta'|\theta_t)$. (The proposal kernel is discussed below.) The second step is to determine whether the proposal should be accepted. The proposal is accepted with probability given by

$$\min\left\{1, \frac{p(D|\theta')\, p(\theta')}{p(D|\theta_t)\, p(\theta_t)}\right\}$$

which is called the Metropolis–Hastings sampling criterion. Finally, if the proposal $\theta'$ is accepted, the location of the sampler is updated using the mapping $\theta_{t+1} \leftarrow \theta'$. Otherwise, $\theta_{t+1} \leftarrow \theta_t$. When the proposal kernel $K$ is designed such that the stochastic process is ergodic (that is, there is non-zero probability to visit any given point in parameter space from any starting point) and Metropolis–Hastings sampling criterion is used to accept/reject move proposals, the long-time statistics of the stochastic process converges to $p(D|\theta)\, p(\theta)$, which is proportional to $p(\theta|D)$.

**The proposal kernel $K$ is critical for efficient MCMC sampling.** During MCMC sampling, the convergence rate to the target (posterior) distribution strongly depends on the choice of the proposal kernel $K$, despite the fact that an ergodic process with any proposal kernel is guaranteed to converge given a sufficiently long trajectory. PyBNF implements a commonly chosen proposal kernel: $K(\theta'|\theta_t) \sim N(\theta_t, \Sigma)$, where $\Sigma$ is the covariance matrix of the multivariate Gaussian distribution. It is known that the optimal choice of $\Sigma$ is inference problem-specific and is related to the covariance matrix of the target density $p(\theta|D)$, i.e., $\Sigma_{\theta|D} \coloneqq \int (\theta_i - \bar{\theta}_i)(\theta_j - \bar{\theta}_j)\, p(\theta|D)\, d\theta$, where the mean of the $i^{\text{th}}$ and $j^{\text{th}}$ parameter are given by $\bar{\theta}_i \coloneqq \int \theta_i\, p(\theta|D)\, d\theta$ and $\bar{\theta}_j \coloneqq$ $\int \theta_j \, p(\theta|D) \, d\theta$, respectively. Ideally, $\Sigma = f \cdot \Sigma_{\theta|D}$ where $f = 2.38^2/N$ (Andrieu and Thoms, 2008). However, $\Sigma_{\theta|D}$ is generally not known *a priori*, and the prefactor $f$ potentially depends on higher-moment statistics of the target distribution $p(\theta|D)$. This situation has motivated development of a family of adaptive methods for learning the optimal prefactor $f$ and $\Sigma_{ij}^*$ (an approximation of $\Sigma_{\theta|D}$) *on-the-fly* from MCMC trajectories (Andrieu and Thomas, 2008).

**How PyBNF's adaptive MCMC sampler works.** In our update of PyBNF to version 1.1.9, we implemented the adaptive MCMC algorithm labeled as Algorithm 4 in Andrieu and Thoms (2008). We found this method to be useful for solving challenging Bayesian inference problems in an epidemiological forecasting study (Lin et al., 2021). For pseudo code, we refer the reader to Andrieu and Thoms (2008). Here, we explain how the adaptive algorithm learns the prefactor $f$ and $\Sigma_{ij}^*$ (the approximation of $\Sigma_{\theta|D}$) on-the-fly as understanding this procedure is important for proper use of PyBNF's am method. Use of this method to solve an inference problem consists of four sampling phases, which are defined by three user-supplied settings for the algorithmic parameters $n_{\text{burn-in}}$, $n_{\text{adaptive}}$, and $n_{\text{stationary}}$. The four phases are described as follows. **(1) *Burn-in phase*.** The MCMC chain is initiated at a point in parameter space corresponding to user-specified initial guesses for the uncertain parameters, $\theta_0$, and move proposals are made on the basis of a multivariate Gaussian kernel having a user-specified covariance matrix, $\Sigma_0$. Sampling during the burn-in phase proceeds until iteration $n_{\text{burn-in}}$. The goal of burn-in is to allow the chain to relax to a high-density region of the posterior distribution $p(\theta|D)$. If $\theta_0$ is known *a priori* to be located within a high-density region, then the burn-in phase can by bypassed (by setting $n_{\text{burn-in}} = 0$). **(2) *Mixing phase*.** MCMC sampling on the basis of $\Sigma_0$ continues until the current iteration reaches $n_{\text{burn-in}} + n_{\text{adaptive}}$. The $n_{\text{adaptive}}$ samples collected during the mixing phase are used to compute an empirical covariance matrix $\hat{\Sigma}$, which is an $N \times N$ positive definite matrix. The user should check to make sure that the move acceptance rate during mixing was large enough to produce enough samples to enable learning the covariance matrix. **(3) *Adaptation phase*.** At iteration $n_{\text{burn-in}} + n_{\text{adaptive}}$, the empirical covariance matrix $\hat{\Sigma}$ (defined on the basis of samples generated during the mixing phase) is used to initialize $\Sigma_{ij}^*$. The prefactor $f$ is also initialized, on the basis of a user-supplied setting. During the adaptation phase, $f$ and $\Sigma_{ij}^*$ are updated on-the-fly as new samples are generated. The algorithm imposes a decaying weight on samples such that the learned settings for $f$ and $\Sigma_{ij}^*$ eventually stabilize. The weight is proportional to $1/n$, where $n$ is the number of iterations after the start of adaptation. (After $1/n$ becomes small, which stabilizes settings for $f$ and $\Sigma_{ij}^*$, the am algorithm behaves like a standard MCMC algorithm.) The purpose of the adaptation phase is to make $\Sigma_{ij}^*$ as close to $\Sigma_{\theta|D}$ as possible and to find a setting for $f$ that yields an acceptance rate close to the optimal acceptance rate of 0.234 (Gelman et al., 1997). *NB:* the adaptation phase may fail to achieve these goals. Importantly, if $\Sigma_{ij}^*$ is deemed ***not*** to be close to $\Sigma_{\theta|D}$ after adaptation (which will manifest as inefficient sampling, which can be detected as discussed below), one should restart sampling at the mixing phase, ideally using a better choice for $\Sigma_0$. *NB:* trying to remedy inefficient sampling because of failure of the adaptation phase by generating a large number of samples in the stationary phase (which is discussed below) will be unnecessarily wasteful of computational resources and is likely to be unhelpful. **(4) *Stationary phase*.** At iteration $n_{\text{stationary}}$, samples will begin to be collected for the purpose of defining the posterior distribution $p(\theta|D)$. We recommend selecting $n_{\text{stationary}}$ such that $\Delta n := n_{\text{stationary}} - (n_{\text{burn-in}} + n_{\text{adaptive}})$ is at least $10^3$. In test applications, we set $n_{\text{stationary}}$ such that $\Delta n > 10^4$. Sampling during the stationary phase should initially be conservative so as to allow efficiency of sampling to be assessed before committing to the generation of a large number of samples (needed for fine resolution of the parameter posterior). PyBNF supports sampling restarts; we refer the reader to the online documentation for details.

**Assessing the efficiency of MCMC sampling.** The quality and efficiency of sampling should always be assessed. There are numerous ways to accomplish this task. Here, for selected chains which were deemed to be long and mixed enough to approximate the posterior distribution, we calculated the effective sample size (ESS), which provides a metric for estimating the number of independent samples generated for a specified estimand (Gelman et al., 2014). We used the TensorFlow Probability Python library (https://www.tensorflow.org/probability) to calculate ESS for parameters follows:

```
import tensorflow_probability as tfp
ESS = tfp.mcmc.effective_sample_size(parLog).numpy()
```

where parLog is the 2-dimensional array of an MCMC chain wherein each column corresponds to a particular parameter and each row is a sample along the MCMC chain. An output file of PyBNF, for example, fileForDegran_mh.txt, can be loaded via the genfromtxt function in the NumPy library, i.e., parLog = np.genfromtxt('fileForDegran_mh.txt'). An ESS is computed for each parameter (i.e., each column in the array), so the returned ESS is a 1-dimensional array with $N$ entries, where $N$ is the number of free/adjustable parameters. Results of ESS calculations are shown in Supplementary Table S1. A small ESS value relative to the total number of samples in the full chain corresponds to a highly temporally correlated chain, which is indicative of inefficient sampling. Larger values allow for better coverage of the parameter posterior. We inspected log-likelihood and parametric trace plots (parameter value vs. iteration/position in the chain), which are shown in panel C of Supplementary Figs. S1–S20. A trace plot should indicate stationarity and good mixing (i.e., exploration of the posterior in the region of high density). We inspected pairs plots, which are shown in panels D and E of Supplementary Figs. S1–S20. A pairs plot is a matrix of 1- and 2-dimensional marginalizations of a multivariate parameter posterior. The 1-dimensional marginalizations (or marginal posteriors) appear on a diagonal (top left to bottom right). The other elements of the matrix are the 2-dimensional marginalizations. These plots, which provide visual insights into the shape of the parameter posterior, may be generated as histograms or as contour plots, which are found through kernel density estimation (KDE). The 2-dimensional marginalizations shown in Supplementary Figs. S1–S20 are contour plots. These plots were made using `seaborn.kdeplot` in the Python seaborn module (https://seaborn.pydata.org/generated/seaborn.kdeplot.html) with default settings, that is, a Gaussian kernel with the bandwidth set by Scott's rule. We used KDE to generate contour plots (vs. showing two-dimensional histograms) because our pairs plots are based on large-size collections of samples. KDE was used as a data dimension reduction technique to pare figure file sizes down to manageable sizes.

**Posterior predictive checking.** Visualizations of posterior predictive distributions (Gelman et al., 2014) are shown in panel A of Supplementary Figs. S1–S20. These visualizations are useful for assessing a model's explanatory power. Here, a posterior predictive is a distribution of model outputs generated by re-sampling from the parameter posterior and using the sampled parameter values to perform simulations that generate predictions (or other model-based calculations of outputs). Thus, the posterior predictive captures uncertainty in predictions arising from uncertainty in parameter values. The posterior predictive may be defined so as to also capture measurement noise. Injecting measurement noise into the posterior predictive allows for assessment of model consistency with data. A 95% credible interval, for example, for such a posterior predictive should cover 95% of the data used in inference.

### Test problems used to evaluate PyBNF's am method

**Test Problem 1.** We contrived a linear regression problem to illustrate the power of PyBNF's adaptive MCMC sampler (am) over PyBNF's conventional MH-MCMC sampler (mh). Consider an affine transformation $f: \mathbb{R}^{10 \times 1} \rightarrow \mathbb{R}^1$ mapping a 10-dimensional input vector $x$ to a 1-dimensional real number $y$:

$$y = f(x) + \varepsilon = w^T \cdot x + b + \varepsilon,$$

with weight vector $w \in \mathbb{R}^{10 \times 1}$ and bias $b \in \mathbb{R}^1$. For each input $x$, we injected normally and independently distributed noise $\varepsilon \sim \mathcal{N}(0, \sigma^2)$. Our goal is to quantify the posterior distribution of $(w, b)$ given a set of $N$ input-output pairs $\{x^{[i]}, y^{[i]}\}_{i=1}^{N}$. We assume that noise statistics, $\varepsilon \sim \mathcal{N}(0, \sigma^2)$, have been characterized, i.e., we know $\sigma$. In this problem, we set $N = 256$ and $\sigma = 1/2$. To facilitate derivation of an analytical expression for the posterior, we take the prior for $(w, b)$ to be everywhere uniform. This prior is improper. In the PyBNF job setup, we specified a proper uniform prior for $(w, b)$ that is functionally equivalent to an everywhere-uniform improper prior. (PyBNF only supports proper uniform priors.) An everywhere-uniform improper prior means that $P(\theta) = $ constant for every $\theta$ in the parameter space. The value of the constant is unimportant for MCMC sampling, as it does not enter the ratio in the Metropolis–Hastings criterion. Because PyBNF only supports proper uniform priors, practically, we can achieve an improper uniform prior by setting bounds on parameters that are far away from the high-density region of parameter space. With this approach, the sampler is unlikely to reach the bounds in a chain of finite length. In this particular problem, bounds were imposed at $x_i = \pm 10$ (cf. the CONF file in Supplementary File S1). These bounds are far from the high-density region of parameter space (cf. Panel B in Supplementary Fig. S1).

We assume that the input distribution, $p(x)$, is a zero-mean multivariate Gaussian with a diagonal covariance matrix $C$. We assign values to the diagonal entries of the covariance matrix that are spread over several multiple scales: $C_{ii} \coloneqq 2^{-i+1}$, $i = 1 \dots 10$. As will be seen below, such a construction leads to a Gaussian posterior distribution with a wide range of variances in the high-dimensional parameter space. We assume a set of ground-truth parameters $w = [1,1, \dots 1]^T$ and $b = 0$ to generate the synthetic dataset. We first generate $N$ multivariate Gaussian $x^{[i]}$'s and univariate Gaussian $\varepsilon^{[i]}$'s, then carry out the affine transformation with the ground-truth parameters to obtain the synthetic $y^{[i]}$'s. The dataset is provided in Supplementary File S1. In the ZIP archive, the inputs can be found in the BNGL file and the outputs can be found in the EXP file.

Because the noise statistics are known—recall that $\varepsilon \sim \mathcal{N}(0, \sigma^2)$—the log-likelihood of an input-output pair has the follow functional form:

$$\log \mathcal{L}\left(w, b | \{x^{[i]}, y^{[i]}\}\right) = \frac{1}{2\sigma^2} \left[y^{[i]} - \left(w^T \cdot x^{[i]} + b\right)\right]^2 + \log \sigma.$$

Note that the choice of $\sigma = 1/2$ leads to the sum-of-squares likelihood implemented in PyBioNetFit. Because each pair in the data set is independently distributed, the full log-likelihood of the dataset is the sum of the log-likelihood of each input-output pair:

$$\log \mathcal{L}\left(w, b | \{x^{[i]}, y^{[i]}\}_{i=1}^{N}\right) = \frac{1}{2\sigma^2} \sum_{i=1}^{N} \left[y^{[i]} - \left(w^T \cdot x^{[i]} + b\right)\right]^2 + N \cdot \log \sigma$$

From the prior being everywhere uniform, it follows that the posterior is also multivariate Gaussian:

$$p\left(w, b | \{x^{[i]}, y^{[i]}\}_{i=1}^{N}\right) \sim \exp \sum_{i=1}^{N} \left[y^{[i]} - \left(w^T \cdot x^{[i]} + b\right)\right]^2 \sim \mathcal{N}\left(0, \sigma^2 = \frac{1}{2}\right).$$

Given the dataset $\{x^{[i]}, y^{[i]}\}_{i=1}^{N}$, an analytical equation for the posterior distribution exists. Let us introduce $\theta$ to represent the augmented model parameters $(w, b)$ and introduce $d^{[i]}$ to represent the augmented input vector (accounting for bias):

$$\theta \coloneqq [w_1, w_2, \dots w_{10}, b]^T \in \mathbb{R}^{11 \times 1}, \quad \text{and} \quad d^{[i]} \coloneqq \left[x_1^{[i]}, x_2^{[i]} \dots x_{10}^{[i]}, 1\right]^T \in \mathbb{R}^{11 \times 1},$$

The sample second moments can then be expressed as follows:

$$\Lambda := \frac{1}{N} \sum_{k=1}^{N} d^{[k]} \cdot \left(d^{[k]}\right)^T \in \mathbb{R}^{11 \times 11} \quad \text{and} \quad \phi := \frac{1}{N} \sum_{k=1}^{N} d^{[k]} \cdot y^{[k]} \in \mathbb{R}^{11 \times 1}.$$

It is straightforward to find the following expression for the posterior density:

$$p\left(\theta \big| \{x^{[i]}, y^{[i]}\}_{i=1}^{N}\right) \sim \mathcal{N}(\text{mean} = \Lambda^{-1} \cdot \phi, \text{covariance} = \Lambda^{-1}\sigma^2).$$

Note that the posterior density inherits the wide spectrum of the input covariance matrix $C$, because the first $10 \times 10$ submatrix of $\Lambda$ carries the sample statistics of the multivariate Gaussian random variable $x$ with the covariance matrix $C$. Consistent with Bayesian reasoning, the maximum *a posteriori* estimate (MAP) of the model parameters $\Lambda^{-1} \cdot \phi$ depends on the given dataset, and are not equal to the ground-truth parameters.

The PyBNF job setup files for Test Problem 1 are provided in Supplementary File S1. Results obtained with PyBNF's am method are compared those obtained with PyBNF's mh method and results from the analytical posterior (Supplementary Fig. S1).

**Test Problem 2.** In this test problem, we analyze patient data from Ho et al. (1995) using a two-phase exponential decay model for HIV dynamics under antiretroviral therapy (Perelson et al., 1996). We consider three sets of patient data, for Patients 303, 403, and 409 (Ho et al., 1995). The model can be expressed as follows:

$$\log V(t) = \begin{cases} \log V_0 & \text{for } 0 < t < \tau \\ \log[V_0(Ae^{-d_1(t-\tau)} + (1-A)e^{-d_2(t-\tau)})] & \text{for } t \geq \tau \end{cases}$$

The model has four uncertain patient-specific parameters: $V_0$, the baseline plasma viral load, which is assumed to be in a steady state from time $t = 0$ (the time at which plasma viral load in each patient began to be monitored) to $t = \tau$ (the start time of therapy); a dimensionless constant $A$, which lies between 0 and 1; and two strictly positive decay constants, $d_1$ and $d_2$. The constants $d_1$ and $d_2$ are taken to characterize a fast phase of viral decay during therapy and a subsequent, slow phase of decay. As a simplification (to break the symmetry of $d_1$ and $d_2$), we constrain $d_2$ to be less than $d_1$, i.e., we designate $d_2$ to be the constant characterizing the slow phase of decay. The model has one known parameter: $\tau = 7$ d, the start time of therapy. Inferences of the parameter posterior for $\theta = (V_0, A, d_1, d_2)$ are conditioned on the model (as usual) and the fixed setting for $\tau$, which can be viewed as part of the model. We assume that errors in measurements of $\log V(t)$ are normally distributed identically and independently across all measurement times. In this problem, we take the standard deviation of the noise kernel, $\sigma$, to be a hyperparameter, which we infer jointly together with the uncertain model parameters. The inference problem can be stated more precisely as follows. The dataset for a given patient consists of $N$ measurements of viral load: $V_i$ (at time $t_i$) for $i = 1, \ldots, N$. From the assumptions stated above, we find the following expression for the log-likelihood of the model parameters $\theta$ and the hyperparameter $\sigma$ $\{V_0, A, d_1, d_2, \sigma\}$ given the data $\{t_i, V_i\}_{i=1}^{N}$:

$$\log \mathcal{L}\big((V_0, A, d_1, d_2, \sigma,)|\{t_i, V_i\}_{i=1}^{N}\big) = \frac{1}{2\sigma^2} \sum_{i=1}^{N} [\log V_i - \log V(t = t_i)]^2 + N \cdot \log \sigma$$

We chose priors for model parameters as follows:

$$V_0 \sim \text{Unif}(0,2000)$$
$$A \sim \text{Unif}(0,1)$$
$$(d_1, d_2) \sim \text{Unif}([0,1]^2) \text{ s.t. } d_1 \geq d_2$$
$$\sigma \sim \text{Unif}(0,10)$$

The units of $V_0$ are viral RNA copies per $\mu$L of plasma as in Fig. 1 of Ho et al. (1995). The units of $d_1$ and $d_2$ are $d^{-1}$ (because time $t$ has units of d). Recall that $A$ is dimensionless.

The PyBNF job setup files for Test Problem 2 are provided in Supplementary File S2. This ZIP archive file includes a subdirectory with setup files for each of the three patients. Results obtained with PyBNF's am method are compared to those obtained with a problem-specific code in Supplementary Figs. S2–S4. These figures show results from inferences based on the datasets available for Patient 303 (Test Problem 2A), Patient 403 (Test Problem 2B), and Patient 409 (Test Problem 2C).

**Test Problem 3.** In this test problem, we analyze measured degranulation responses of mast cells to stimulation of IgE receptor signaling using a model for IgE receptor signaling (Harmon et al., 2017). This problem was previously used to evaluate PyBNF's mh method (Mitra et al., 2019). We refer the reader to Harmon et al. (2017) and Mitra et al. (2019) for additional details.

The PyBNF job setup files for Test Problem 3 are provided in Supplementary File S3. Results obtained with PyBNF's am method are compared to those obtained with PyBNF's mh method in Supplementary File S5.

**Test Problem 4.** In this test problem, we analyze regional COVID-19 surveillance data (daily reports of new cases) using a compartmental model (Lin et al., 2021). We consider 15 regions, the metropolitan statistical areas (MSAs) encompassing the following cities (listed in order of population size, from largest to smallest): A) New York City, New York; B), Los Angeles, California; C) Chicago, Illinois; D) Dallas, Texas; E) Houston, Texas; F) Washington, DC; G) Miami, Florida; H) Philadelphia, Pennsylvania; I) Atlanta, Georgia; J) Phoenix, Arizona; K) Boston, Massachusetts; L) San Francisco, California; M) Riverside, California; N) Detroit, Michigan; and O) Seattle, Washington. We refer the reader to Lin et al. (2021) for a full description of the inference problems: the compartmental model used for all regions, the fixed and adjustable model parameters, the prior distributions for adjustable parameters, the probabilistic model of case detection, and the datasets. It should be noted that each inference is MSA-specific and that the number of uncertain/free/adjustable parameters in a model depends on the setting for $n$, the number of distinct social-distancing periods considered beyond the initial social-distancing period. Each social-distancing period is characterized by three parameters. The setting for $n$ was determined through a model selection procedure (Lin et al., 2021). Inferences are conditioned on the model, the setting for $n$, and several fixed parameter estimates taken to be certain and applicable for all MSAs (Lin et al., 2021).

The PyBNF job setup files for Test Problem 4 are provided in Supplementary File S4. This ZIP archive file includes a subdirectory with setup files for each of the 15 MSAs. Results obtained with PyBNF's am method are compared to those obtained using the problem-specific code of Lin et al. (2021) in Supplementary Figs. S6–S20. These figures show results from inferences based on MSA-specific datasets for New York City (Test Problem 4A), Los Angeles (Test Problem 4B), Chicago (Test Problem 4C), Dallas (Test Problem 4D), Houston (Test Problem 4E), Washington DC (Test Problem 4F), Miami (Test Problem 4G), Philadelphia (Test Problem 4H), Atlanta (Test Problem 4I), Phoenix (Test Problem 4J), Boston (Test Problem 4K), San Francisco (Test Problem 4L), Riverside (Test Problem 4M), Detroit (Test Problem 4N), and Seattle (Test Problem 4O).

**Supplementary Table S1.** Effective sample sizes (ESSs)[a] of chains generated using PyBNF's am and mh methods.

| Parameter[b] | $ESS_{am}$ | $ESS_{mh}$ | Ratio ($ESS_{am}/ESS_{mh}$) |
|---|---|---|---|
| X_tot | 422 | 61.4 | 6.9 |
| k_Xoff | 780 | 72.6 | 10.7 |
| k_Xon | 655 | 41.1 | 15.9 |
| kase | 1050 | 52.0 | 20.1 |
| kdegX | 243 | 57.6 | 4.2 |
| kdegran | 2010 | 72.6 | 27.7 |
| km_Ship1 | 388 | 74.4 | 5.2 |
| km_Syk | 817 | 76.0 | 10.8 |
| km_x | 738 | 43.2 | 17.1 |
| koff | 435 | 42.7 | 10.2 |
| kp_Ship1 | 473 | 136 | 3.5 |
| kp_Syk | 566 | 74.6 | 7.6 |
| kp_x | 214 | 70.6 | 3.0 |
| kpten | 584 | 58.7 | 10.0 |
| ksynth1 | 1080 | 68.7 | 15.7 |
| pase | 818 | 55.4 | 14.7 |

[a]Effective sample size was calculated for each adjustable model parameter as described in Supplementary Methods.
[b]Parameters are identified by their names in the BNGL file included in Supplementary File S3. The model is that of Harmon et al. (2017). The inference problem considered here was previously used to evaluate the mh method (Mitra et al., 2019).

**SUPPLEMENTARY FIGURE S1 – PANEL A**

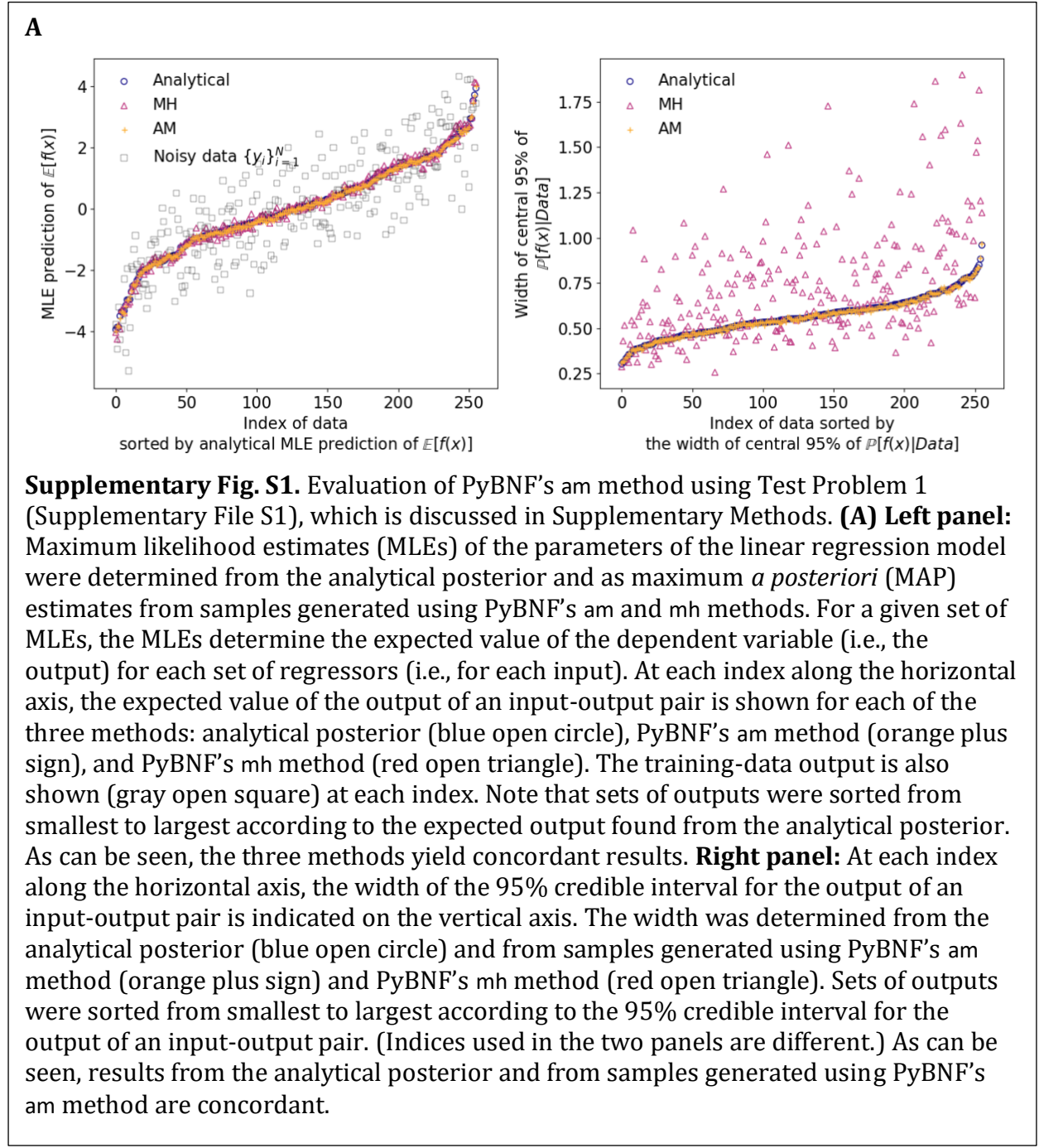

**Supplementary Fig. S1.** Evaluation of PyBNF's am method using Test Problem 1 (Supplementary File S1), which is discussed in Supplementary Methods. **(A) Left panel:** Maximum likelihood estimates (MLEs) of the parameters of the linear regression model were determined from the analytical posterior and as maximum *a posteriori* (MAP) estimates from samples generated using PyBNF's am and mh methods. For a given set of MLEs, the MLEs determine the expected value of the dependent variable (i.e., the output) for each set of regressors (i.e., for each input). At each index along the horizontal axis, the expected value of the output of an input-output pair is shown for each of the three methods: analytical posterior (blue open circle), PyBNF's am method (orange plus sign), and PyBNF's mh method (red open triangle). The training-data output is also shown (gray open square) at each index. Note that sets of outputs were sorted from smallest to largest according to the expected output found from the analytical posterior. As can be seen, the three methods yield concordant results. **Right panel:** At each index along the horizontal axis, the width of the 95% credible interval for the output of an input-output pair is indicated on the vertical axis. The width was determined from the analytical posterior (blue open circle) and from samples generated using PyBNF's am method (orange plus sign) and PyBNF's mh method (red open triangle). Sets of outputs were sorted from smallest to largest according to the 95% credible interval for the output of an input-output pair. (Indices used in the two panels are different.) As can be seen, results from the analytical posterior and from samples generated using PyBNF's am method are concordant.

**SUPPLEMENTARY FIGURE S1 – PANEL B**

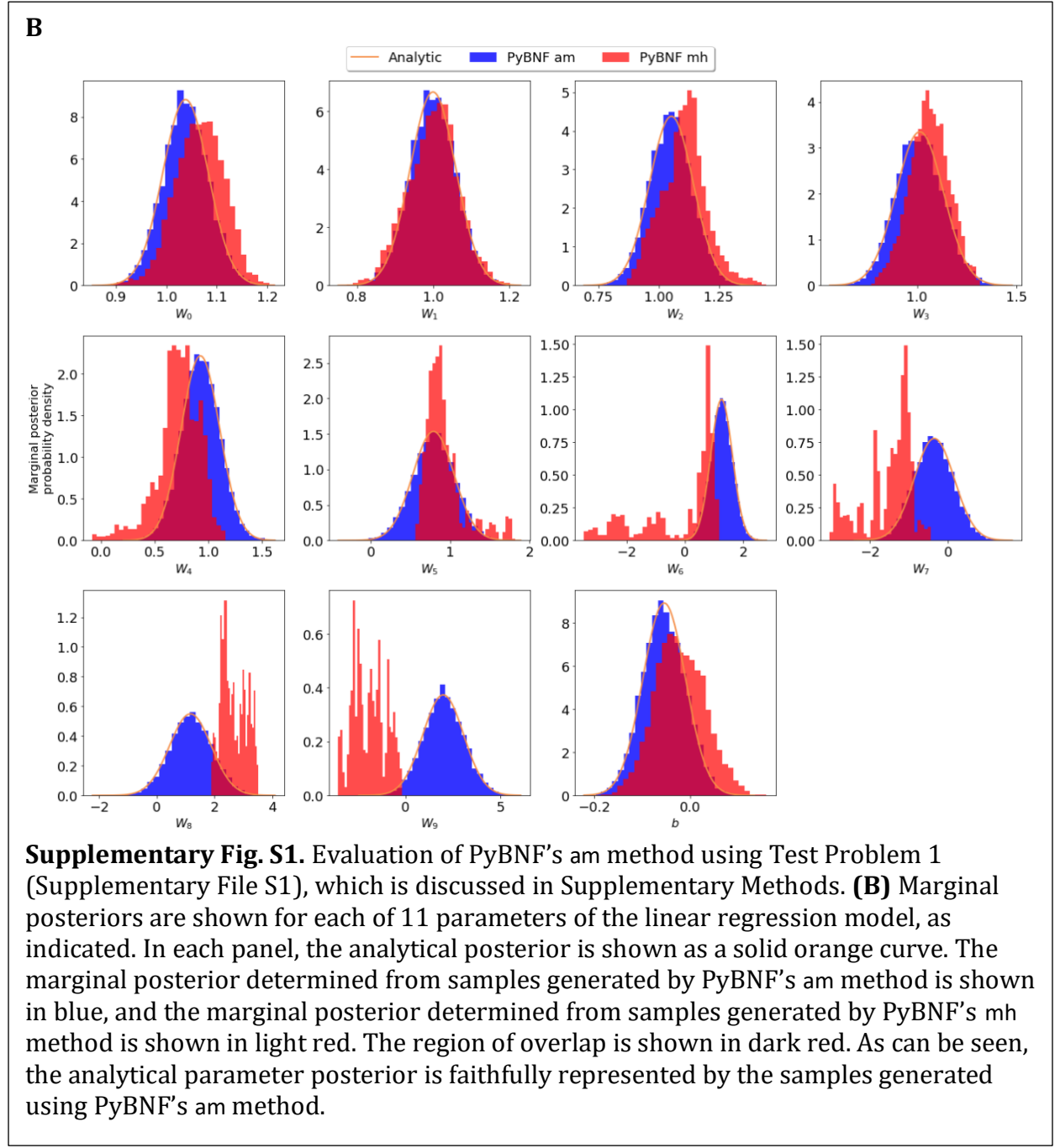

**Supplementary Fig. S1.** Evaluation of PyBNF's am method using Test Problem 1 (Supplementary File S1), which is discussed in Supplementary Methods. **(B)** Marginal posteriors are shown for each of 11 parameters of the linear regression model, as indicated. In each panel, the analytical posterior is shown as a solid orange curve. The marginal posterior determined from samples generated by PyBNF's am method is shown in blue, and the marginal posterior determined from samples generated by PyBNF's mh method is shown in light red. The region of overlap is shown in dark red. As can be seen, the analytical parameter posterior is faithfully represented by the samples generated using PyBNF's am method.

**SUPPLEMENTARY FIGURE S1 – PANEL C**

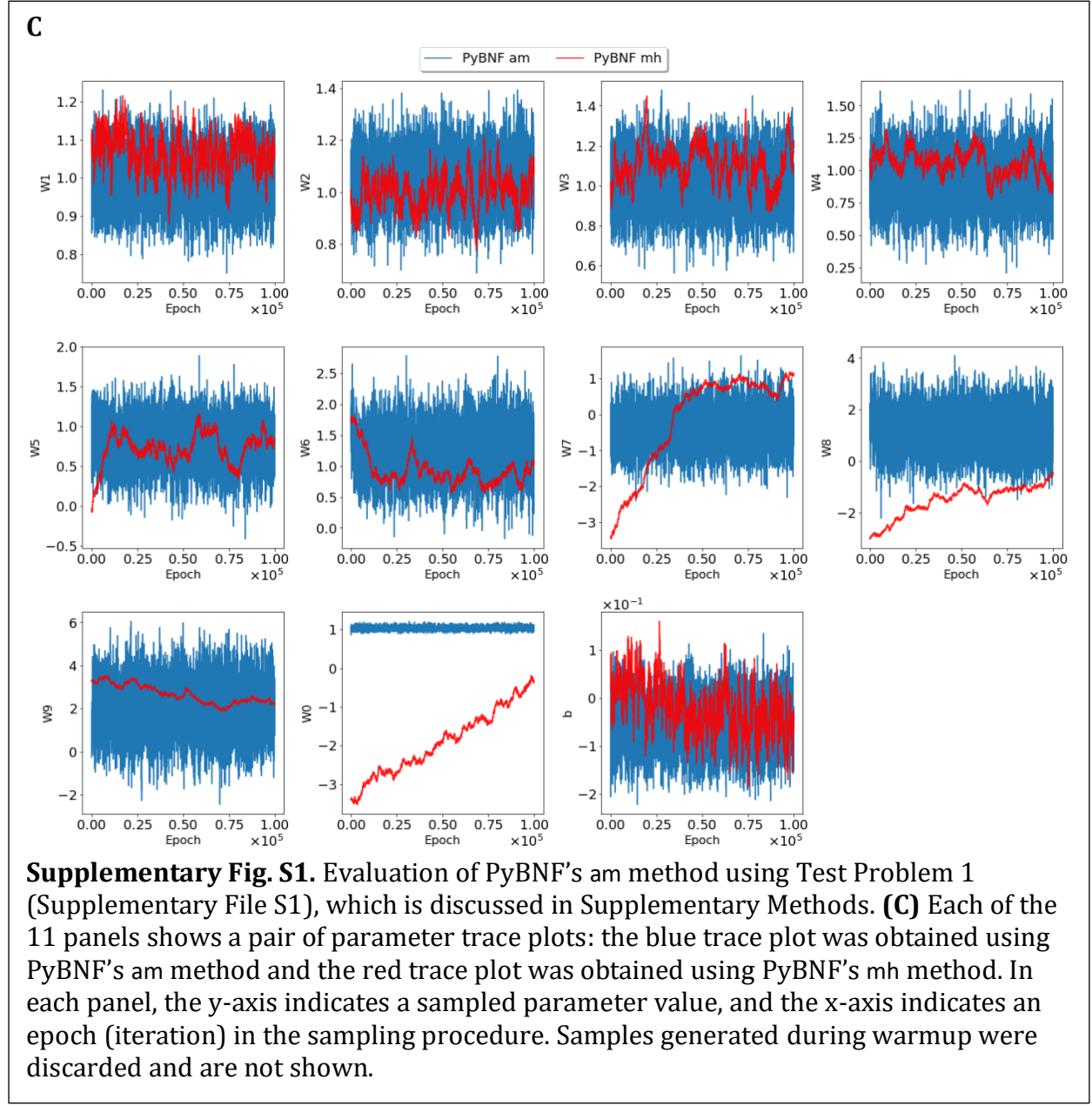

**Supplementary Fig. S1.** Evaluation of PyBNF's am method using Test Problem 1 (Supplementary File S1), which is discussed in Supplementary Methods. **(C)** Each of the 11 panels shows a pair of parameter trace plots: the blue trace plot was obtained using PyBNF's am method and the red trace plot was obtained using PyBNF's mh method. In each panel, the y-axis indicates a sampled parameter value, and the x-axis indicates an epoch (iteration) in the sampling procedure. Samples generated during warmup were discarded and are not shown.

**SUPPLEMENTARY FIGURE S1 – PANEL D**

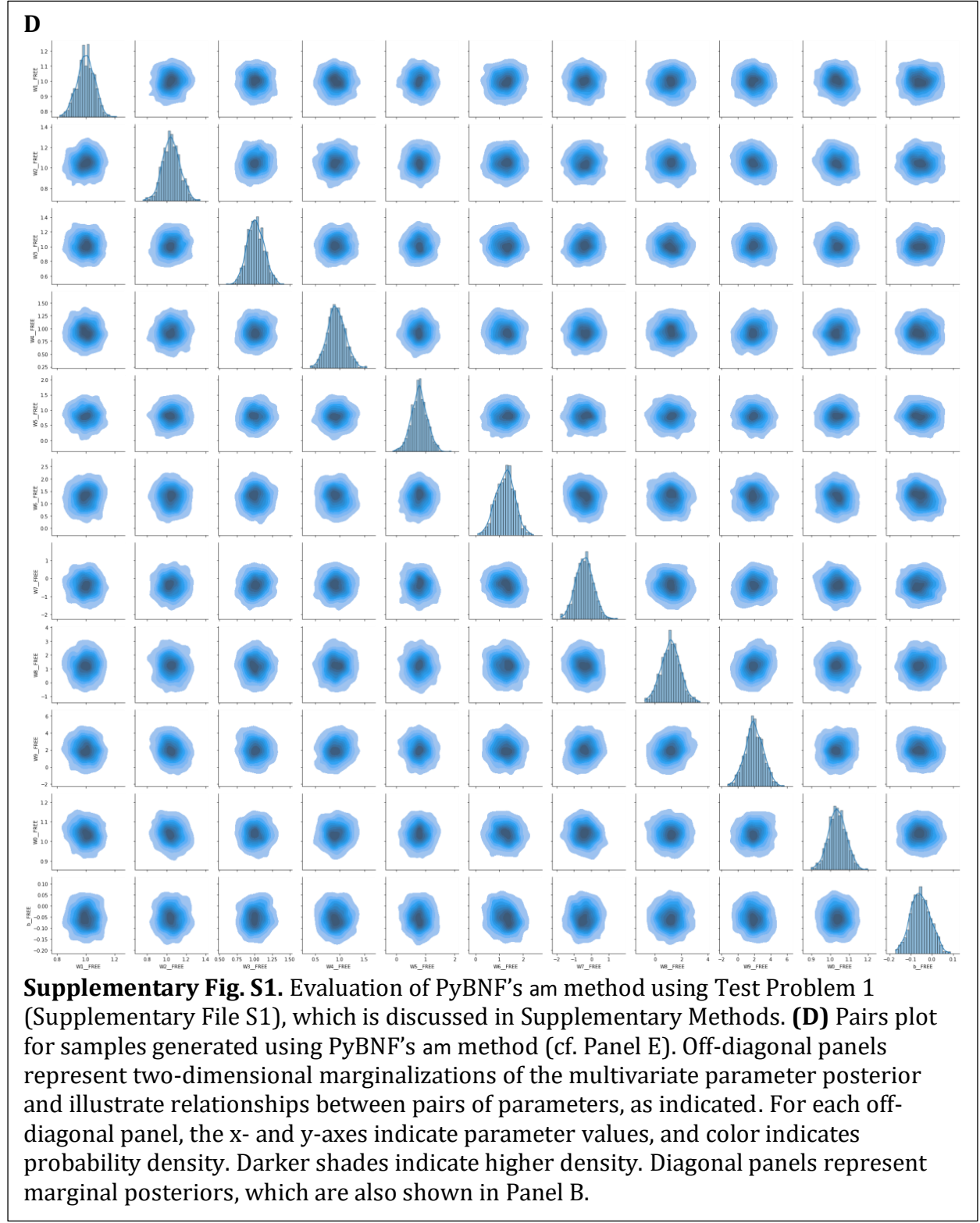

**Supplementary Fig. S1.** Evaluation of PyBNF's ɑm method using Test Problem 1 (Supplementary File S1), which is discussed in Supplementary Methods. **(D)** Pairs plot for samples generated using PyBNF's ɑm method (cf. Panel E). Off-diagonal panels represent two-dimensional marginalizations of the multivariate parameter posterior and illustrate relationships between pairs of parameters, as indicated. For each off-diagonal panel, the x- and y-axes indicate parameter values, and color indicates probability density. Darker shades indicate higher density. Diagonal panels represent marginal posteriors, which are also shown in Panel B.

**SUPPLEMENTARY FIGURE S1 – PANEL E**

**E**

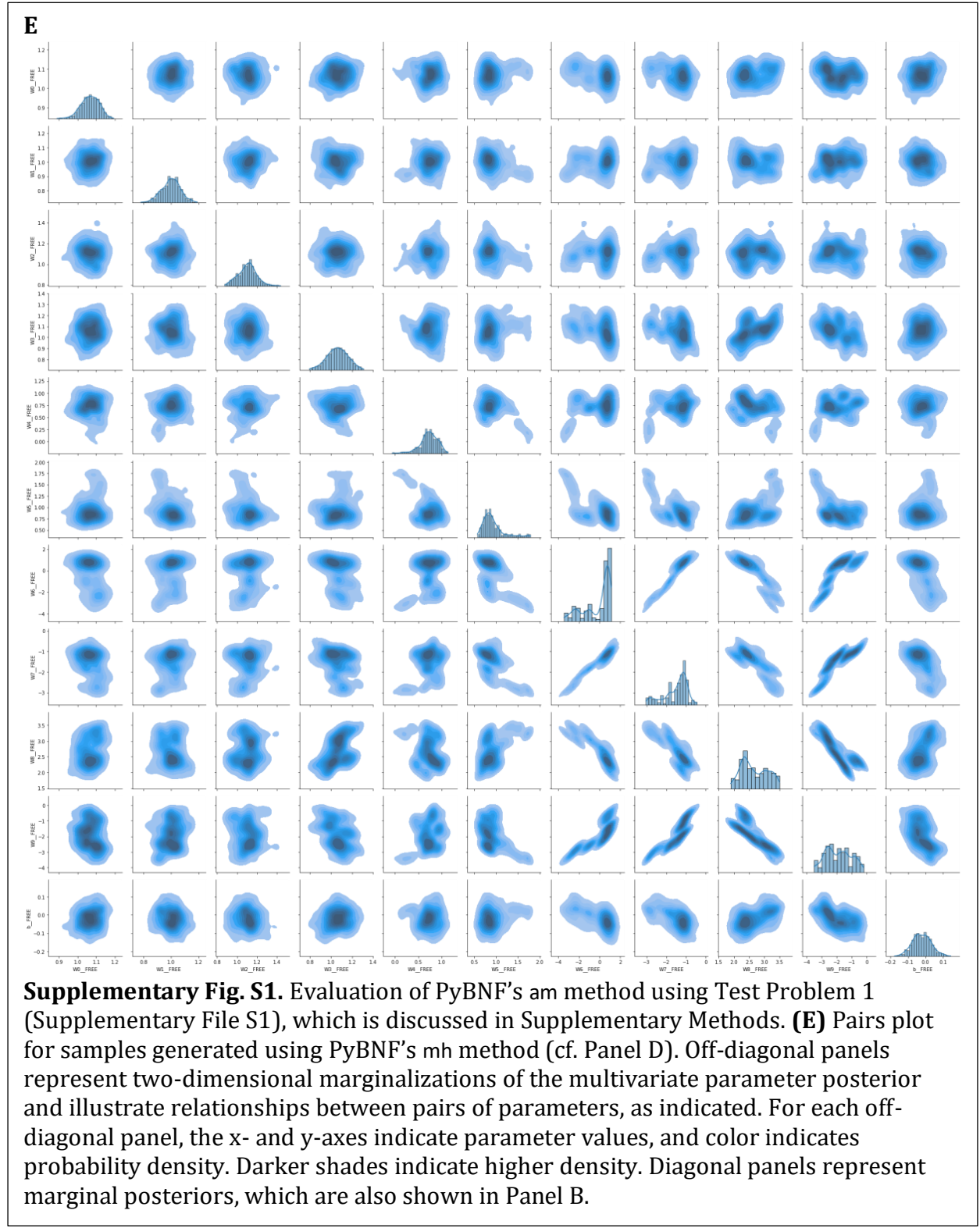

**Supplementary Fig. S1.** Evaluation of PyBNF's am method using Test Problem 1 (Supplementary File S1), which is discussed in Supplementary Methods. **(E)** Pairs plot for samples generated using PyBNF's mh method (cf. Panel D). Off-diagonal panels represent two-dimensional marginalizations of the multivariate parameter posterior and illustrate relationships between pairs of parameters, as indicated. For each off-diagonal panel, the x- and y-axes indicate parameter values, and color indicates probability density. Darker shades indicate higher density. Diagonal panels represent marginal posteriors, which are also shown in Panel B.

**SUPPLEMENTARY FIGURE S2 – PANEL A**

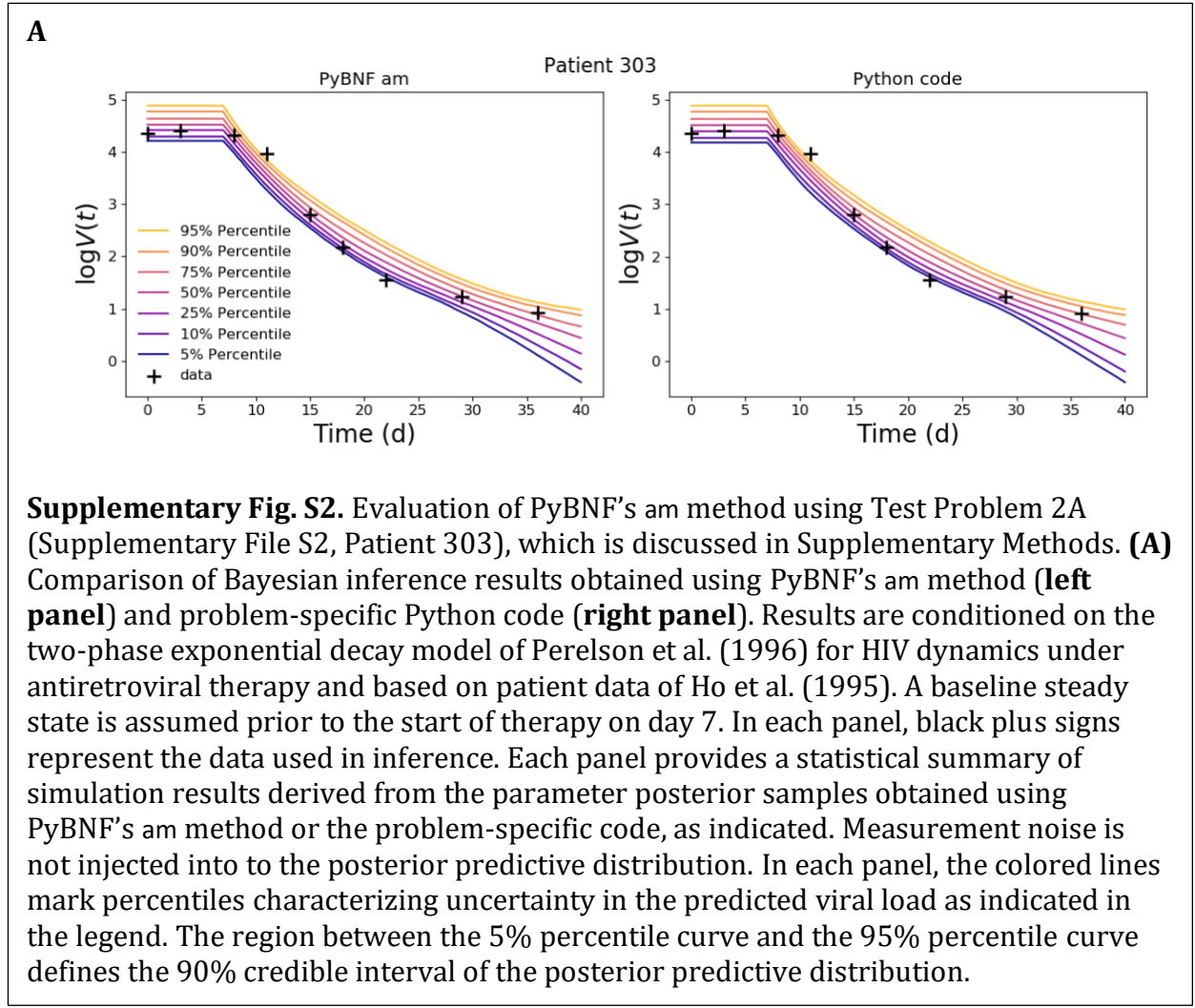

**Supplementary Fig. S2.** Evaluation of PyBNF's am method using Test Problem 2A (Supplementary File S2, Patient 303), which is discussed in Supplementary Methods. **(A)** Comparison of Bayesian inference results obtained using PyBNF's am method (**left panel**) and problem-specific Python code (**right panel**). Results are conditioned on the two-phase exponential decay model of Perelson et al. (1996) for HIV dynamics under antiretroviral therapy and based on patient data of Ho et al. (1995). A baseline steady state is assumed prior to the start of therapy on day 7. In each panel, black plus signs represent the data used in inference. Each panel provides a statistical summary of simulation results derived from the parameter posterior samples obtained using PyBNF's am method or the problem-specific code, as indicated. Measurement noise is not injected into to the posterior predictive distribution. In each panel, the colored lines mark percentiles characterizing uncertainty in the predicted viral load as indicated in the legend. The region between the 5% percentile curve and the 95% percentile curve defines the 90% credible interval of the posterior predictive distribution.

**SUPPLEMENTARY FIGURE S2 – PANEL B**

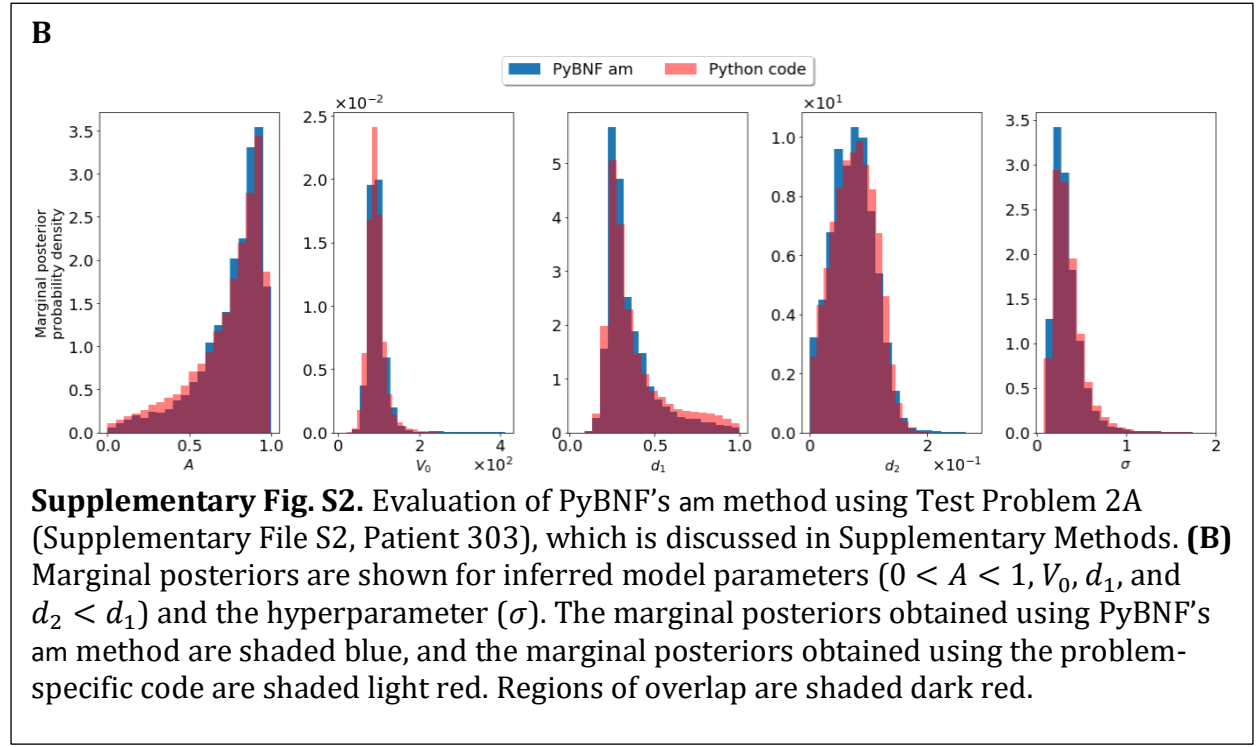

**Supplementary Fig. S2.** Evaluation of PyBNF's am method using Test Problem 2A (Supplementary File S2, Patient 303), which is discussed in Supplementary Methods. **(B)** Marginal posteriors are shown for inferred model parameters ($0 < A < 1$, $V_0$, $d_1$, and $d_2 < d_1$) and the hyperparameter ($\sigma$). The marginal posteriors obtained using PyBNF's am method are shaded blue, and the marginal posteriors obtained using the problem-specific code are shaded light red. Regions of overlap are shaded dark red.

**SUPPLEMENTARY FIGURE S2 – PANEL C**

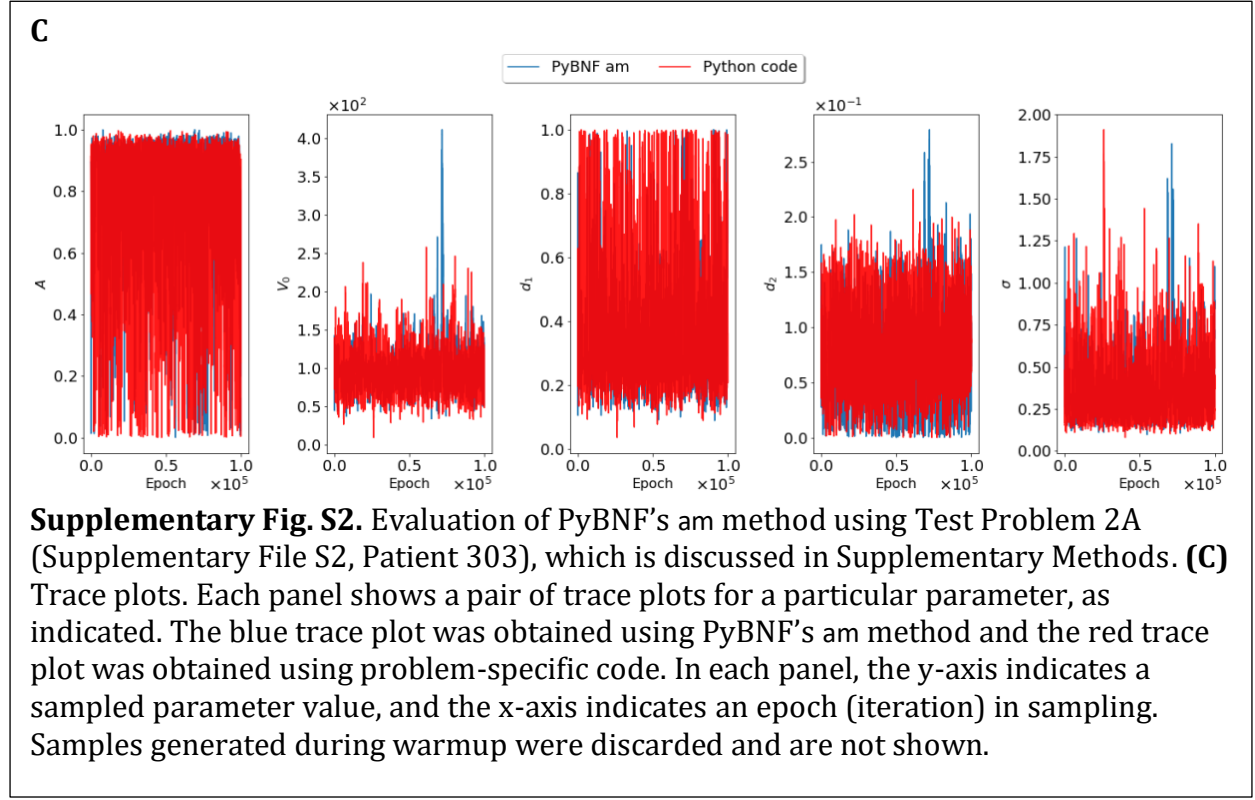

**Supplementary Fig. S2.** Evaluation of PyBNF's am method using Test Problem 2A (Supplementary File S2, Patient 303), which is discussed in Supplementary Methods. **(C)** Trace plots. Each panel shows a pair of trace plots for a particular parameter, as indicated. The blue trace plot was obtained using PyBNF's am method and the red trace plot was obtained using problem-specific code. In each panel, the y-axis indicates a sampled parameter value, and the x-axis indicates an epoch (iteration) in sampling. Samples generated during warmup were discarded and are not shown.

**SUPPLEMENTARY FIGURE S2 – PANEL D**

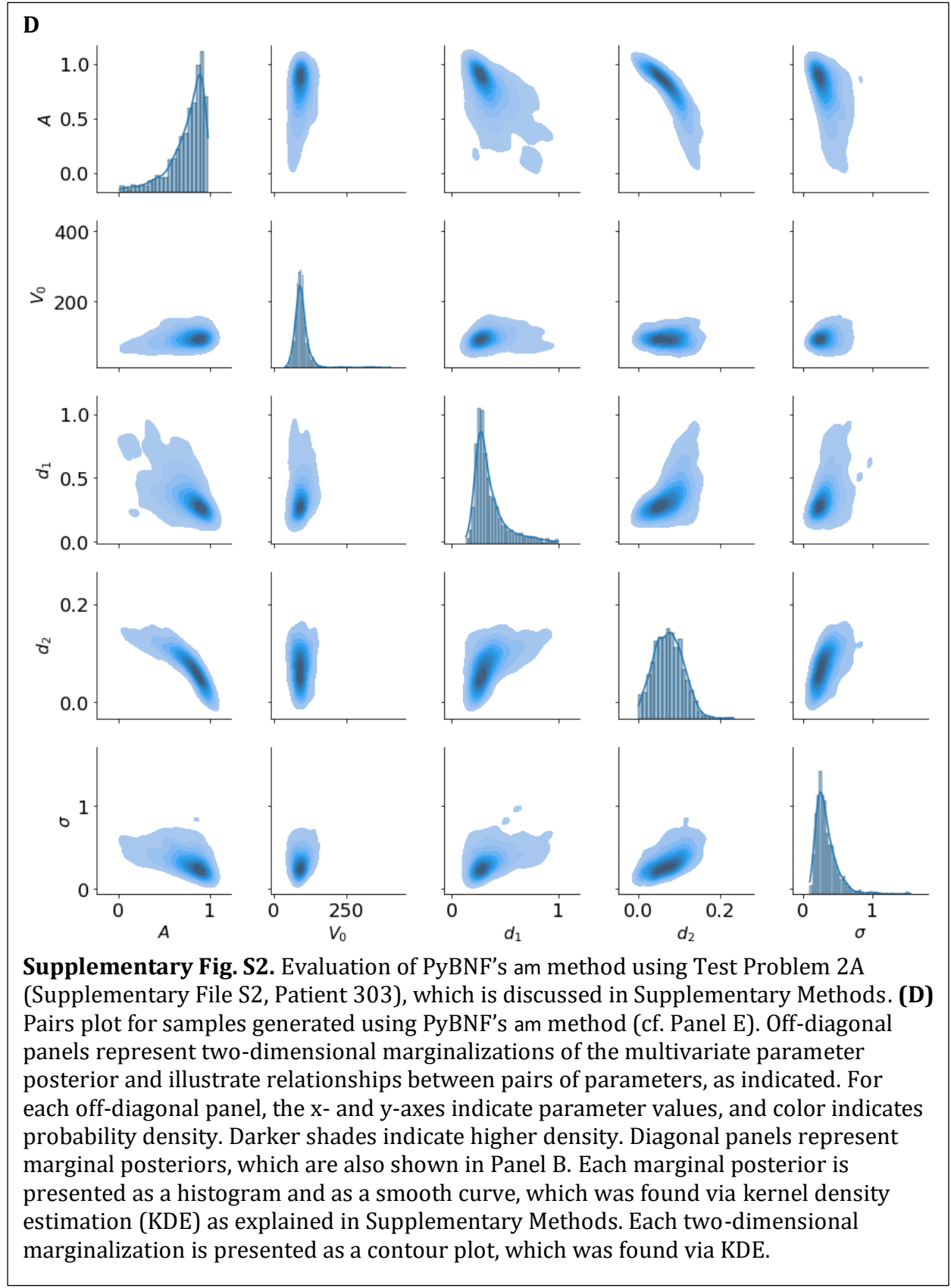

**Supplementary Fig. S2.** Evaluation of PyBNF's am method using Test Problem 2A (Supplementary File S2, Patient 303), which is discussed in Supplementary Methods. **(D)** Pairs plot for samples generated using PyBNF's am method (cf. Panel E). Off-diagonal panels represent two-dimensional marginalizations of the multivariate parameter posterior and illustrate relationships between pairs of parameters, as indicated. For each off-diagonal panel, the x- and y-axes indicate parameter values, and color indicates probability density. Darker shades indicate higher density. Diagonal panels represent marginal posteriors, which are also shown in Panel B. Each marginal posterior is presented as a histogram and as a smooth curve, which was found via kernel density estimation (KDE) as explained in Supplementary Methods. Each two-dimensional marginalization is presented as a contour plot, which was found via KDE.

**SUPPLEMENTARY FIGURE S2 – PANEL E**

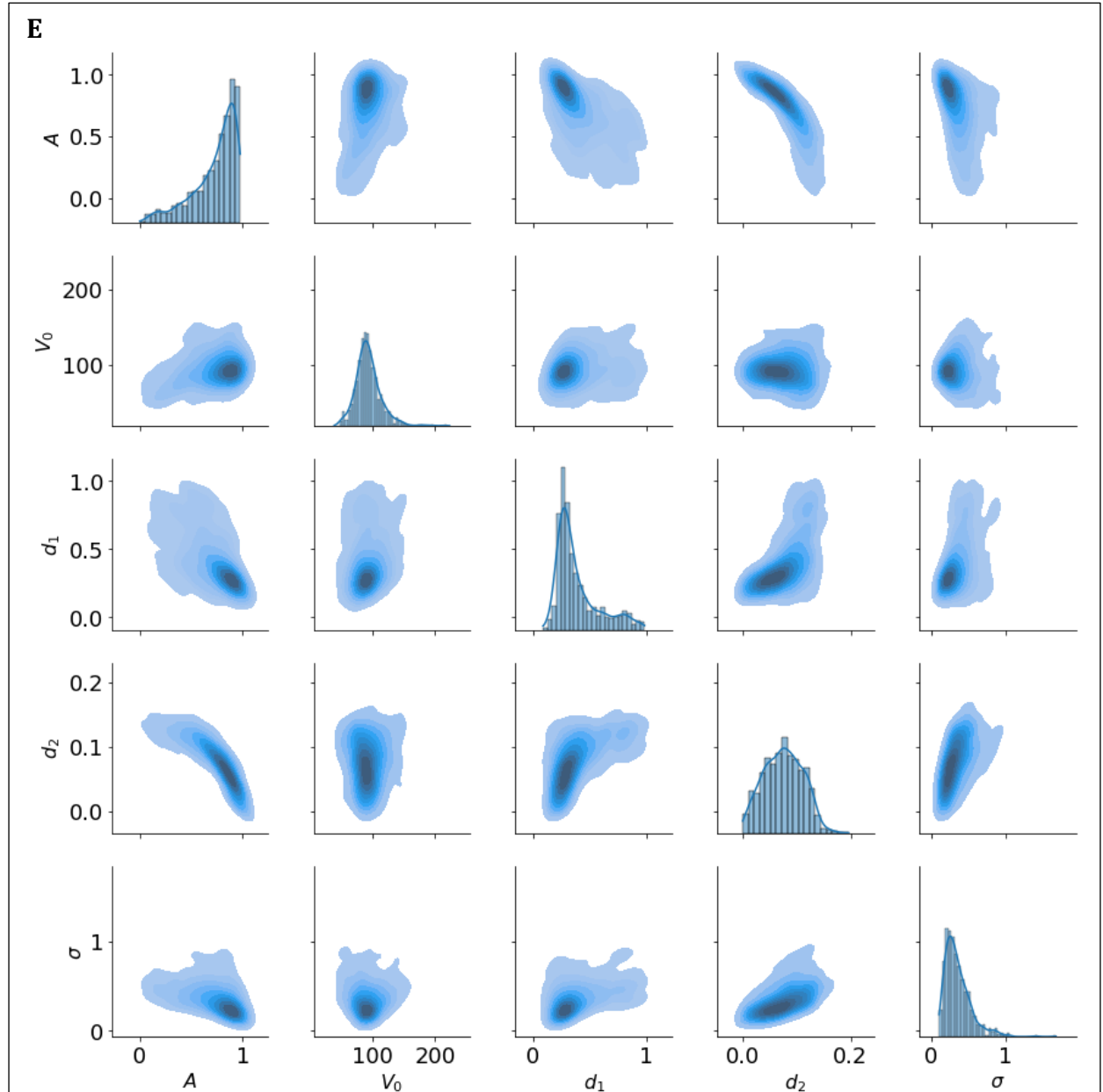

**Supplementary Fig. S2.** Evaluation of PyBNF's am method using Test Problem 2A (Supplementary File S2, Patient 303), which is discussed in Supplementary Methods. **(E)** Pairs plot for samples generated using problem-specific code (cf. Panel D). Off-diagonal panels represent two-dimensional marginalizations of the multivariate parameter posterior and illustrate relationships between pairs of parameters, as indicated. For each off-diagonal panel, the x- and y-axes indicate parameter values, and color indicates probability density. Darker shades indicate higher density. Diagonal panels represent marginal posteriors, which are also shown in Panel B. Each marginal posterior is presented as a histogram and as a smooth curve, which was found via kernel density estimation (KDE) as explained in Supplementary Methods. Each two-dimensional marginalization is presented as a contour plot, which was found via KDE.

**SUPPLEMENTARY FIGURE S3 – PANEL A**

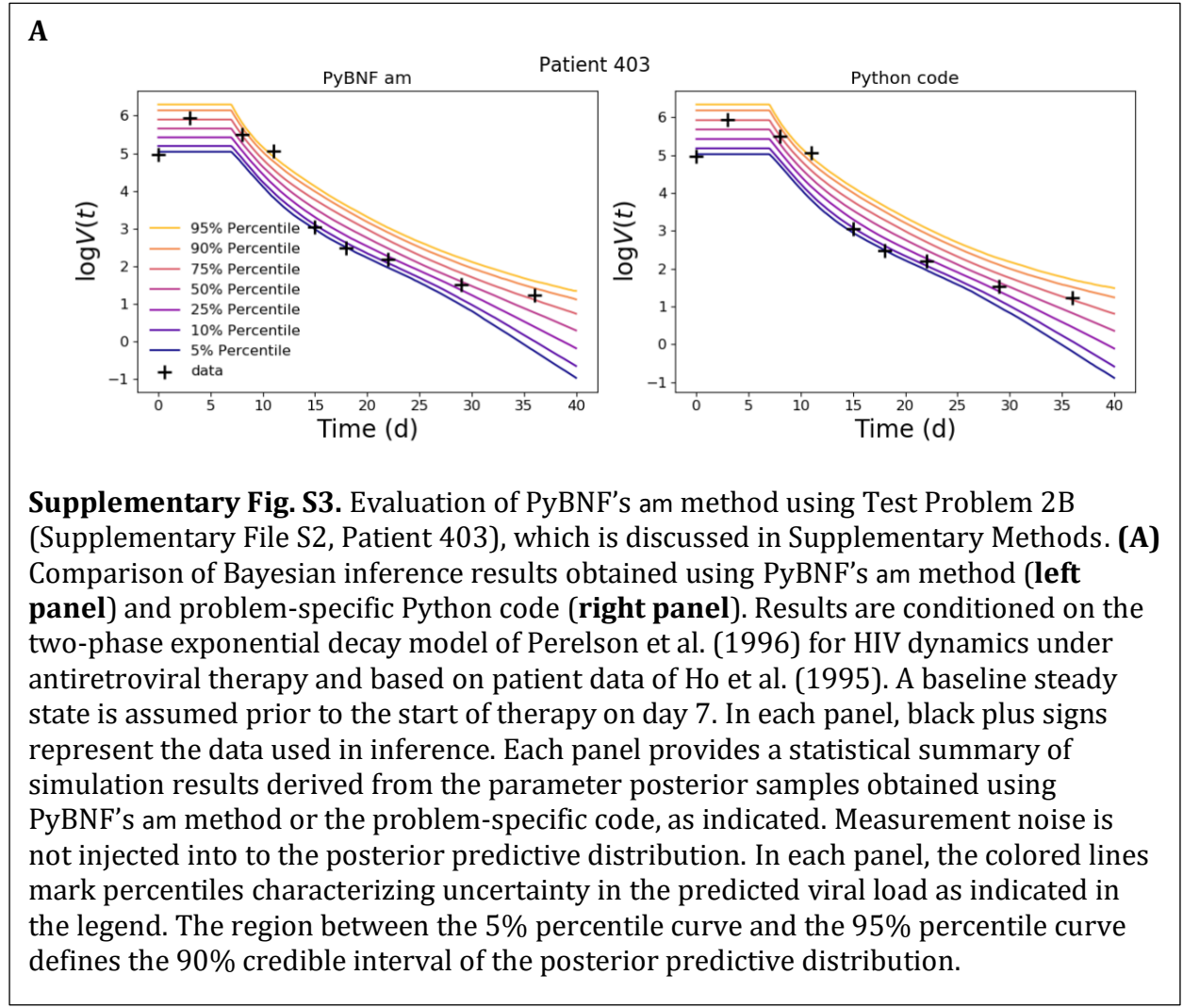

**Supplementary Fig. S3.** Evaluation of PyBNF's am method using Test Problem 2B (Supplementary File S2, Patient 403), which is discussed in Supplementary Methods. **(A)** Comparison of Bayesian inference results obtained using PyBNF's am method (**left panel**) and problem-specific Python code (**right panel**). Results are conditioned on the two-phase exponential decay model of Perelson et al. (1996) for HIV dynamics under antiretroviral therapy and based on patient data of Ho et al. (1995). A baseline steady state is assumed prior to the start of therapy on day 7. In each panel, black plus signs represent the data used in inference. Each panel provides a statistical summary of simulation results derived from the parameter posterior samples obtained using PyBNF's am method or the problem-specific code, as indicated. Measurement noise is not injected into to the posterior predictive distribution. In each panel, the colored lines mark percentiles characterizing uncertainty in the predicted viral load as indicated in the legend. The region between the 5% percentile curve and the 95% percentile curve defines the 90% credible interval of the posterior predictive distribution.

**SUPPLEMENTARY FIGURE S3 – PANEL B**

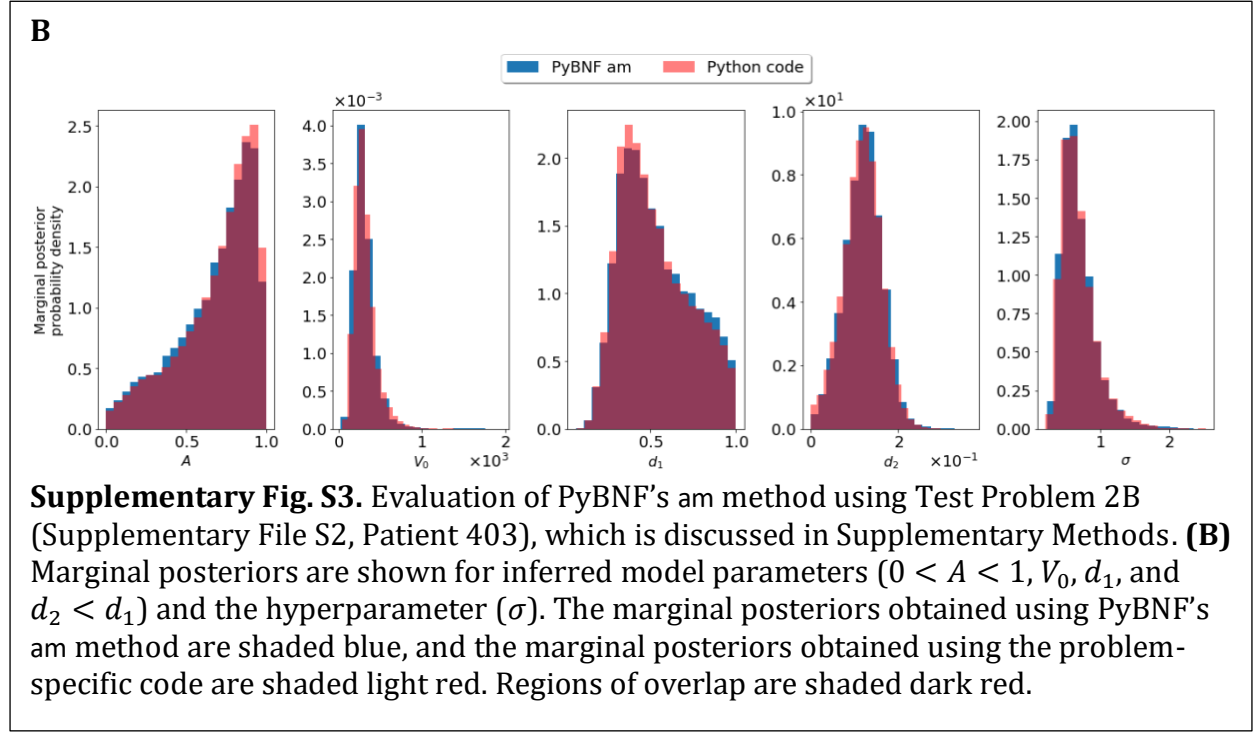

**Supplementary Fig. S3.** Evaluation of PyBNF's am method using Test Problem 2B (Supplementary File S2, Patient 403), which is discussed in Supplementary Methods. **(B)** Marginal posteriors are shown for inferred model parameters ($0 < A < 1$, $V_0$, $d_1$, and $d_2 < d_1$) and the hyperparameter ($\sigma$). The marginal posteriors obtained using PyBNF's am method are shaded blue, and the marginal posteriors obtained using the problem-specific code are shaded light red. Regions of overlap are shaded dark red.

**SUPPLEMENTARY FIGURE S3 – PANEL C**

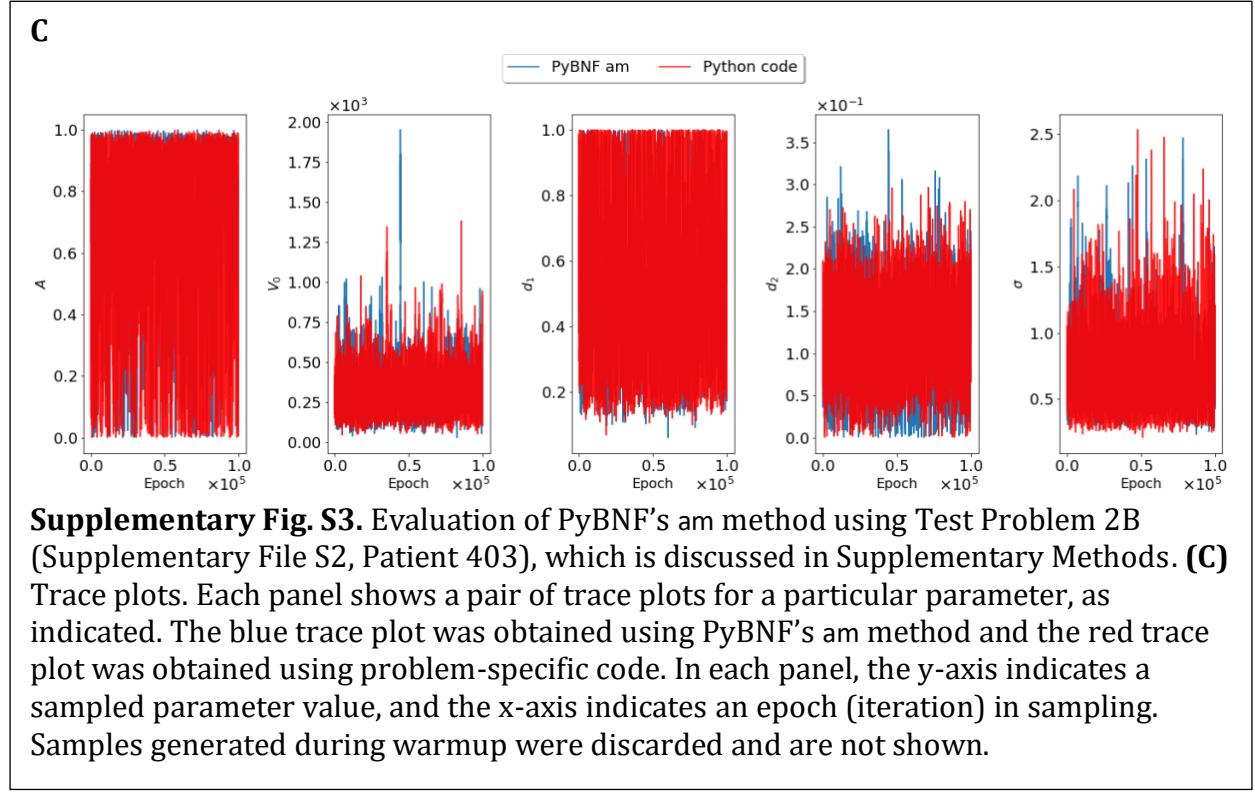

**Supplementary Fig. S3.** Evaluation of PyBNF's am method using Test Problem 2B (Supplementary File S2, Patient 403), which is discussed in Supplementary Methods. **(C)** Trace plots. Each panel shows a pair of trace plots for a particular parameter, as indicated. The blue trace plot was obtained using PyBNF's am method and the red trace plot was obtained using problem-specific code. In each panel, the y-axis indicates a sampled parameter value, and the x-axis indicates an epoch (iteration) in sampling. Samples generated during warmup were discarded and are not shown.

**SUPPLEMENTARY FIGURE S3 – PANEL D**

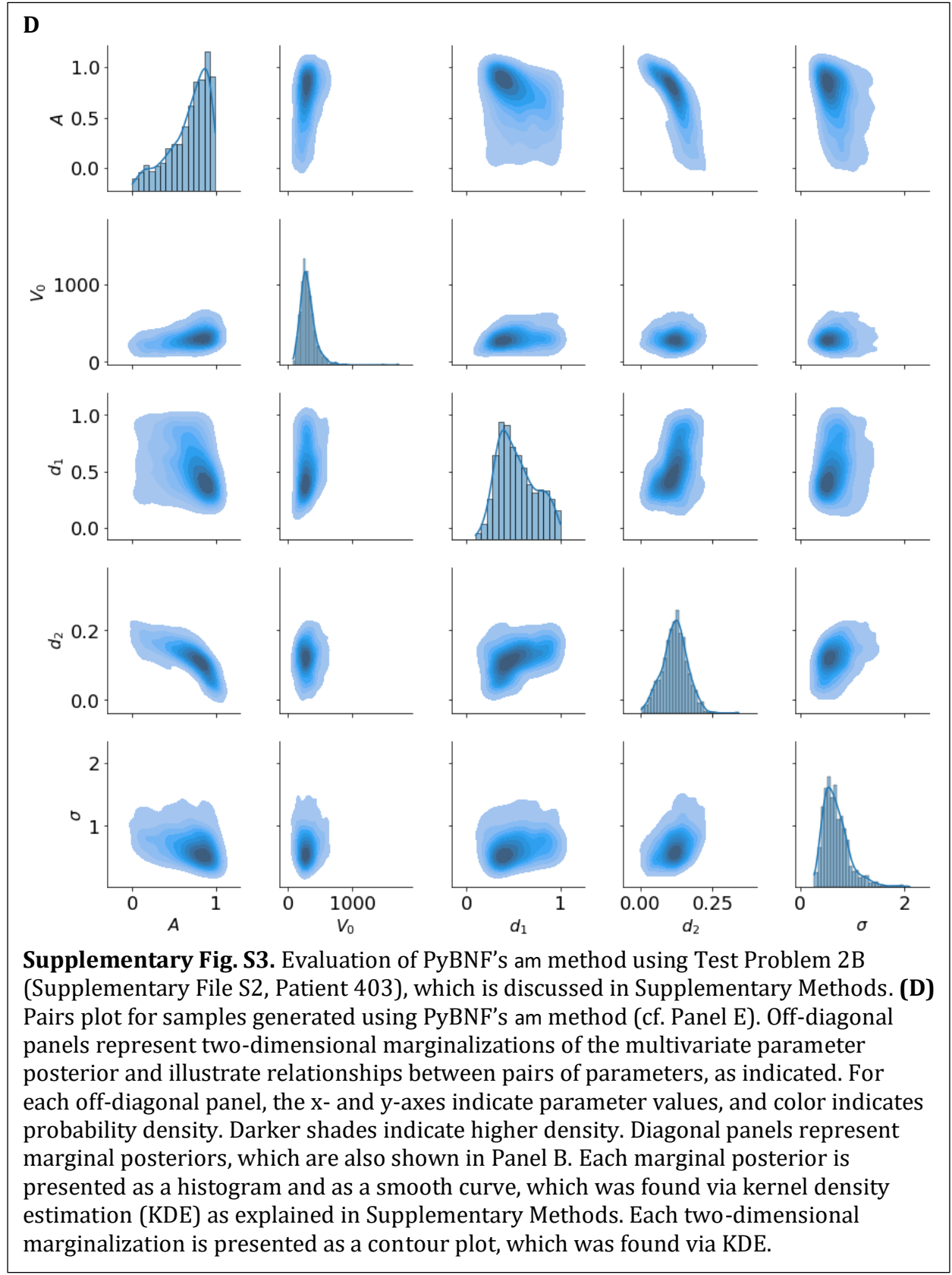

**Supplementary Fig. S3.** Evaluation of PyBNF's am method using Test Problem 2B (Supplementary File S2, Patient 403), which is discussed in Supplementary Methods. **(D)** Pairs plot for samples generated using PyBNF's am method (cf. Panel E). Off-diagonal panels represent two-dimensional marginalizations of the multivariate parameter posterior and illustrate relationships between pairs of parameters, as indicated. For each off-diagonal panel, the x- and y-axes indicate parameter values, and color indicates probability density. Darker shades indicate higher density. Diagonal panels represent marginal posteriors, which are also shown in Panel B. Each marginal posterior is presented as a histogram and as a smooth curve, which was found via kernel density estimation (KDE) as explained in Supplementary Methods. Each two-dimensional marginalization is presented as a contour plot, which was found via KDE.

**SUPPLEMENTARY FIGURE S3 – PANEL E**

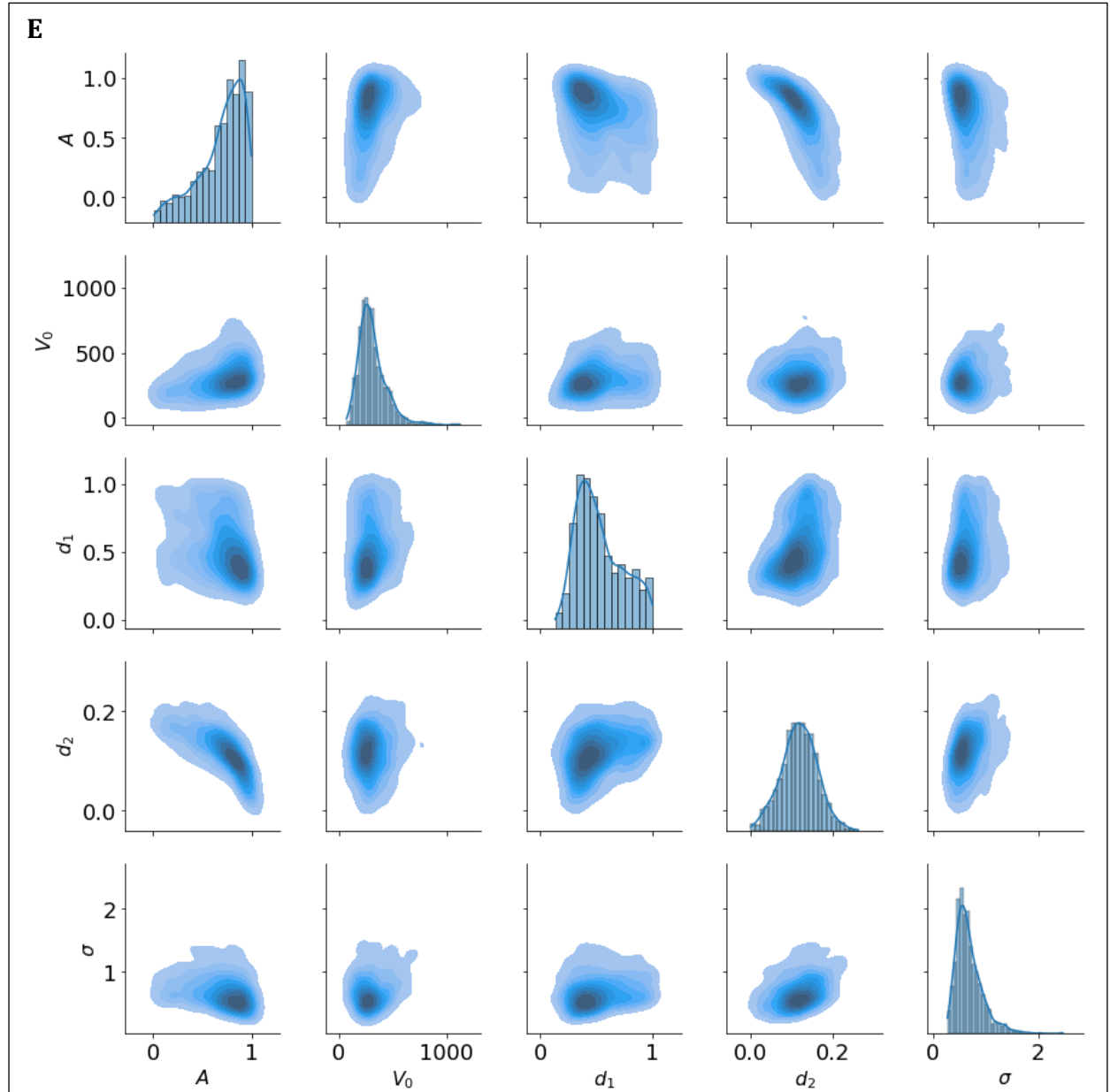

**Supplementary Fig. S3.** Evaluation of PyBNF's am method using Test Problem 2B (Supplementary File S2, Patient 403), which is discussed in Supplementary Methods. **(E)** Pairs plot for samples generated using problem-specific code (cf. Panel D). Off-diagonal panels represent two-dimensional marginalizations of the multivariate parameter posterior and illustrate relationships between pairs of parameters, as indicated. For each off-diagonal panel, the x- and y-axes indicate parameter values, and color indicates probability density. Darker shades indicate higher density. Diagonal panels represent marginal posteriors, which are also shown in Panel B. Each marginal posterior is presented as a histogram and as a smooth curve, which was found via kernel density estimation (KDE) as explained in Supplementary Methods. Each two-dimensional marginalization is presented as a contour plot, which was found via KDE.

**SUPPLEMENTARY FIGURE S4 – PANEL A**

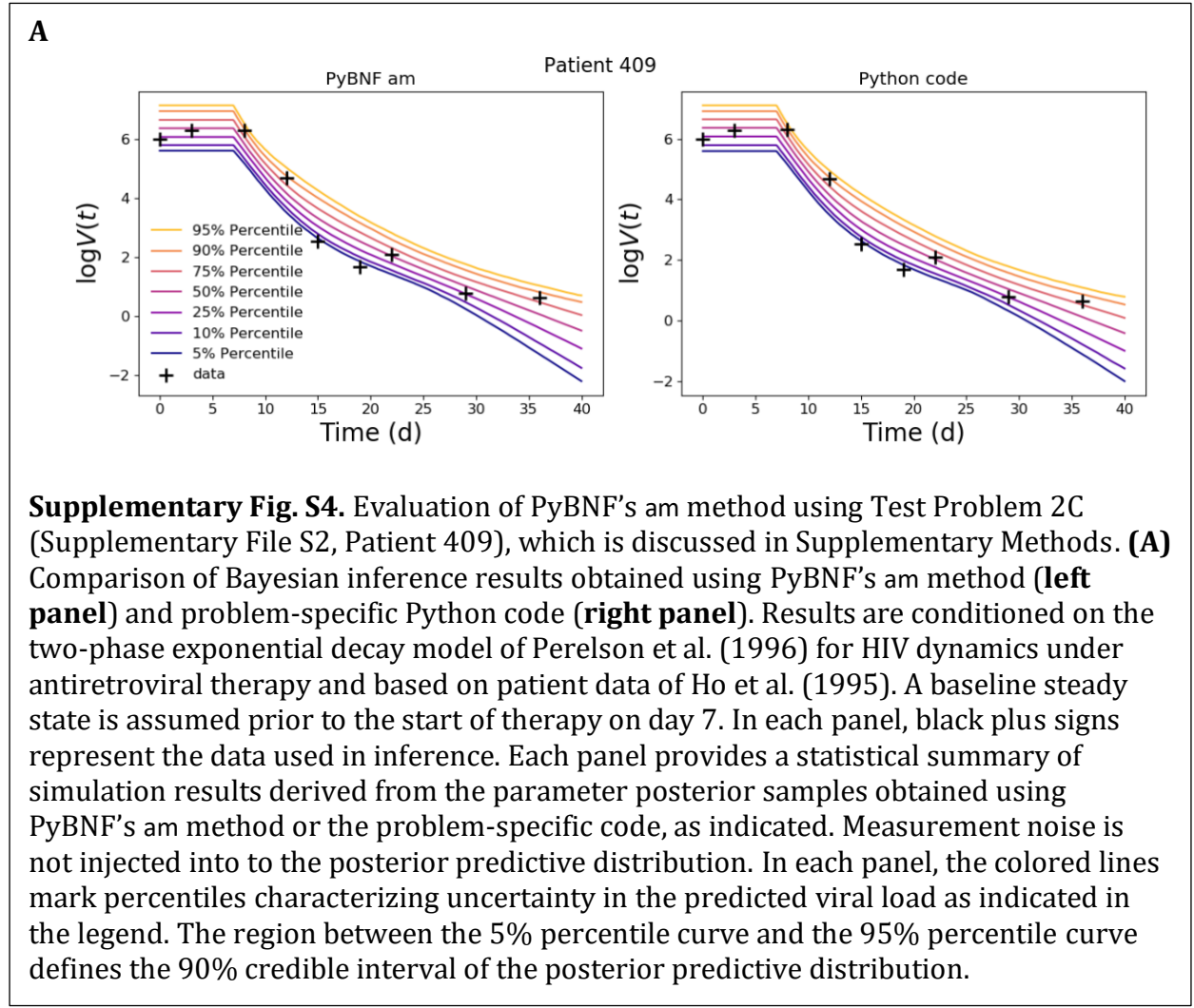

**Supplementary Fig. S4.** Evaluation of PyBNF's am method using Test Problem 2C (Supplementary File S2, Patient 409), which is discussed in Supplementary Methods. **(A)** Comparison of Bayesian inference results obtained using PyBNF's am method (**left panel**) and problem-specific Python code (**right panel**). Results are conditioned on the two-phase exponential decay model of Perelson et al. (1996) for HIV dynamics under antiretroviral therapy and based on patient data of Ho et al. (1995). A baseline steady state is assumed prior to the start of therapy on day 7. In each panel, black plus signs represent the data used in inference. Each panel provides a statistical summary of simulation results derived from the parameter posterior samples obtained using PyBNF's am method or the problem-specific code, as indicated. Measurement noise is not injected into to the posterior predictive distribution. In each panel, the colored lines mark percentiles characterizing uncertainty in the predicted viral load as indicated in the legend. The region between the 5% percentile curve and the 95% percentile curve defines the 90% credible interval of the posterior predictive distribution.

**SUPPLEMENTARY FIGURE S4 – PANEL B**

**Supplementary Fig. S4.** Evaluation of PyBNF's am method using Test Problem 2C (Supplementary File S2, Patient 409), which is discussed in Supplementary Methods. **(B)** Marginal posteriors are shown for inferred model parameters ($0 < A < 1$, $V_0$, $d_1$, and $d_2 < d_1$) and the hyperparameter ($\sigma$). The marginal posteriors obtained using PyBNF's am method are shaded blue, and the marginal posteriors obtained using the problem-specific code are shaded light red. Regions of overlap are shaded dark red.

**SUPPLEMENTARY FIGURE S4 – PANEL C**

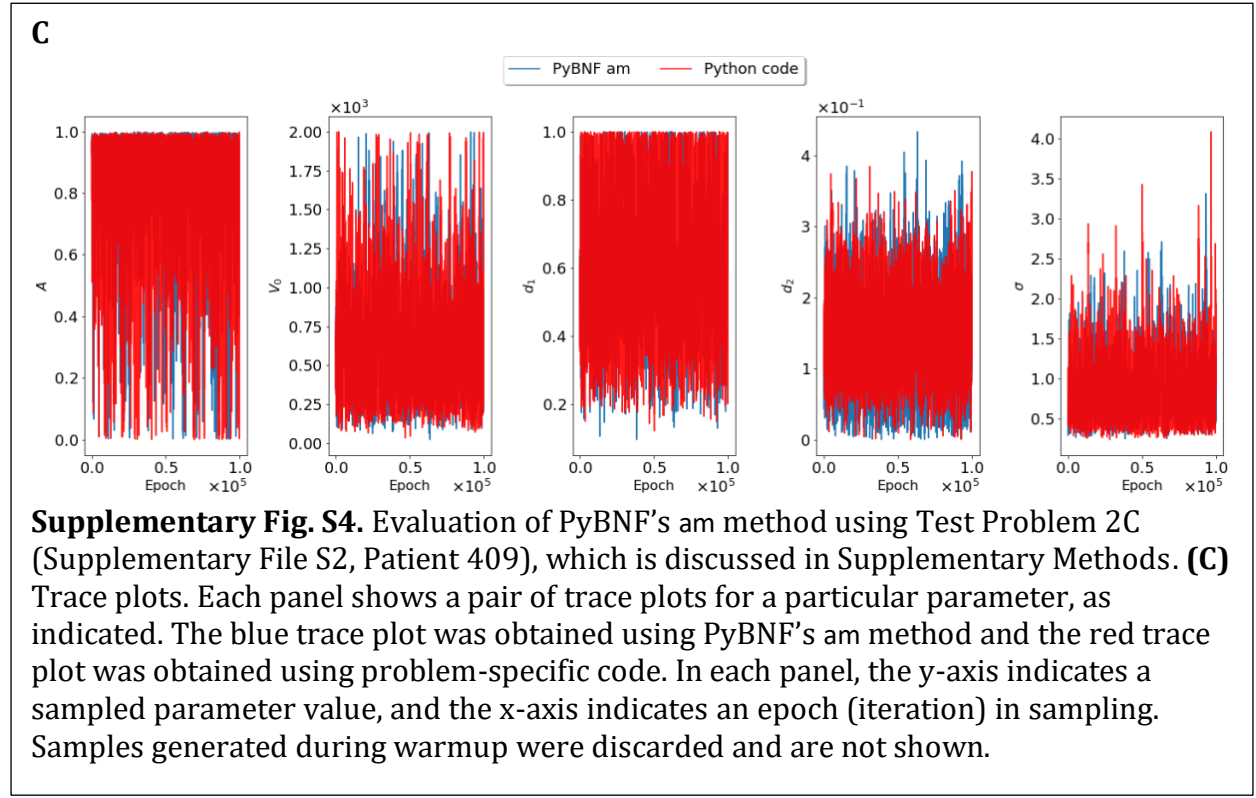

**Supplementary Fig. S4.** Evaluation of PyBNF's am method using Test Problem 2C (Supplementary File S2, Patient 409), which is discussed in Supplementary Methods. **(C)** Trace plots. Each panel shows a pair of trace plots for a particular parameter, as indicated. The blue trace plot was obtained using PyBNF's am method and the red trace plot was obtained using problem-specific code. In each panel, the y-axis indicates a sampled parameter value, and the x-axis indicates an epoch (iteration) in sampling. Samples generated during warmup were discarded and are not shown.

**SUPPLEMENTARY FIGURE S4 – PANEL D**

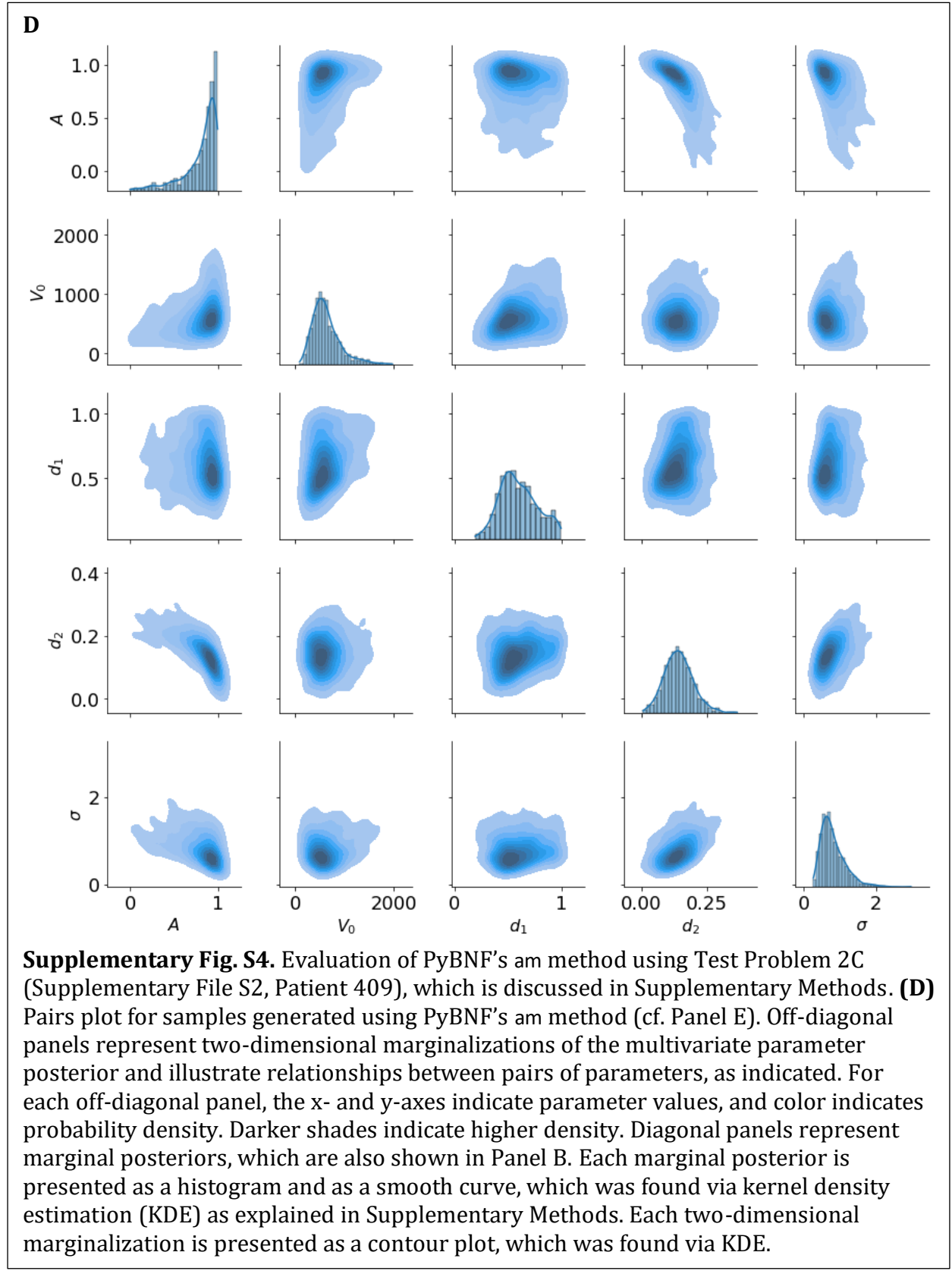

**Supplementary Fig. S4.** Evaluation of PyBNF's am method using Test Problem 2C (Supplementary File S2, Patient 409), which is discussed in Supplementary Methods. **(D)** Pairs plot for samples generated using PyBNF's am method (cf. Panel E). Off-diagonal panels represent two-dimensional marginalizations of the multivariate parameter posterior and illustrate relationships between pairs of parameters, as indicated. For each off-diagonal panel, the x- and y-axes indicate parameter values, and color indicates probability density. Darker shades indicate higher density. Diagonal panels represent marginal posteriors, which are also shown in Panel B. Each marginal posterior is presented as a histogram and as a smooth curve, which was found via kernel density estimation (KDE) as explained in Supplementary Methods. Each two-dimensional marginalization is presented as a contour plot, which was found via KDE.

**SUPPLEMENTARY FIGURE S4 – PANEL E**

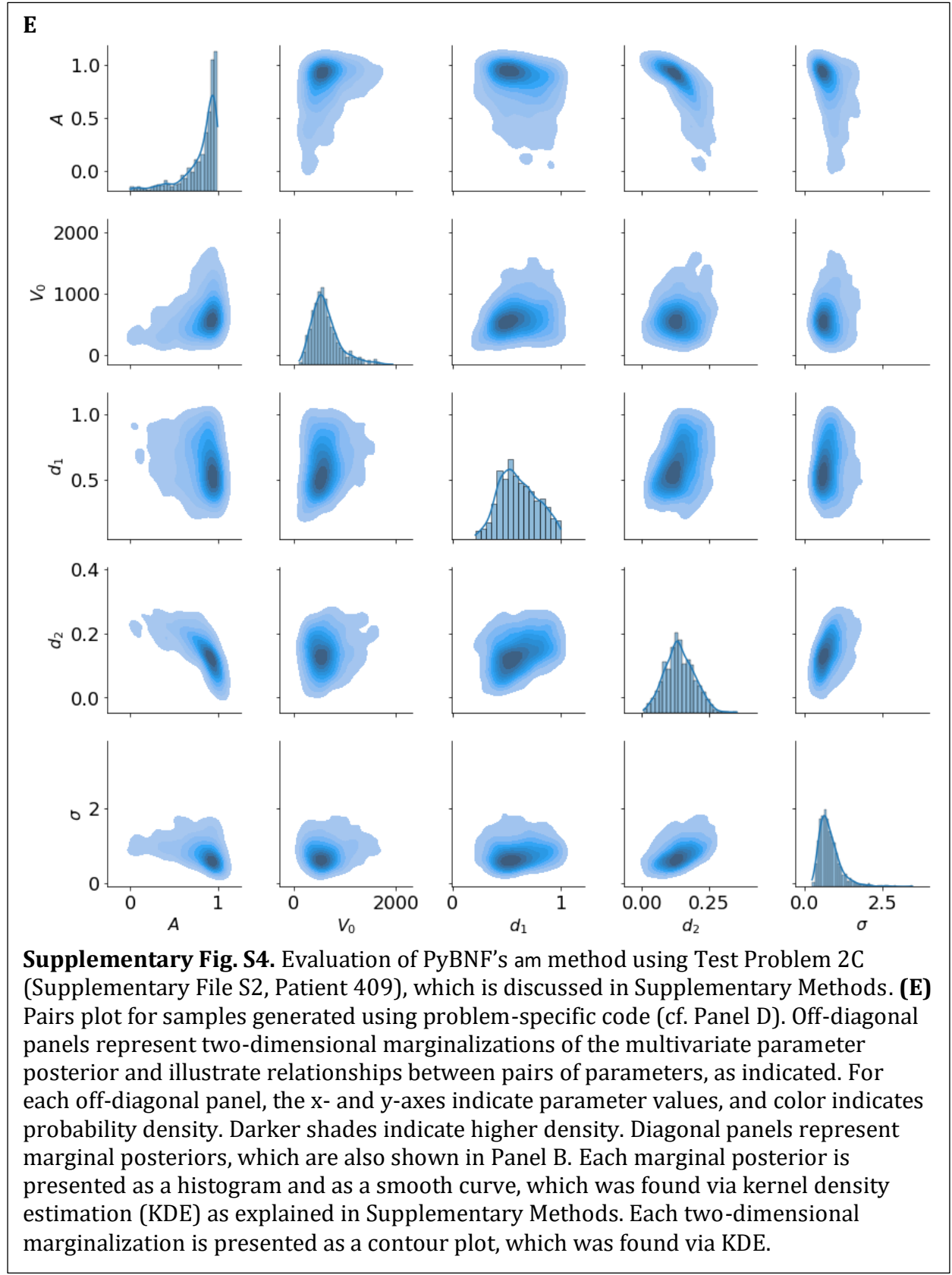

**Supplementary Fig. S4.** Evaluation of PyBNF's am method using Test Problem 2C (Supplementary File S2, Patient 409), which is discussed in Supplementary Methods. **(E)** Pairs plot for samples generated using problem-specific code (cf. Panel D). Off-diagonal panels represent two-dimensional marginalizations of the multivariate parameter posterior and illustrate relationships between pairs of parameters, as indicated. For each off-diagonal panel, the x- and y-axes indicate parameter values, and color indicates probability density. Darker shades indicate higher density. Diagonal panels represent marginal posteriors, which are also shown in Panel B. Each marginal posterior is presented as a histogram and as a smooth curve, which was found via kernel density estimation (KDE) as explained in Supplementary Methods. Each two-dimensional marginalization is presented as a contour plot, which was found via KDE.

**SUPPLEMENTARY FIGURE S5 – PANEL A**

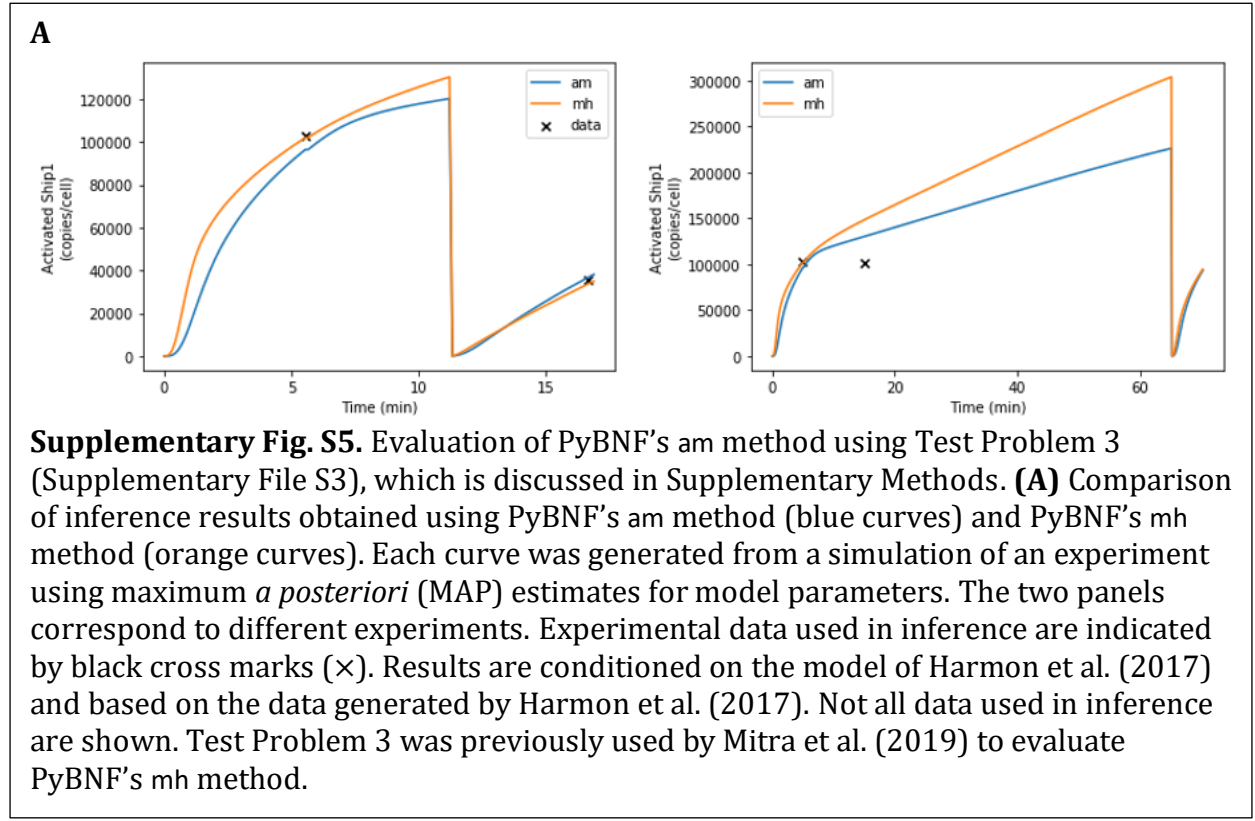

**Supplementary Fig. S5.** Evaluation of PyBNF's am method using Test Problem 3 (Supplementary File S3), which is discussed in Supplementary Methods. **(A)** Comparison of inference results obtained using PyBNF's am method (blue curves) and PyBNF's mh method (orange curves). Each curve was generated from a simulation of an experiment using maximum *a posteriori* (MAP) estimates for model parameters. The two panels correspond to different experiments. Experimental data used in inference are indicated by black cross marks (×). Results are conditioned on the model of Harmon et al. (2017) and based on the data generated by Harmon et al. (2017). Not all data used in inference are shown. Test Problem 3 was previously used by Mitra et al. (2019) to evaluate PyBNF's mh method.

**SUPPLEMENTARY FIGURE S5 – PANEL B**

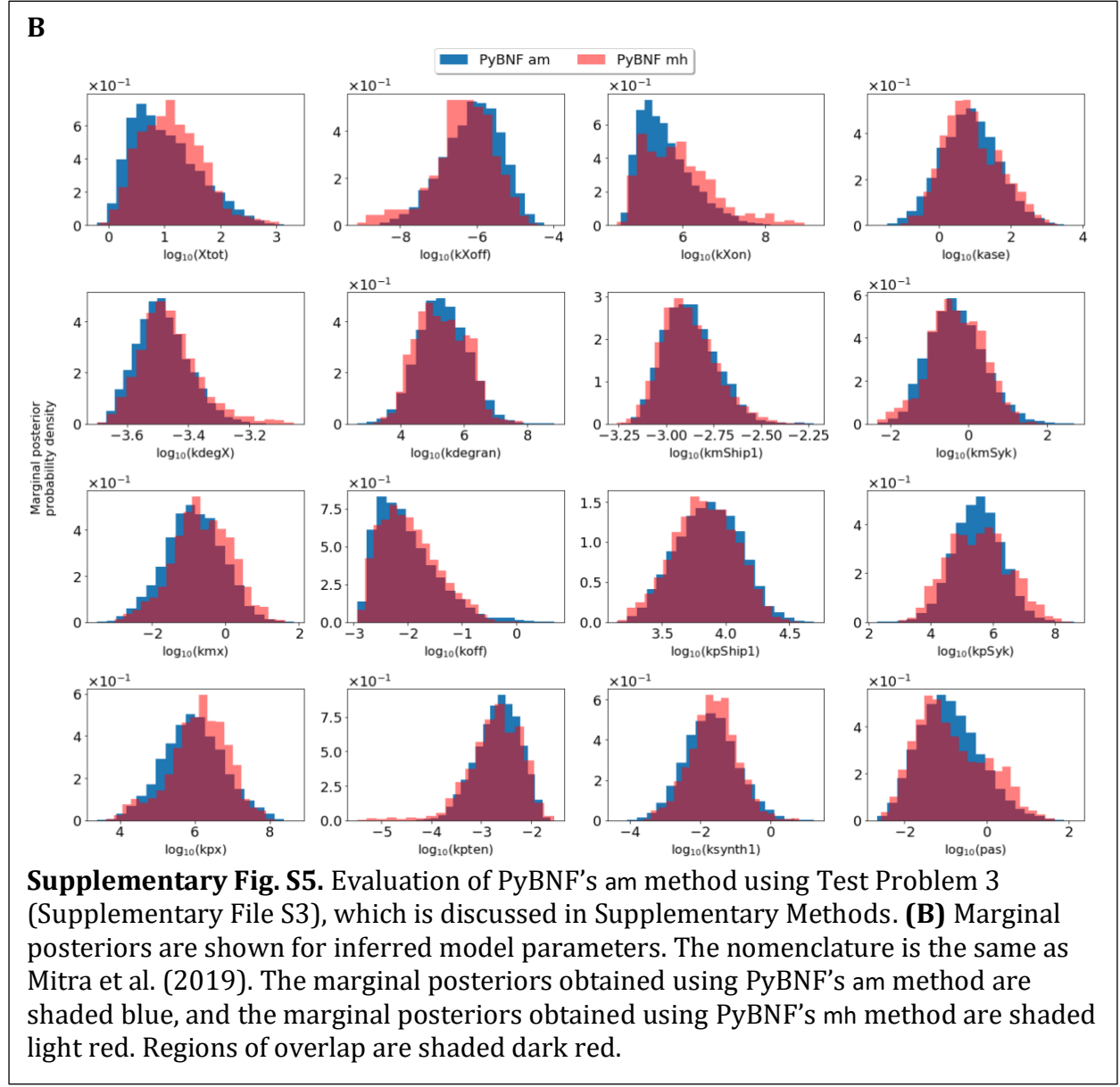

**Supplementary Fig. S5.** Evaluation of PyBNF's am method using Test Problem 3 (Supplementary File S3), which is discussed in Supplementary Methods. **(B)** Marginal posteriors are shown for inferred model parameters. The nomenclature is the same as Mitra et al. (2019). The marginal posteriors obtained using PyBNF's am method are shaded blue, and the marginal posteriors obtained using PyBNF's mh method are shaded light red. Regions of overlap are shaded dark red.

**SUPPLEMENTARY FIGURE S5 – PANEL C**

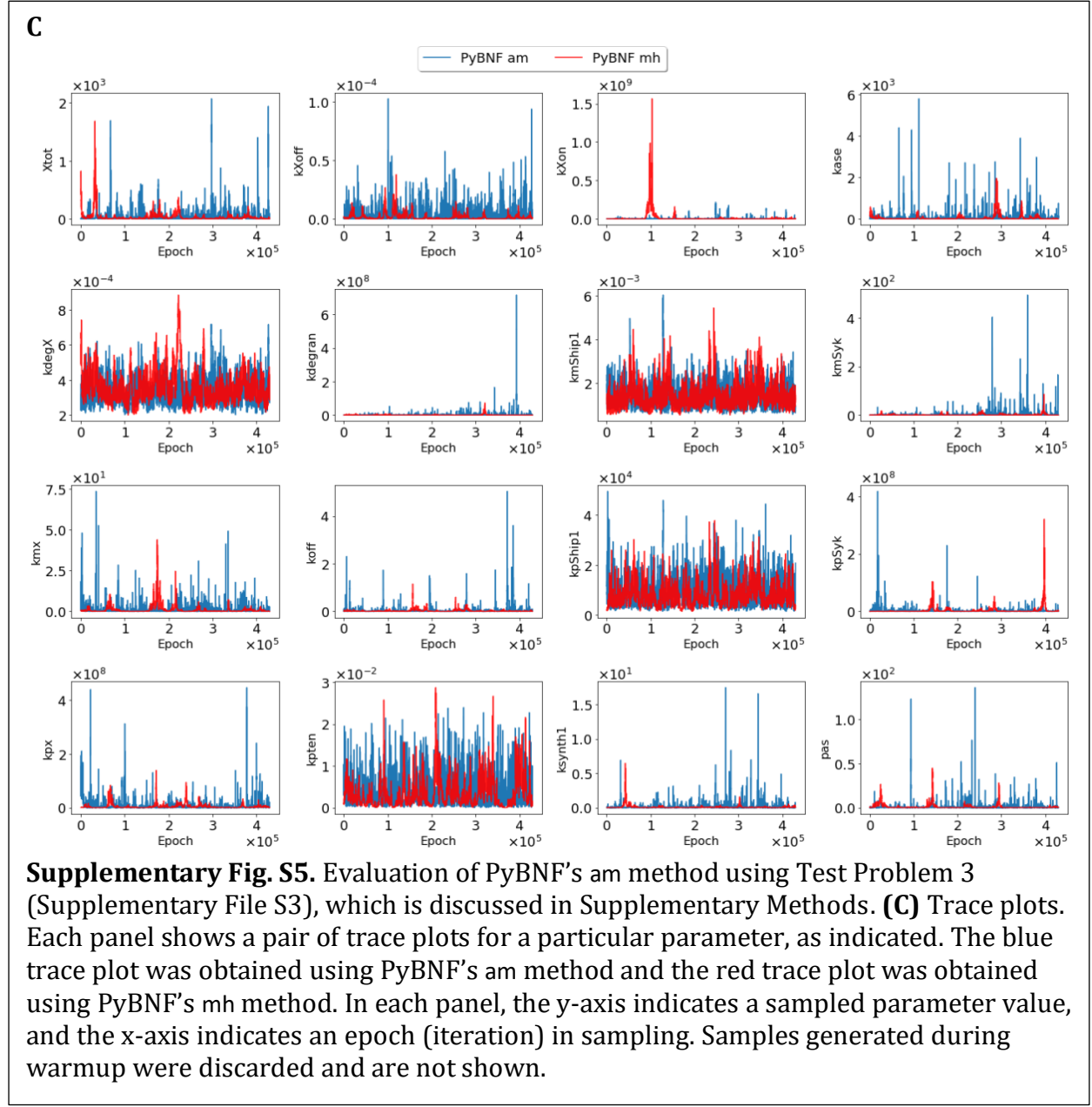

**Supplementary Fig. S5.** Evaluation of PyBNF's am method using Test Problem 3 (Supplementary File S3), which is discussed in Supplementary Methods. **(C)** Trace plots. Each panel shows a pair of trace plots for a particular parameter, as indicated. The blue trace plot was obtained using PyBNF's am method and the red trace plot was obtained using PyBNF's mh method. In each panel, the y-axis indicates a sampled parameter value, and the x-axis indicates an epoch (iteration) in sampling. Samples generated during warmup were discarded and are not shown.

**SUPPLEMENTARY FIGURE S5 – PANEL D**

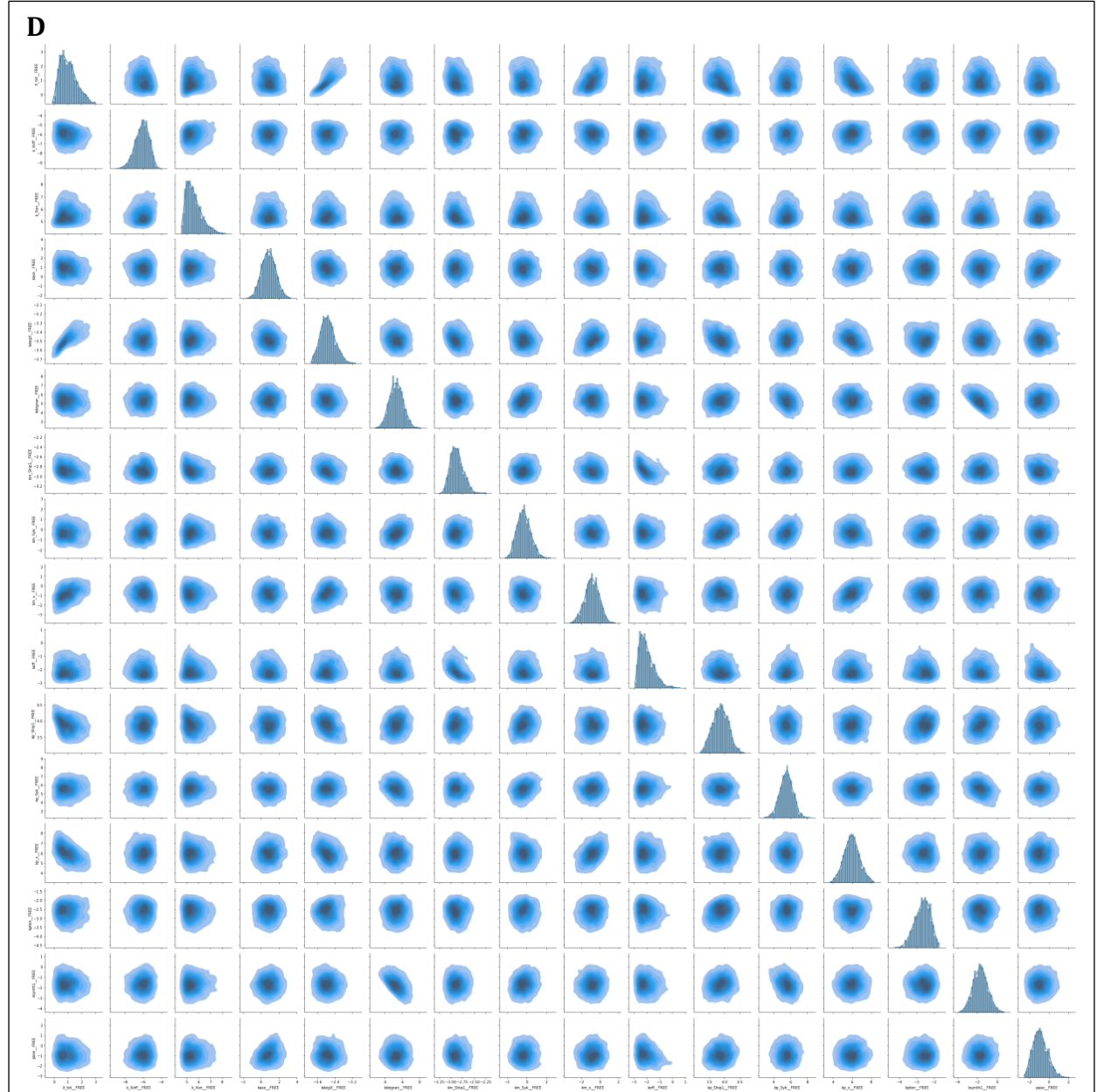

**Supplementary Fig. S5.** Evaluation of PyBNF's am method using Test Problem 3 (Supplementary File S3), which is discussed in Supplementary Methods. **(D)** Pairs plot for samples generated using PyBNF's am method (cf. Panel E). Off-diagonal panels represent two-dimensional marginalizations of the multivariate parameter posterior and illustrate relationships between pairs of parameters, as indicated. For each off-diagonal panel, the x- and y-axes indicate parameter values, and color indicates probability density. Darker shades indicate higher density. Diagonal panels represent marginal posteriors, which are also shown in Panel B. Each marginal posterior is presented as a histogram and as a smooth curve, which was found via kernel density estimation (KDE) as explained in Supplementary Methods. Each two-dimensional marginalization is presented as a contour plot, which was found via KDE.

**SUPPLEMENTARY FIGURE S5 – PANEL E**

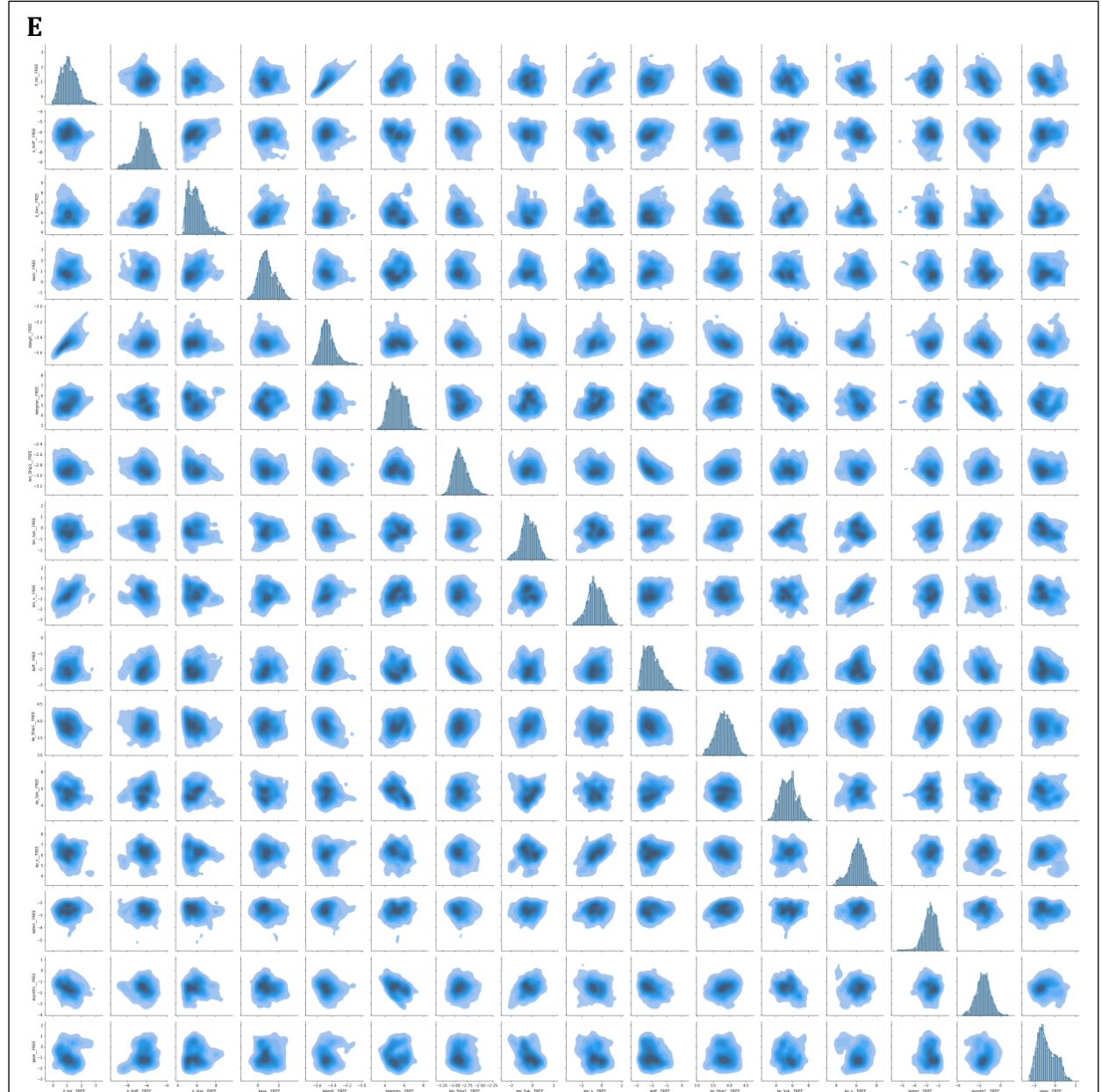

**Supplementary Fig. S5.** Evaluation of PyBNF's am method using Test Problem 3 (Supplementary File S3), which is discussed in Supplementary Methods. **(E)** Pairs plot for samples generated using PyBNF's mh method (cf. Panel E). Off-diagonal panels represent two-dimensional marginalizations of the multivariate parameter posterior and illustrate relationships between pairs of parameters, as indicated. For each off-diagonal panel, the x- and y-axes indicate parameter values, and color indicates probability density. Darker shades indicate higher density. Diagonal panels represent marginal posteriors, which are also shown in Panel B. Each marginal posterior is presented as a histogram and as a smooth curve, which was found via kernel density estimation (KDE) as explained in Supplementary Methods. Each two-dimensional marginalization is presented as a contour plot, which was found via KDE.

**SUPPLEMENTARY FIGURE S6 – PANEL A**

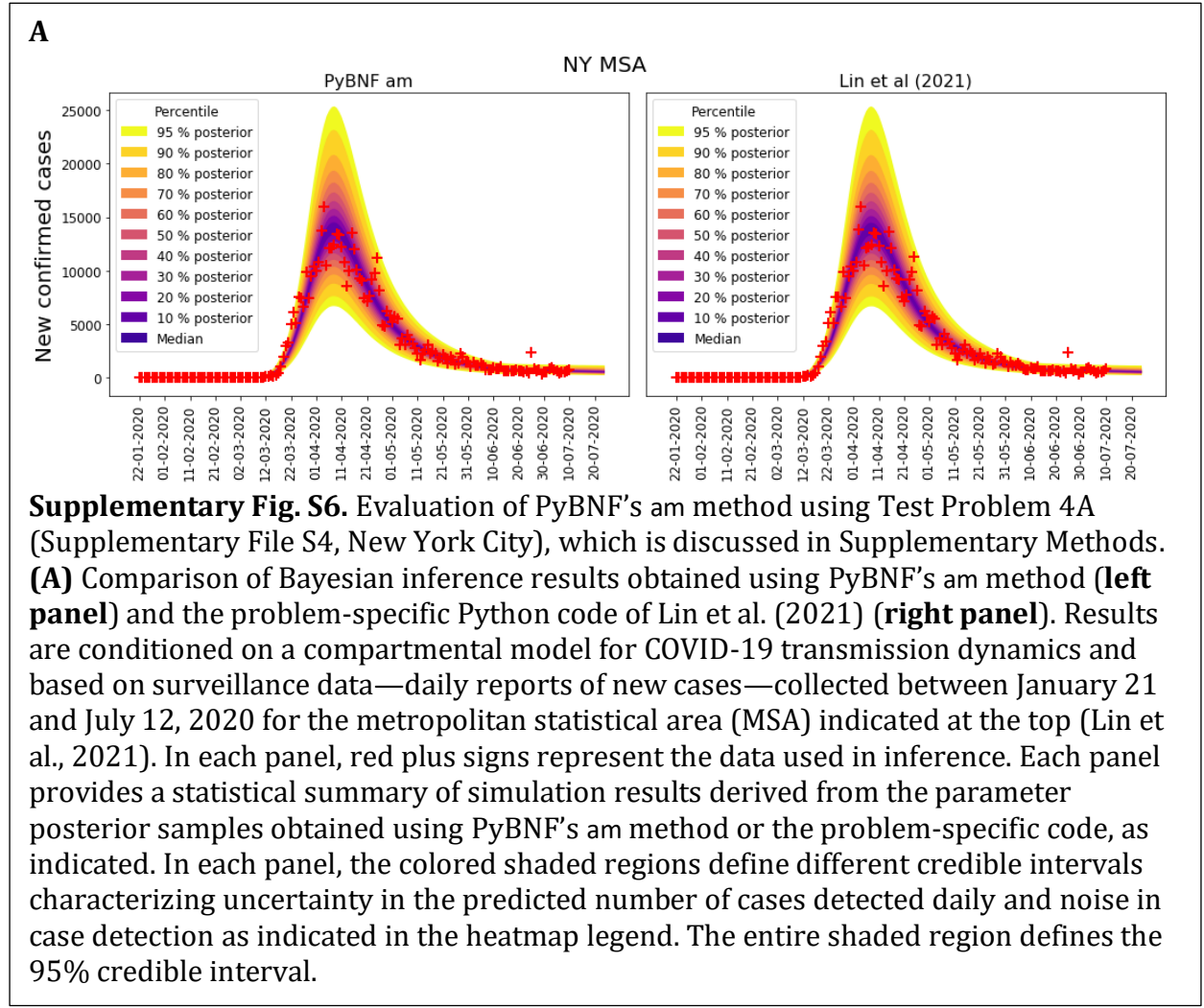

**Supplementary Fig. S6.** Evaluation of PyBNF's ᴀᴍ method using Test Problem 4A (Supplementary File S4, New York City), which is discussed in Supplementary Methods. **(A)** Comparison of Bayesian inference results obtained using PyBNF's ᴀᴍ method (**left panel**) and the problem-specific Python code of Lin et al. (2021) (**right panel**). Results are conditioned on a compartmental model for COVID-19 transmission dynamics and based on surveillance data—daily reports of new cases—collected between January 21 and July 12, 2020 for the metropolitan statistical area (MSA) indicated at the top (Lin et al., 2021). In each panel, red plus signs represent the data used in inference. Each panel provides a statistical summary of simulation results derived from the parameter posterior samples obtained using PyBNF's ᴀᴍ method or the problem-specific code, as indicated. In each panel, the colored shaded regions define different credible intervals characterizing uncertainty in the predicted number of cases detected daily and noise in case detection as indicated in the heatmap legend. The entire shaded region defines the 95% credible interval.

**SUPPLEMENTARY FIGURE S6 – PANEL B**

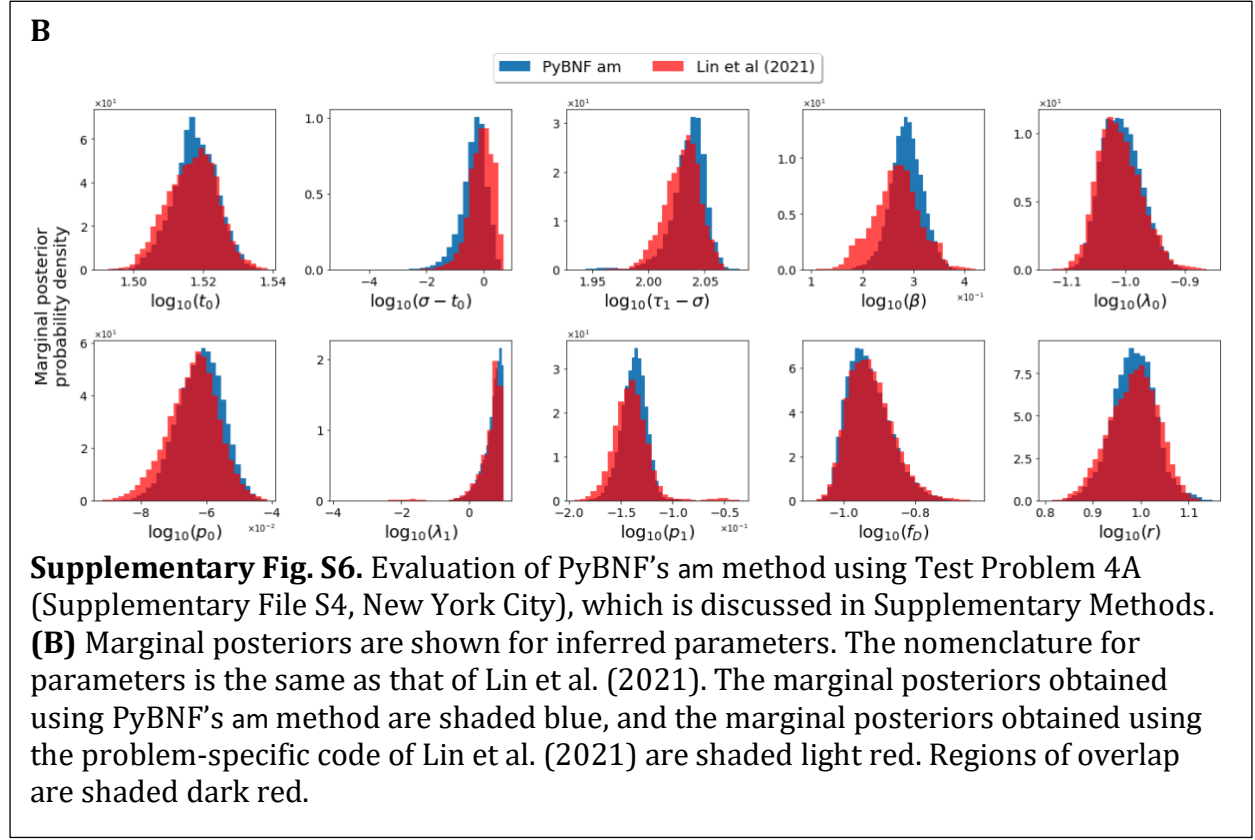

**Supplementary Fig. S6.** Evaluation of PyBNF's am method using Test Problem 4A (Supplementary File S4, New York City), which is discussed in Supplementary Methods. **(B)** Marginal posteriors are shown for inferred parameters. The nomenclature for parameters is the same as that of Lin et al. (2021). The marginal posteriors obtained using PyBNF's am method are shaded blue, and the marginal posteriors obtained using the problem-specific code of Lin et al. (2021) are shaded light red. Regions of overlap are shaded dark red.

**SUPPLEMENTARY FIGURE S6 – PANEL C**

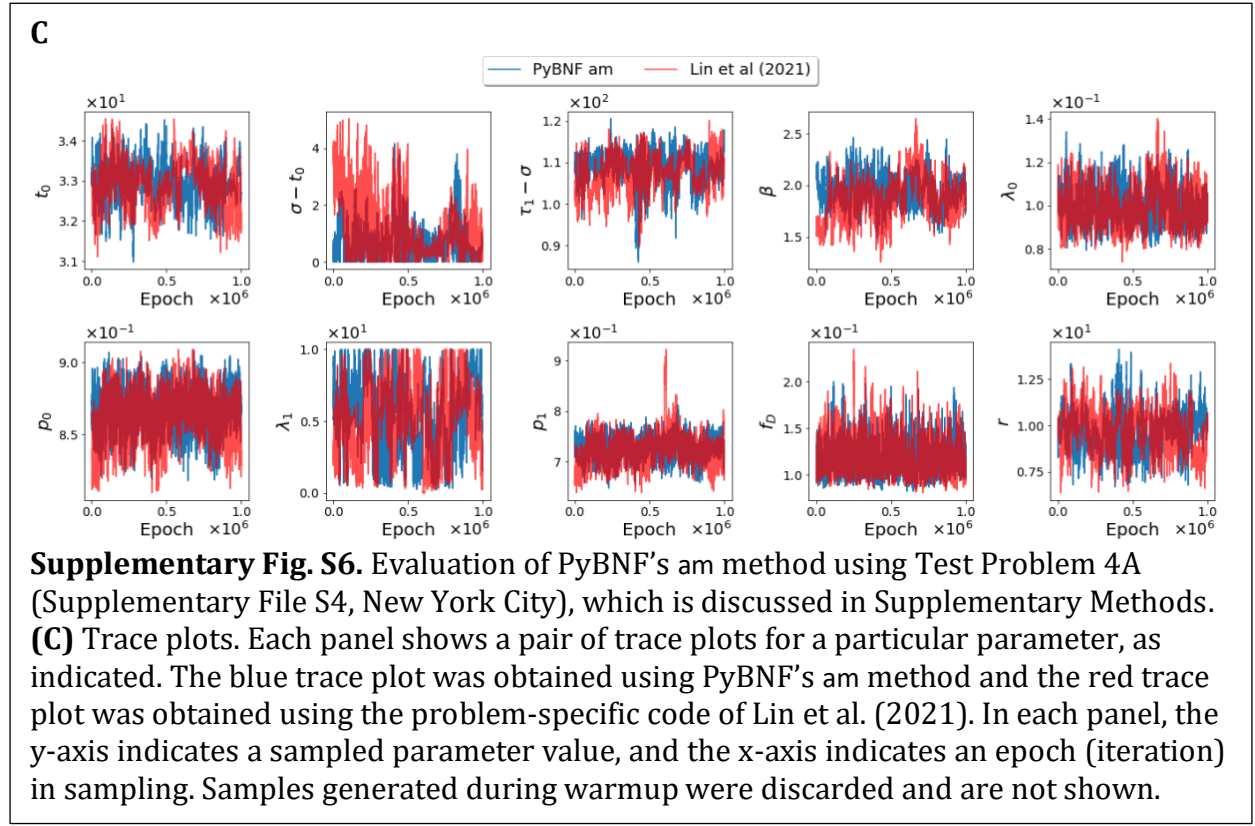

**Supplementary Fig. S6.** Evaluation of PyBNF's am method using Test Problem 4A (Supplementary File S4, New York City), which is discussed in Supplementary Methods. **(C)** Trace plots. Each panel shows a pair of trace plots for a particular parameter, as indicated. The blue trace plot was obtained using PyBNF's am method and the red trace plot was obtained using the problem-specific code of Lin et al. (2021). In each panel, the y-axis indicates a sampled parameter value, and the x-axis indicates an epoch (iteration) in sampling. Samples generated during warmup were discarded and are not shown.

**SUPPLEMENTARY FIGURE S6 – PANEL D**

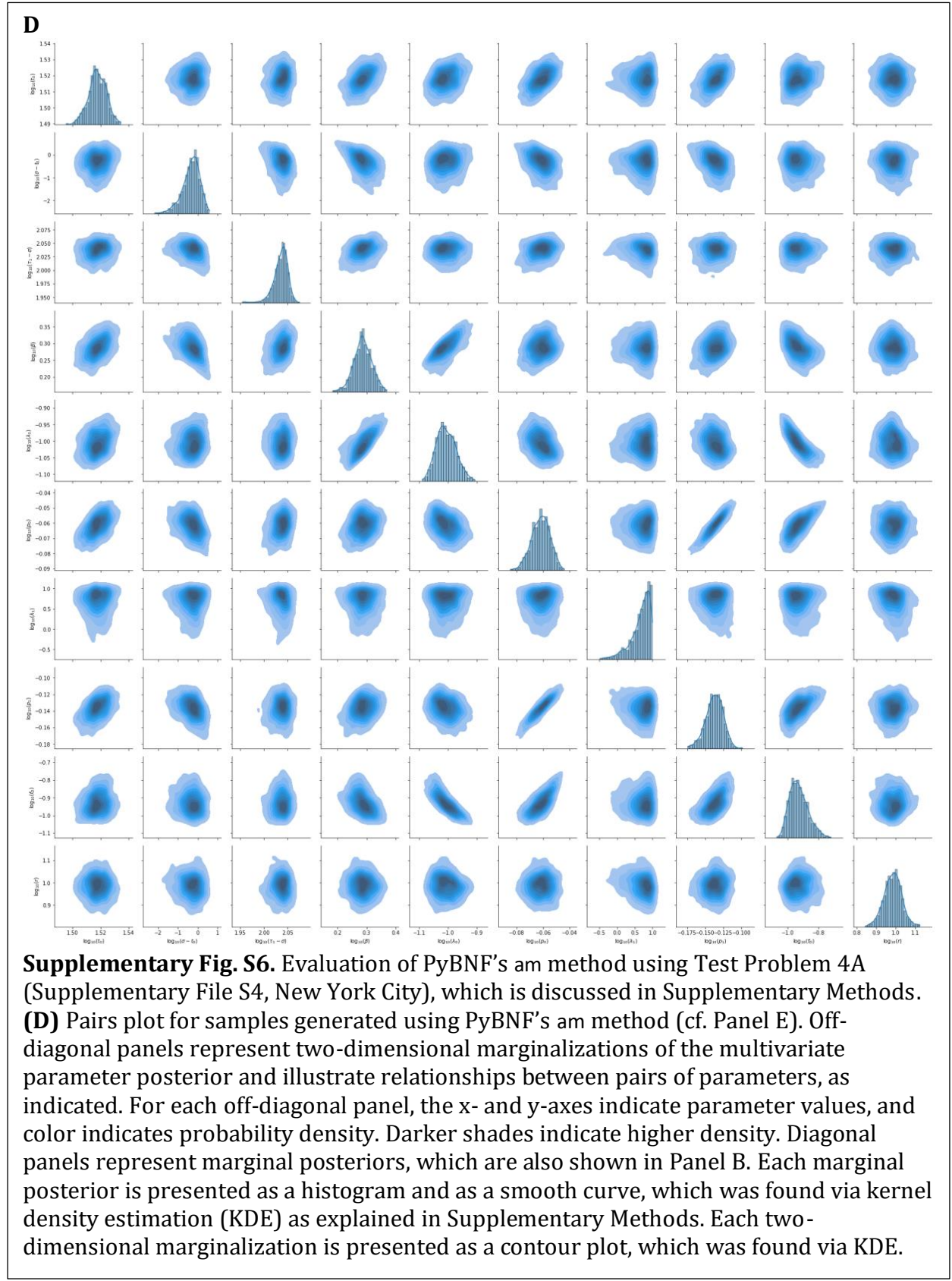

**Supplementary Fig. S6.** Evaluation of PyBNF's am method using Test Problem 4A (Supplementary File S4, New York City), which is discussed in Supplementary Methods. **(D)** Pairs plot for samples generated using PyBNF's am method (cf. Panel E). Off-diagonal panels represent two-dimensional marginalizations of the multivariate parameter posterior and illustrate relationships between pairs of parameters, as indicated. For each off-diagonal panel, the x- and y-axes indicate parameter values, and color indicates probability density. Darker shades indicate higher density. Diagonal panels represent marginal posteriors, which are also shown in Panel B. Each marginal posterior is presented as a histogram and as a smooth curve, which was found via kernel density estimation (KDE) as explained in Supplementary Methods. Each two-dimensional marginalization is presented as a contour plot, which was found via KDE.

**SUPPLEMENTARY FIGURE S6 – PANEL E**

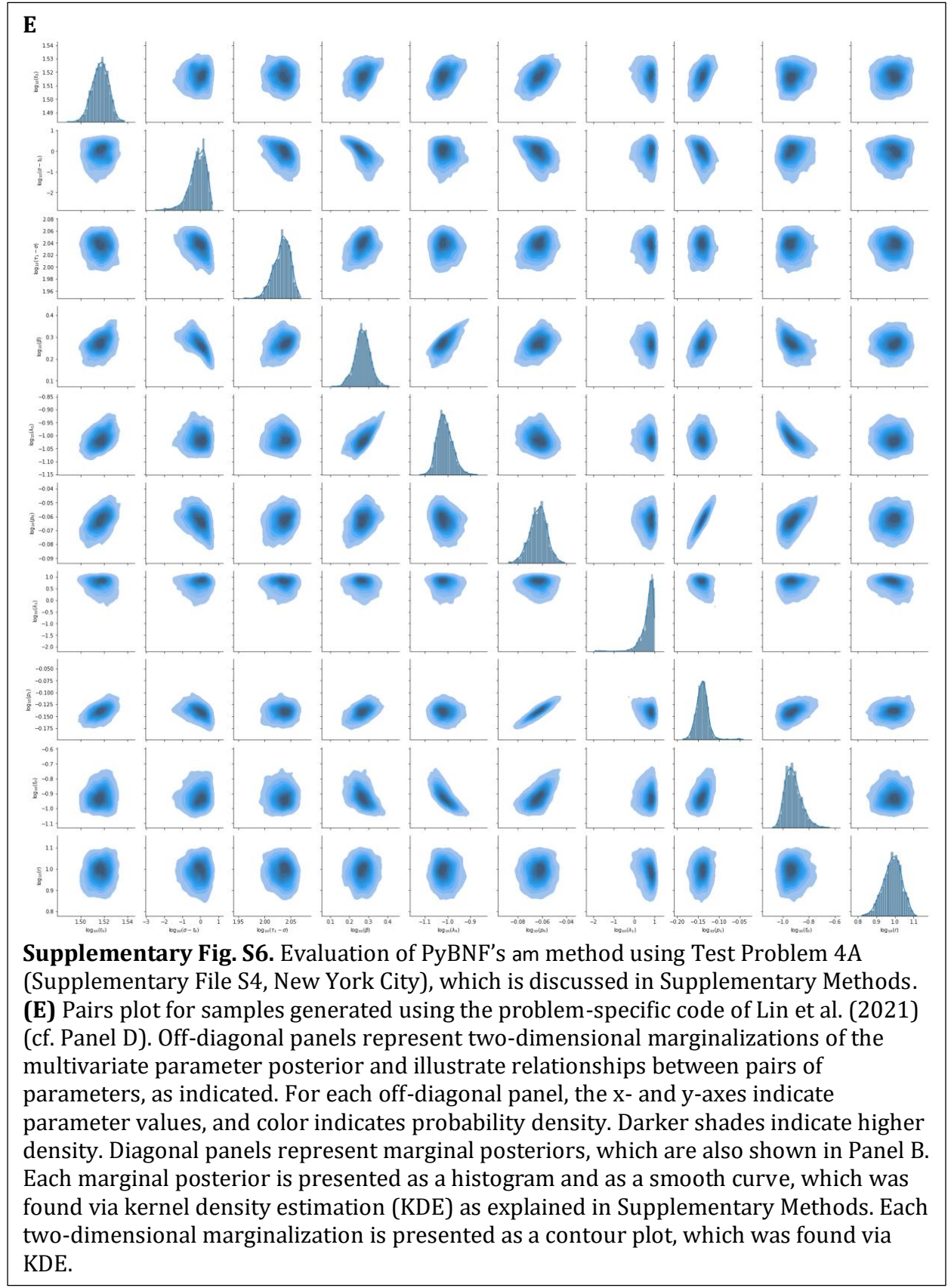

**Supplementary Fig. S6.** Evaluation of PyBNF's am method using Test Problem 4A (Supplementary File S4, New York City), which is discussed in Supplementary Methods. **(E)** Pairs plot for samples generated using the problem-specific code of Lin et al. (2021) (cf. Panel D). Off-diagonal panels represent two-dimensional marginalizations of the multivariate parameter posterior and illustrate relationships between pairs of parameters, as indicated. For each off-diagonal panel, the x- and y-axes indicate parameter values, and color indicates probability density. Darker shades indicate higher density. Diagonal panels represent marginal posteriors, which are also shown in Panel B. Each marginal posterior is presented as a histogram and as a smooth curve, which was found via kernel density estimation (KDE) as explained in Supplementary Methods. Each two-dimensional marginalization is presented as a contour plot, which was found via KDE.

**SUPPLEMENTARY FIGURE S7 – PANEL A**

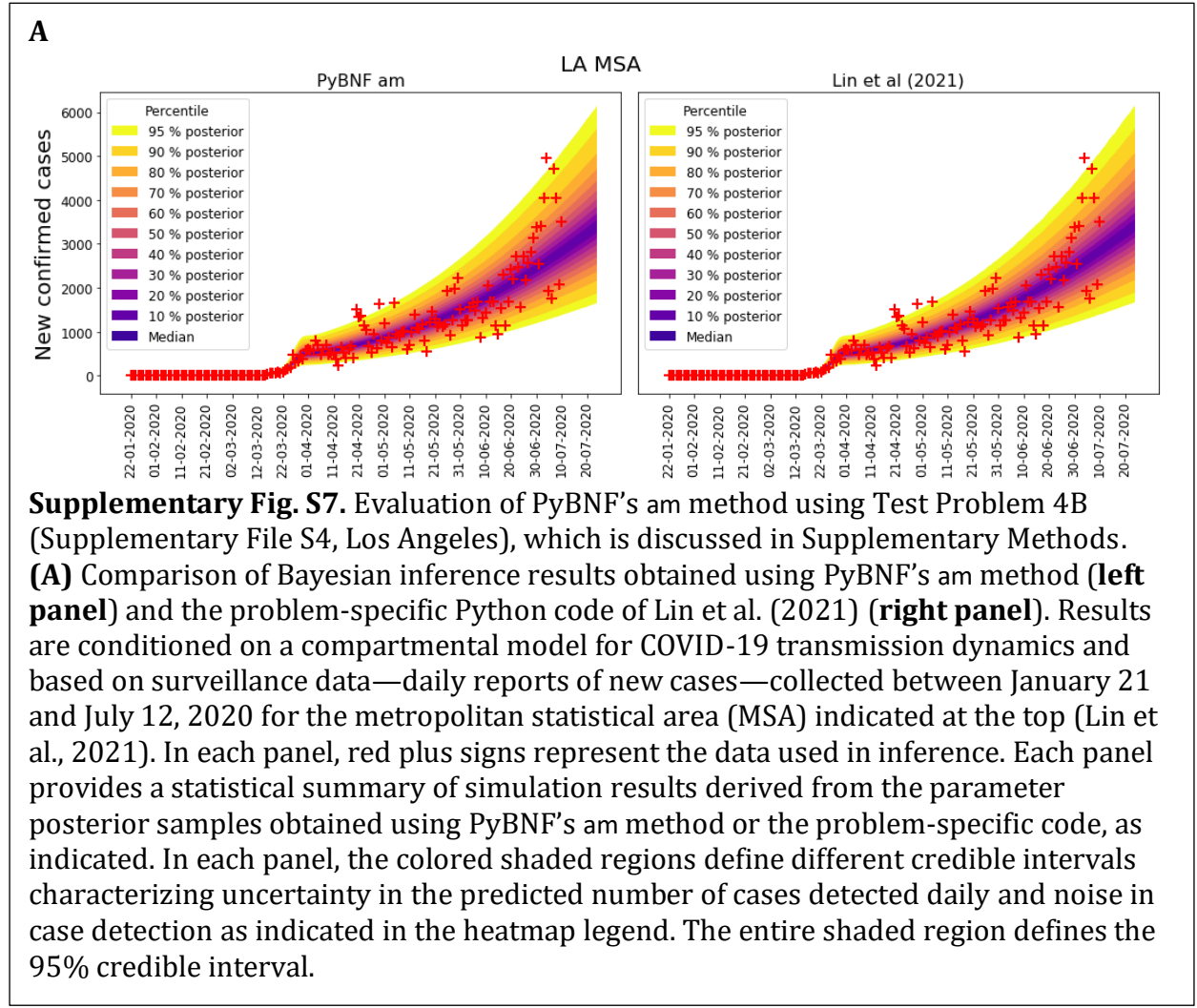

**Supplementary Fig. S7.** Evaluation of PyBNF's am method using Test Problem 4B (Supplementary File S4, Los Angeles), which is discussed in Supplementary Methods. **(A)** Comparison of Bayesian inference results obtained using PyBNF's method (**left panel**) and the problem-specific Python code of Lin et al. (2021) (**right panel**). Results are conditioned on a compartmental model for COVID-19 transmission dynamics and based on surveillance data—daily reports of new cases—collected between January 21 and July 12, 2020 for the metropolitan statistical area (MSA) indicated at the top (Lin et al., 2021). In each panel, red plus signs represent the data used in inference. Each panel provides a statistical summary of simulation results derived from the parameter posterior samples obtained using PyBNF's am method or the problem-specific code, as indicated. In each panel, the colored shaded regions define different credible intervals characterizing uncertainty in the predicted number of cases detected daily and noise in case detection as indicated in the heatmap legend. The entire shaded region defines the 95% credible interval.

**SUPPLEMENTARY FIGURE S7 – PANEL B**

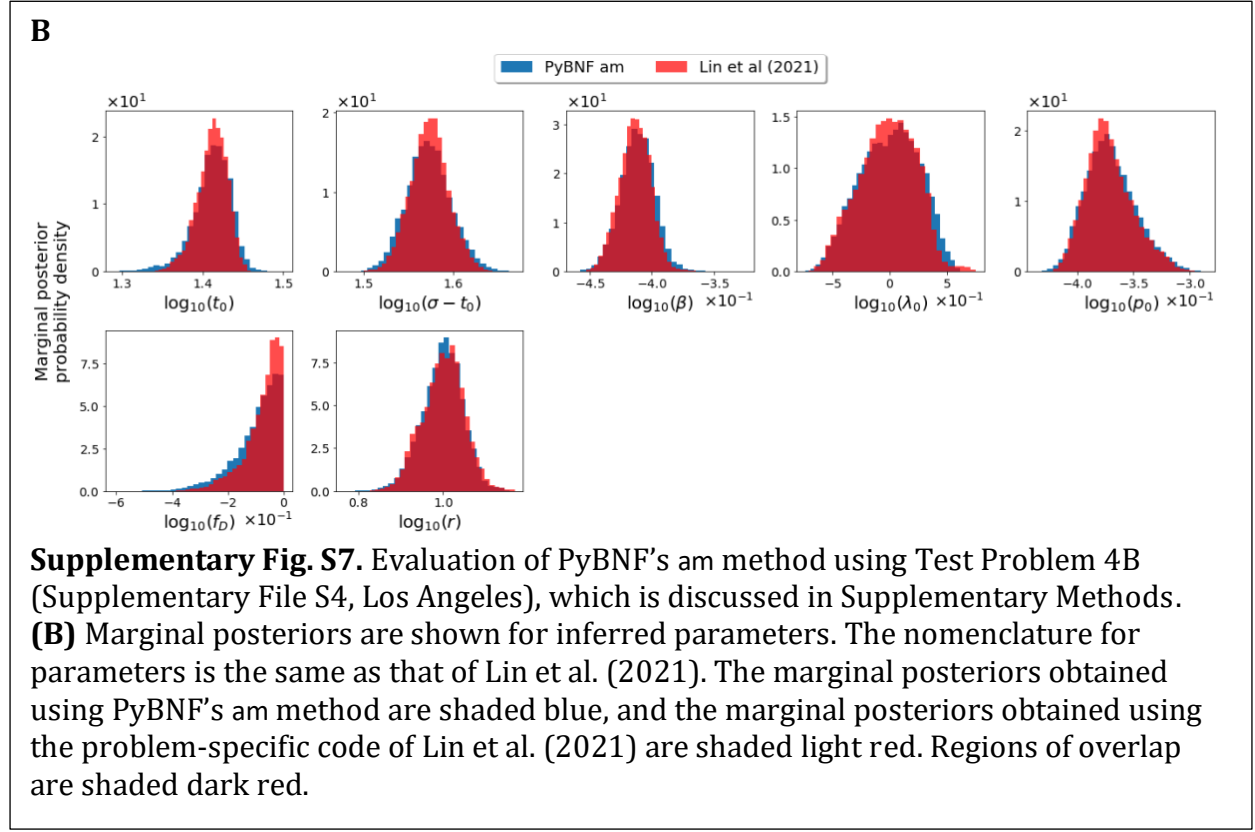

**Supplementary Fig. S7.** Evaluation of PyBNF's am method using Test Problem 4B (Supplementary File S4, Los Angeles), which is discussed in Supplementary Methods. **(B)** Marginal posteriors are shown for inferred parameters. The nomenclature for parameters is the same as that of Lin et al. (2021). The marginal posteriors obtained using PyBNF's am method are shaded blue, and the marginal posteriors obtained using the problem-specific code of Lin et al. (2021) are shaded light red. Regions of overlap are shaded dark red.

**SUPPLEMENTARY FIGURE S7 – PANEL C**

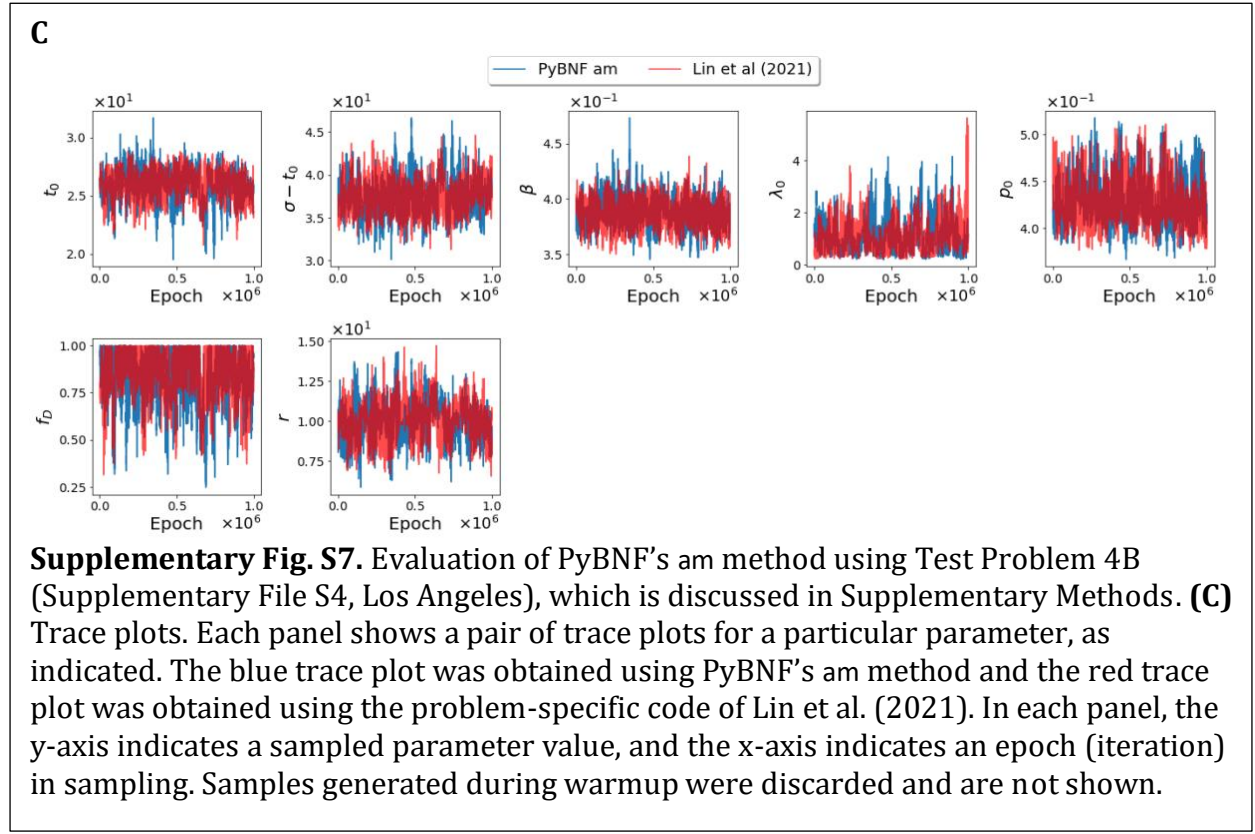

**Supplementary Fig. S7.** Evaluation of PyBNF's am method using Test Problem 4B (Supplementary File S4, Los Angeles), which is discussed in Supplementary Methods. **(C)** Trace plots. Each panel shows a pair of trace plots for a particular parameter, as indicated. The blue trace plot was obtained using PyBNF's am method and the red trace plot was obtained using the problem-specific code of Lin et al. (2021). In each panel, the y-axis indicates a sampled parameter value, and the x-axis indicates an epoch (iteration) in sampling. Samples generated during warmup were discarded and are not shown.

**SUPPLEMENTARY FIGURE S7 – PANEL D**

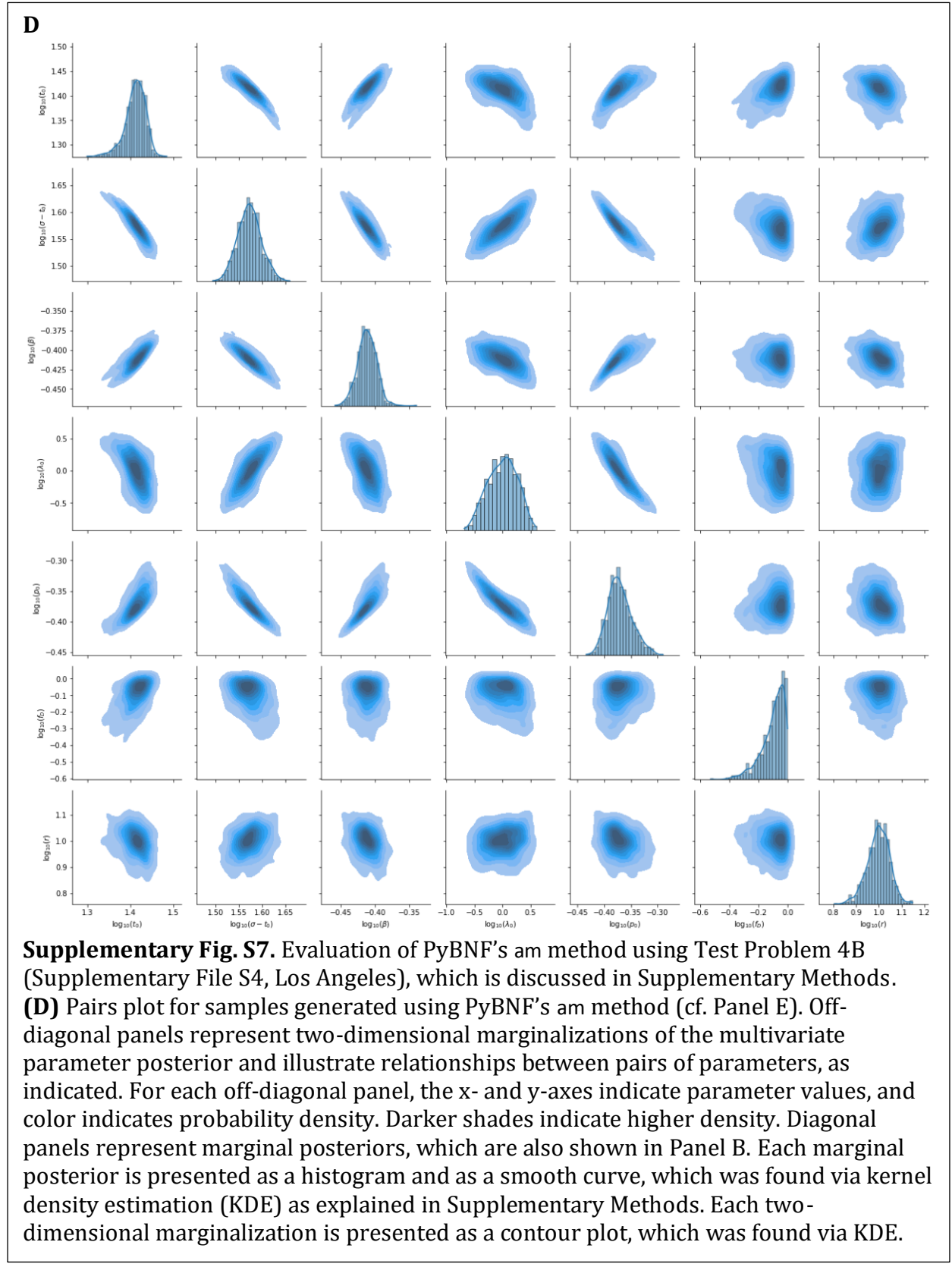

**Supplementary Fig. S7.** Evaluation of PyBNF's am method using Test Problem 4B (Supplementary File S4, Los Angeles), which is discussed in Supplementary Methods. **(D)** Pairs plot for samples generated using PyBNF's am method (cf. Panel E). Off-diagonal panels represent two-dimensional marginalizations of the multivariate parameter posterior and illustrate relationships between pairs of parameters, as indicated. For each off-diagonal panel, the x- and y-axes indicate parameter values, and color indicates probability density. Darker shades indicate higher density. Diagonal panels represent marginal posteriors, which are also shown in Panel B. Each marginal posterior is presented as a histogram and as a smooth curve, which was found via kernel density estimation (KDE) as explained in Supplementary Methods. Each two-dimensional marginalization is presented as a contour plot, which was found via KDE.

**SUPPLEMENTARY FIGURE S7 – PANEL E**

**E**

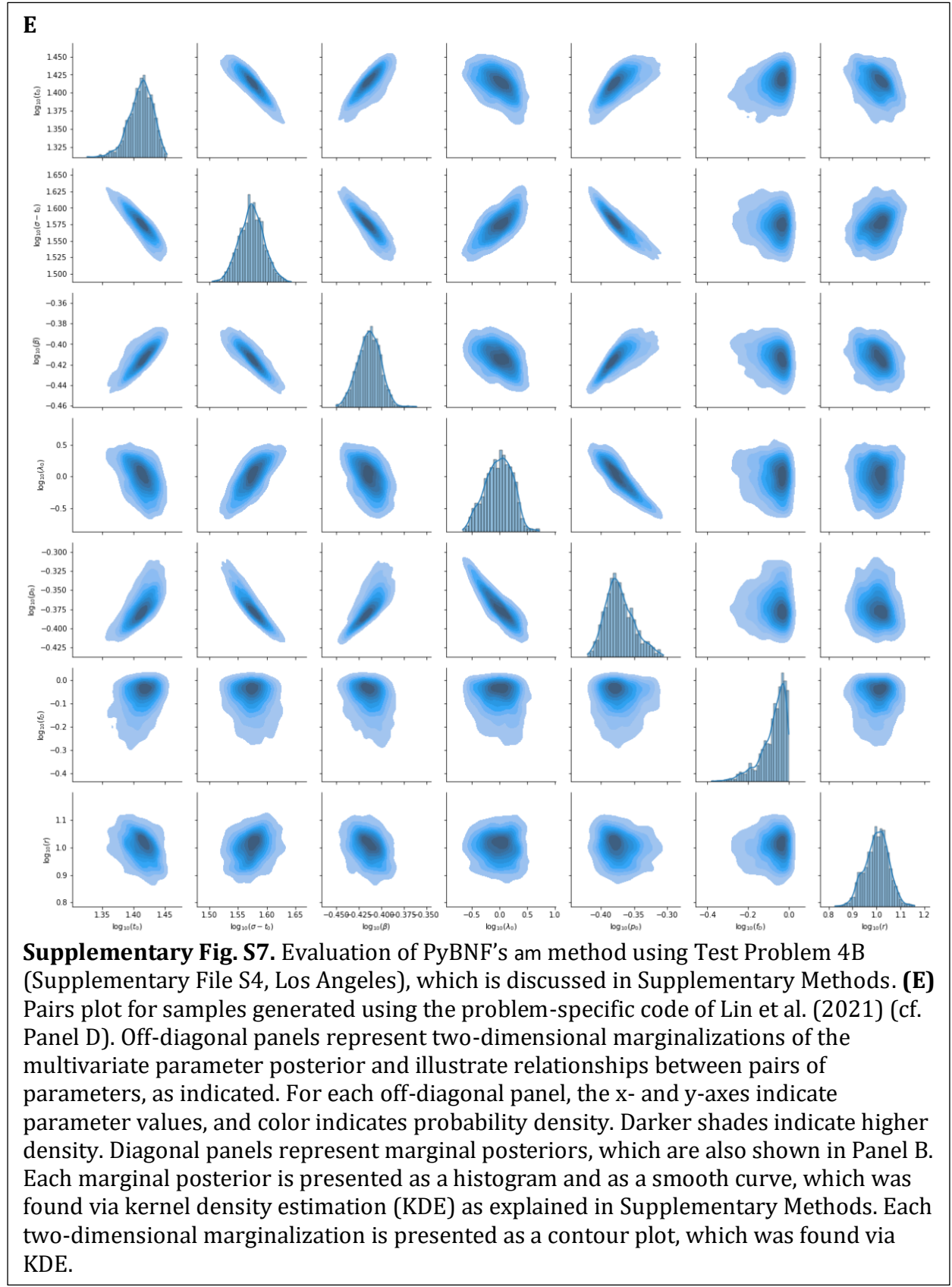

**Supplementary Fig. S7.** Evaluation of PyBNF's am method using Test Problem 4B (Supplementary File S4, Los Angeles), which is discussed in Supplementary Methods. **(E)** Pairs plot for samples generated using the problem-specific code of Lin et al. (2021) (cf. Panel D). Off-diagonal panels represent two-dimensional marginalizations of the multivariate parameter posterior and illustrate relationships between pairs of parameters, as indicated. For each off-diagonal panel, the x- and y-axes indicate parameter values, and color indicates probability density. Darker shades indicate higher density. Diagonal panels represent marginal posteriors, which are also shown in Panel B. Each marginal posterior is presented as a histogram and as a smooth curve, which was found via kernel density estimation (KDE) as explained in Supplementary Methods. Each two-dimensional marginalization is presented as a contour plot, which was found via KDE.

**SUPPLEMENTARY FIGURE S8 – PANEL A**

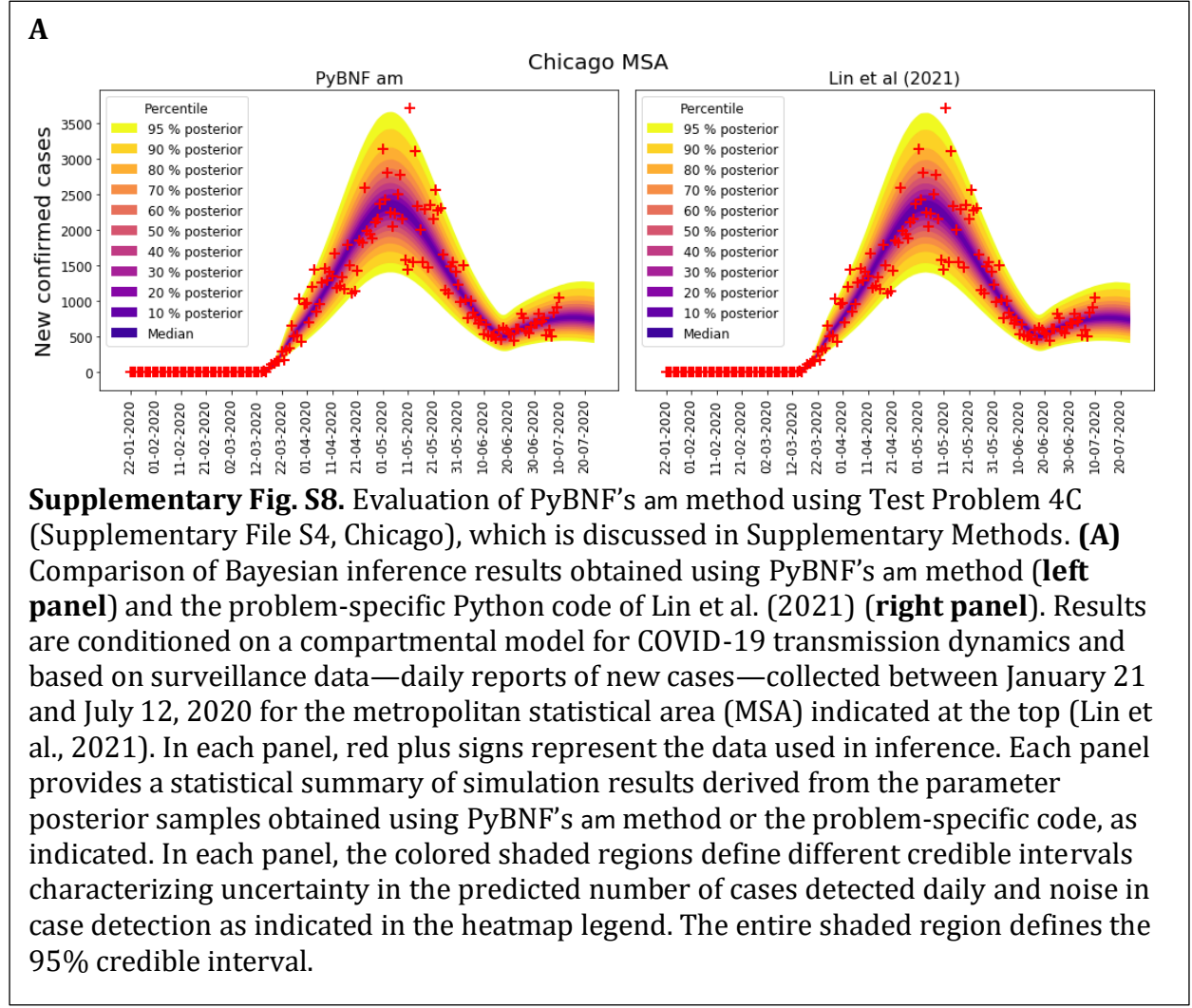

**Supplementary Fig. S8.** Evaluation of PyBNF's am method using Test Problem 4C (Supplementary File S4, Chicago), which is discussed in Supplementary Methods. **(A)** Comparison of Bayesian inference results obtained using PyBNF's am method (**left panel**) and the problem-specific Python code of Lin et al. (2021) (**right panel**). Results are conditioned on a compartmental model for COVID-19 transmission dynamics and based on surveillance data—daily reports of new cases—collected between January 21 and July 12, 2020 for the metropolitan statistical area (MSA) indicated at the top (Lin et al., 2021). In each panel, red plus signs represent the data used in inference. Each panel provides a statistical summary of simulation results derived from the parameter posterior samples obtained using PyBNF's am method or the problem-specific code, as indicated. In each panel, the colored shaded regions define different credible intervals characterizing uncertainty in the predicted number of cases detected daily and noise in case detection as indicated in the heatmap legend. The entire shaded region defines the 95% credible interval.

**SUPPLEMENTARY FIGURE S8 – PANEL B**

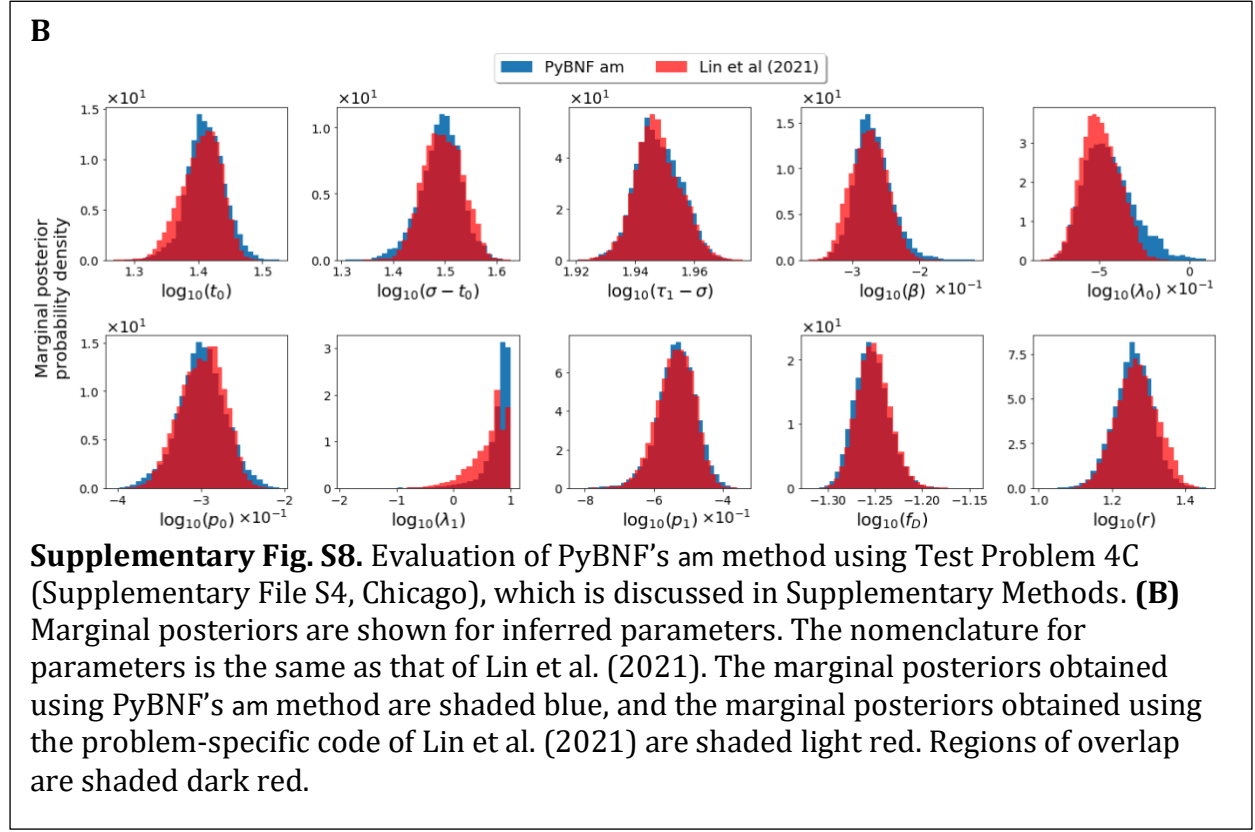

**Supplementary Fig. S8.** Evaluation of PyBNF's am method using Test Problem 4C (Supplementary File S4, Chicago), which is discussed in Supplementary Methods. **(B)** Marginal posteriors are shown for inferred parameters. The nomenclature for parameters is the same as that of Lin et al. (2021). The marginal posteriors obtained using PyBNF's am method are shaded blue, and the marginal posteriors obtained using the problem-specific code of Lin et al. (2021) are shaded light red. Regions of overlap are shaded dark red.

**SUPPLEMENTARY FIGURE S8 – PANEL C**

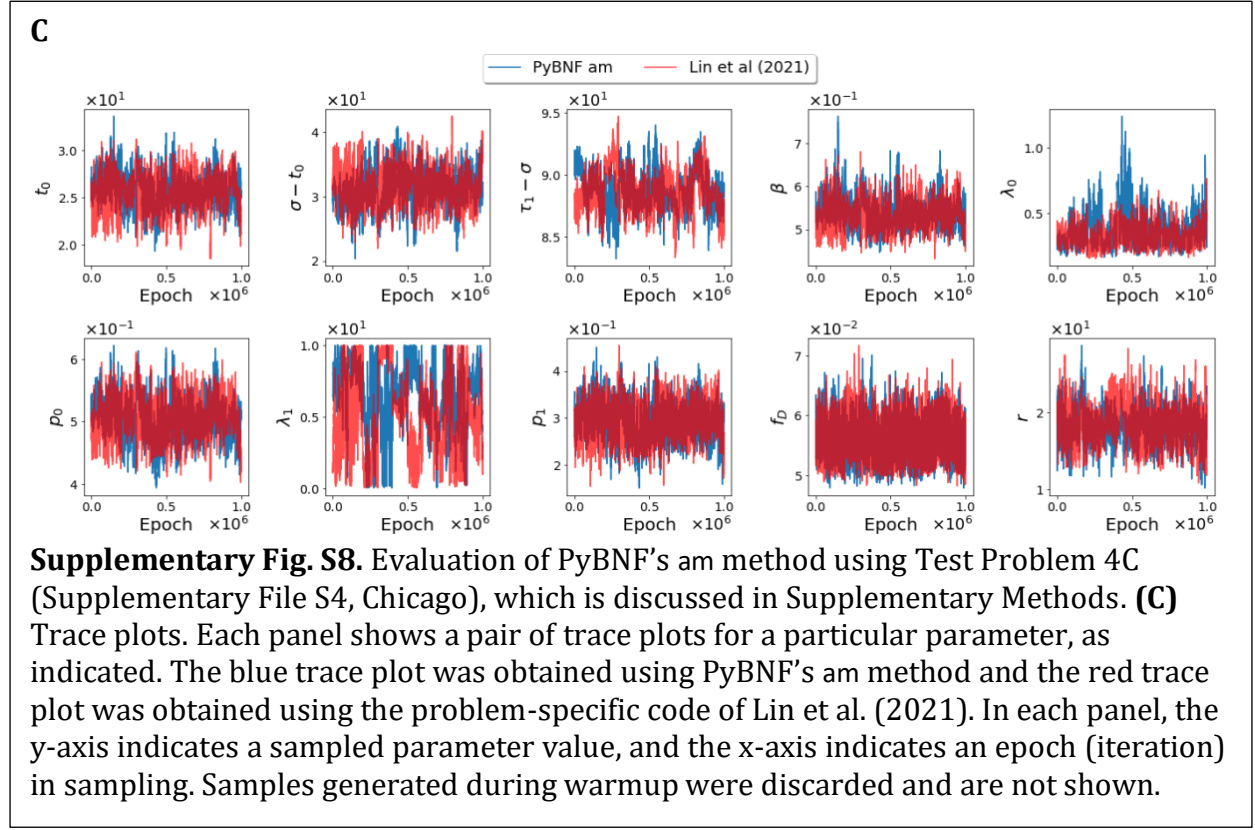

**Supplementary Fig. S8.** Evaluation of PyBNF's am method using Test Problem 4C (Supplementary File S4, Chicago), which is discussed in Supplementary Methods. **(C)** Trace plots. Each panel shows a pair of trace plots for a particular parameter, as indicated. The blue trace plot was obtained using PyBNF's am method and the red trace plot was obtained using the problem-specific code of Lin et al. (2021). In each panel, the y-axis indicates a sampled parameter value, and the x-axis indicates an epoch (iteration) in sampling. Samples generated during warmup were discarded and are not shown.

**SUPPLEMENTARY FIGURE S8 – PANEL D**

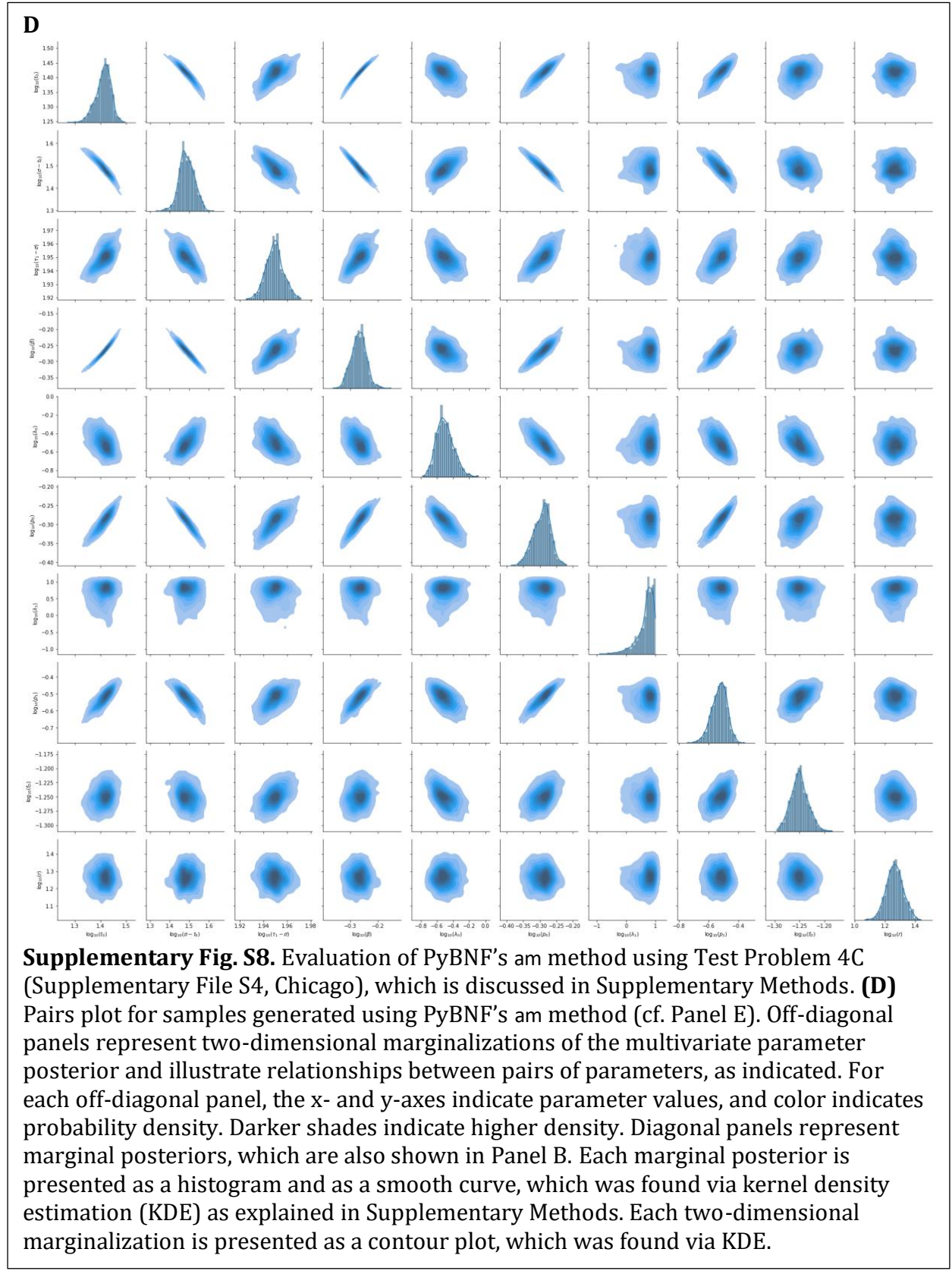

**Supplementary Fig. S8.** Evaluation of PyBNF's am method using Test Problem 4C (Supplementary File S4, Chicago), which is discussed in Supplementary Methods. **(D)** Pairs plot for samples generated using PyBNF's am method (cf. Panel E). Off-diagonal panels represent two-dimensional marginalizations of the multivariate parameter posterior and illustrate relationships between pairs of parameters, as indicated. For each off-diagonal panel, the x- and y-axes indicate parameter values, and color indicates probability density. Darker shades indicate higher density. Diagonal panels represent marginal posteriors, which are also shown in Panel B. Each marginal posterior is presented as a histogram and as a smooth curve, which was found via kernel density estimation (KDE) as explained in Supplementary Methods. Each two-dimensional marginalization is presented as a contour plot, which was found via KDE.

**SUPPLEMENTARY FIGURE S8 – PANEL E**

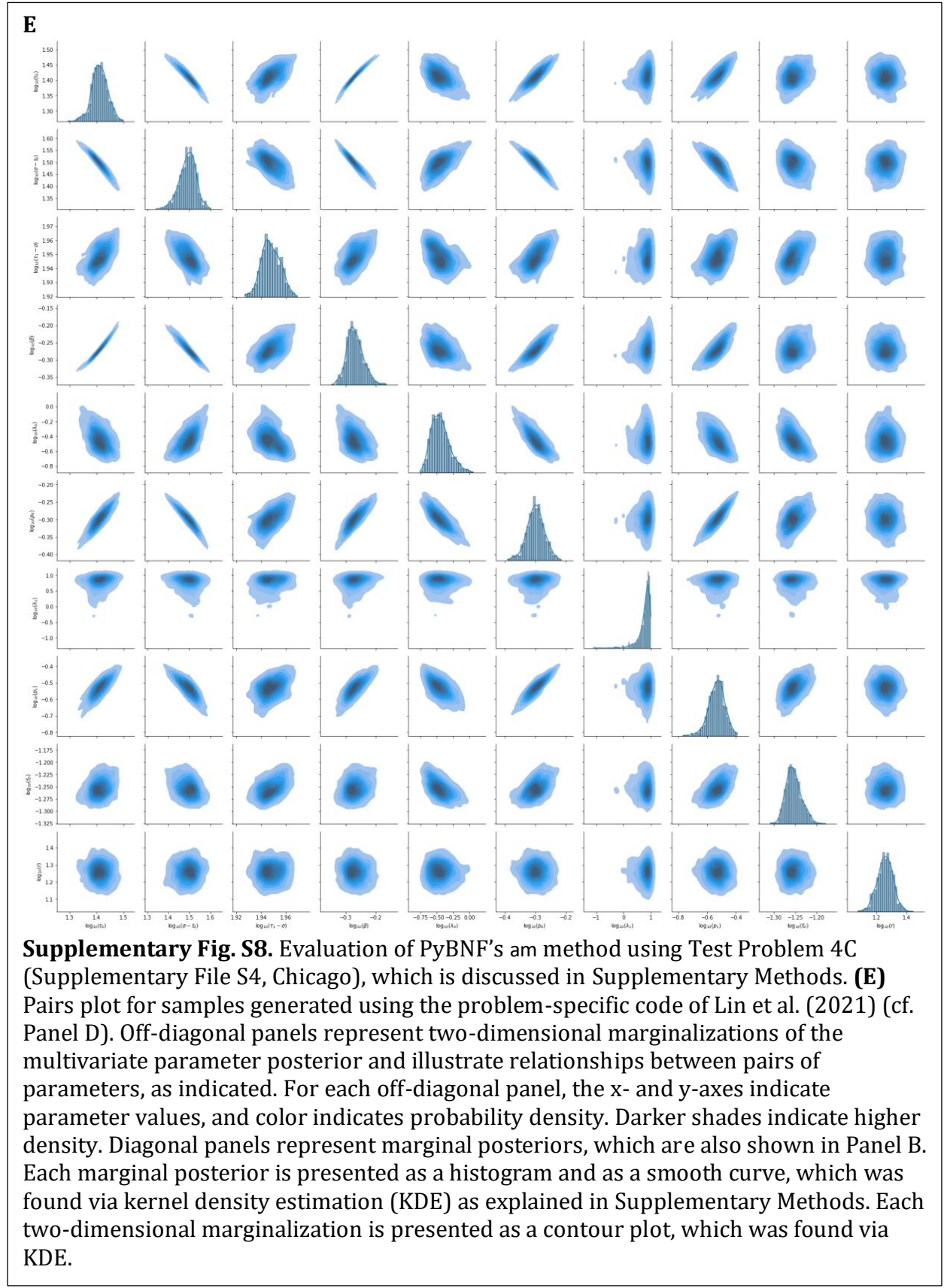

**Supplementary Fig. S8.** Evaluation of PyBNF's am method using Test Problem 4C (Supplementary File S4, Chicago), which is discussed in Supplementary Methods. **(E)** Pairs plot for samples generated using the problem-specific code of Lin et al. (2021) (cf. Panel D). Off-diagonal panels represent two-dimensional marginalizations of the multivariate parameter posterior and illustrate relationships between pairs of parameters, as indicated. For each off-diagonal panel, the x- and y-axes indicate parameter values, and color indicates probability density. Darker shades indicate higher density. Diagonal panels represent marginal posteriors, which are also shown in Panel B. Each marginal posterior is presented as a histogram and as a smooth curve, which was found via kernel density estimation (KDE) as explained in Supplementary Methods. Each two-dimensional marginalization is presented as a contour plot, which was found via KDE.

**SUPPLEMENTARY FIGURE S9 – PANEL A**

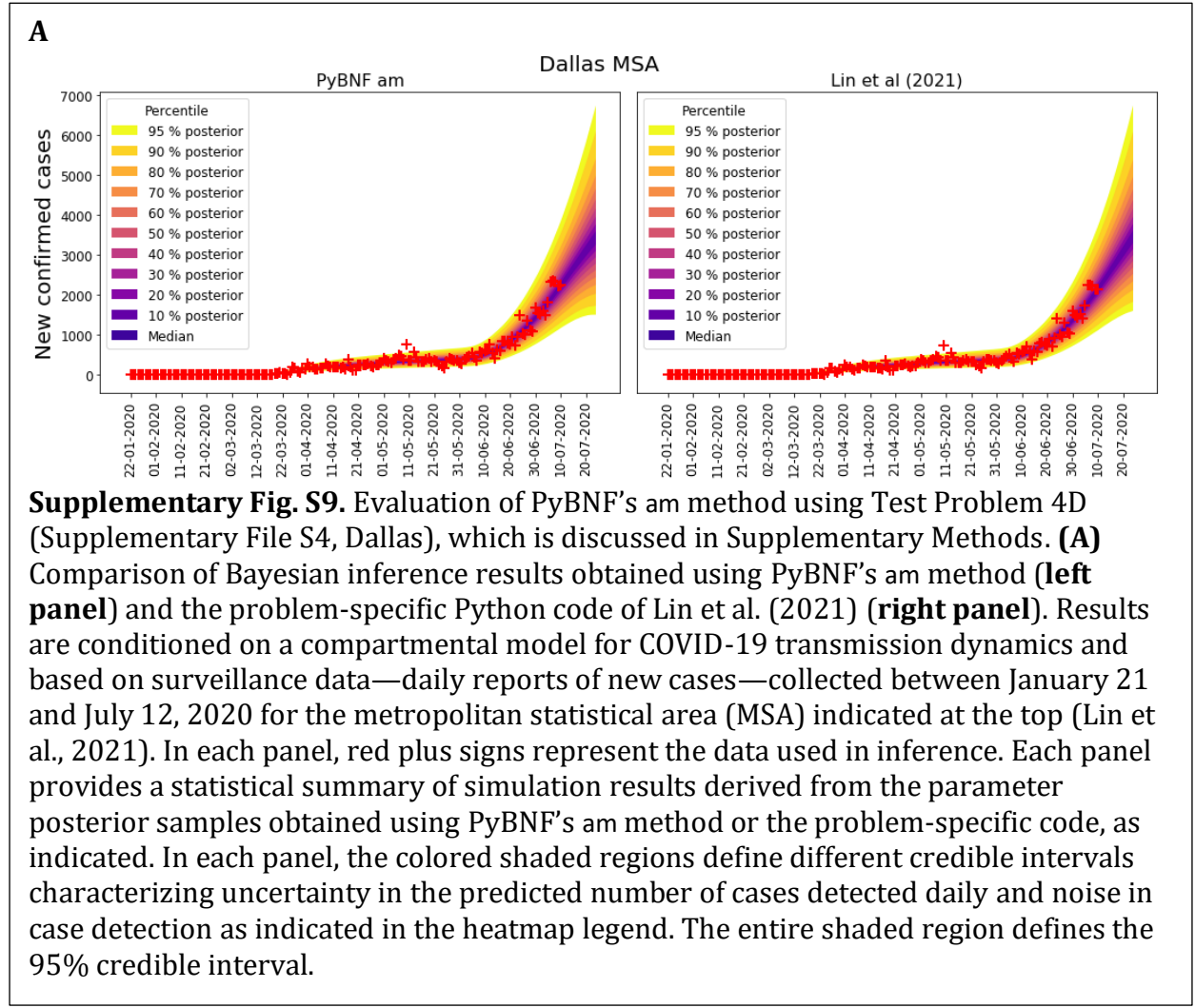

**Supplementary Fig. S9.** Evaluation of PyBNF's am method using Test Problem 4D (Supplementary File S4, Dallas), which is discussed in Supplementary Methods. **(A)** Comparison of Bayesian inference results obtained using PyBNF's am method (**left panel**) and the problem-specific Python code of Lin et al. (2021) (**right panel**). Results are conditioned on a compartmental model for COVID-19 transmission dynamics and based on surveillance data—daily reports of new cases—collected between January 21 and July 12, 2020 for the metropolitan statistical area (MSA) indicated at the top (Lin et al., 2021). In each panel, red plus signs represent the data used in inference. Each panel provides a statistical summary of simulation results derived from the parameter posterior samples obtained using PyBNF's am method or the problem-specific code, as indicated. In each panel, the colored shaded regions define different credible intervals characterizing uncertainty in the predicted number of cases detected daily and noise in case detection as indicated in the heatmap legend. The entire shaded region defines the 95% credible interval.

**SUPPLEMENTARY FIGURE S9 – PANEL B**

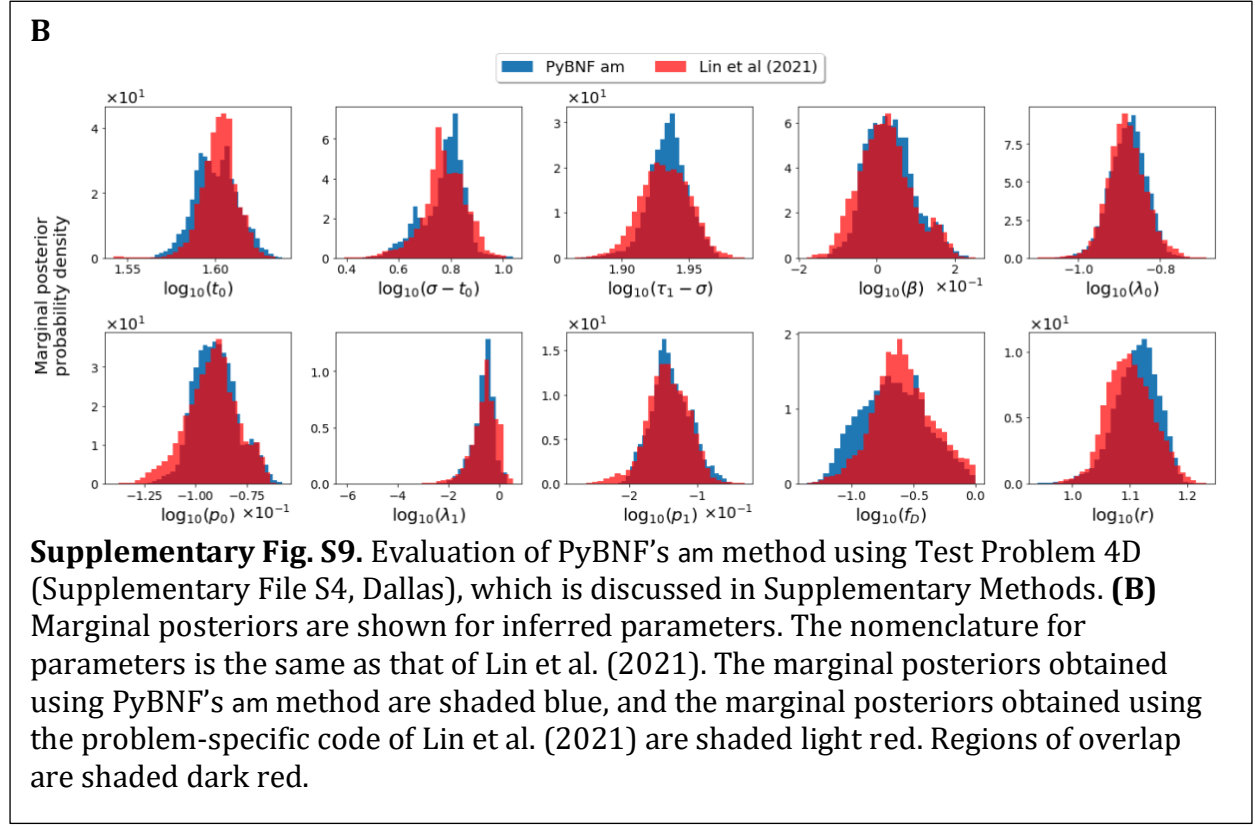

**Supplementary Fig. S9.** Evaluation of PyBNF's am method using Test Problem 4D (Supplementary File S4, Dallas), which is discussed in Supplementary Methods. **(B)** Marginal posteriors are shown for inferred parameters. The nomenclature for parameters is the same as that of Lin et al. (2021). The marginal posteriors obtained using PyBNF's am method are shaded blue, and the marginal posteriors obtained using the problem-specific code of Lin et al. (2021) are shaded light red. Regions of overlap are shaded dark red.

**SUPPLEMENTARY FIGURE S9 – PANEL C**

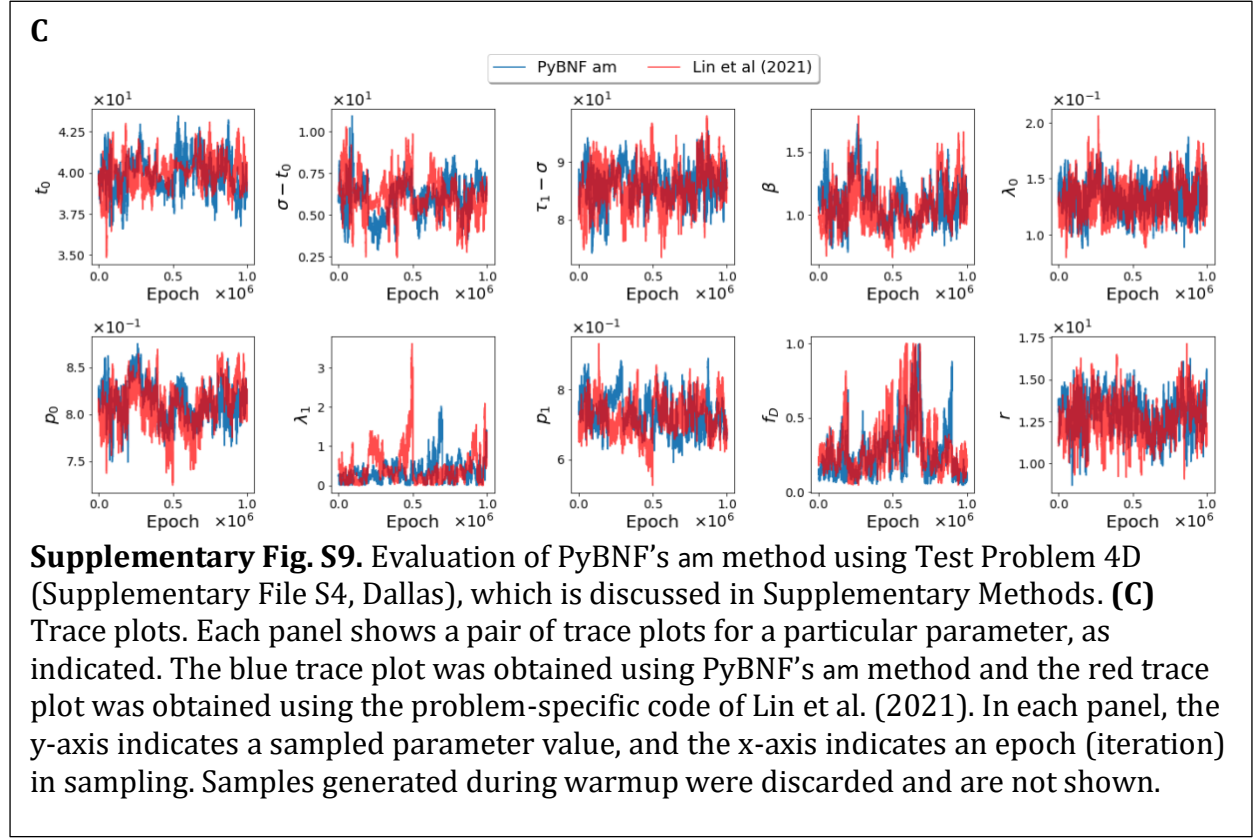

**Supplementary Fig. S9.** Evaluation of PyBNF's am method using Test Problem 4D (Supplementary File S4, Dallas), which is discussed in Supplementary Methods. **(C)** Trace plots. Each panel shows a pair of trace plots for a particular parameter, as indicated. The blue trace plot was obtained using PyBNF's am method and the red trace plot was obtained using the problem-specific code of Lin et al. (2021). In each panel, the y-axis indicates a sampled parameter value, and the x-axis indicates an epoch (iteration) in sampling. Samples generated during warmup were discarded and are not shown.

**SUPPLEMENTARY FIGURE S9 – PANEL D**

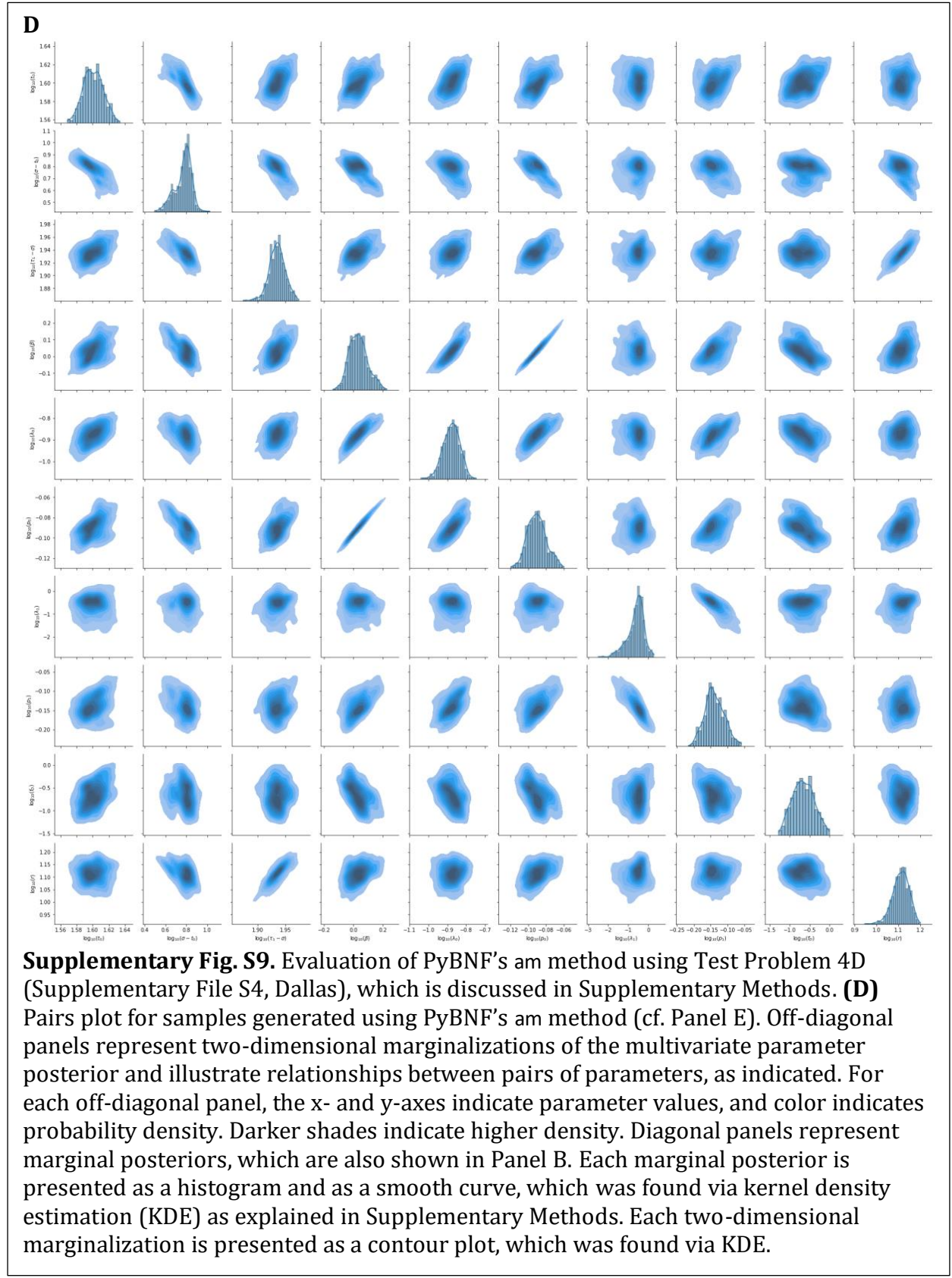

**Supplementary Fig. S9.** Evaluation of PyBNF's am method using Test Problem 4D (Supplementary File S4, Dallas), which is discussed in Supplementary Methods. **(D)** Pairs plot for samples generated using PyBNF's am method (cf. Panel E). Off-diagonal panels represent two-dimensional marginalizations of the multivariate parameter posterior and illustrate relationships between pairs of parameters, as indicated. For each off-diagonal panel, the x- and y-axes indicate parameter values, and color indicates probability density. Darker shades indicate higher density. Diagonal panels represent marginal posteriors, which are also shown in Panel B. Each marginal posterior is presented as a histogram and as a smooth curve, which was found via kernel density estimation (KDE) as explained in Supplementary Methods. Each two-dimensional marginalization is presented as a contour plot, which was found via KDE.

**SUPPLEMENTARY FIGURE S9 – PANEL E**

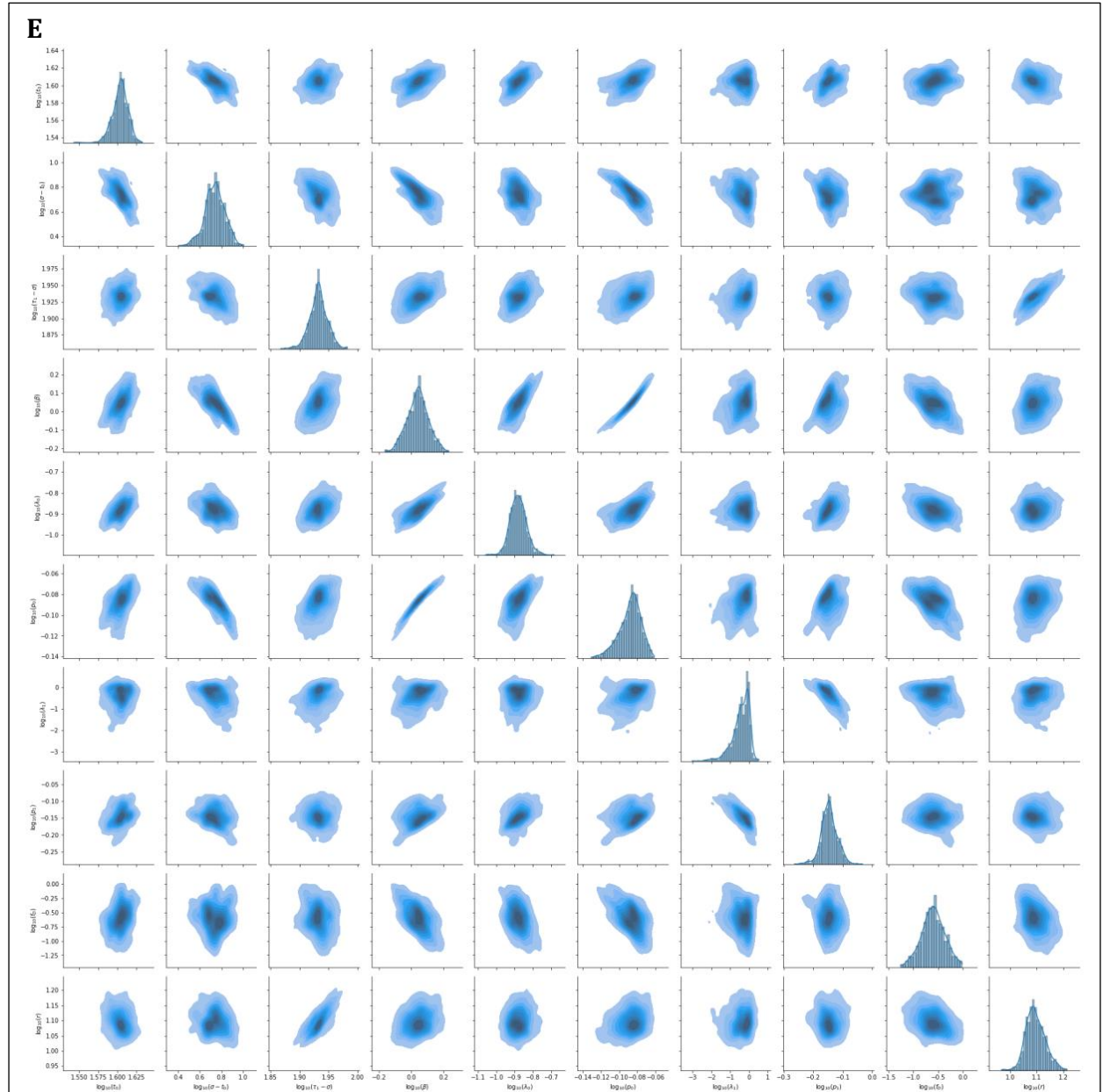

**Supplementary Fig. S9.** Evaluation of PyBNF's am method using Test Problem 4D (Supplementary File S4, Dallas), which is discussed in Supplementary Methods. **(E)** Pairs plot for samples generated using the problem-specific code of Lin et al. (2021) (cf. Panel D). Off-diagonal panels represent two-dimensional marginalizations of the multivariate parameter posterior and illustrate relationships between pairs of parameters, as indicated. For each off-diagonal panel, the x- and y-axes indicate parameter values, and color indicates probability density. Darker shades indicate higher density. Diagonal panels represent marginal posteriors, which are also shown in Panel B. Each marginal posterior is presented as a histogram and as a smooth curve, which was found via kernel density estimation (KDE) as explained in Supplementary Methods. Each two-dimensional marginalization is presented as a contour plot, which was found via KDE.

**SUPPLEMENTARY FIGURE S10 – PANEL A**

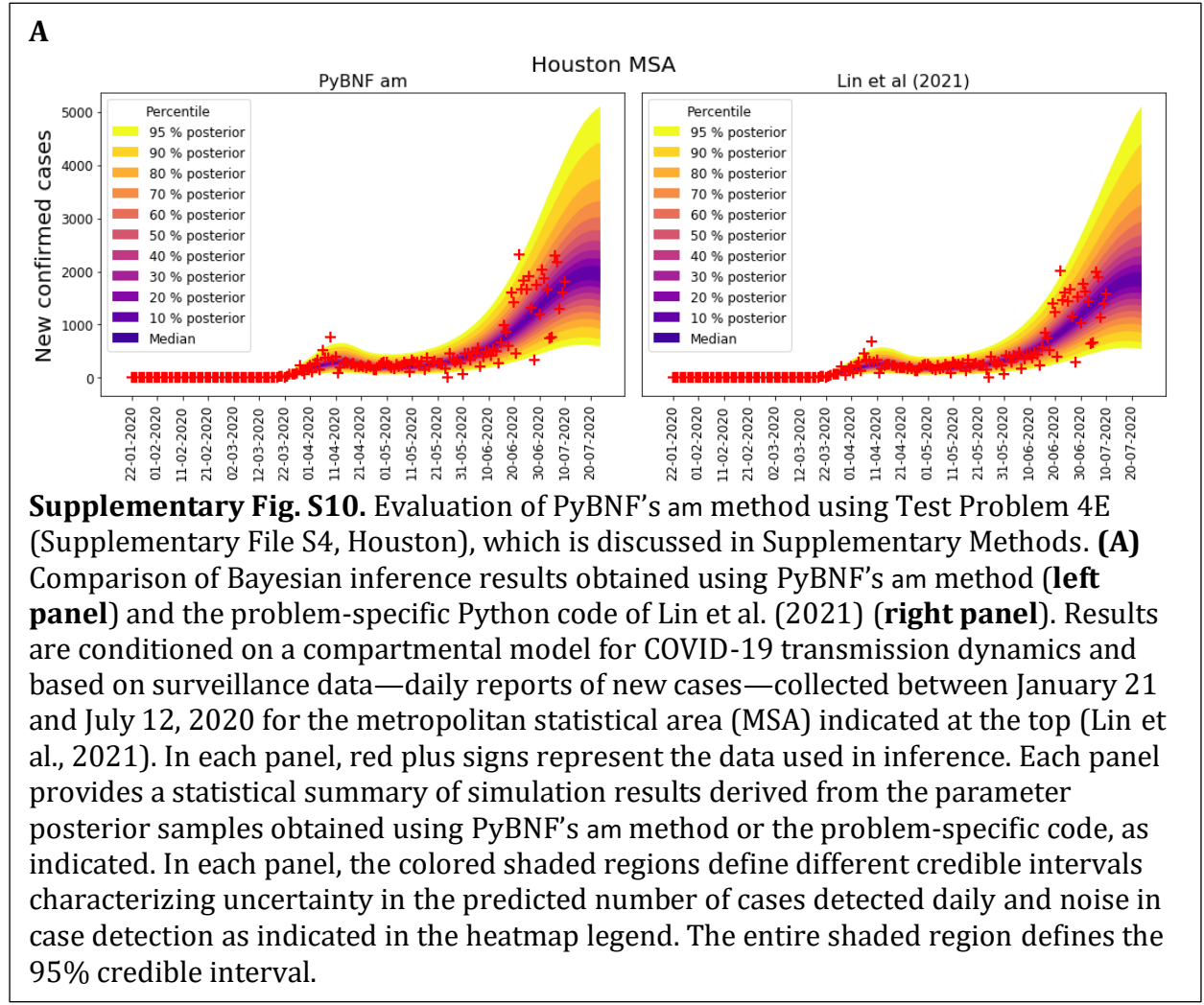

**Supplementary Fig. S10.** Evaluation of PyBNF's ам method using Test Problem 4E (Supplementary File S4, Houston), which is discussed in Supplementary Methods. **(A)** Comparison of Bayesian inference results obtained using PyBNF's ам method (**left panel**) and the problem-specific Python code of Lin et al. (2021) (**right panel**). Results are conditioned on a compartmental model for COVID-19 transmission dynamics and based on surveillance data—daily reports of new cases—collected between January 21 and July 12, 2020 for the metropolitan statistical area (MSA) indicated at the top (Lin et al., 2021). In each panel, red plus signs represent the data used in inference. Each panel provides a statistical summary of simulation results derived from the parameter posterior samples obtained using PyBNF's ам method or the problem-specific code, as indicated. In each panel, the colored shaded regions define different credible intervals characterizing uncertainty in the predicted number of cases detected daily and noise in case detection as indicated in the heatmap legend. The entire shaded region defines the 95% credible interval.

**SUPPLEMENTARY FIGURE S10 – PANEL B**

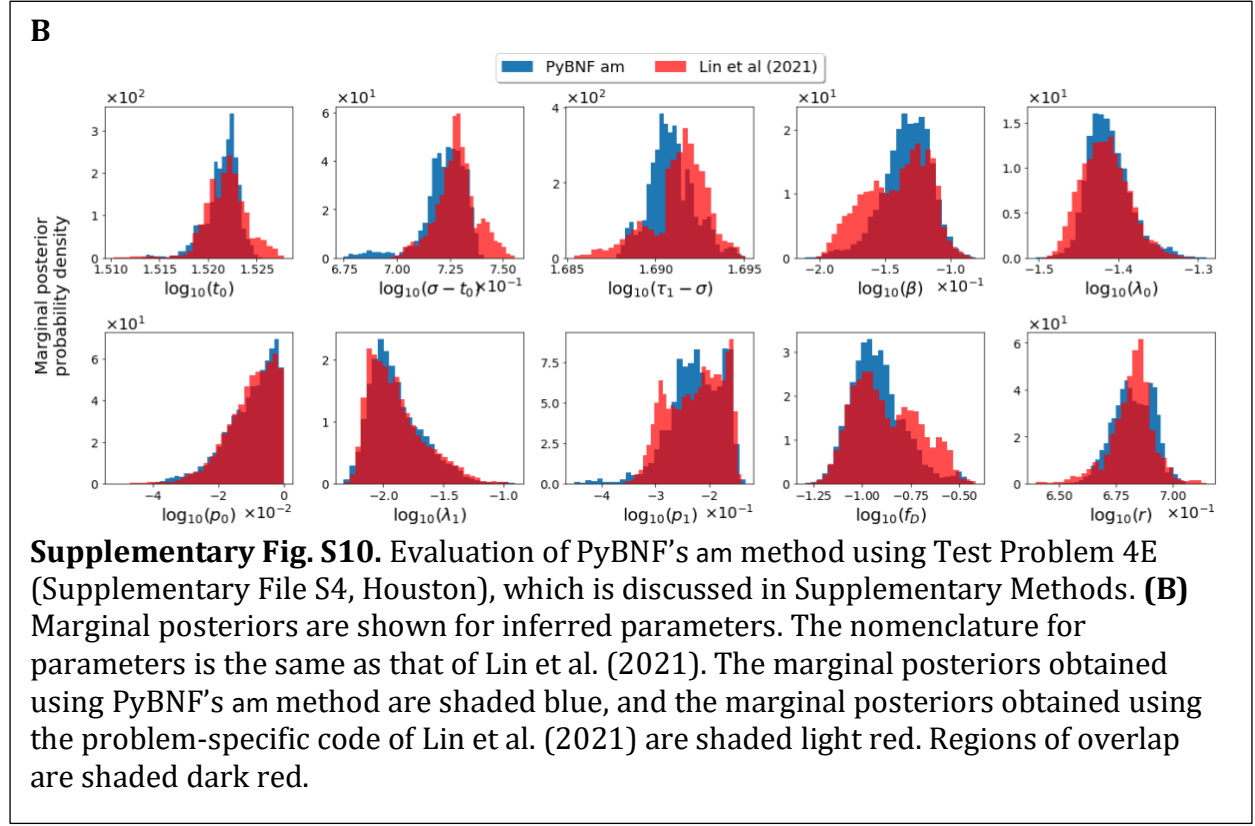

**Supplementary Fig. S10.** Evaluation of PyBNF's am method using Test Problem 4E (Supplementary File S4, Houston), which is discussed in Supplementary Methods. **(B)** Marginal posteriors are shown for inferred parameters. The nomenclature for parameters is the same as that of Lin et al. (2021). The marginal posteriors obtained using PyBNF's am method are shaded blue, and the marginal posteriors obtained using the problem-specific code of Lin et al. (2021) are shaded light red. Regions of overlap are shaded dark red.

**SUPPLEMENTARY FIGURE S10 – PANEL C**

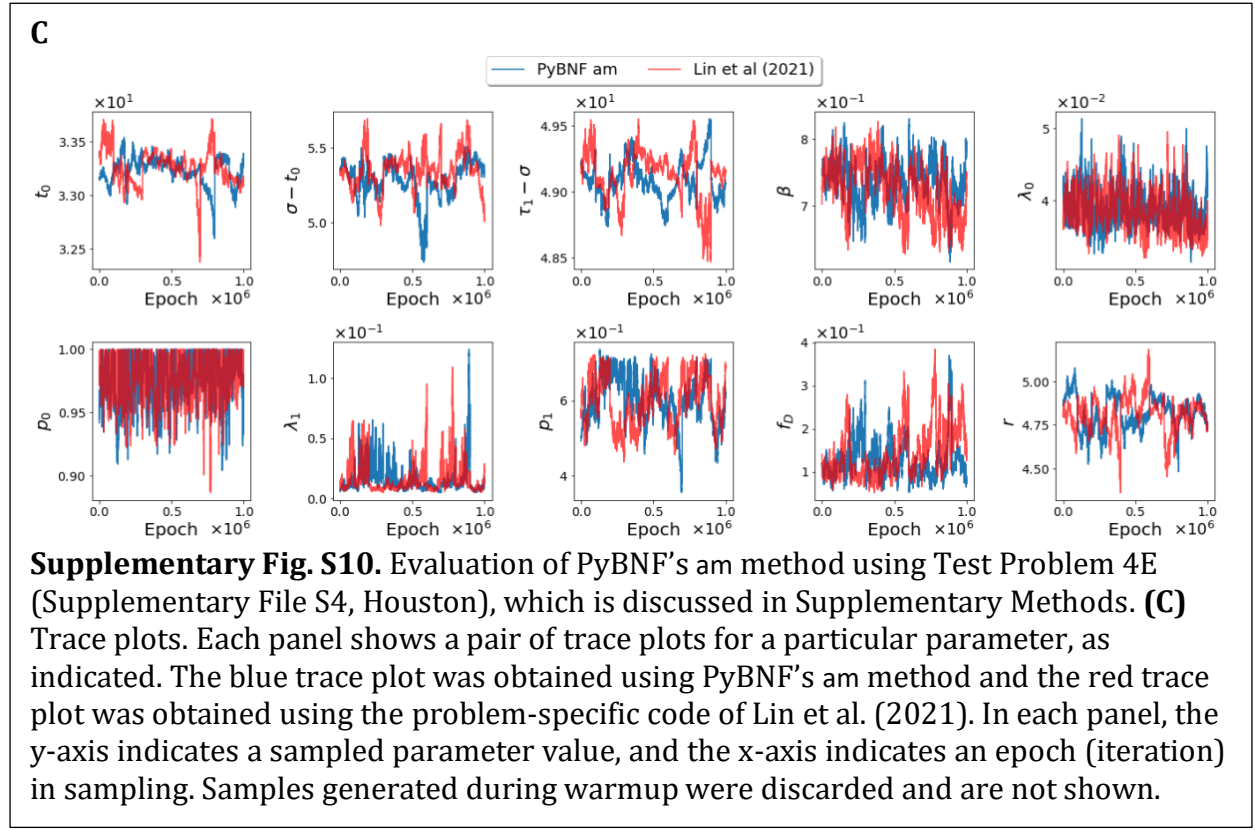

**Supplementary Fig. S10.** Evaluation of PyBNF's am method using Test Problem 4E (Supplementary File S4, Houston), which is discussed in Supplementary Methods. **(C)** Trace plots. Each panel shows a pair of trace plots for a particular parameter, as indicated. The blue trace plot was obtained using PyBNF's am method and the red trace plot was obtained using the problem-specific code of Lin et al. (2021). In each panel, the y-axis indicates a sampled parameter value, and the x-axis indicates an epoch (iteration) in sampling. Samples generated during warmup were discarded and are not shown.

**SUPPLEMENTARY FIGURE S10 – PANEL D**

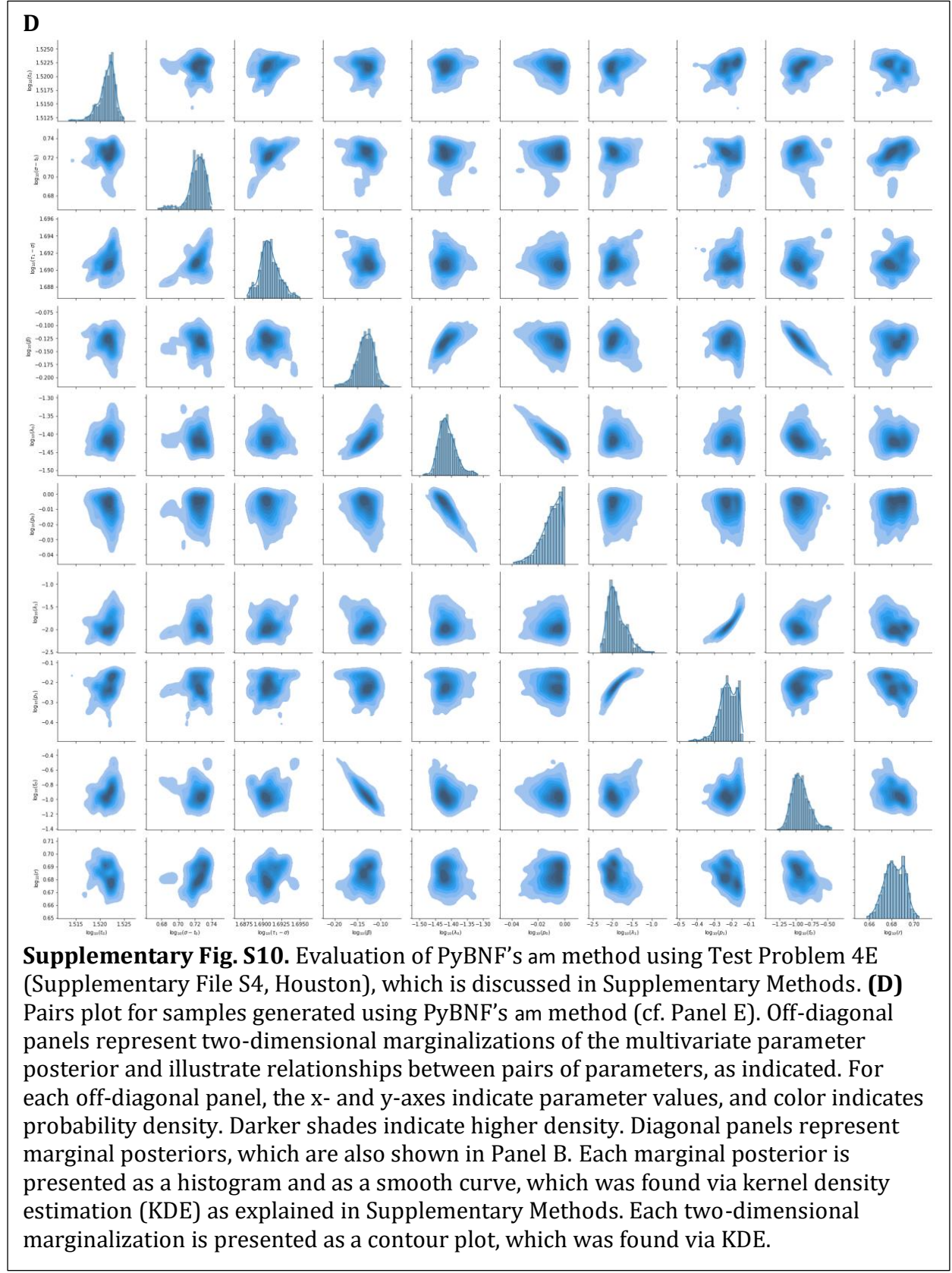

**Supplementary Fig. S10.** Evaluation of PyBNF's am method using Test Problem 4E (Supplementary File S4, Houston), which is discussed in Supplementary Methods. **(D)** Pairs plot for samples generated using PyBNF's am method (cf. Panel E). Off-diagonal panels represent two-dimensional marginalizations of the multivariate parameter posterior and illustrate relationships between pairs of parameters, as indicated. For each off-diagonal panel, the x- and y-axes indicate parameter values, and color indicates probability density. Darker shades indicate higher density. Diagonal panels represent marginal posteriors, which are also shown in Panel B. Each marginal posterior is presented as a histogram and as a smooth curve, which was found via kernel density estimation (KDE) as explained in Supplementary Methods. Each two-dimensional marginalization is presented as a contour plot, which was found via KDE.

**SUPPLEMENTARY FIGURE S10 – PANEL E**

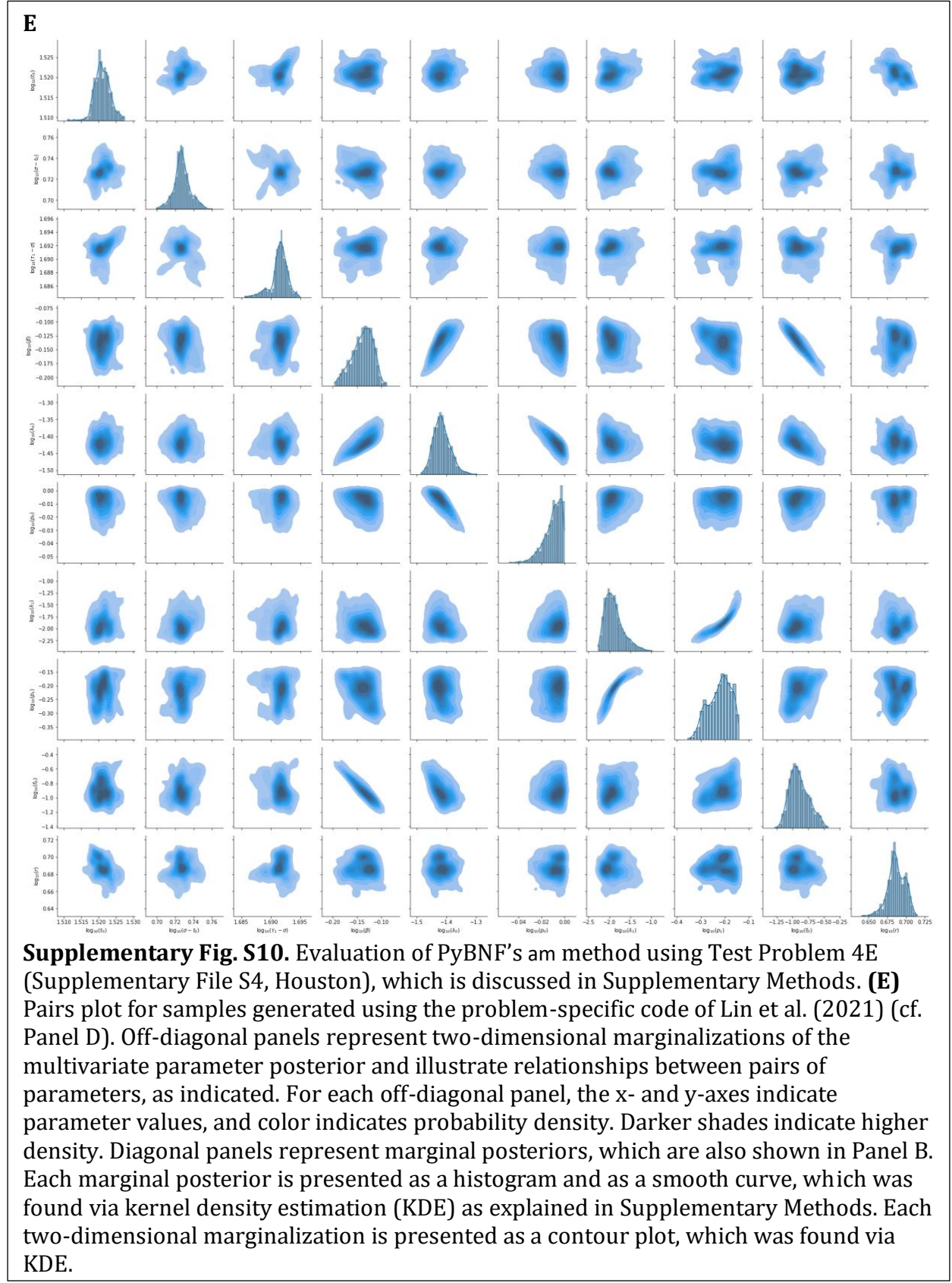

**Supplementary Fig. S10.** Evaluation of PyBNF's am method using Test Problem 4E (Supplementary File S4, Houston), which is discussed in Supplementary Methods. **(E)** Pairs plot for samples generated using the problem-specific code of Lin et al. (2021) (cf. Panel D). Off-diagonal panels represent two-dimensional marginalizations of the multivariate parameter posterior and illustrate relationships between pairs of parameters, as indicated. For each off-diagonal panel, the x- and y-axes indicate parameter values, and color indicates probability density. Darker shades indicate higher density. Diagonal panels represent marginal posteriors, which are also shown in Panel B. Each marginal posterior is presented as a histogram and as a smooth curve, which was found via kernel density estimation (KDE) as explained in Supplementary Methods. Each two-dimensional marginalization is presented as a contour plot, which was found via KDE.

**SUPPLEMENTARY FIGURE S11 – PANEL A**

A

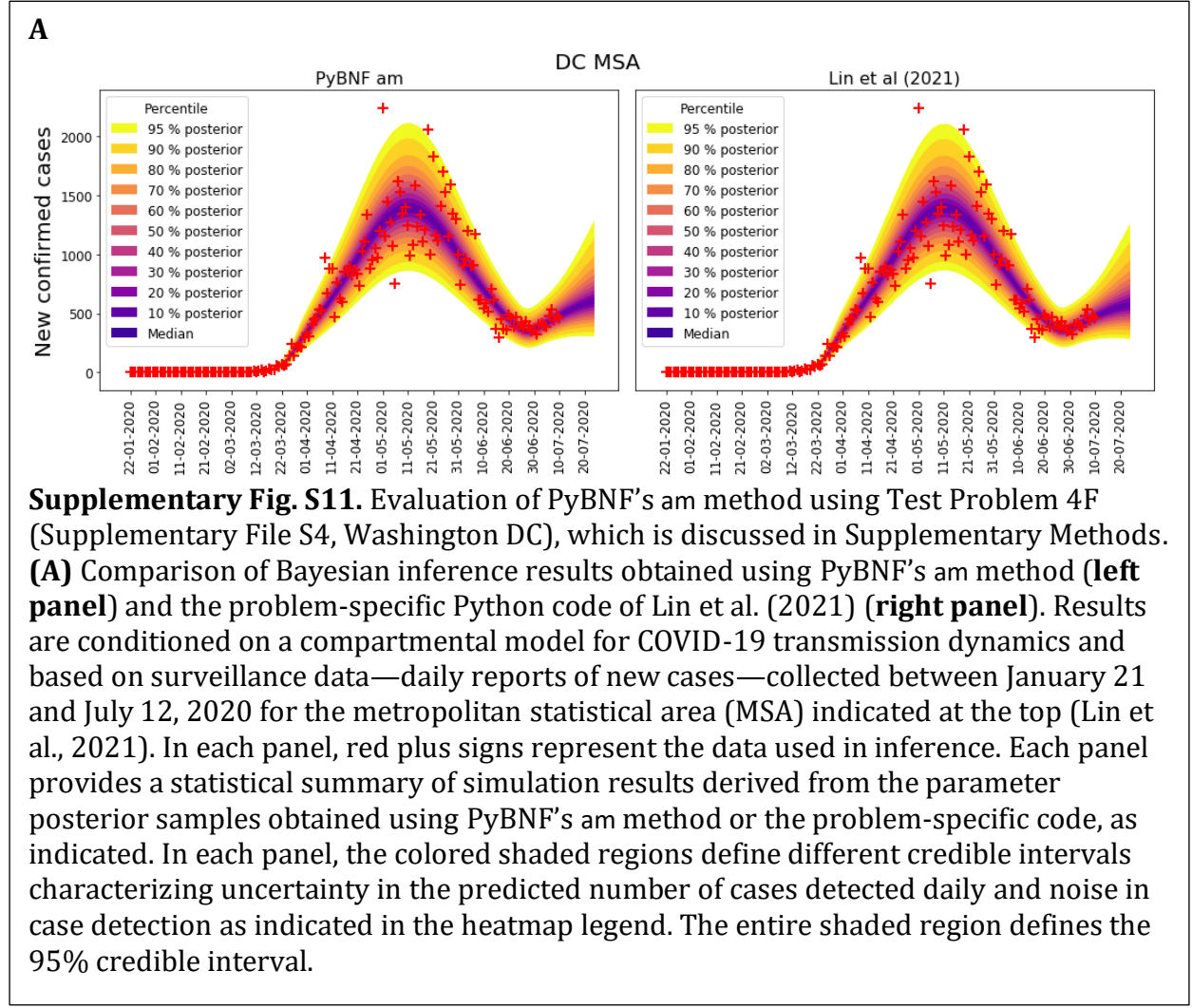

**Supplementary Fig. S11.** Evaluation of PyBNF's ʌм method using Test Problem 4F (Supplementary File S4, Washington DC), which is discussed in Supplementary Methods. **(A)** Comparison of Bayesian inference results obtained using PyBNF's ʌм method (**left panel**) and the problem-specific Python code of Lin et al. (2021) (**right panel**). Results are conditioned on a compartmental model for COVID-19 transmission dynamics and based on surveillance data—daily reports of new cases—collected between January 21 and July 12, 2020 for the metropolitan statistical area (MSA) indicated at the top (Lin et al., 2021). In each panel, red plus signs represent the data used in inference. Each panel provides a statistical summary of simulation results derived from the parameter posterior samples obtained using PyBNF's ʌм method or the problem-specific code, as indicated. In each panel, the colored shaded regions define different credible intervals characterizing uncertainty in the predicted number of cases detected daily and noise in case detection as indicated in the heatmap legend. The entire shaded region defines the 95% credible interval.

**SUPPLEMENTARY FIGURE S11 – PANEL B**

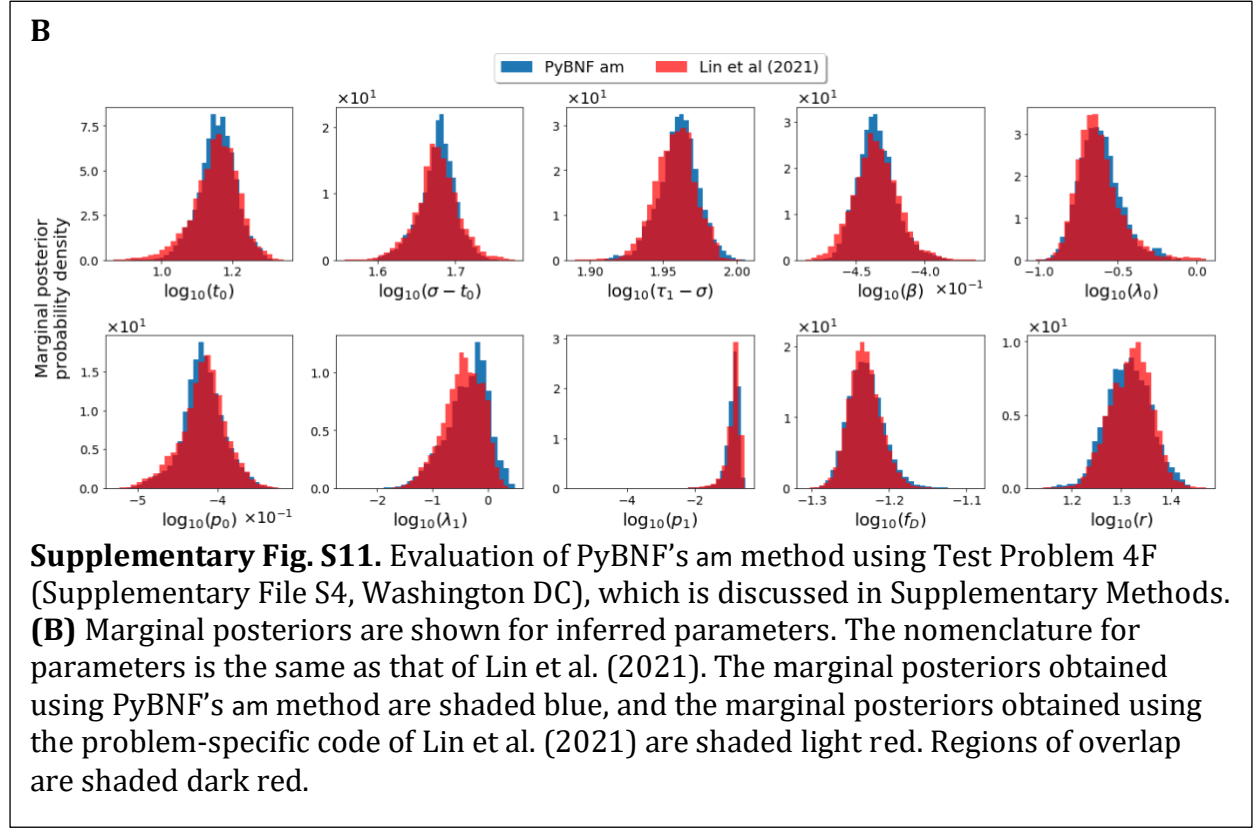

**Supplementary Fig. S11.** Evaluation of PyBNF's am method using Test Problem 4F (Supplementary File S4, Washington DC), which is discussed in Supplementary Methods. **(B)** Marginal posteriors are shown for inferred parameters. The nomenclature for parameters is the same as that of Lin et al. (2021). The marginal posteriors obtained using PyBNF's am method are shaded blue, and the marginal posteriors obtained using the problem-specific code of Lin et al. (2021) are shaded light red. Regions of overlap are shaded dark red.

**SUPPLEMENTARY FIGURE S11 – PANEL C**

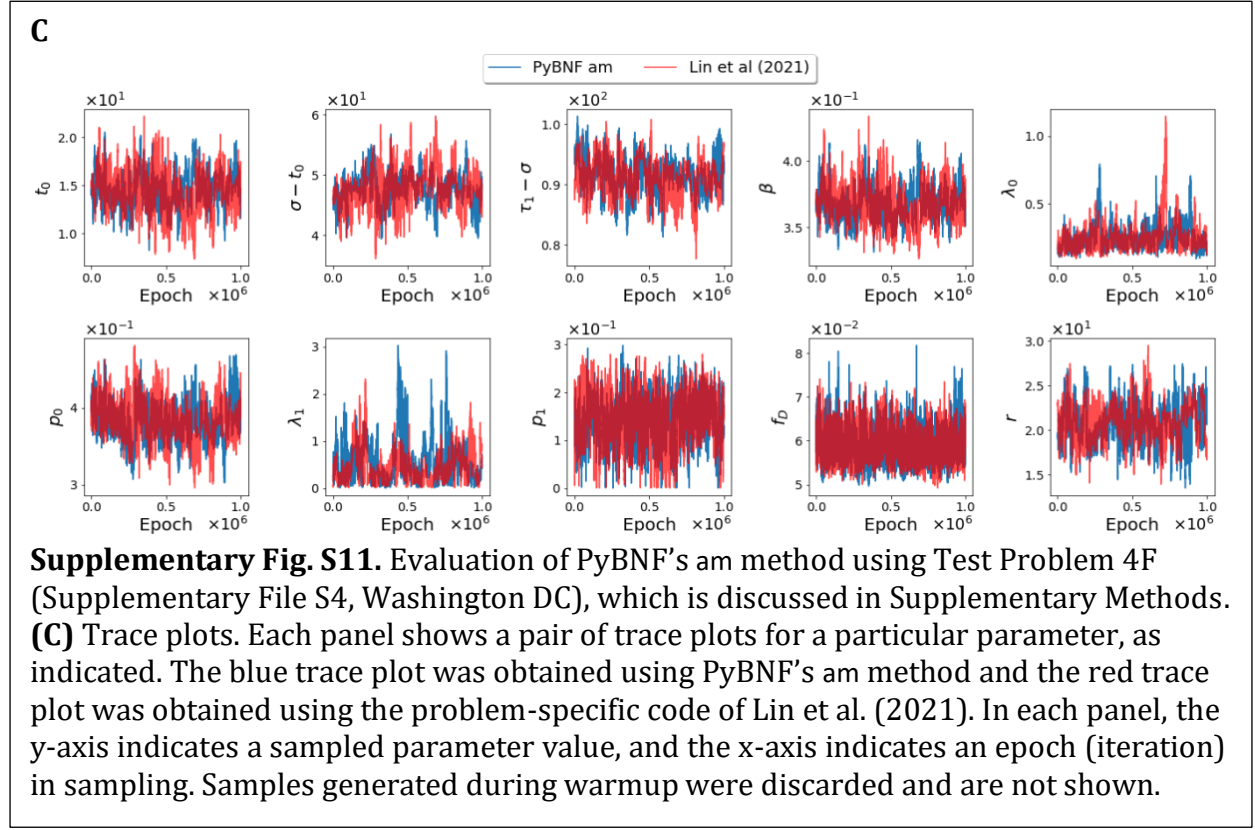

**Supplementary Fig. S11.** Evaluation of PyBNF's am method using Test Problem 4F (Supplementary File S4, Washington DC), which is discussed in Supplementary Methods. **(C)** Trace plots. Each panel shows a pair of trace plots for a particular parameter, as indicated. The blue trace plot was obtained using PyBNF's am method and the red trace plot was obtained using the problem-specific code of Lin et al. (2021). In each panel, the y-axis indicates a sampled parameter value, and the x-axis indicates an epoch (iteration) in sampling. Samples generated during warmup were discarded and are not shown.

**SUPPLEMENTARY FIGURE S11 – PANEL D**

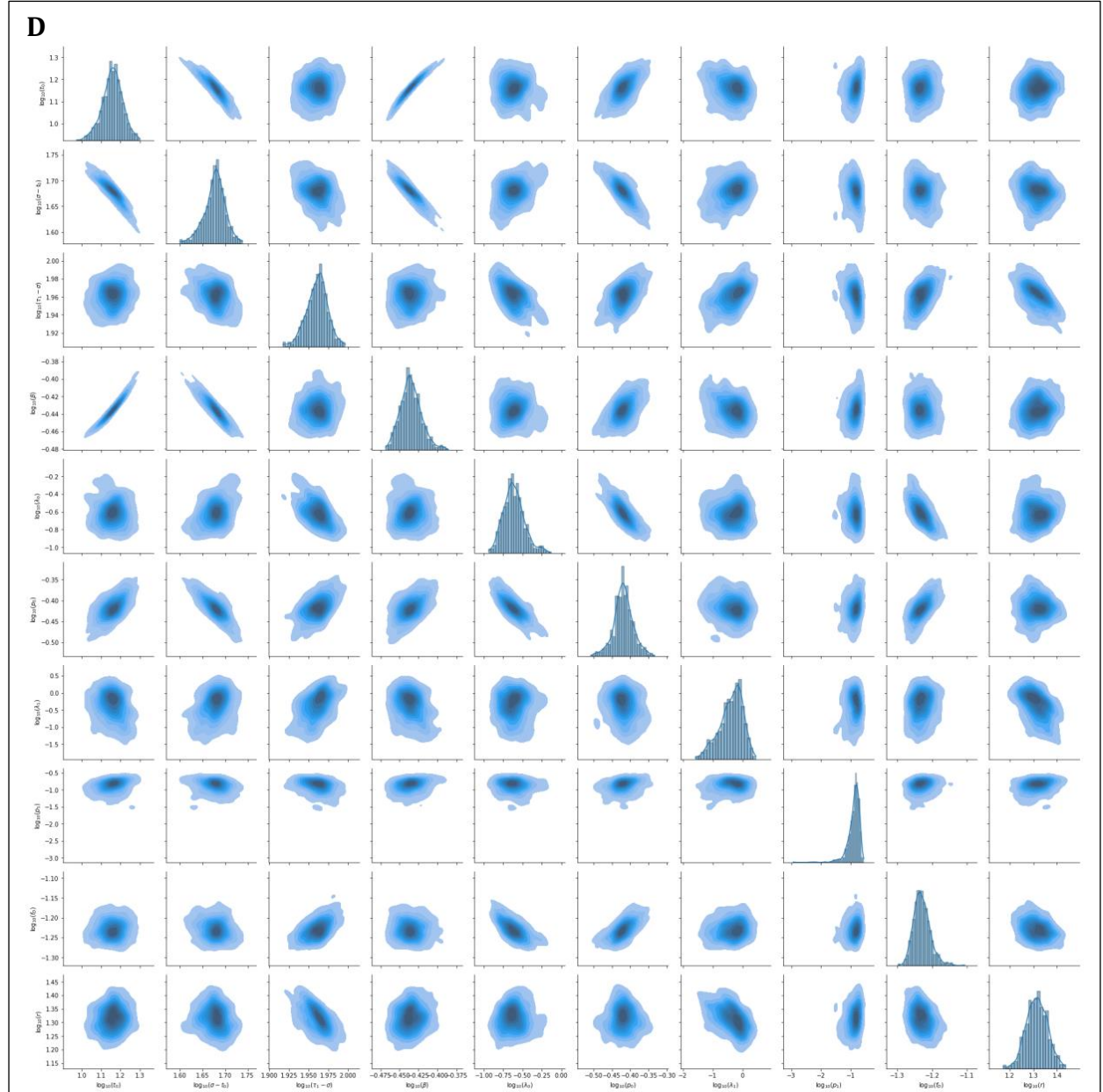

**Supplementary Fig. S11.** Evaluation of PyBNF's am method using Test Problem 4F (Supplementary File S4, Washington DC), which is discussed in Supplementary Methods. **(D)** Pairs plot for samples generated using PyBNF's am method (cf. Panel E). Off-diagonal panels represent two-dimensional marginalizations of the multivariate parameter posterior and illustrate relationships between pairs of parameters, as indicated. For each off-diagonal panel, the x- and y-axes indicate parameter values, and color indicates probability density. Darker shades indicate higher density. Diagonal panels represent marginal posteriors, which are also shown in Panel B. Each marginal posterior is presented as a histogram and as a smooth curve, which was found via kernel density estimation (KDE) as explained in Supplementary Methods. Each two-dimensional marginalization is presented as a contour plot, which was found via KDE.

**SUPPLEMENTARY FIGURE S11 – PANEL E**

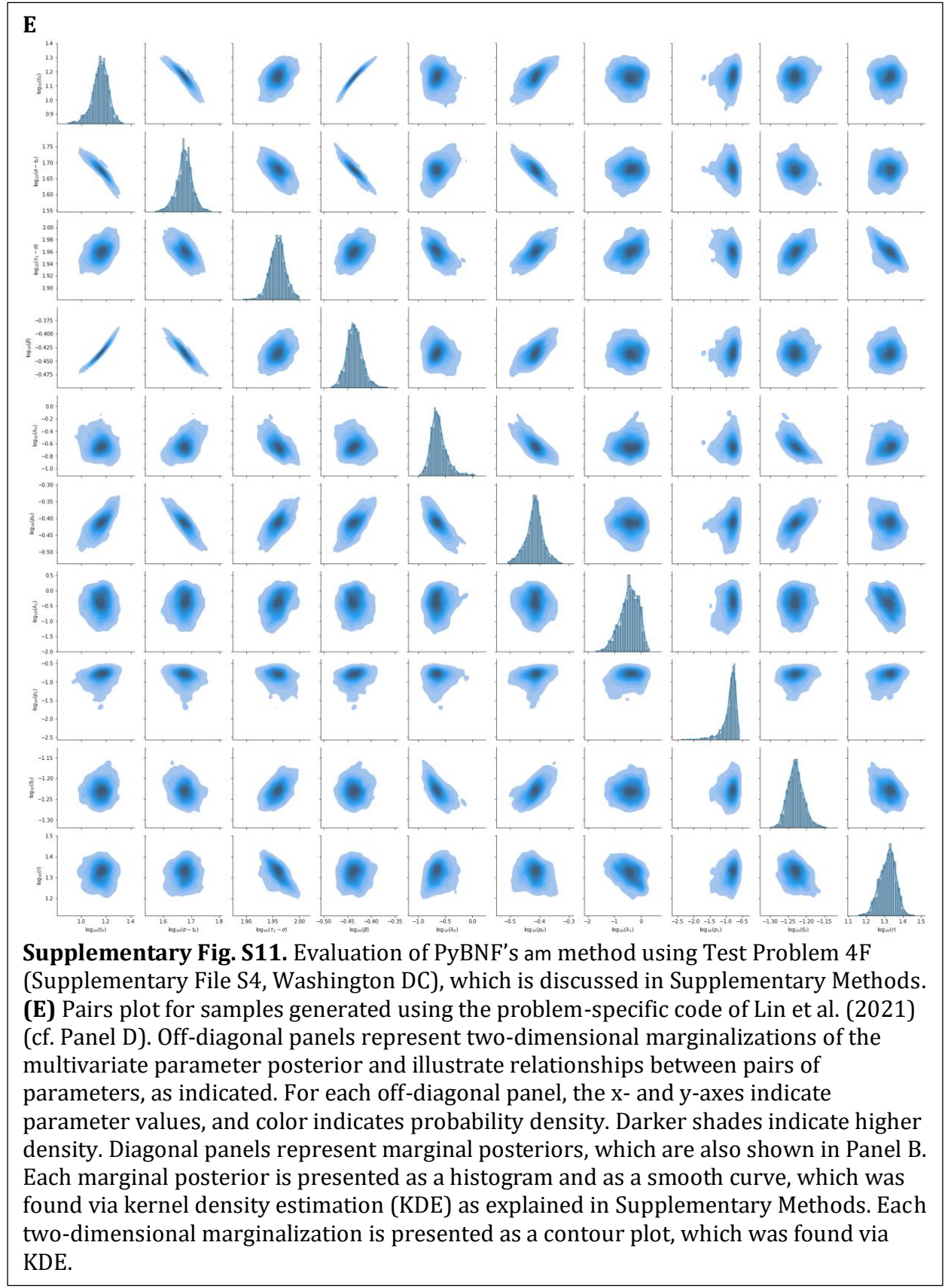

**Supplementary Fig. S11.** Evaluation of PyBNF's am method using Test Problem 4F (Supplementary File S4, Washington DC), which is discussed in Supplementary Methods. **(E)** Pairs plot for samples generated using the problem-specific code of Lin et al. (2021) (cf. Panel D). Off-diagonal panels represent two-dimensional marginalizations of the multivariate parameter posterior and illustrate relationships between pairs of parameters, as indicated. For each off-diagonal panel, the x- and y-axes indicate parameter values, and color indicates probability density. Darker shades indicate higher density. Diagonal panels represent marginal posteriors, which are also shown in Panel B. Each marginal posterior is presented as a histogram and as a smooth curve, which was found via kernel density estimation (KDE) as explained in Supplementary Methods. Each two-dimensional marginalization is presented as a contour plot, which was found via KDE.

**SUPPLEMENTARY FIGURE S12 – PANEL A**

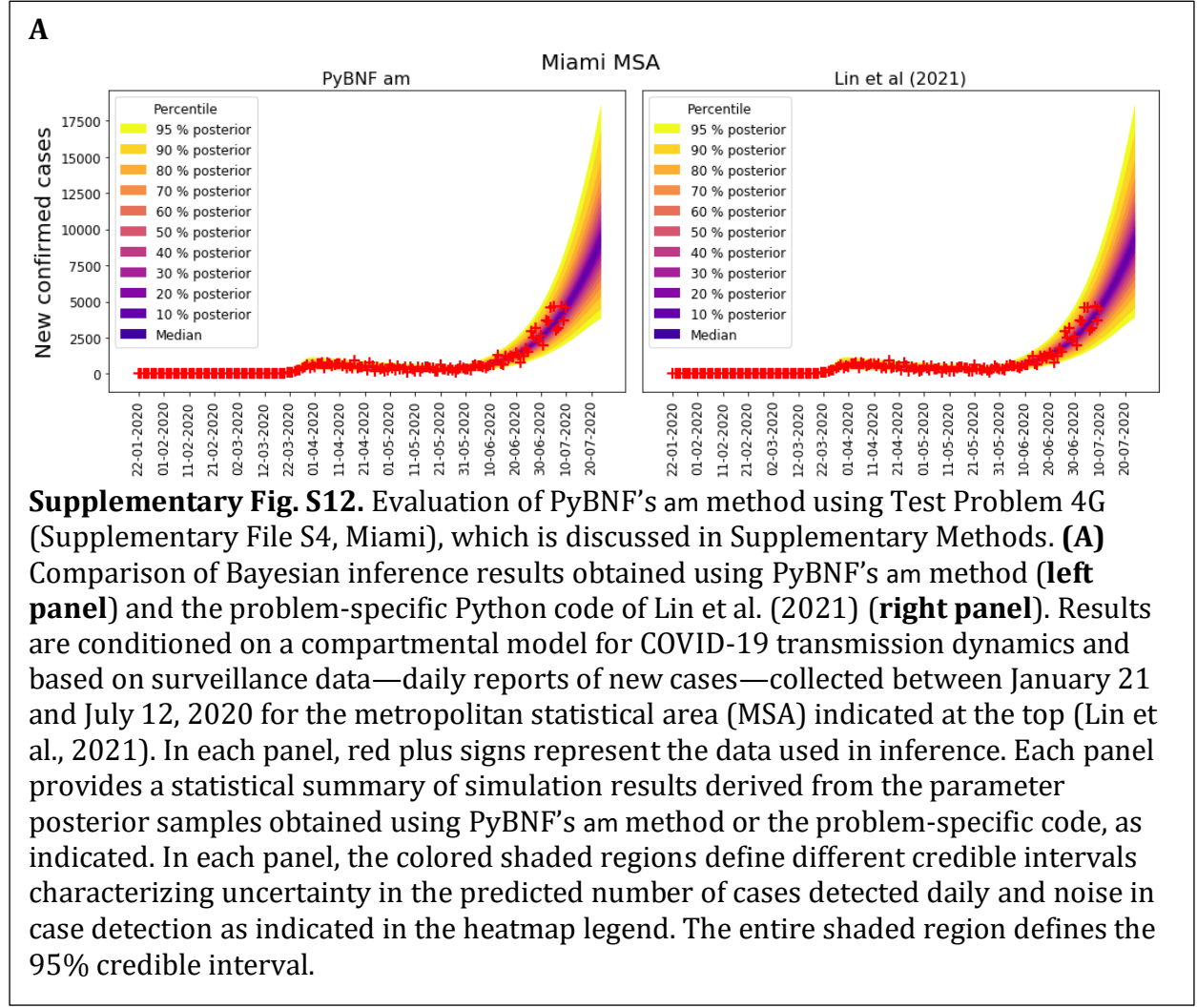

**Supplementary Fig. S12.** Evaluation of PyBNF's aᴍ method using Test Problem 4G (Supplementary File S4, Miami), which is discussed in Supplementary Methods. **(A)** Comparison of Bayesian inference results obtained using PyBNF's aᴍ method (**left panel**) and the problem-specific Python code of Lin et al. (2021) (**right panel**). Results are conditioned on a compartmental model for COVID-19 transmission dynamics and based on surveillance data—daily reports of new cases—collected between January 21 and July 12, 2020 for the metropolitan statistical area (MSA) indicated at the top (Lin et al., 2021). In each panel, red plus signs represent the data used in inference. Each panel provides a statistical summary of simulation results derived from the parameter posterior samples obtained using PyBNF's aᴍ method or the problem-specific code, as indicated. In each panel, the colored shaded regions define different credible intervals characterizing uncertainty in the predicted number of cases detected daily and noise in case detection as indicated in the heatmap legend. The entire shaded region defines the 95% credible interval.

**SUPPLEMENTARY FIGURE S12 – PANEL B**

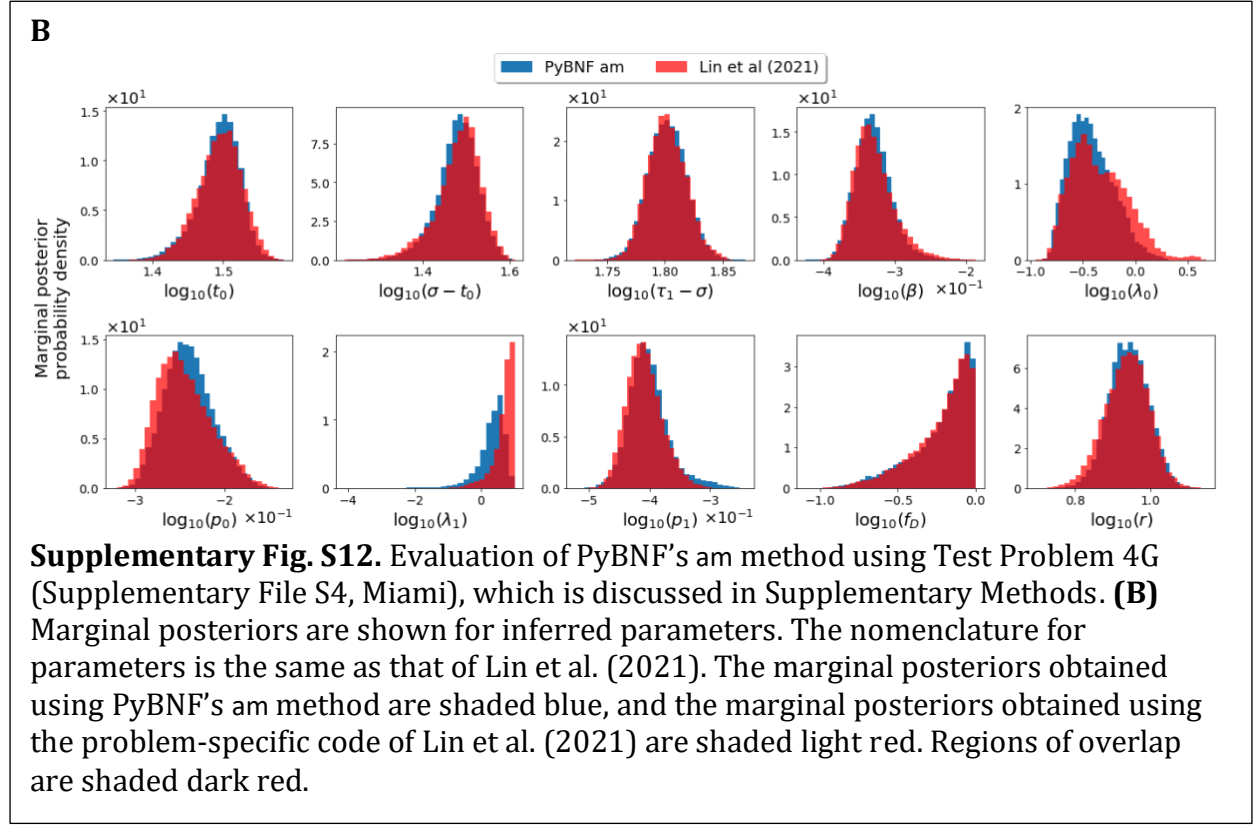

**Supplementary Fig. S12.** Evaluation of PyBNF's am method using Test Problem 4G (Supplementary File S4, Miami), which is discussed in Supplementary Methods. **(B)** Marginal posteriors are shown for inferred parameters. The nomenclature for parameters is the same as that of Lin et al. (2021). The marginal posteriors obtained using PyBNF's am method are shaded blue, and the marginal posteriors obtained using the problem-specific code of Lin et al. (2021) are shaded light red. Regions of overlap are shaded dark red.

**SUPPLEMENTARY FIGURE S12 – PANEL C**

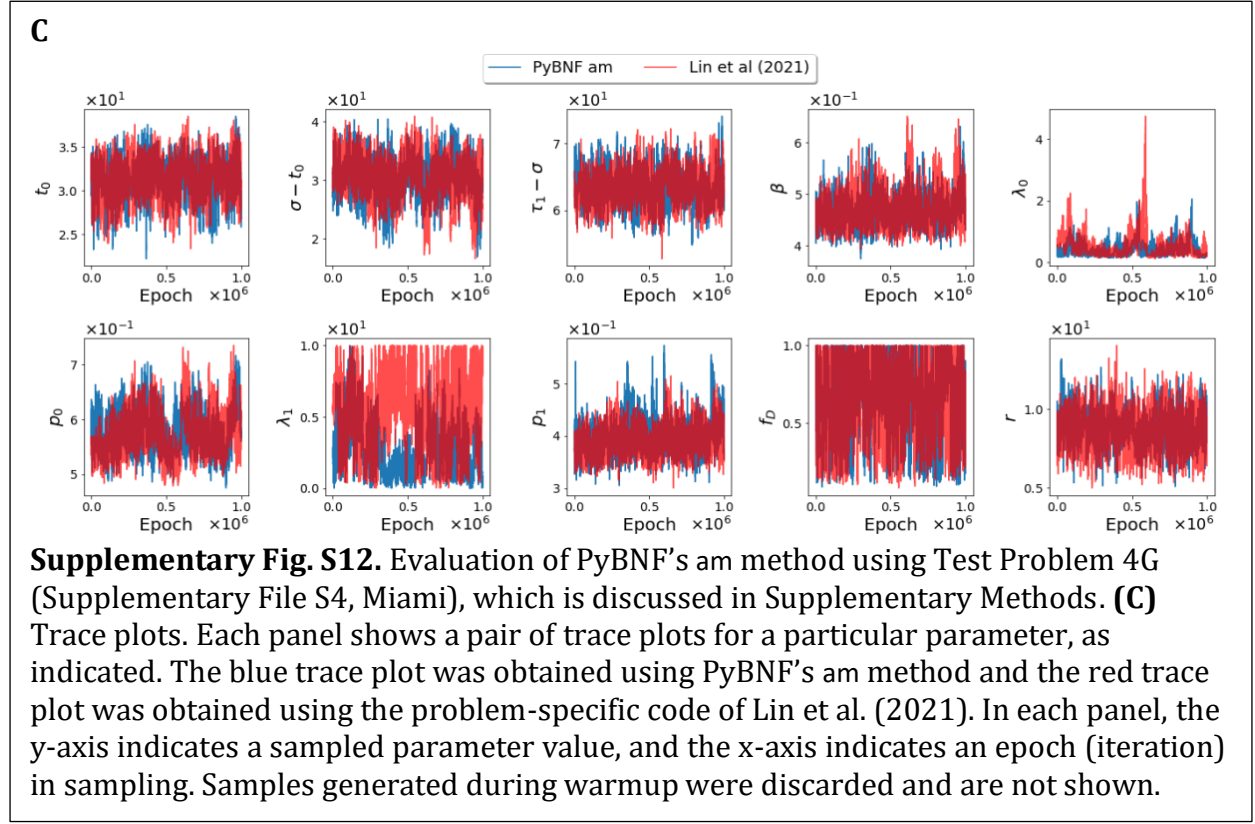

**Supplementary Fig. S12.** Evaluation of PyBNF's am method using Test Problem 4G (Supplementary File S4, Miami), which is discussed in Supplementary Methods. **(C)** Trace plots. Each panel shows a pair of trace plots for a particular parameter, as indicated. The blue trace plot was obtained using PyBNF's am method and the red trace plot was obtained using the problem-specific code of Lin et al. (2021). In each panel, the y-axis indicates a sampled parameter value, and the x-axis indicates an epoch (iteration) in sampling. Samples generated during warmup were discarded and are not shown.

**SUPPLEMENTARY FIGURE S12 – PANEL D**

**D**

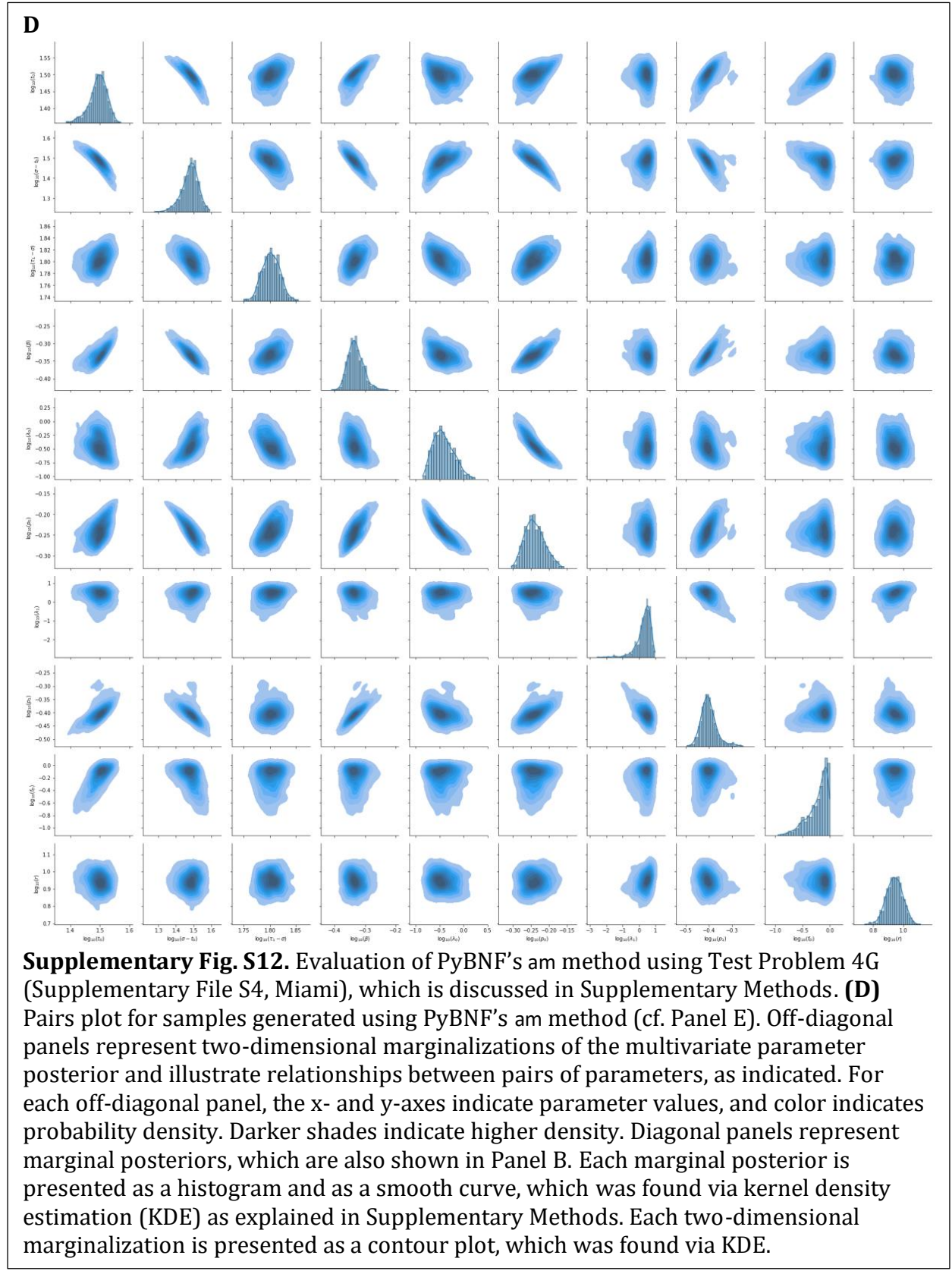

**Supplementary Fig. S12.** Evaluation of PyBNF's am method using Test Problem 4G (Supplementary File S4, Miami), which is discussed in Supplementary Methods. **(D)** Pairs plot for samples generated using PyBNF's am method (cf. Panel E). Off-diagonal panels represent two-dimensional marginalizations of the multivariate parameter posterior and illustrate relationships between pairs of parameters, as indicated. For each off-diagonal panel, the x- and y-axes indicate parameter values, and color indicates probability density. Darker shades indicate higher density. Diagonal panels represent marginal posteriors, which are also shown in Panel B. Each marginal posterior is presented as a histogram and as a smooth curve, which was found via kernel density estimation (KDE) as explained in Supplementary Methods. Each two-dimensional marginalization is presented as a contour plot, which was found via KDE.

**SUPPLEMENTARY FIGURE S12 – PANEL E**

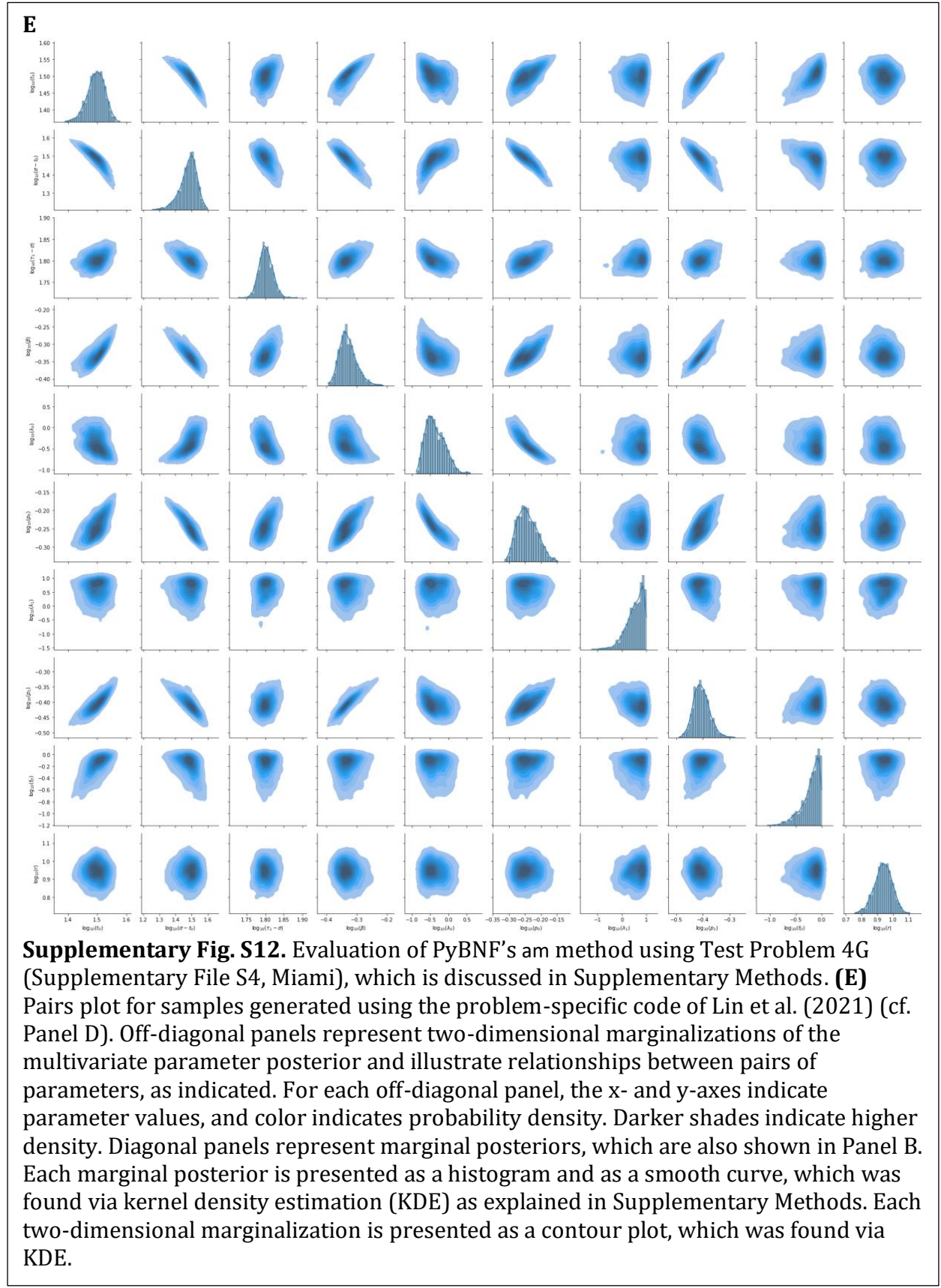

**Supplementary Fig. S12.** Evaluation of PyBNF's am method using Test Problem 4G (Supplementary File S4, Miami), which is discussed in Supplementary Methods. **(E)** Pairs plot for samples generated using the problem-specific code of Lin et al. (2021) (cf. Panel D). Off-diagonal panels represent two-dimensional marginalizations of the multivariate parameter posterior and illustrate relationships between pairs of parameters, as indicated. For each off-diagonal panel, the x- and y-axes indicate parameter values, and color indicates probability density. Darker shades indicate higher density. Diagonal panels represent marginal posteriors, which are also shown in Panel B. Each marginal posterior is presented as a histogram and as a smooth curve, which was found via kernel density estimation (KDE) as explained in Supplementary Methods. Each two-dimensional marginalization is presented as a contour plot, which was found via KDE.

**SUPPLEMENTARY FIGURE S13 – PANEL A**

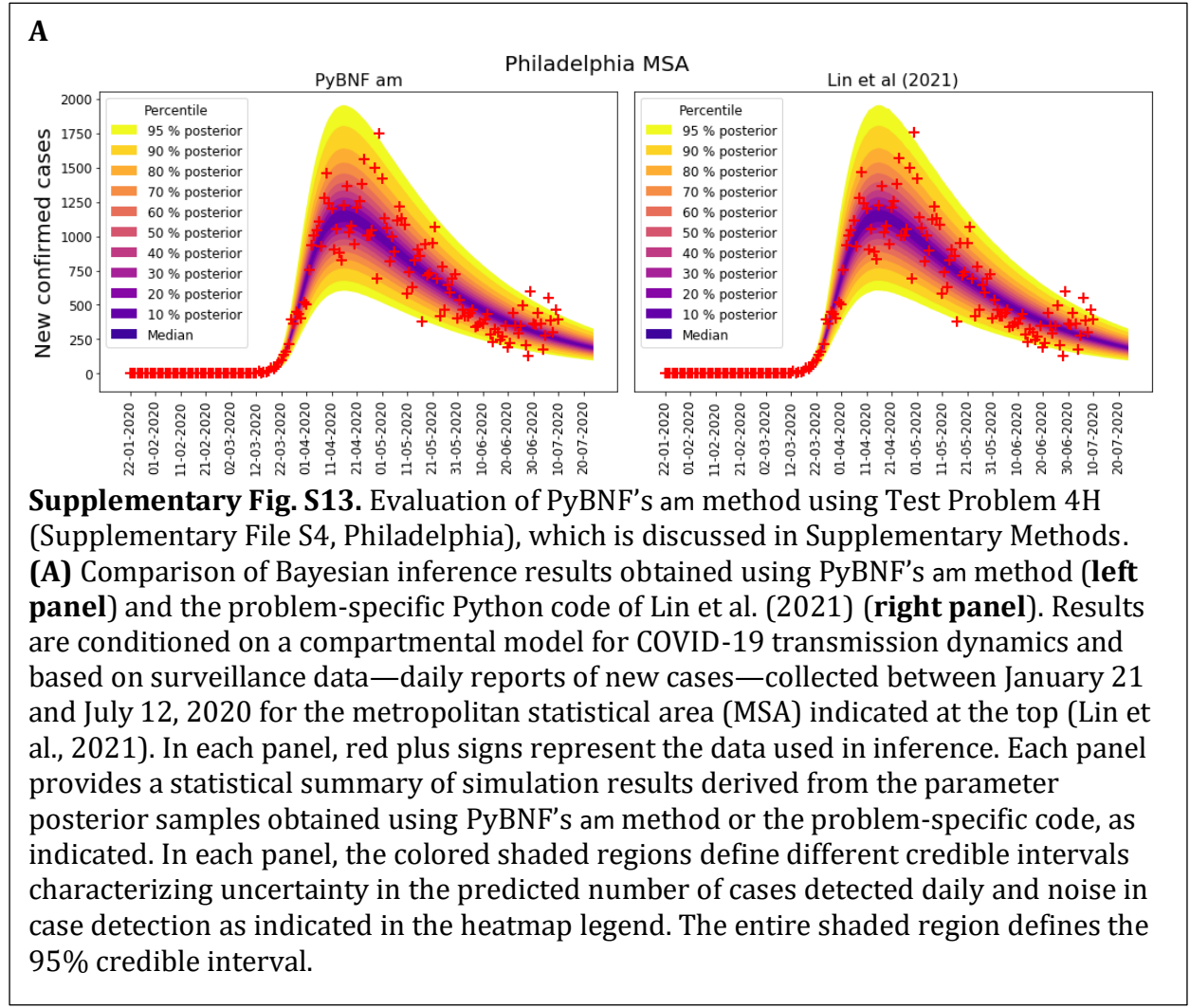

**Supplementary Fig. S13.** Evaluation of PyBNF's am method using Test Problem 4H (Supplementary File S4, Philadelphia), which is discussed in Supplementary Methods. **(A)** Comparison of Bayesian inference results obtained using PyBNF's am method (**left panel**) and the problem-specific Python code of Lin et al. (2021) (**right panel**). Results are conditioned on a compartmental model for COVID-19 transmission dynamics and based on surveillance data—daily reports of new cases—collected between January 21 and July 12, 2020 for the metropolitan statistical area (MSA) indicated at the top (Lin et al., 2021). In each panel, red plus signs represent the data used in inference. Each panel provides a statistical summary of simulation results derived from the parameter posterior samples obtained using PyBNF's am method or the problem-specific code, as indicated. In each panel, the colored shaded regions define different credible intervals characterizing uncertainty in the predicted number of cases detected daily and noise in case detection as indicated in the heatmap legend. The entire shaded region defines the 95% credible interval.

**SUPPLEMENTARY FIGURE S13 – PANEL B**

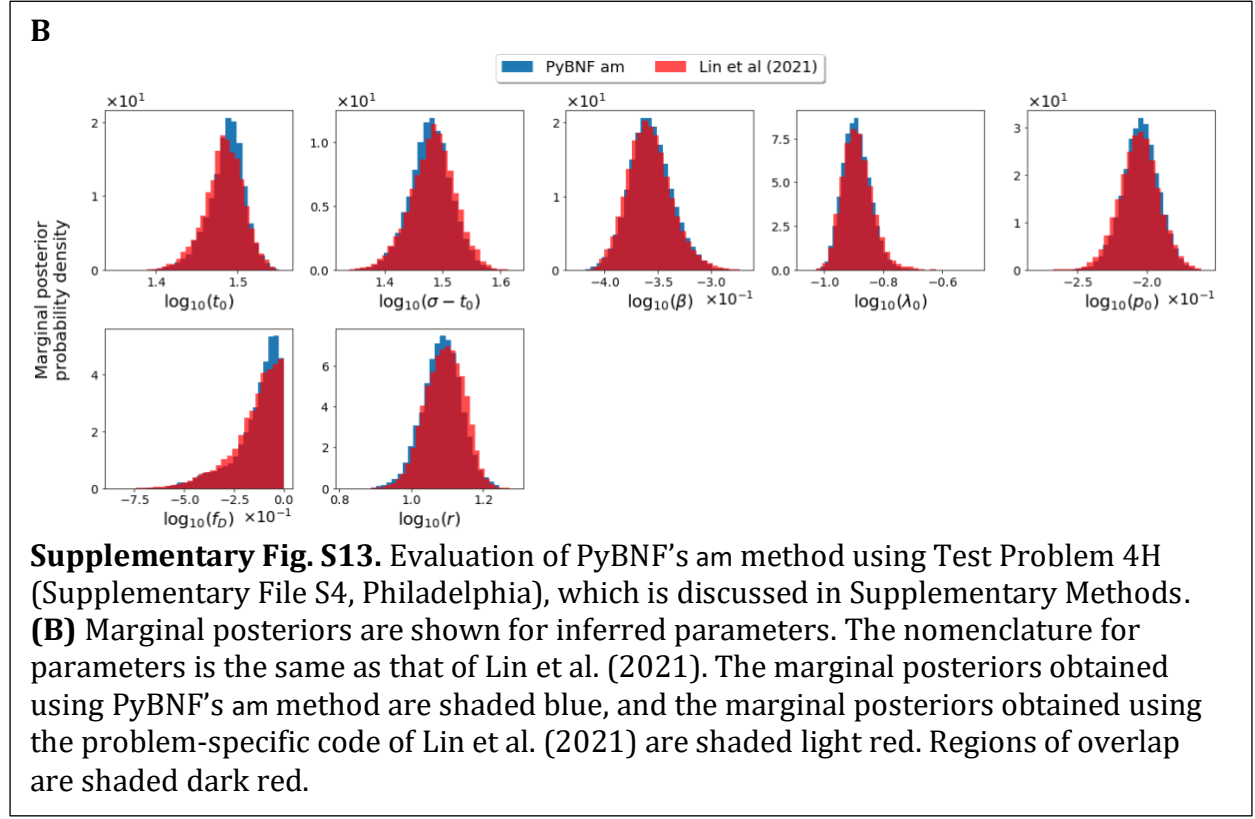

**Supplementary Fig. S13.** Evaluation of PyBNF's am method using Test Problem 4H (Supplementary File S4, Philadelphia), which is discussed in Supplementary Methods. **(B)** Marginal posteriors are shown for inferred parameters. The nomenclature for parameters is the same as that of Lin et al. (2021). The marginal posteriors obtained using PyBNF's am method are shaded blue, and the marginal posteriors obtained using the problem-specific code of Lin et al. (2021) are shaded light red. Regions of overlap are shaded dark red.

**SUPPLEMENTARY FIGURE S13 – PANEL C**

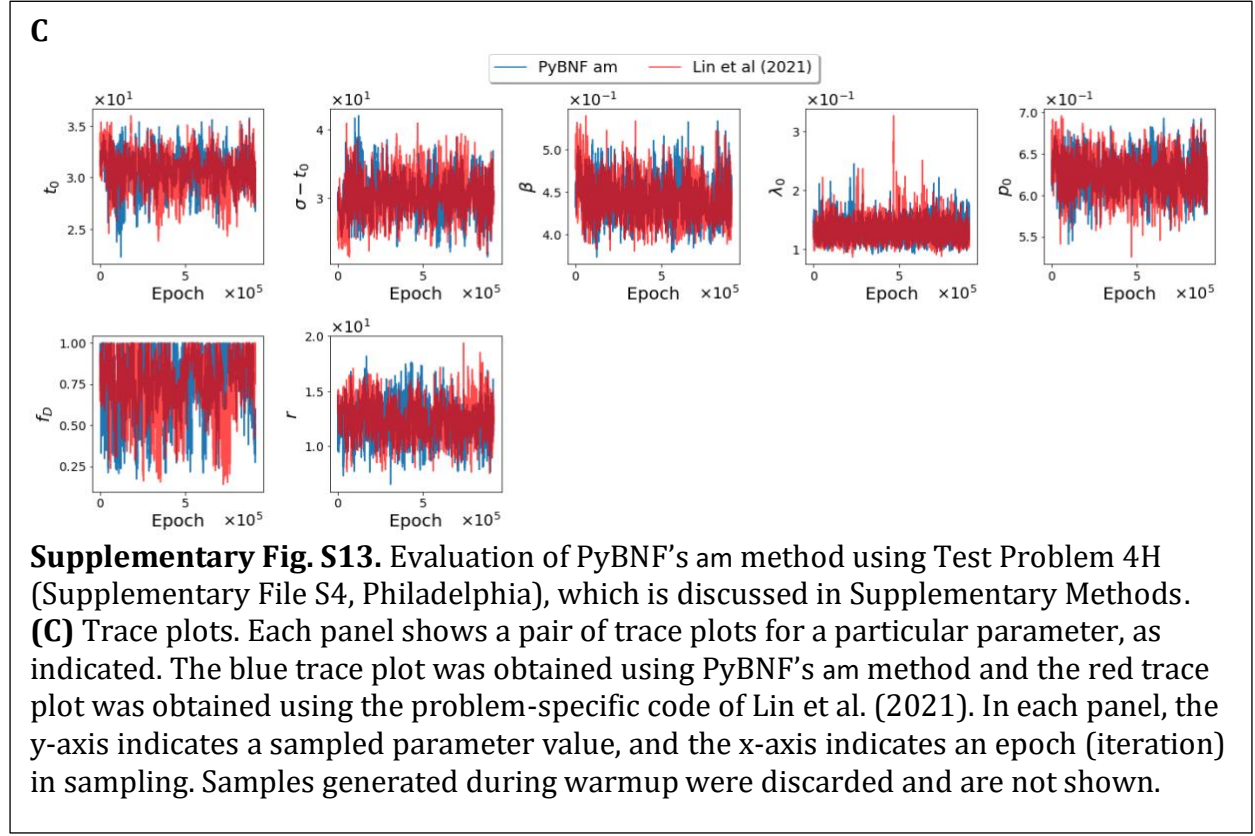

**Supplementary Fig. S13.** Evaluation of PyBNF's am method using Test Problem 4H (Supplementary File S4, Philadelphia), which is discussed in Supplementary Methods. **(C)** Trace plots. Each panel shows a pair of trace plots for a particular parameter, as indicated. The blue trace plot was obtained using PyBNF's am method and the red trace plot was obtained using the problem-specific code of Lin et al. (2021). In each panel, the y-axis indicates a sampled parameter value, and the x-axis indicates an epoch (iteration) in sampling. Samples generated during warmup were discarded and are not shown.

**SUPPLEMENTARY FIGURE S13 – PANEL D**

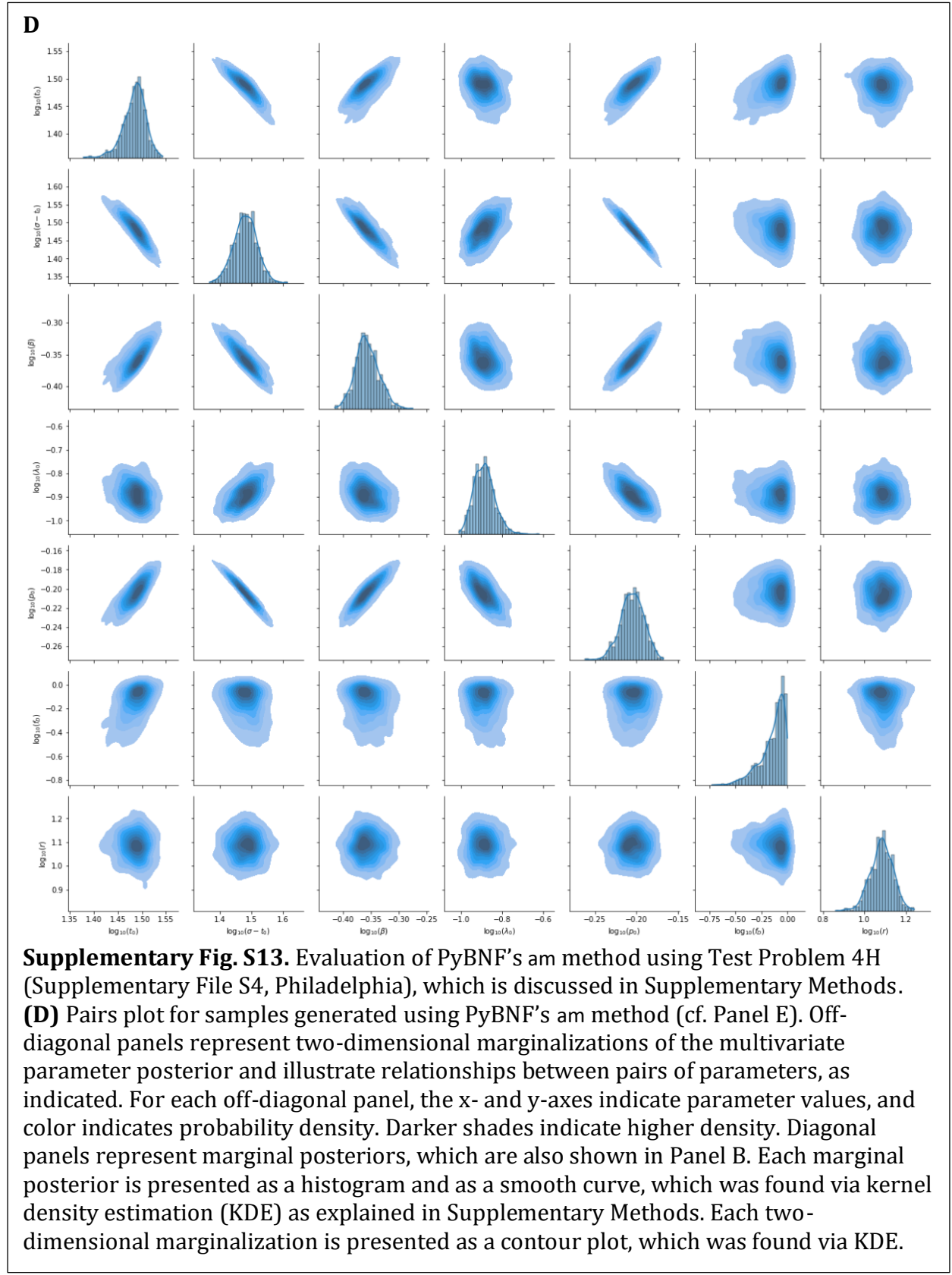

**Supplementary Fig. S13.** Evaluation of PyBNF's ᴀᴍ method using Test Problem 4H (Supplementary File S4, Philadelphia), which is discussed in Supplementary Methods. **(D)** Pairs plot for samples generated using PyBNF's ᴀᴍ method (cf. Panel E). Off-diagonal panels represent two-dimensional marginalizations of the multivariate parameter posterior and illustrate relationships between pairs of parameters, as indicated. For each off-diagonal panel, the x- and y-axes indicate parameter values, and color indicates probability density. Darker shades indicate higher density. Diagonal panels represent marginal posteriors, which are also shown in Panel B. Each marginal posterior is presented as a histogram and as a smooth curve, which was found via kernel density estimation (KDE) as explained in Supplementary Methods. Each two-dimensional marginalization is presented as a contour plot, which was found via KDE.

**SUPPLEMENTARY FIGURE S13 – PANEL E**

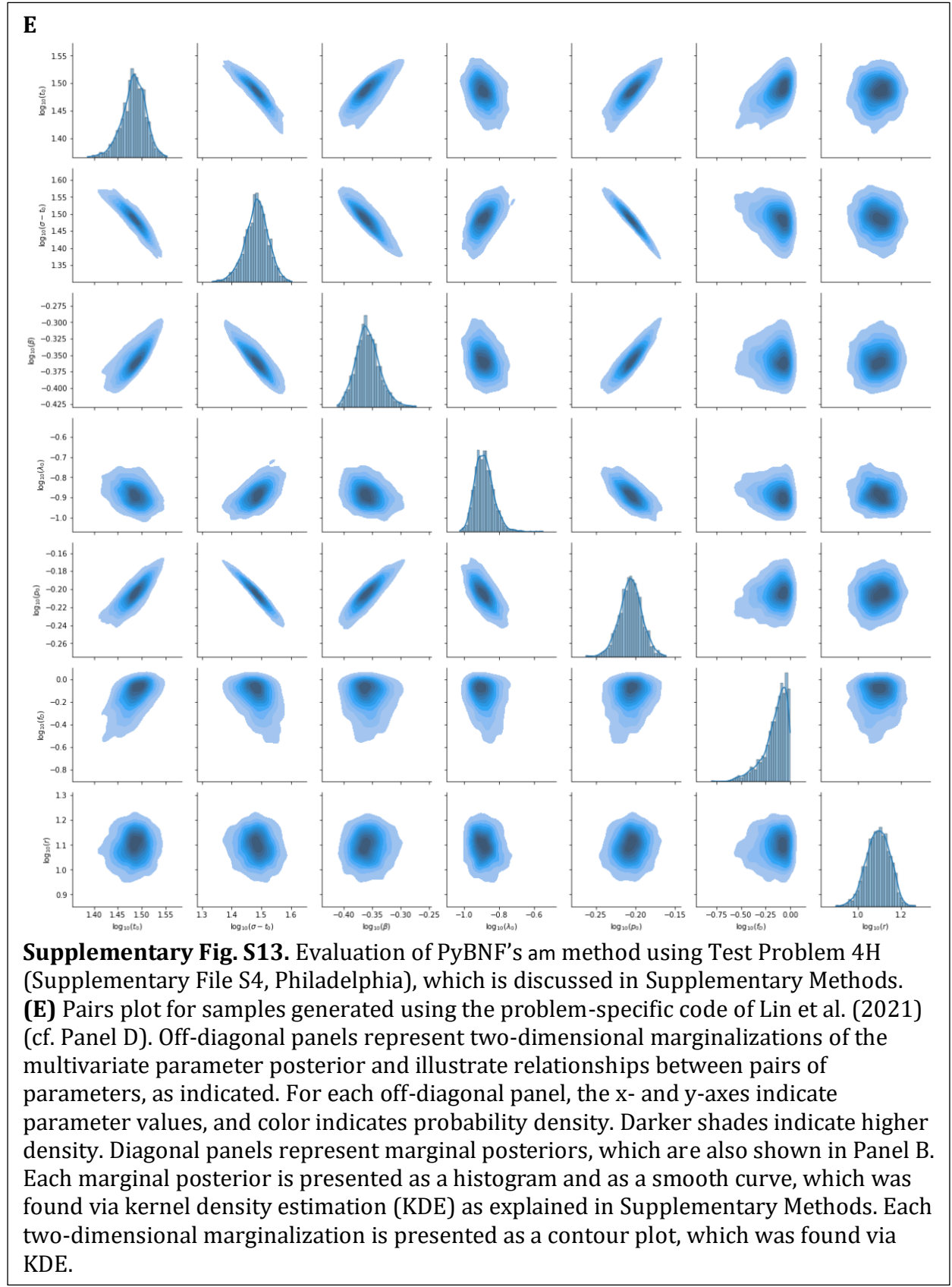

**Supplementary Fig. S13.** Evaluation of PyBNF's am method using Test Problem 4H (Supplementary File S4, Philadelphia), which is discussed in Supplementary Methods. **(E)** Pairs plot for samples generated using the problem-specific code of Lin et al. (2021) (cf. Panel D). Off-diagonal panels represent two-dimensional marginalizations of the multivariate parameter posterior and illustrate relationships between pairs of parameters, as indicated. For each off-diagonal panel, the x- and y-axes indicate parameter values, and color indicates probability density. Darker shades indicate higher density. Diagonal panels represent marginal posteriors, which are also shown in Panel B. Each marginal posterior is presented as a histogram and as a smooth curve, which was found via kernel density estimation (KDE) as explained in Supplementary Methods. Each two-dimensional marginalization is presented as a contour plot, which was found via KDE.

**SUPPLEMENTARY FIGURE S14 – PANEL A**

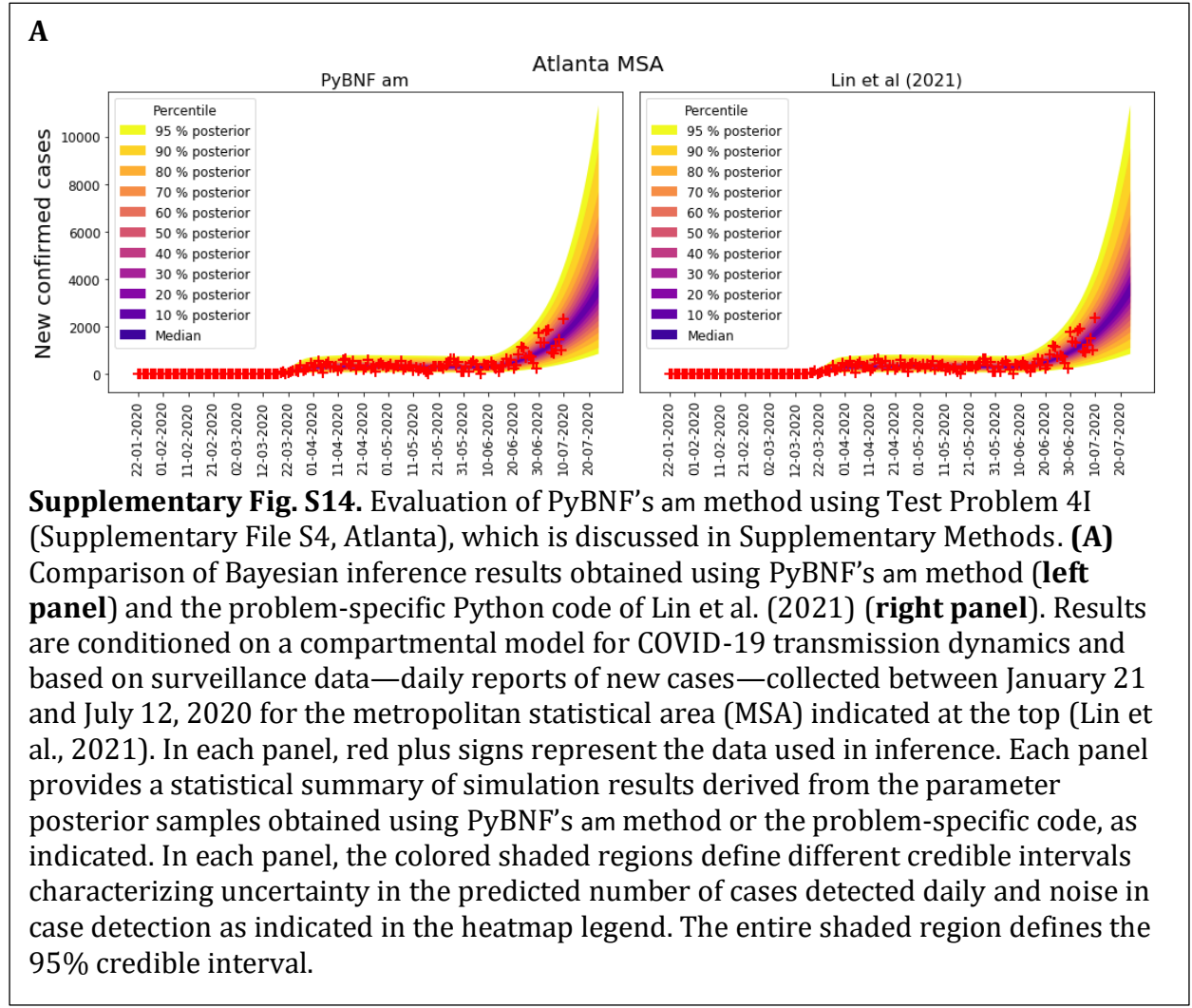

**Supplementary Fig. S14.** Evaluation of PyBNF's ᴀᴍ method using Test Problem 4I (Supplementary File S4, Atlanta), which is discussed in Supplementary Methods. **(A)** Comparison of Bayesian inference results obtained using PyBNF's ᴀᴍ method (**left panel**) and the problem-specific Python code of Lin et al. (2021) (**right panel**). Results are conditioned on a compartmental model for COVID-19 transmission dynamics and based on surveillance data—daily reports of new cases—collected between January 21 and July 12, 2020 for the metropolitan statistical area (MSA) indicated at the top (Lin et al., 2021). In each panel, red plus signs represent the data used in inference. Each panel provides a statistical summary of simulation results derived from the parameter posterior samples obtained using PyBNF's ᴀᴍ method or the problem-specific code, as indicated. In each panel, the colored shaded regions define different credible intervals characterizing uncertainty in the predicted number of cases detected daily and noise in case detection as indicated in the heatmap legend. The entire shaded region defines the 95% credible interval.

**SUPPLEMENTARY FIGURE S14 – PANEL B**

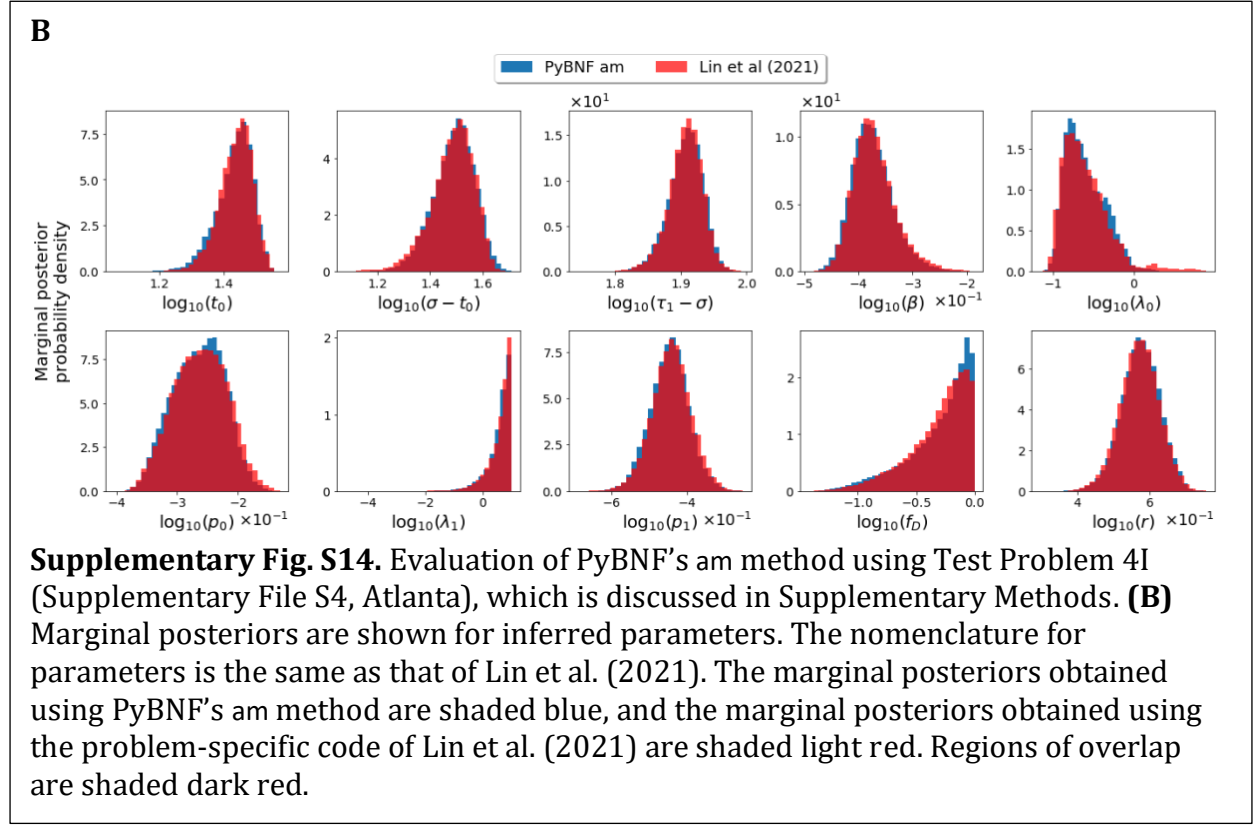

**Supplementary Fig. S14.** Evaluation of PyBNF's am method using Test Problem 4I (Supplementary File S4, Atlanta), which is discussed in Supplementary Methods. **(B)** Marginal posteriors are shown for inferred parameters. The nomenclature for parameters is the same as that of Lin et al. (2021). The marginal posteriors obtained using PyBNF's am method are shaded blue, and the marginal posteriors obtained using the problem-specific code of Lin et al. (2021) are shaded light red. Regions of overlap are shaded dark red.

**SUPPLEMENTARY FIGURE S14 – PANEL C**

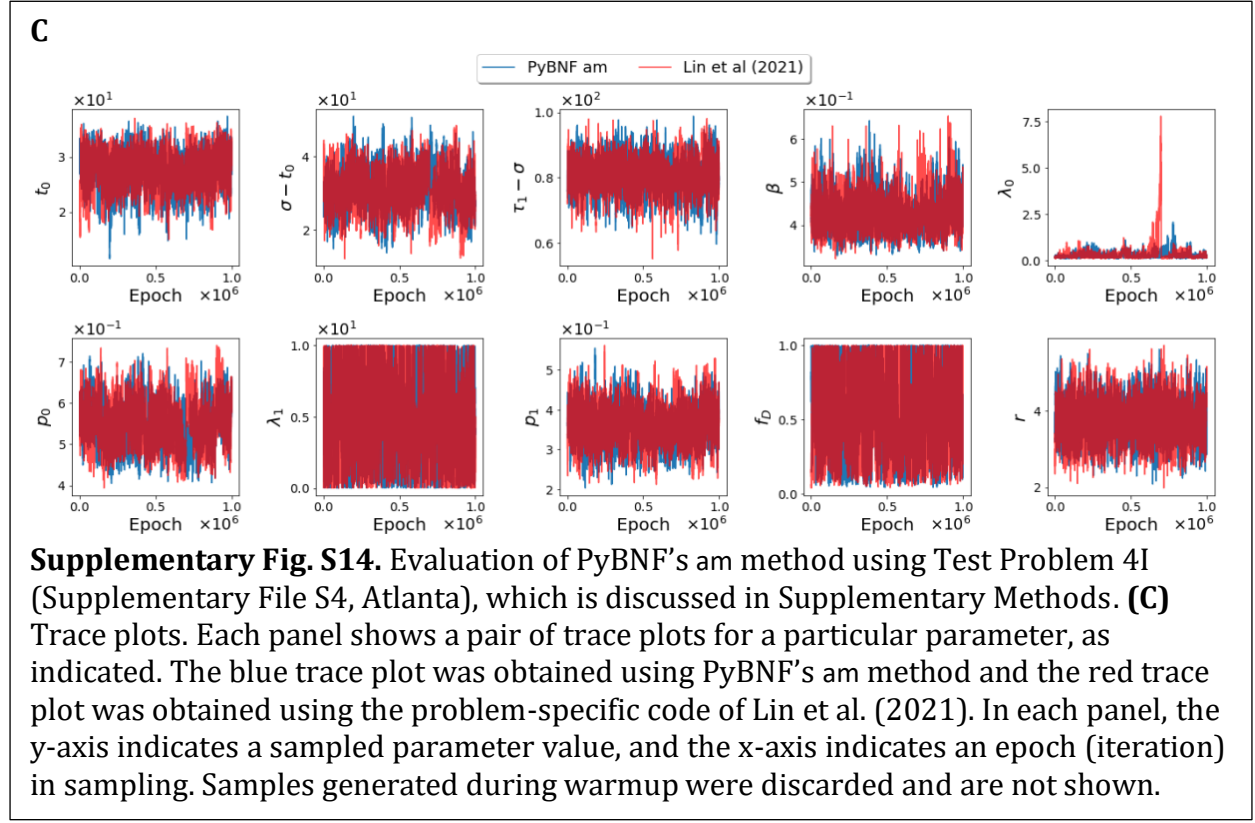

**Supplementary Fig. S14.** Evaluation of PyBNF's am method using Test Problem 4I (Supplementary File S4, Atlanta), which is discussed in Supplementary Methods. **(C)** Trace plots. Each panel shows a pair of trace plots for a particular parameter, as indicated. The blue trace plot was obtained using PyBNF's am method and the red trace plot was obtained using the problem-specific code of Lin et al. (2021). In each panel, the y-axis indicates a sampled parameter value, and the x-axis indicates an epoch (iteration) in sampling. Samples generated during warmup were discarded and are not shown.

**SUPPLEMENTARY FIGURE S14 – PANEL D**

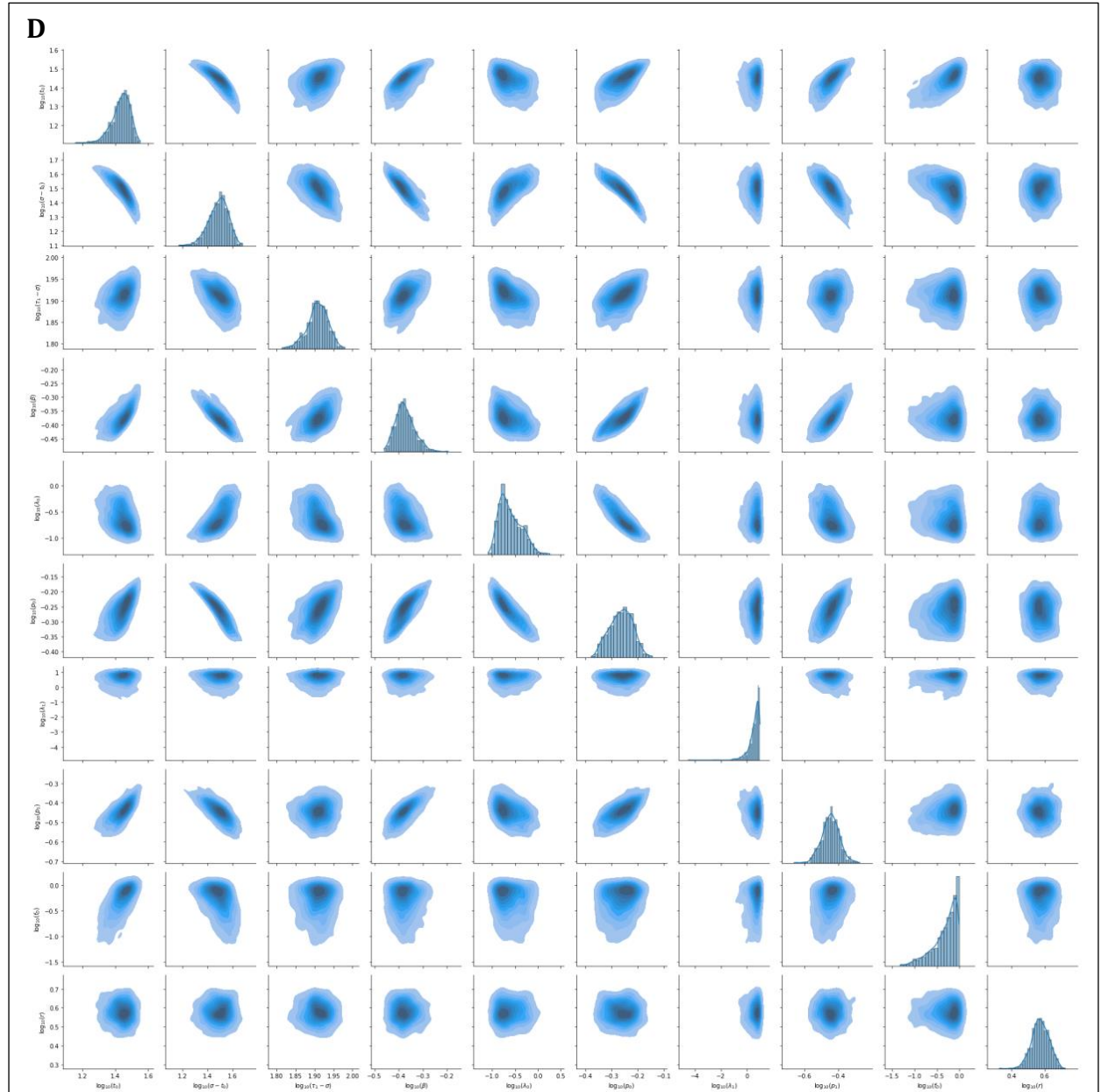

**Supplementary Fig. S14.** Evaluation of PyBNF's am method using Test Problem 4I (Supplementary File S4, Atlanta), which is discussed in Supplementary Methods. **(D)** Pairs plot for samples generated using PyBNF's am method (cf. Panel E). Off-diagonal panels represent two-dimensional marginalizations of the multivariate parameter posterior and illustrate relationships between pairs of parameters, as indicated. For each off-diagonal panel, the x- and y-axes indicate parameter values, and color indicates probability density. Darker shades indicate higher density. Diagonal panels represent marginal posteriors, which are also shown in Panel B. Each marginal posterior is presented as a histogram and as a smooth curve, which was found via kernel density estimation (KDE) as explained in Supplementary Methods. Each two-dimensional marginalization is presented as a contour plot, which was found via KDE.

**SUPPLEMENTARY FIGURE S14 – PANEL E**

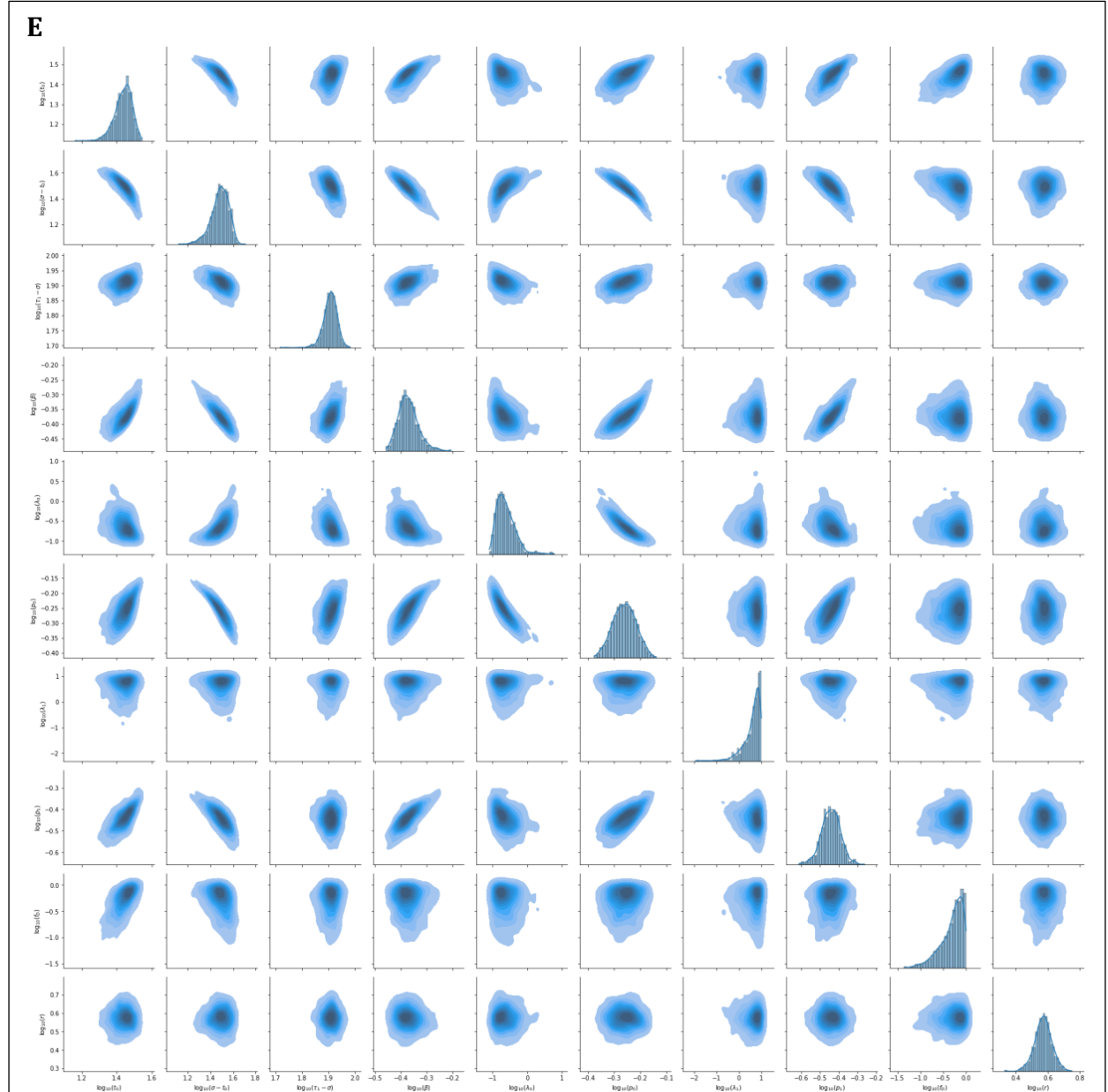

**Supplementary Fig. S14.** Evaluation of PyBNF's am method using Test Problem 4I (Supplementary File S4, Atlanta), which is discussed in Supplementary Methods. **(E)** Pairs plot for samples generated using the problem-specific code of Lin et al. (2021) (cf. Panel D). Off-diagonal panels represent two-dimensional marginalizations of the multivariate parameter posterior and illustrate relationships between pairs of parameters, as indicated. For each off-diagonal panel, the x- and y-axes indicate parameter values, and color indicates probability density. Darker shades indicate higher density. Diagonal panels represent marginal posteriors, which are also shown in Panel B. Each marginal posterior is presented as a histogram and as a smooth curve, which was found via kernel density estimation (KDE) as explained in Supplementary Methods. Each two-dimensional marginalization is presented as a contour plot, which was found via KDE.

**SUPPLEMENTARY FIGURE S15 – PANEL A**

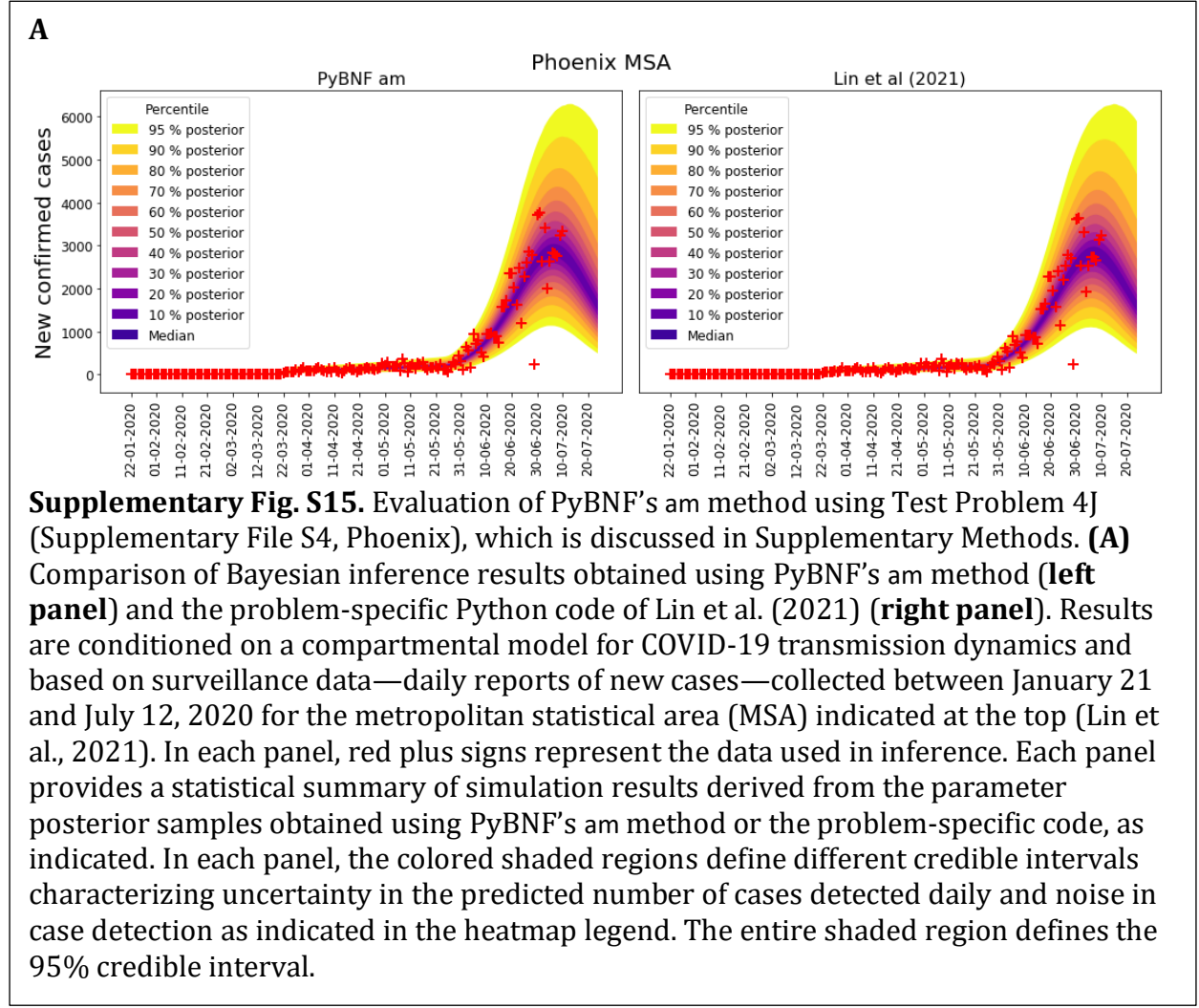

**Supplementary Fig. S15.** Evaluation of PyBNF's ᴀᴍ method using Test Problem 4J (Supplementary File S4, Phoenix), which is discussed in Supplementary Methods. **(A)** Comparison of Bayesian inference results obtained using PyBNF's ᴀᴍ method (**left panel**) and the problem-specific Python code of Lin et al. (2021) (**right panel**). Results are conditioned on a compartmental model for COVID-19 transmission dynamics and based on surveillance data—daily reports of new cases—collected between January 21 and July 12, 2020 for the metropolitan statistical area (MSA) indicated at the top (Lin et al., 2021). In each panel, red plus signs represent the data used in inference. Each panel provides a statistical summary of simulation results derived from the parameter posterior samples obtained using PyBNF's ᴀᴍ method or the problem-specific code, as indicated. In each panel, the colored shaded regions define different credible intervals characterizing uncertainty in the predicted number of cases detected daily and noise in case detection as indicated in the heatmap legend. The entire shaded region defines the 95% credible interval.

**SUPPLEMENTARY FIGURE S15 – PANEL B**

**Supplementary Fig. S15.** Evaluation of PyBNF's am method using Test Problem 4J (Supplementary File S4, Phoenix), which is discussed in Supplementary Methods. **(B)** Marginal posteriors are shown for inferred parameters. The nomenclature for parameters is the same as that of Lin et al. (2021). The marginal posteriors obtained using PyBNF's am method are shaded blue, and the marginal posteriors obtained using the problem-specific code of Lin et al. (2021) are shaded light red. Regions of overlap are shaded dark red.

**SUPPLEMENTARY FIGURE S15 – PANEL C**

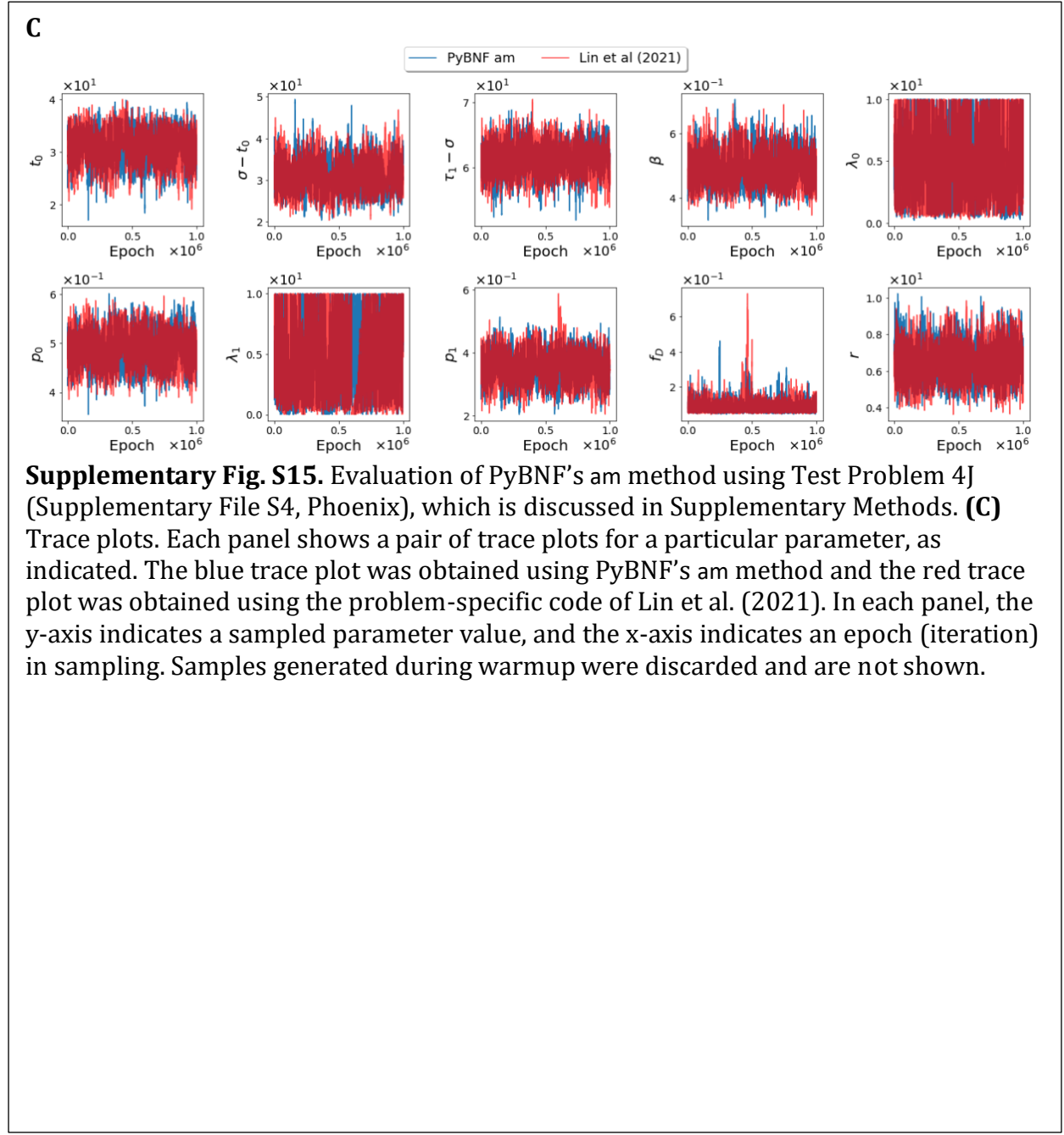

**Supplementary Fig. S15.** Evaluation of PyBNF's am method using Test Problem 4J (Supplementary File S4, Phoenix), which is discussed in Supplementary Methods. **(C)** Trace plots. Each panel shows a pair of trace plots for a particular parameter, as indicated. The blue trace plot was obtained using PyBNF's am method and the red trace plot was obtained using the problem-specific code of Lin et al. (2021). In each panel, the y-axis indicates a sampled parameter value, and the x-axis indicates an epoch (iteration) in sampling. Samples generated during warmup were discarded and are not shown.

**SUPPLEMENTARY FIGURE S15 – PANEL D**

**D**

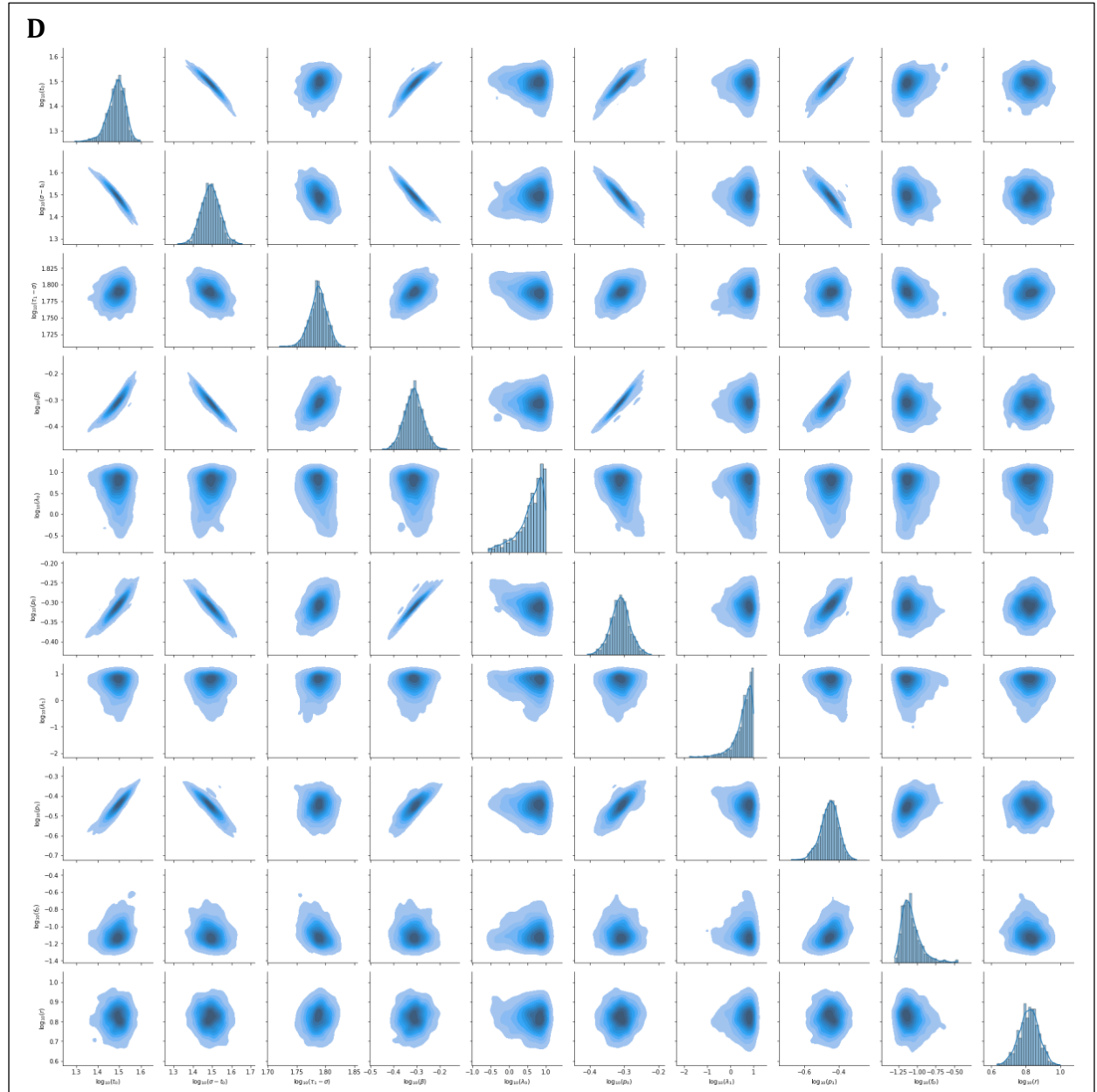

**Supplementary Fig. S15.** Evaluation of PyBNF's am method using Test Problem 4J (Supplementary File S4, Phoenix), which is discussed in Supplementary Methods. **(D)** Pairs plot for samples generated using PyBNF's am method (cf. Panel E). Off-diagonal panels represent two-dimensional marginalizations of the multivariate parameter posterior and illustrate relationships between pairs of parameters, as indicated. For each off-diagonal panel, the x- and y-axes indicate parameter values, and color indicates probability density. Darker shades indicate higher density. Diagonal panels represent marginal posteriors, which are also shown in Panel B. Each marginal posterior is presented as a histogram and as a smooth curve, which was found via kernel density estimation (KDE) as explained in Supplementary Methods. Each two-dimensional marginalization is presented as a contour plot, which was found via KDE.

**SUPPLEMENTARY FIGURE S15 – PANEL E**

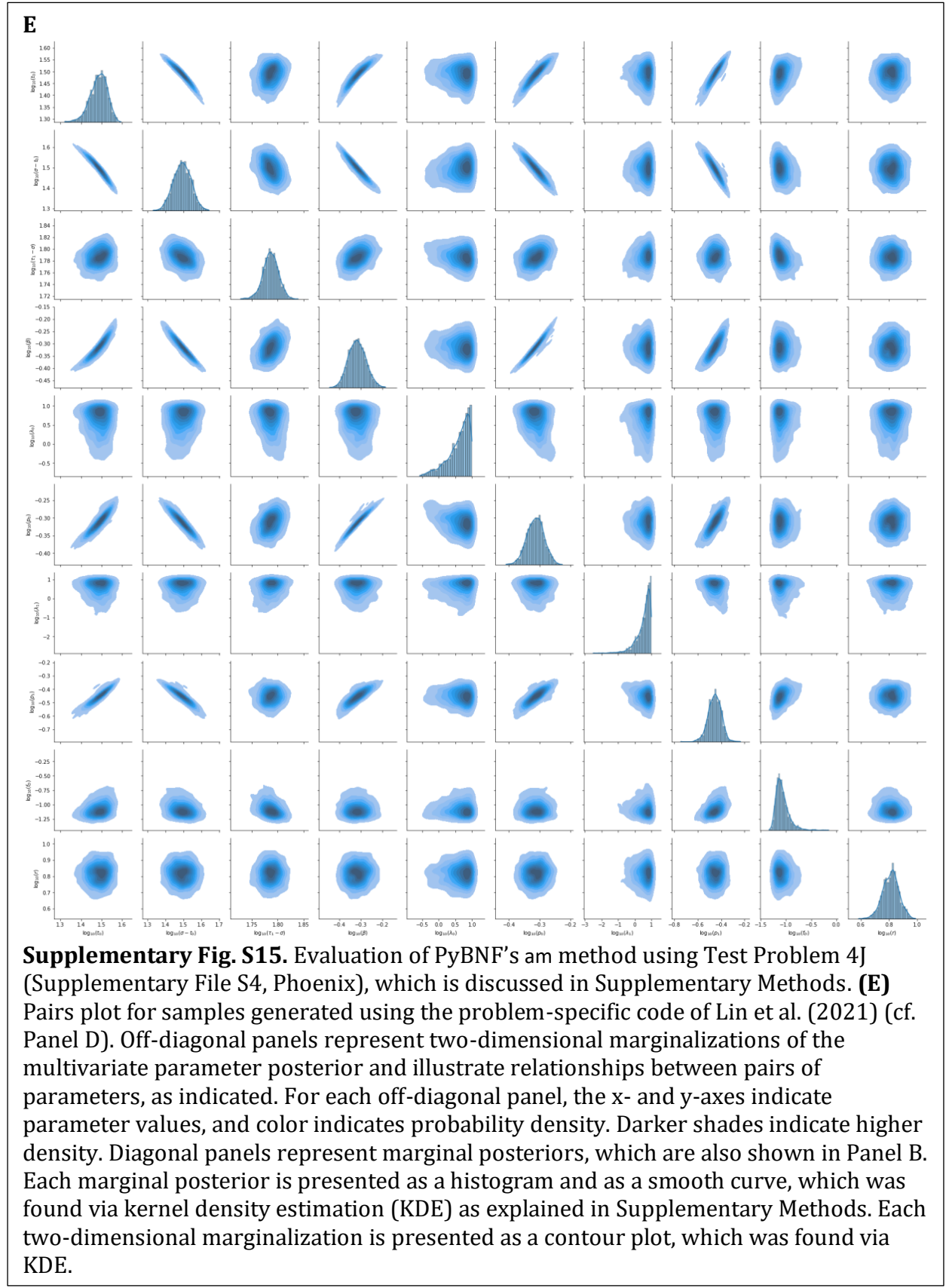

**Supplementary Fig. S15.** Evaluation of PyBNF's am method using Test Problem 4J (Supplementary File S4, Phoenix), which is discussed in Supplementary Methods. **(E)** Pairs plot for samples generated using the problem-specific code of Lin et al. (2021) (cf. Panel D). Off-diagonal panels represent two-dimensional marginalizations of the multivariate parameter posterior and illustrate relationships between pairs of parameters, as indicated. For each off-diagonal panel, the x- and y-axes indicate parameter values, and color indicates probability density. Darker shades indicate higher density. Diagonal panels represent marginal posteriors, which are also shown in Panel B. Each marginal posterior is presented as a histogram and as a smooth curve, which was found via kernel density estimation (KDE) as explained in Supplementary Methods. Each two-dimensional marginalization is presented as a contour plot, which was found via KDE.

**SUPPLEMENTARY FIGURE S16 – PANEL A**

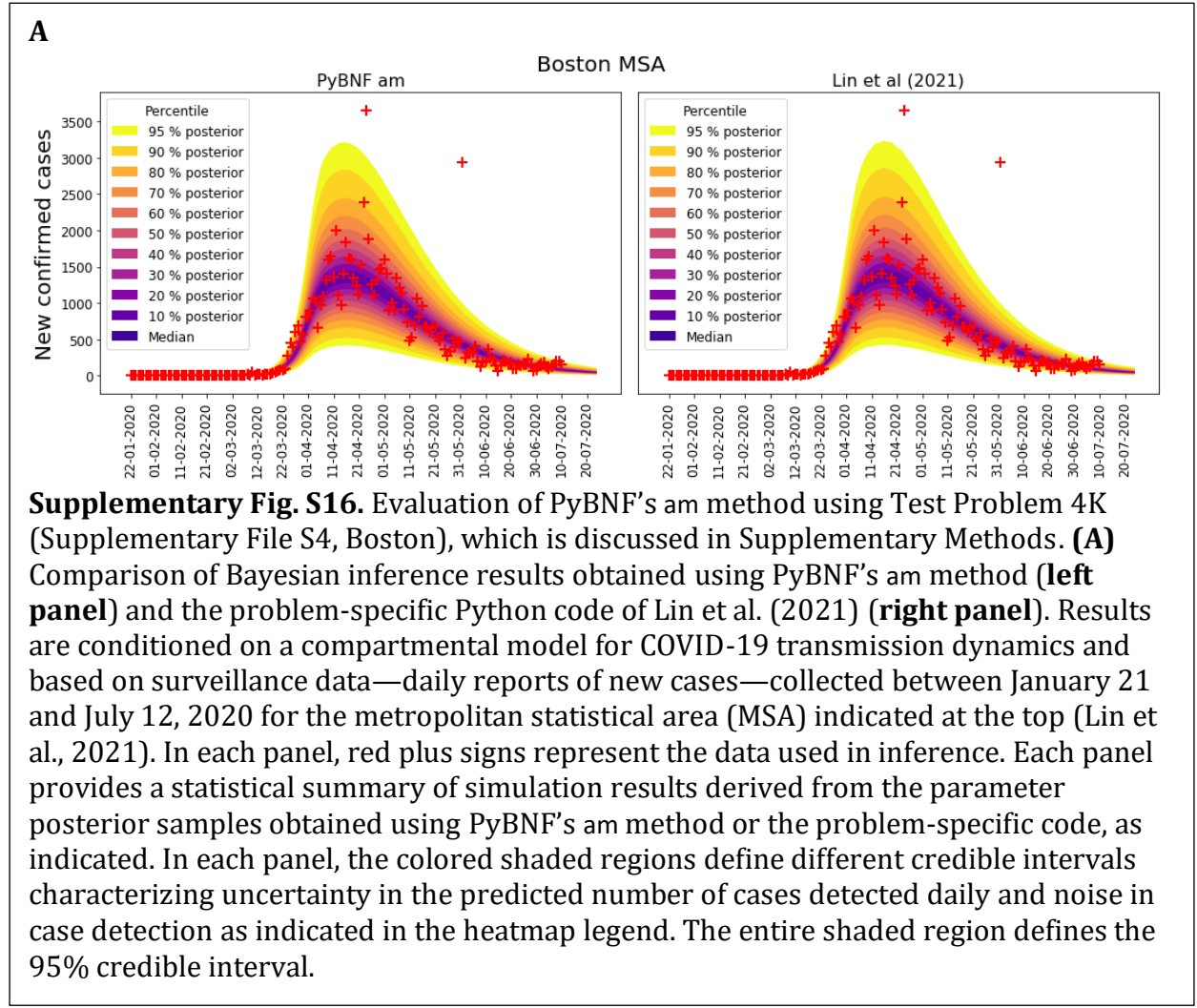

**Supplementary Fig. S16.** Evaluation of PyBNF's ᴀᴍ method using Test Problem 4K (Supplementary File S4, Boston), which is discussed in Supplementary Methods. **(A)** Comparison of Bayesian inference results obtained using PyBNF's ᴀᴍ method (**left panel**) and the problem-specific Python code of Lin et al. (2021) (**right panel**). Results are conditioned on a compartmental model for COVID-19 transmission dynamics and based on surveillance data—daily reports of new cases—collected between January 21 and July 12, 2020 for the metropolitan statistical area (MSA) indicated at the top (Lin et al., 2021). In each panel, red plus signs represent the data used in inference. Each panel provides a statistical summary of simulation results derived from the parameter posterior samples obtained using PyBNF's ᴀᴍ method or the problem-specific code, as indicated. In each panel, the colored shaded regions define different credible intervals characterizing uncertainty in the predicted number of cases detected daily and noise in case detection as indicated in the heatmap legend. The entire shaded region defines the 95% credible interval.

**SUPPLEMENTARY FIGURE S16 – PANEL B**

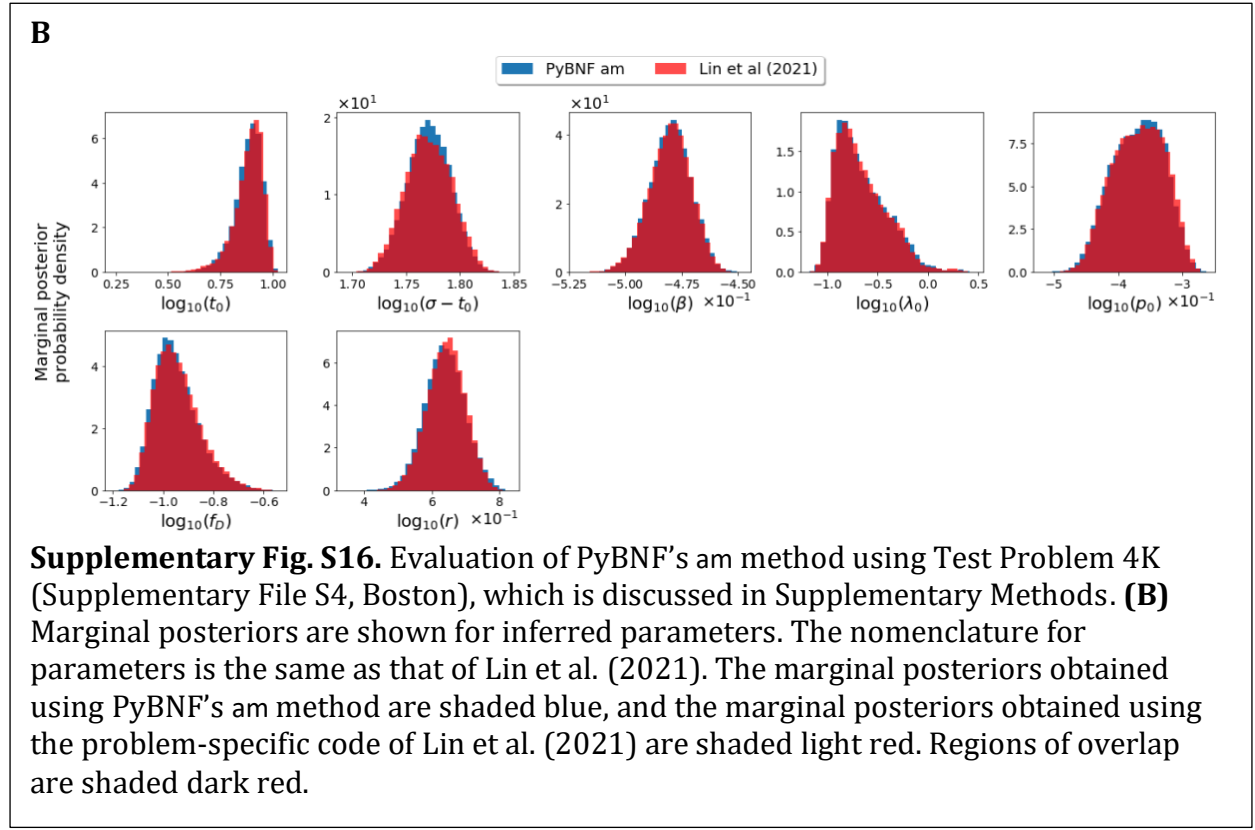

**Supplementary Fig. S16.** Evaluation of PyBNF's am method using Test Problem 4K (Supplementary File S4, Boston), which is discussed in Supplementary Methods. **(B)** Marginal posteriors are shown for inferred parameters. The nomenclature for parameters is the same as that of Lin et al. (2021). The marginal posteriors obtained using PyBNF's am method are shaded blue, and the marginal posteriors obtained using the problem-specific code of Lin et al. (2021) are shaded light red. Regions of overlap are shaded dark red.

**SUPPLEMENTARY FIGURE S16 – PANEL C**

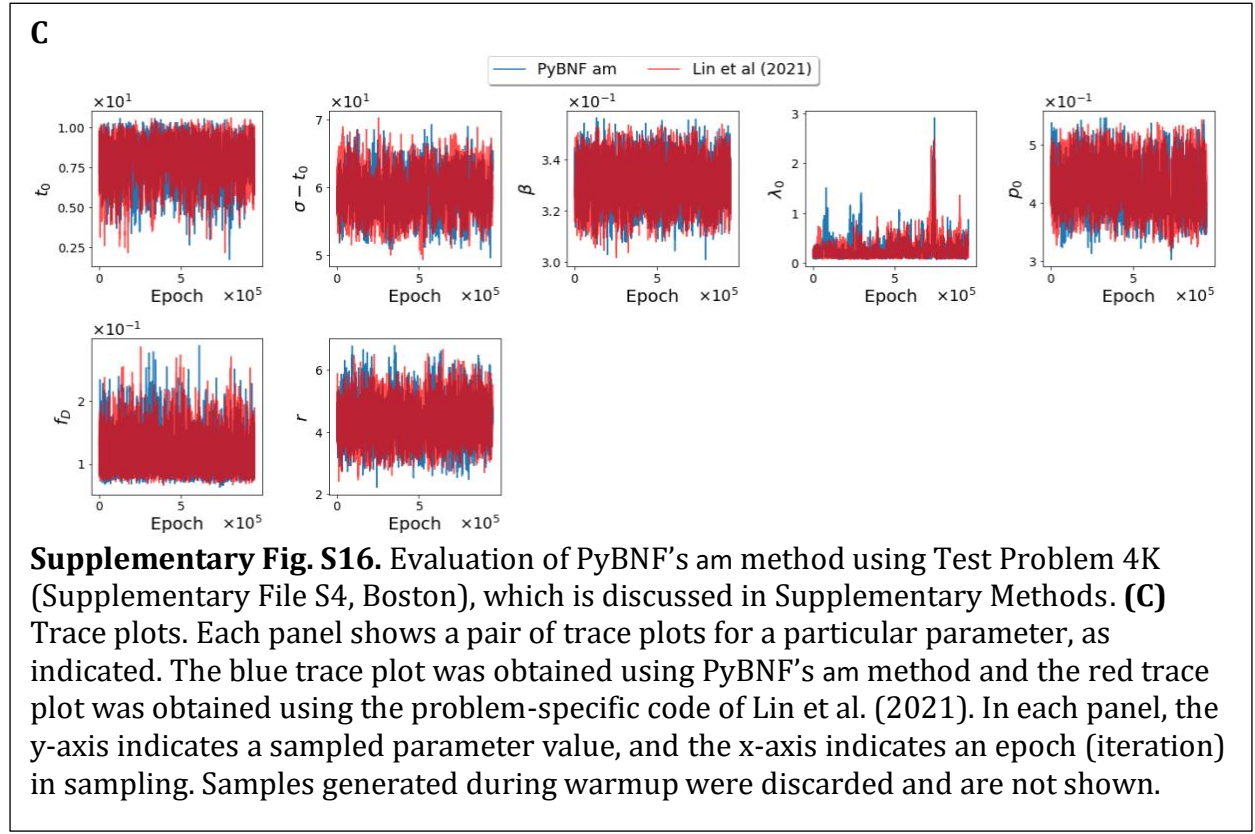

**Supplementary Fig. S16.** Evaluation of PyBNF's am method using Test Problem 4K (Supplementary File S4, Boston), which is discussed in Supplementary Methods. **(C)** Trace plots. Each panel shows a pair of trace plots for a particular parameter, as indicated. The blue trace plot was obtained using PyBNF's am method and the red trace plot was obtained using the problem-specific code of Lin et al. (2021). In each panel, the y-axis indicates a sampled parameter value, and the x-axis indicates an epoch (iteration) in sampling. Samples generated during warmup were discarded and are not shown.

**SUPPLEMENTARY FIGURE S16 – PANEL D**

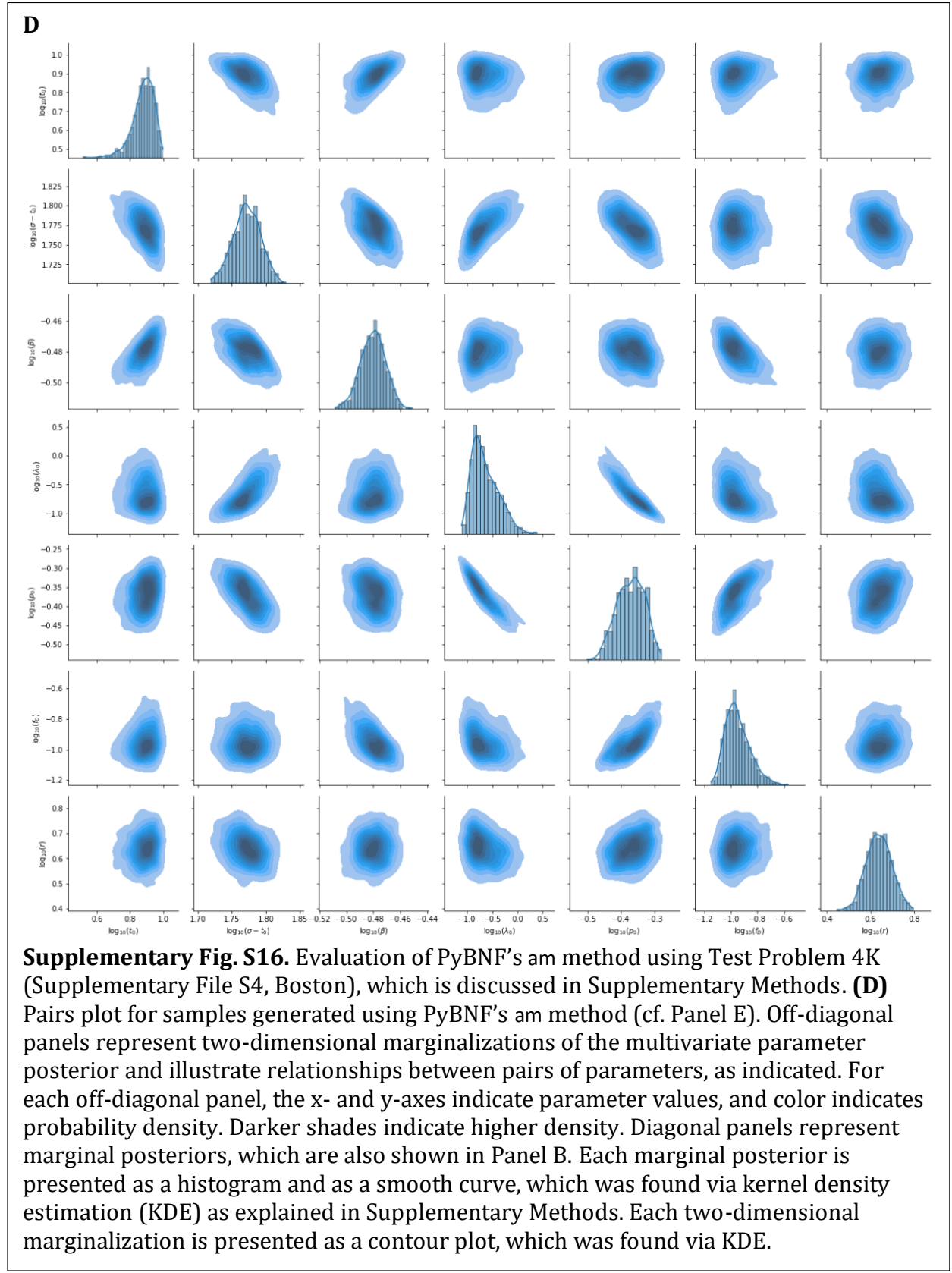

**Supplementary Fig. S16.** Evaluation of PyBNF's am method using Test Problem 4K (Supplementary File S4, Boston), which is discussed in Supplementary Methods. **(D)** Pairs plot for samples generated using PyBNF's am method (cf. Panel E). Off-diagonal panels represent two-dimensional marginalizations of the multivariate parameter posterior and illustrate relationships between pairs of parameters, as indicated. For each off-diagonal panel, the x- and y-axes indicate parameter values, and color indicates probability density. Darker shades indicate higher density. Diagonal panels represent marginal posteriors, which are also shown in Panel B. Each marginal posterior is presented as a histogram and as a smooth curve, which was found via kernel density estimation (KDE) as explained in Supplementary Methods. Each two-dimensional marginalization is presented as a contour plot, which was found via KDE.

**SUPPLEMENTARY FIGURE S16 – PANEL E**

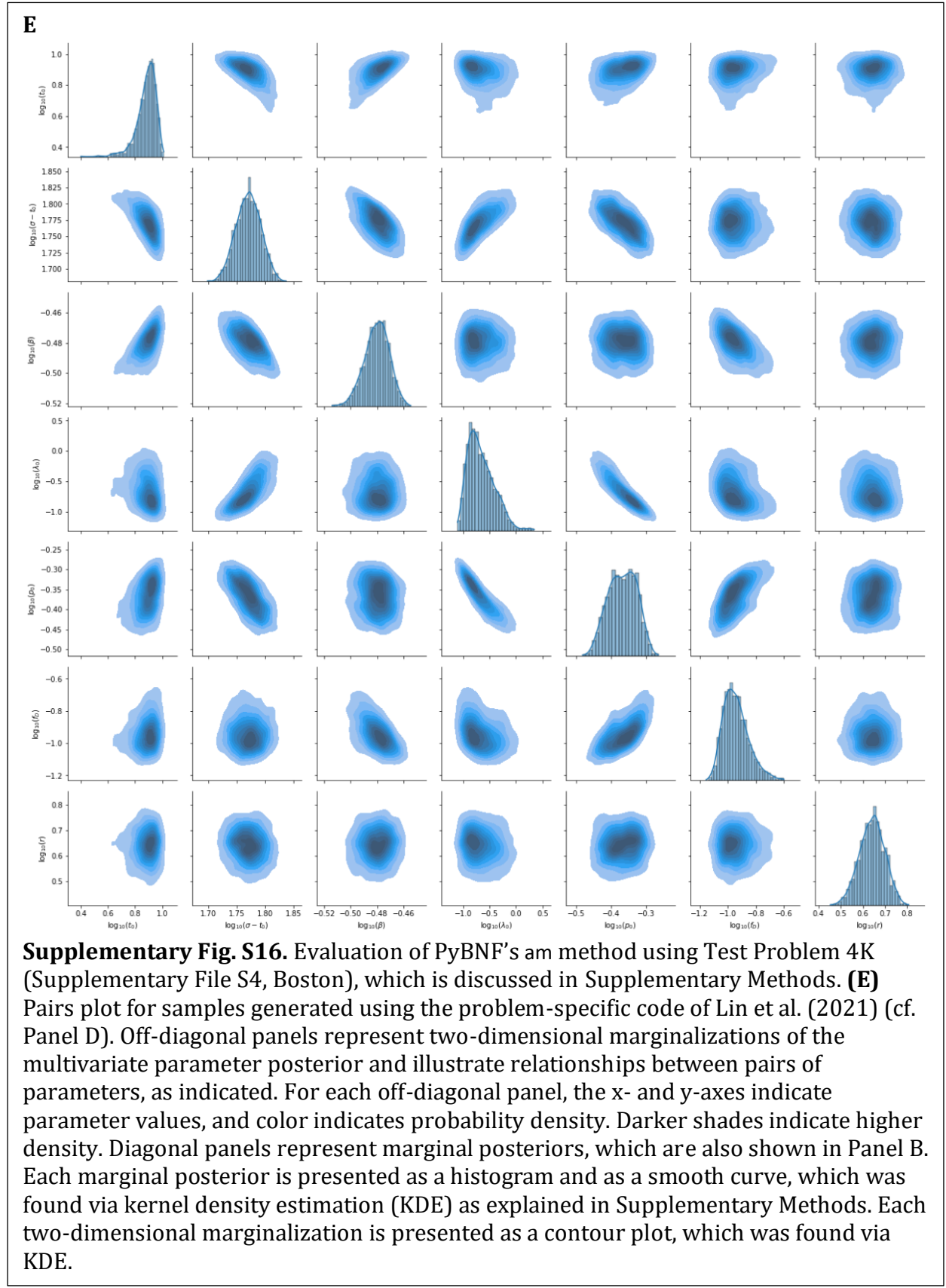

**Supplementary Fig. S16.** Evaluation of PyBNF's am method using Test Problem 4K (Supplementary File S4, Boston), which is discussed in Supplementary Methods. **(E)** Pairs plot for samples generated using the problem-specific code of Lin et al. (2021) (cf. Panel D). Off-diagonal panels represent two-dimensional marginalizations of the multivariate parameter posterior and illustrate relationships between pairs of parameters, as indicated. For each off-diagonal panel, the x- and y-axes indicate parameter values, and color indicates probability density. Darker shades indicate higher density. Diagonal panels represent marginal posteriors, which are also shown in Panel B. Each marginal posterior is presented as a histogram and as a smooth curve, which was found via kernel density estimation (KDE) as explained in Supplementary Methods. Each two-dimensional marginalization is presented as a contour plot, which was found via KDE.

**SUPPLEMENTARY FIGURE S17 – PANEL A**

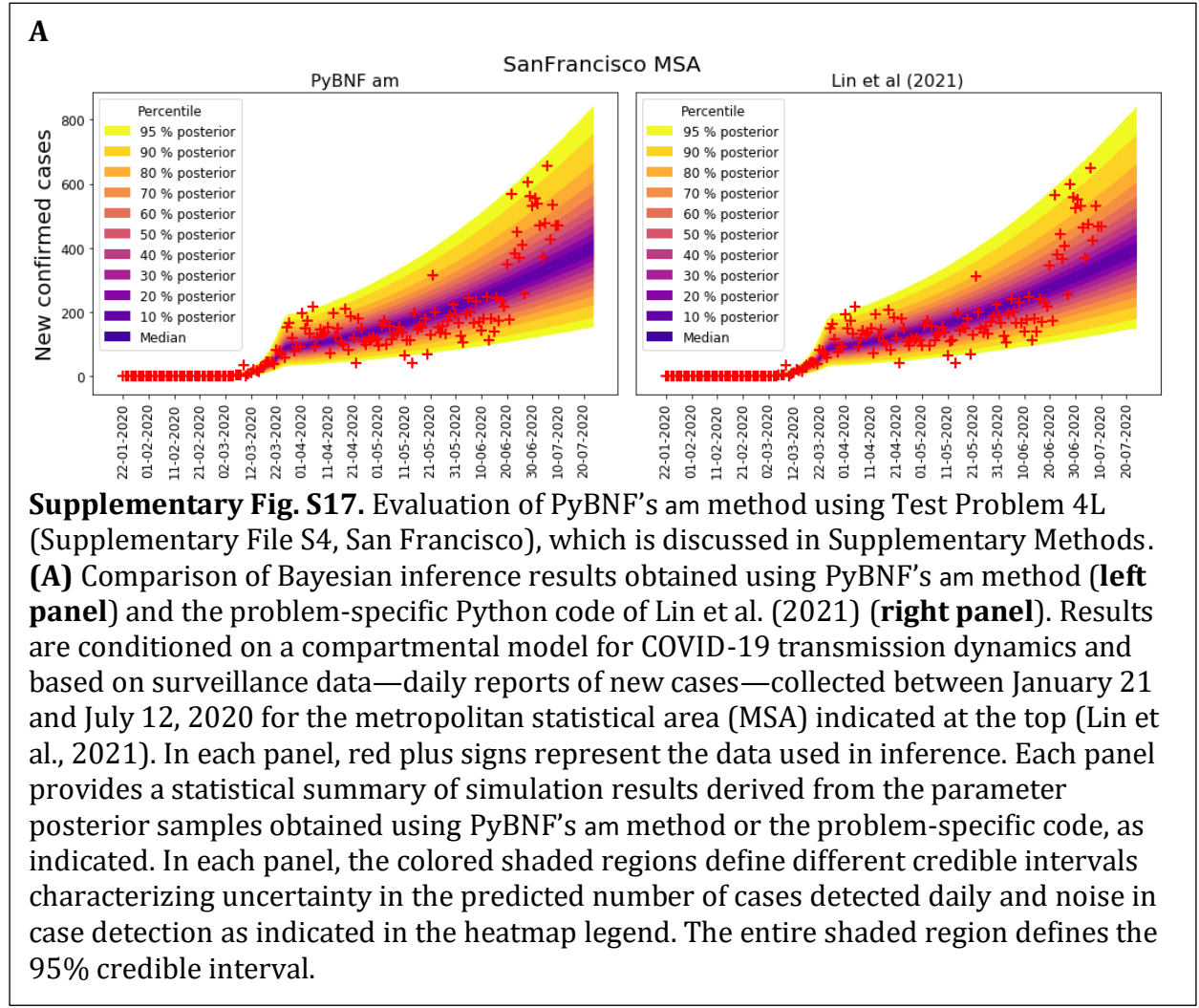

**Supplementary Fig. S17.** Evaluation of PyBNF's ᴀᴍ method using Test Problem 4L (Supplementary File S4, San Francisco), which is discussed in Supplementary Methods. **(A)** Comparison of Bayesian inference results obtained using PyBNF's ᴀᴍ method (**left panel**) and the problem-specific Python code of Lin et al. (2021) (**right panel**). Results are conditioned on a compartmental model for COVID-19 transmission dynamics and based on surveillance data—daily reports of new cases—collected between January 21 and July 12, 2020 for the metropolitan statistical area (MSA) indicated at the top (Lin et al., 2021). In each panel, red plus signs represent the data used in inference. Each panel provides a statistical summary of simulation results derived from the parameter posterior samples obtained using PyBNF's ᴀᴍ method or the problem-specific code, as indicated. In each panel, the colored shaded regions define different credible intervals characterizing uncertainty in the predicted number of cases detected daily and noise in case detection as indicated in the heatmap legend. The entire shaded region defines the 95% credible interval.

**SUPPLEMENTARY FIGURE S17 – PANEL B**

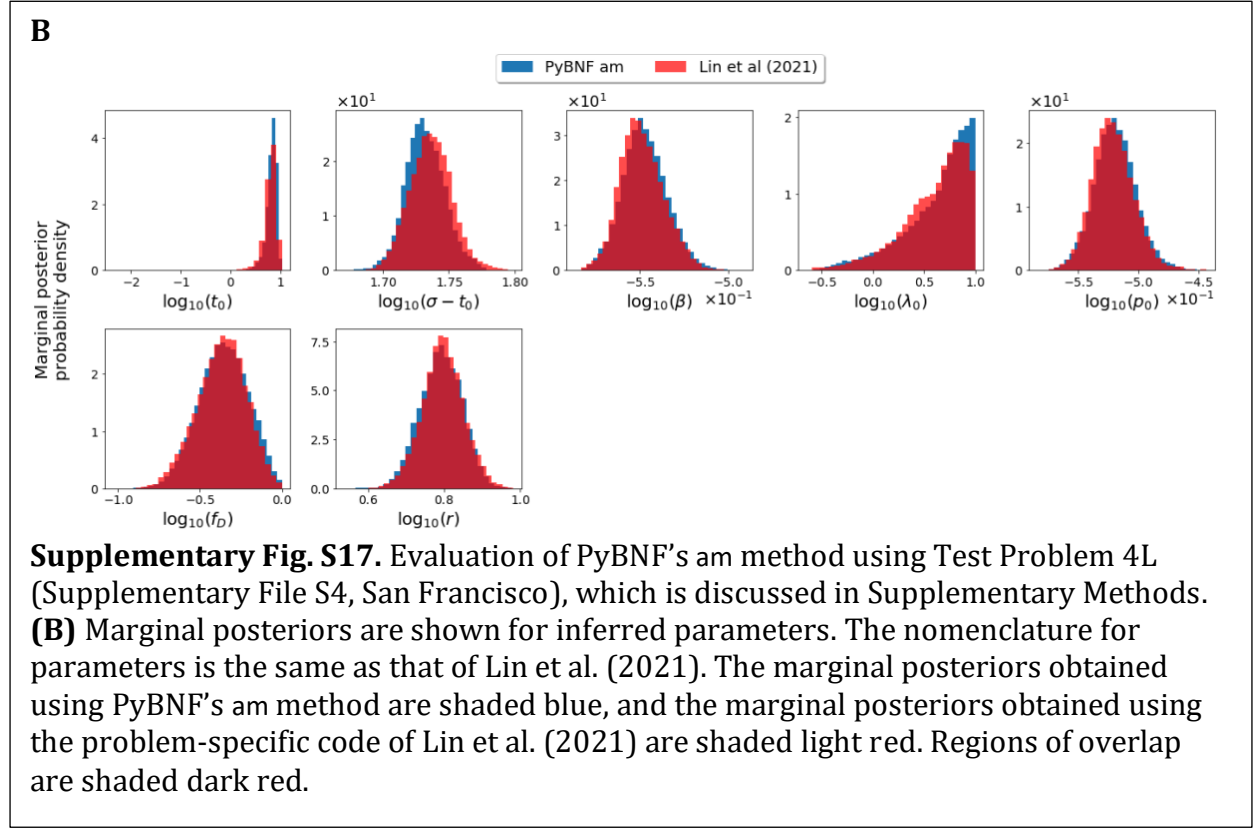

**Supplementary Fig. S17.** Evaluation of PyBNF's am method using Test Problem 4L (Supplementary File S4, San Francisco), which is discussed in Supplementary Methods. **(B)** Marginal posteriors are shown for inferred parameters. The nomenclature for parameters is the same as that of Lin et al. (2021). The marginal posteriors obtained using PyBNF's am method are shaded blue, and the marginal posteriors obtained using the problem-specific code of Lin et al. (2021) are shaded light red. Regions of overlap are shaded dark red.

**SUPPLEMENTARY FIGURE S17 – PANEL C**

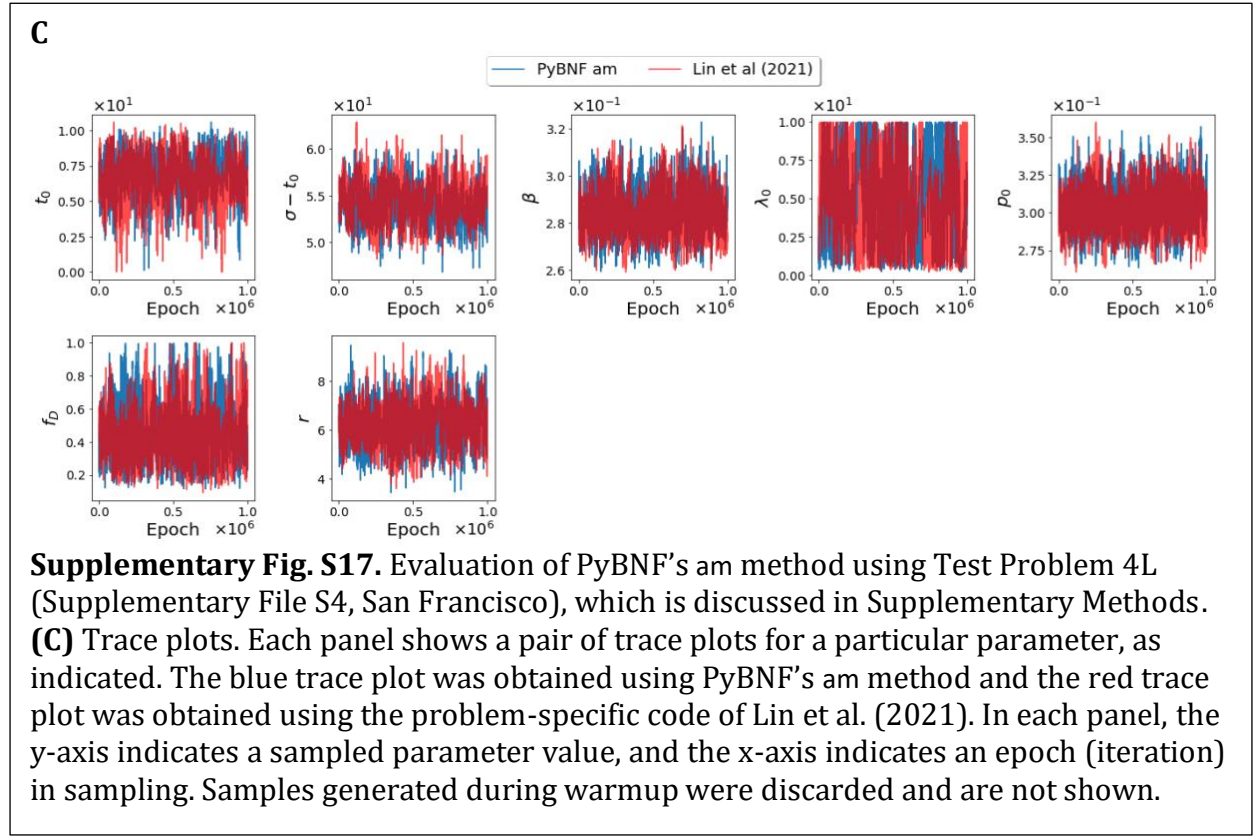

**Supplementary Fig. S17.** Evaluation of PyBNF's am method using Test Problem 4L (Supplementary File S4, San Francisco), which is discussed in Supplementary Methods. **(C)** Trace plots. Each panel shows a pair of trace plots for a particular parameter, as indicated. The blue trace plot was obtained using PyBNF's am method and the red trace plot was obtained using the problem-specific code of Lin et al. (2021). In each panel, the y-axis indicates a sampled parameter value, and the x-axis indicates an epoch (iteration) in sampling. Samples generated during warmup were discarded and are not shown.

**SUPPLEMENTARY FIGURE S17 – PANEL D**

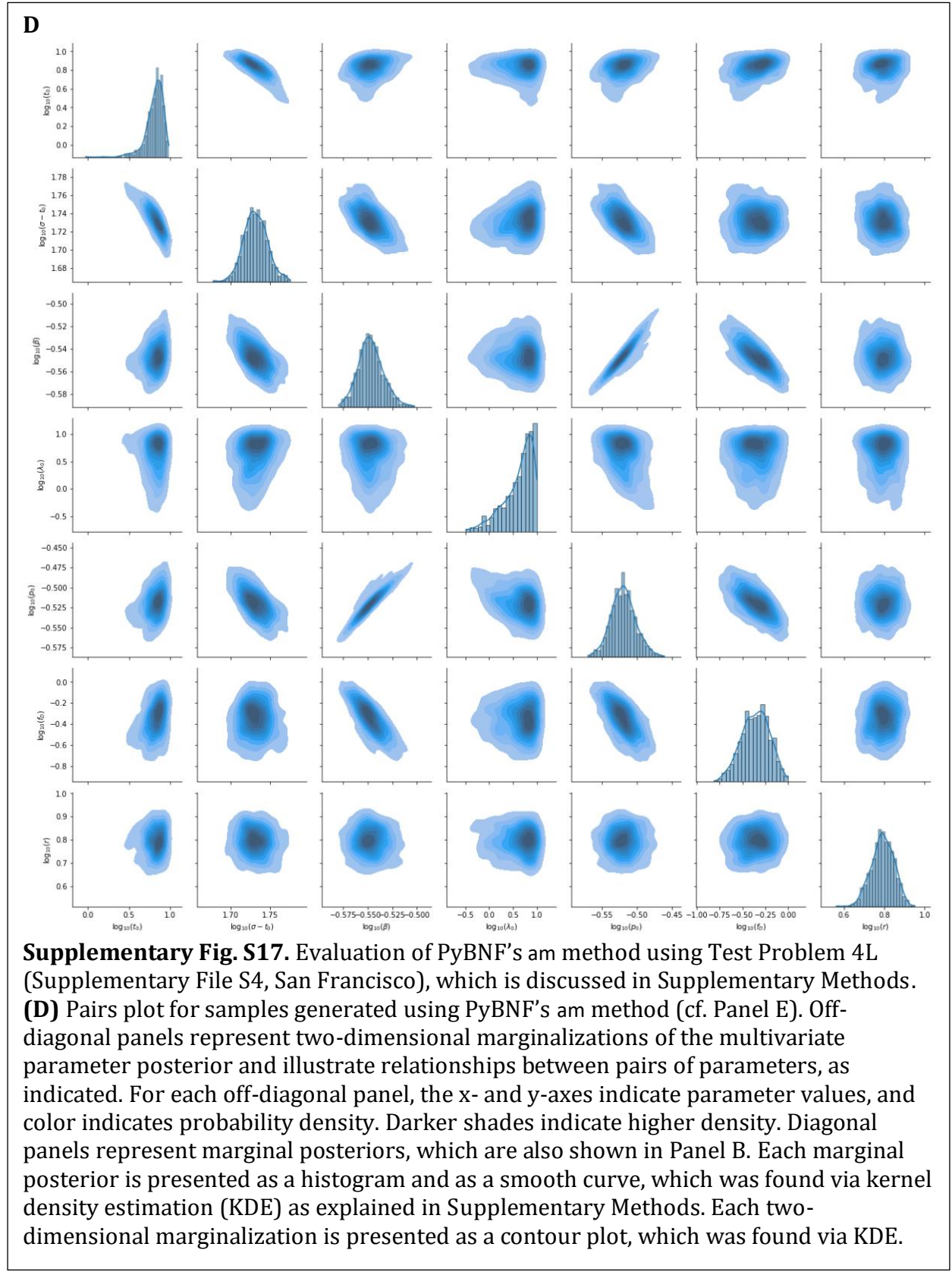

**Supplementary Fig. S17.** Evaluation of PyBNF's am method using Test Problem 4L (Supplementary File S4, San Francisco), which is discussed in Supplementary Methods. **(D)** Pairs plot for samples generated using PyBNF's am method (cf. Panel E). Off-diagonal panels represent two-dimensional marginalizations of the multivariate parameter posterior and illustrate relationships between pairs of parameters, as indicated. For each off-diagonal panel, the x- and y-axes indicate parameter values, and color indicates probability density. Darker shades indicate higher density. Diagonal panels represent marginal posteriors, which are also shown in Panel B. Each marginal posterior is presented as a histogram and as a smooth curve, which was found via kernel density estimation (KDE) as explained in Supplementary Methods. Each two-dimensional marginalization is presented as a contour plot, which was found via KDE.

**SUPPLEMENTARY FIGURE S17 – PANEL E**

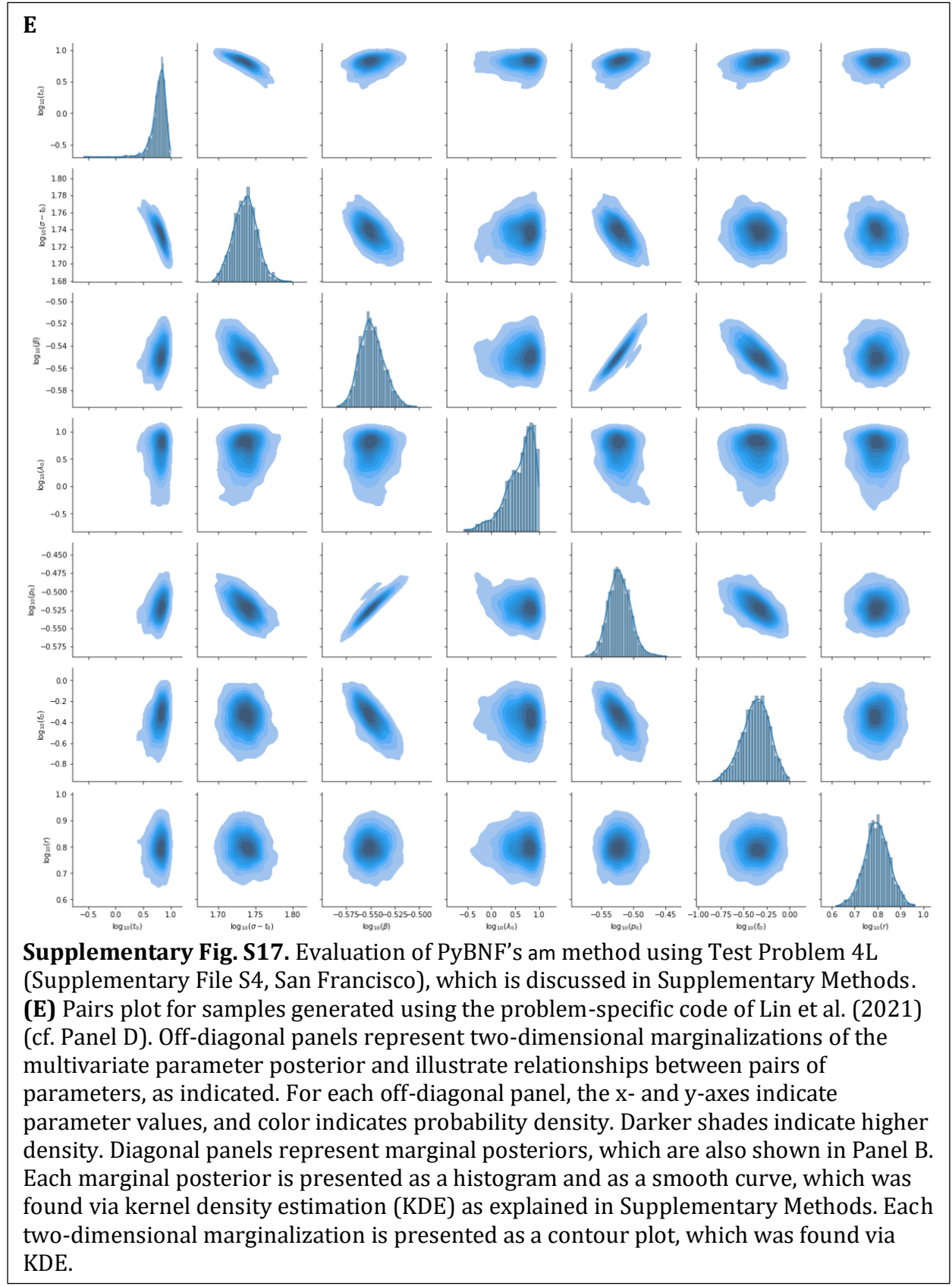

**Supplementary Fig. S17.** Evaluation of PyBNF's am method using Test Problem 4L (Supplementary File S4, San Francisco), which is discussed in Supplementary Methods. **(E)** Pairs plot for samples generated using the problem-specific code of Lin et al. (2021) (cf. Panel D). Off-diagonal panels represent two-dimensional marginalizations of the multivariate parameter posterior and illustrate relationships between pairs of parameters, as indicated. For each off-diagonal panel, the x- and y-axes indicate parameter values, and color indicates probability density. Darker shades indicate higher density. Diagonal panels represent marginal posteriors, which are also shown in Panel B. Each marginal posterior is presented as a histogram and as a smooth curve, which was found via kernel density estimation (KDE) as explained in Supplementary Methods. Each two-dimensional marginalization is presented as a contour plot, which was found via KDE.

**SUPPLEMENTARY FIGURE S18 – PANEL A**

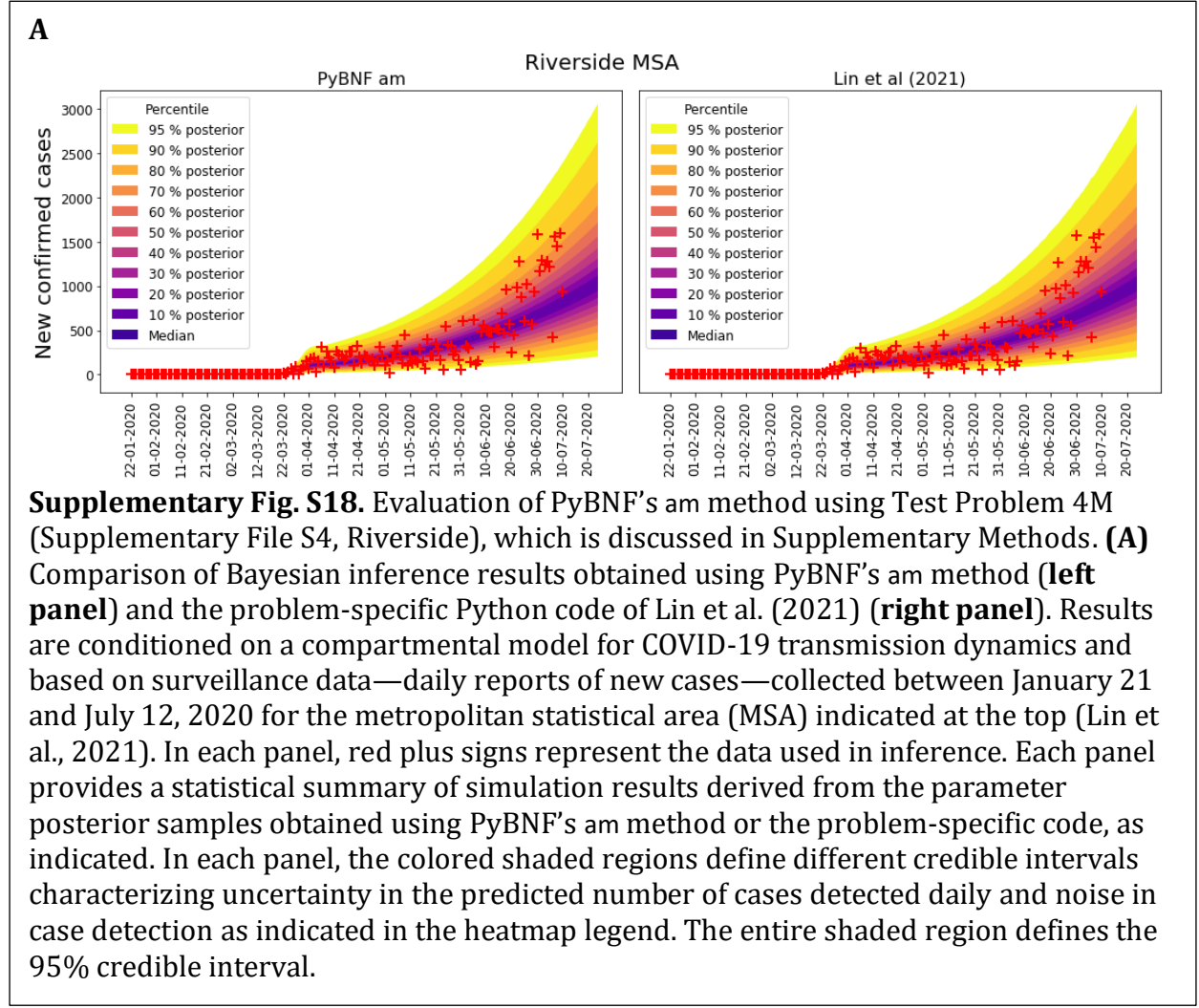

**Supplementary Fig. S18.** Evaluation of PyBNF's am method using Test Problem 4M (Supplementary File S4, Riverside), which is discussed in Supplementary Methods. **(A)** Comparison of Bayesian inference results obtained using PyBNF's am method (**left panel**) and the problem-specific Python code of Lin et al. (2021) (**right panel**). Results are conditioned on a compartmental model for COVID-19 transmission dynamics and based on surveillance data—daily reports of new cases—collected between January 21 and July 12, 2020 for the metropolitan statistical area (MSA) indicated at the top (Lin et al., 2021). In each panel, red plus signs represent the data used in inference. Each panel provides a statistical summary of simulation results derived from the parameter posterior samples obtained using PyBNF's am method or the problem-specific code, as indicated. In each panel, the colored shaded regions define different credible intervals characterizing uncertainty in the predicted number of cases detected daily and noise in case detection as indicated in the heatmap legend. The entire shaded region defines the 95% credible interval.

**SUPPLEMENTARY FIGURE S18 – PANEL B**

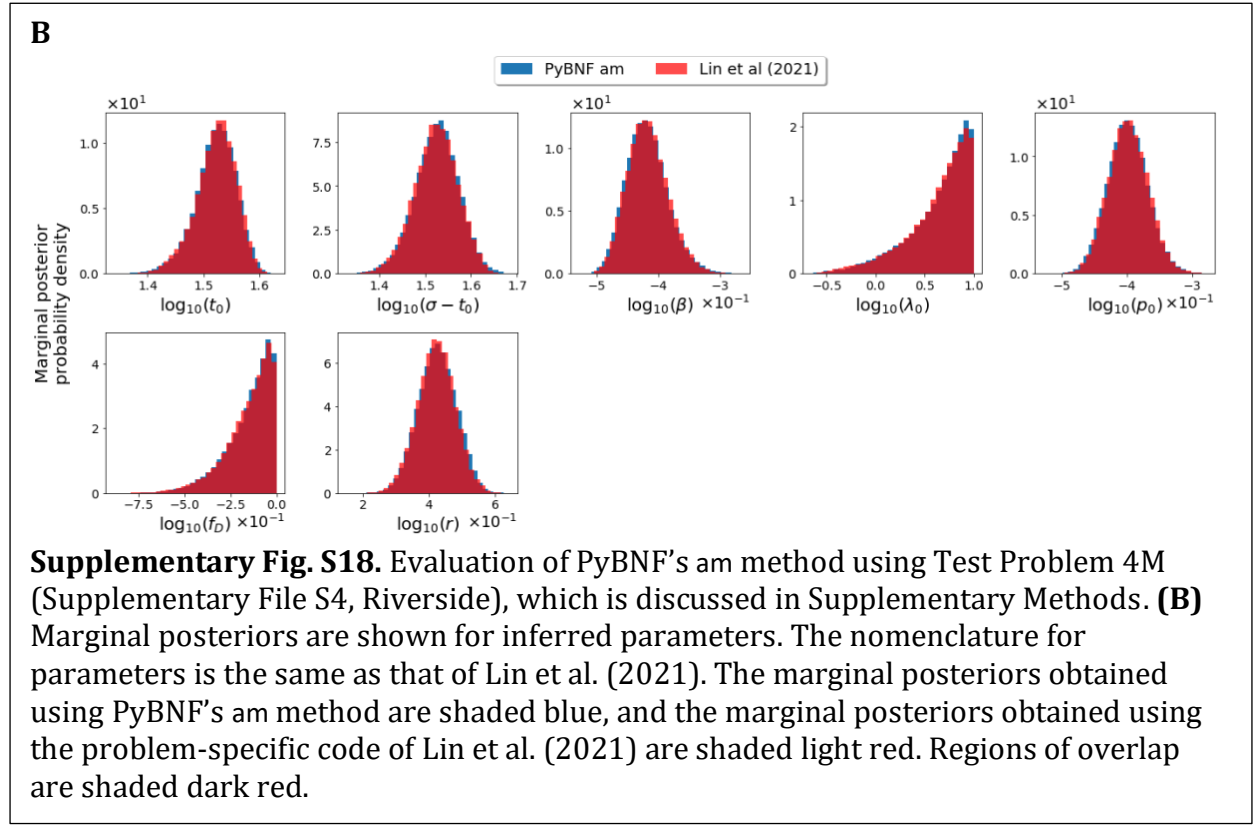

**Supplementary Fig. S18.** Evaluation of PyBNF's am method using Test Problem 4M (Supplementary File S4, Riverside), which is discussed in Supplementary Methods. **(B)** Marginal posteriors are shown for inferred parameters. The nomenclature for parameters is the same as that of Lin et al. (2021). The marginal posteriors obtained using PyBNF's am method are shaded blue, and the marginal posteriors obtained using the problem-specific code of Lin et al. (2021) are shaded light red. Regions of overlap are shaded dark red.

**SUPPLEMENTARY FIGURE S18 – PANEL C**

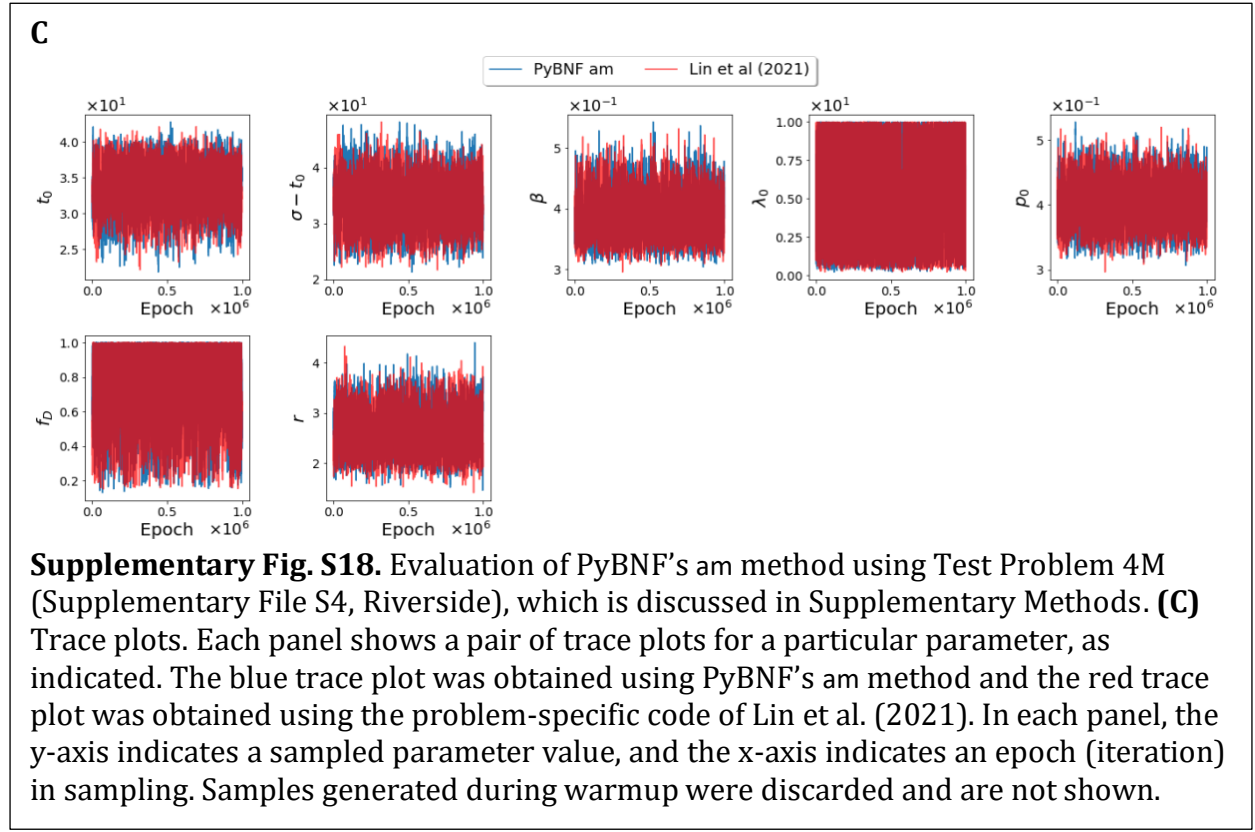

**Supplementary Fig. S18.** Evaluation of PyBNF's am method using Test Problem 4M (Supplementary File S4, Riverside), which is discussed in Supplementary Methods. **(C)** Trace plots. Each panel shows a pair of trace plots for a particular parameter, as indicated. The blue trace plot was obtained using PyBNF's am method and the red trace plot was obtained using the problem-specific code of Lin et al. (2021). In each panel, the y-axis indicates a sampled parameter value, and the x-axis indicates an epoch (iteration) in sampling. Samples generated during warmup were discarded and are not shown.

**SUPPLEMENTARY FIGURE S18 – PANEL D**

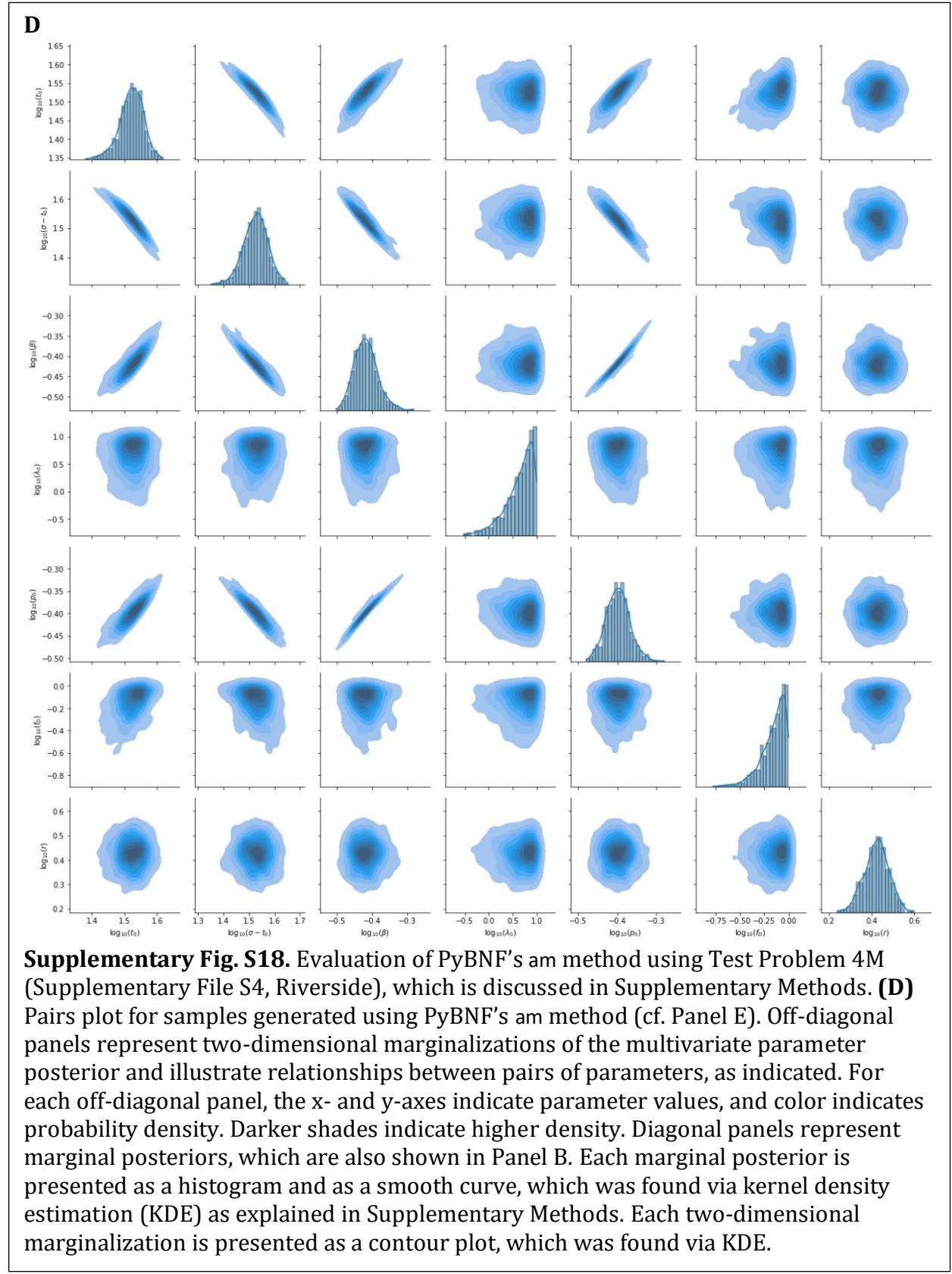

**Supplementary Fig. S18.** Evaluation of PyBNF's am method using Test Problem 4M (Supplementary File S4, Riverside), which is discussed in Supplementary Methods. **(D)** Pairs plot for samples generated using PyBNF's am method (cf. Panel E). Off-diagonal panels represent two-dimensional marginalizations of the multivariate parameter posterior and illustrate relationships between pairs of parameters, as indicated. For each off-diagonal panel, the x- and y-axes indicate parameter values, and color indicates probability density. Darker shades indicate higher density. Diagonal panels represent marginal posteriors, which are also shown in Panel B. Each marginal posterior is presented as a histogram and as a smooth curve, which was found via kernel density estimation (KDE) as explained in Supplementary Methods. Each two-dimensional marginalization is presented as a contour plot, which was found via KDE.

**SUPPLEMENTARY FIGURE S18 – PANEL E**

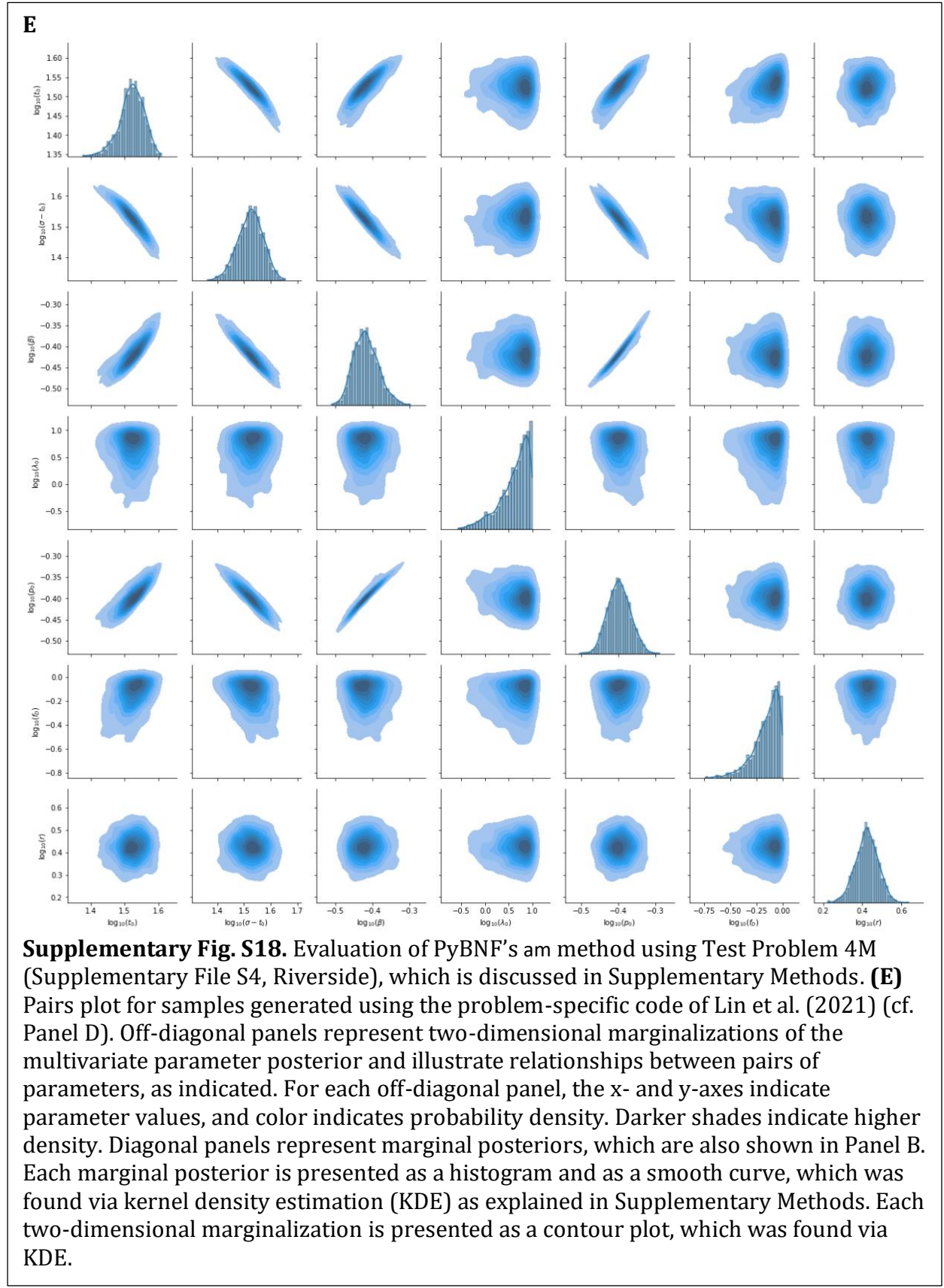

**Supplementary Fig. S18.** Evaluation of PyBNF's am method using Test Problem 4M (Supplementary File S4, Riverside), which is discussed in Supplementary Methods. **(E)** Pairs plot for samples generated using the problem-specific code of Lin et al. (2021) (cf. Panel D). Off-diagonal panels represent two-dimensional marginalizations of the multivariate parameter posterior and illustrate relationships between pairs of parameters, as indicated. For each off-diagonal panel, the x- and y-axes indicate parameter values, and color indicates probability density. Darker shades indicate higher density. Diagonal panels represent marginal posteriors, which are also shown in Panel B. Each marginal posterior is presented as a histogram and as a smooth curve, which was found via kernel density estimation (KDE) as explained in Supplementary Methods. Each two-dimensional marginalization is presented as a contour plot, which was found via KDE.

**SUPPLEMENTARY FIGURE S19 – PANEL A**

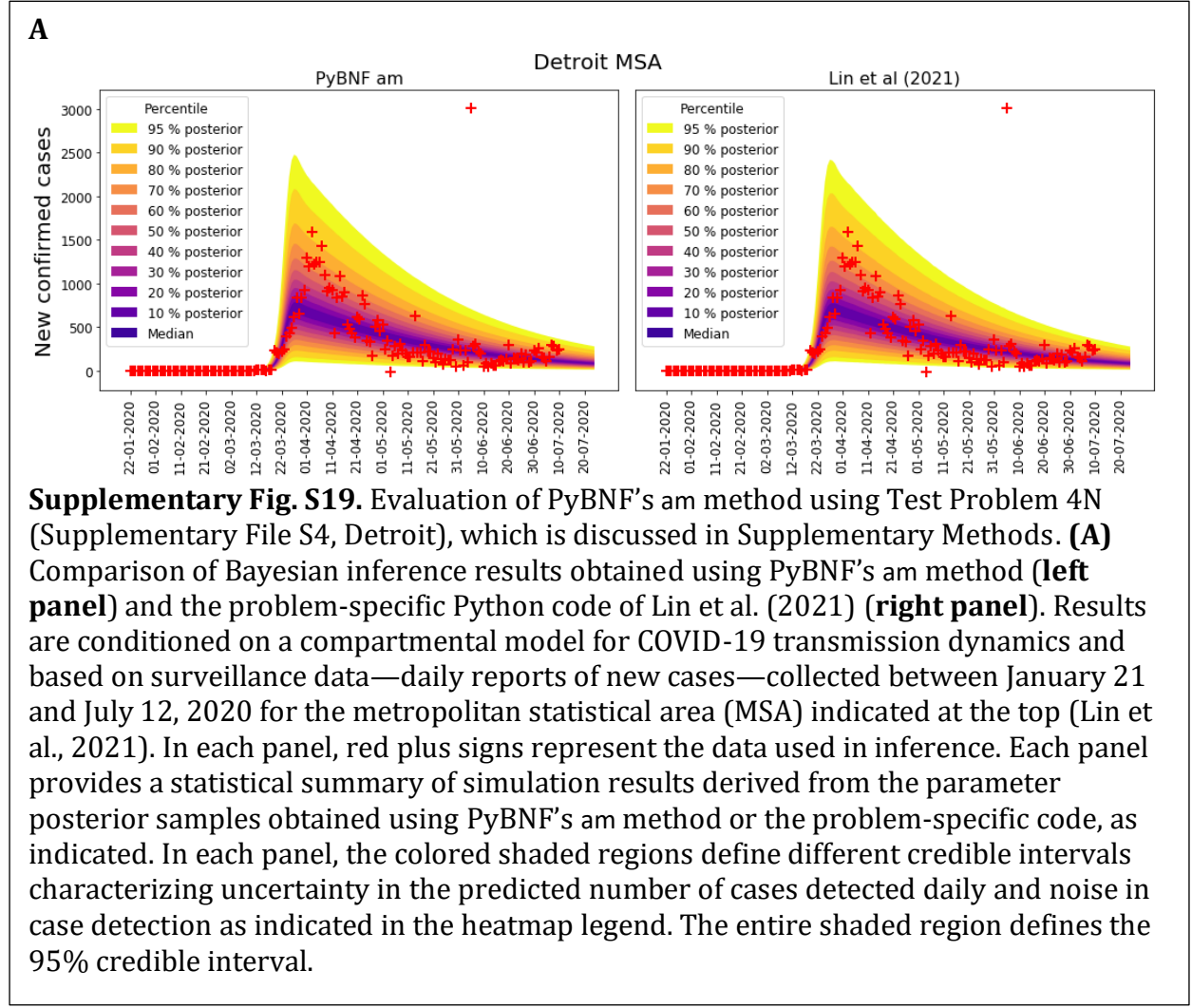

**Supplementary Fig. S19.** Evaluation of PyBNF's ᴀᴍ method using Test Problem 4N (Supplementary File S4, Detroit), which is discussed in Supplementary Methods. **(A)** Comparison of Bayesian inference results obtained using PyBNF's ᴀᴍ method (**left panel**) and the problem-specific Python code of Lin et al. (2021) (**right panel**). Results are conditioned on a compartmental model for COVID-19 transmission dynamics and based on surveillance data—daily reports of new cases—collected between January 21 and July 12, 2020 for the metropolitan statistical area (MSA) indicated at the top (Lin et al., 2021). In each panel, red plus signs represent the data used in inference. Each panel provides a statistical summary of simulation results derived from the parameter posterior samples obtained using PyBNF's ᴀᴍ method or the problem-specific code, as indicated. In each panel, the colored shaded regions define different credible intervals characterizing uncertainty in the predicted number of cases detected daily and noise in case detection as indicated in the heatmap legend. The entire shaded region defines the 95% credible interval.

**SUPPLEMENTARY FIGURE S19 – PANEL B**

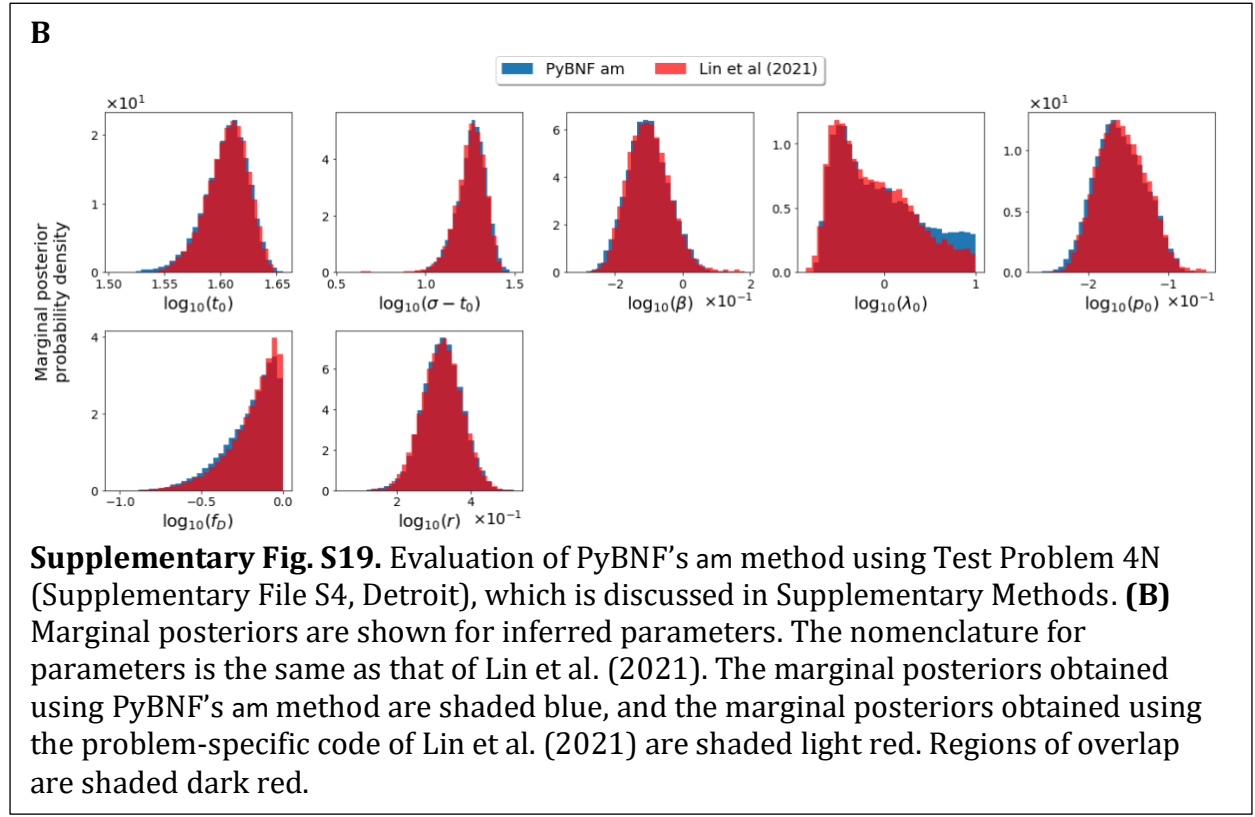

**Supplementary Fig. S19.** Evaluation of PyBNF's am method using Test Problem 4N (Supplementary File S4, Detroit), which is discussed in Supplementary Methods. **(B)** Marginal posteriors are shown for inferred parameters. The nomenclature for parameters is the same as that of Lin et al. (2021). The marginal posteriors obtained using PyBNF's am method are shaded blue, and the marginal posteriors obtained using the problem-specific code of Lin et al. (2021) are shaded light red. Regions of overlap are shaded dark red.

**SUPPLEMENTARY FIGURE S19 – PANEL C**

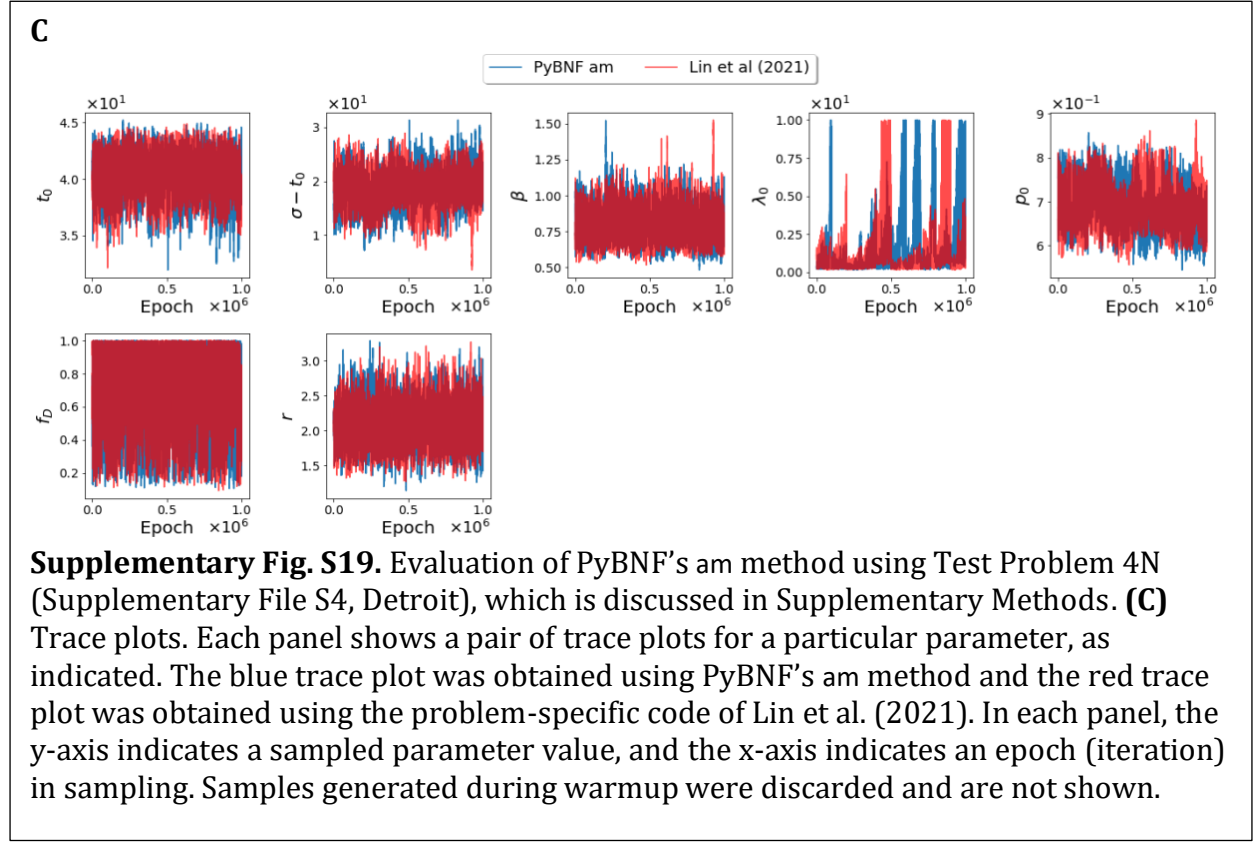

**Supplementary Fig. S19.** Evaluation of PyBNF's am method using Test Problem 4N (Supplementary File S4, Detroit), which is discussed in Supplementary Methods. **(C)** Trace plots. Each panel shows a pair of trace plots for a particular parameter, as indicated. The blue trace plot was obtained using PyBNF's am method and the red trace plot was obtained using the problem-specific code of Lin et al. (2021). In each panel, the y-axis indicates a sampled parameter value, and the x-axis indicates an epoch (iteration) in sampling. Samples generated during warmup were discarded and are not shown.

**SUPPLEMENTARY FIGURE S19 – PANEL D**

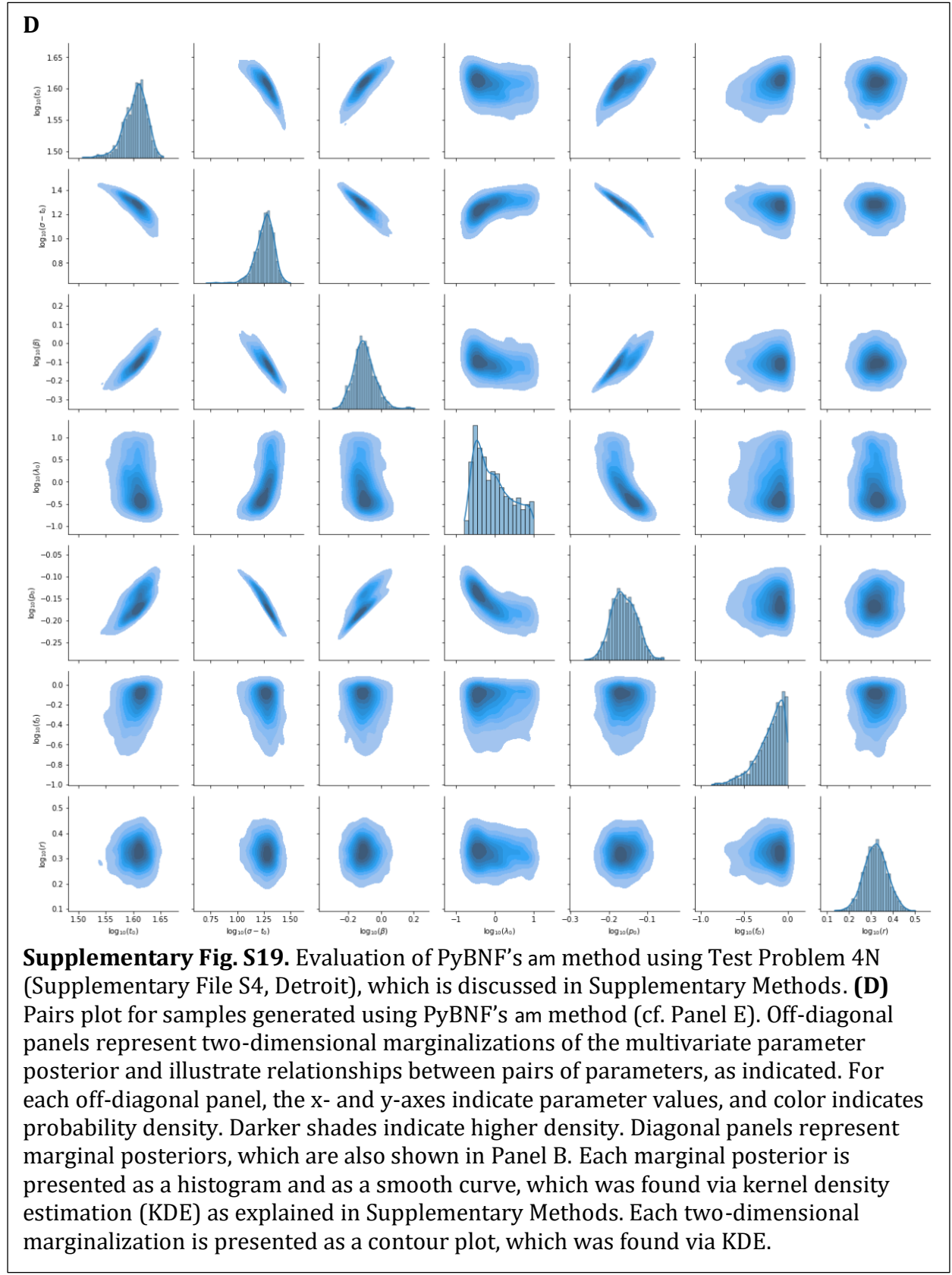

**Supplementary Fig. S19.** Evaluation of PyBNF's am method using Test Problem 4N (Supplementary File S4, Detroit), which is discussed in Supplementary Methods. **(D)** Pairs plot for samples generated using PyBNF's am method (cf. Panel E). Off-diagonal panels represent two-dimensional marginalizations of the multivariate parameter posterior and illustrate relationships between pairs of parameters, as indicated. For each off-diagonal panel, the x- and y-axes indicate parameter values, and color indicates probability density. Darker shades indicate higher density. Diagonal panels represent marginal posteriors, which are also shown in Panel B. Each marginal posterior is presented as a histogram and as a smooth curve, which was found via kernel density estimation (KDE) as explained in Supplementary Methods. Each two-dimensional marginalization is presented as a contour plot, which was found via KDE.

**SUPPLEMENTARY FIGURE S19 – PANEL E**

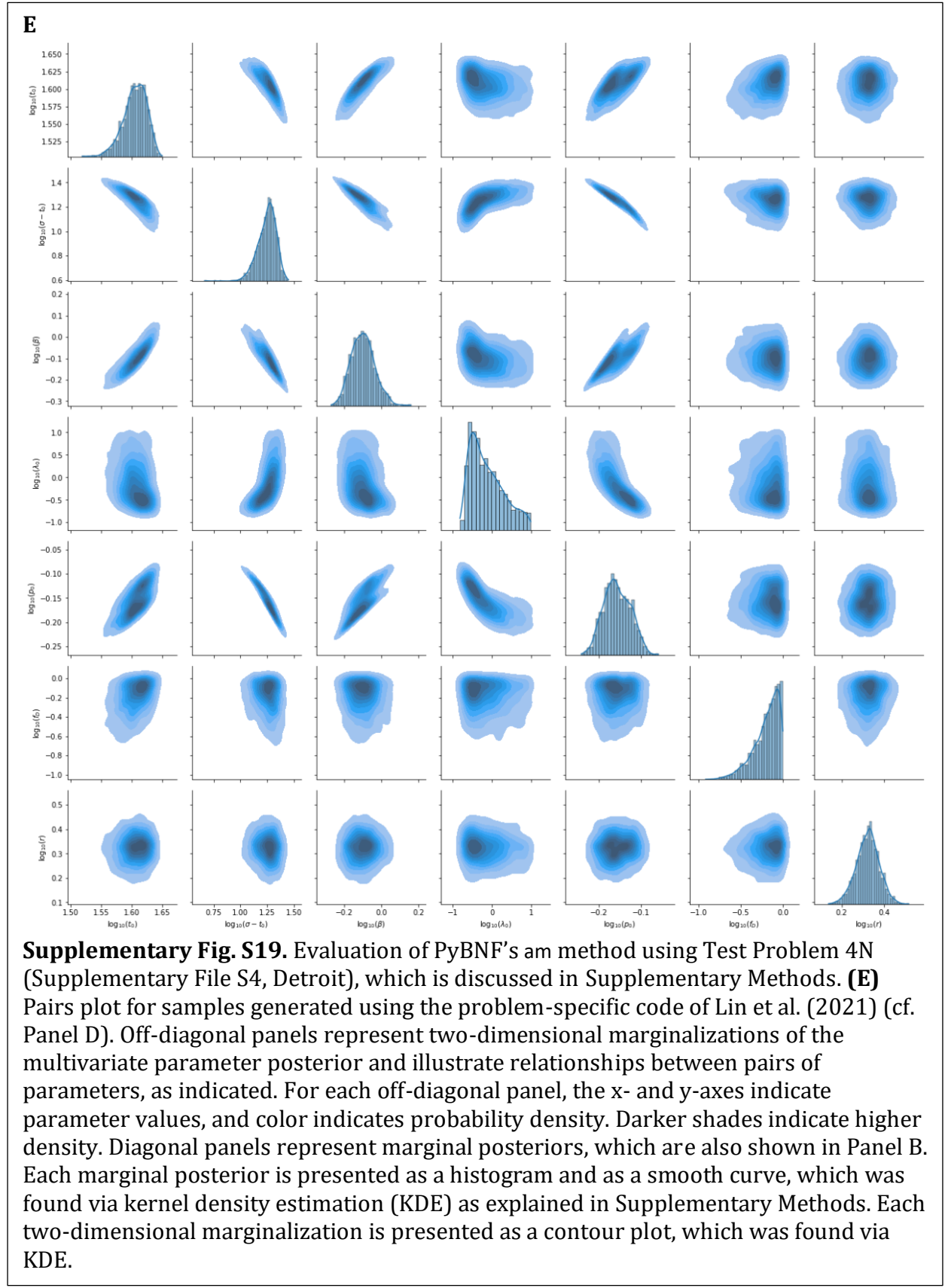

**Supplementary Fig. S19.** Evaluation of PyBNF's am method using Test Problem 4N (Supplementary File S4, Detroit), which is discussed in Supplementary Methods. **(E)** Pairs plot for samples generated using the problem-specific code of Lin et al. (2021) (cf. Panel D). Off-diagonal panels represent two-dimensional marginalizations of the multivariate parameter posterior and illustrate relationships between pairs of parameters, as indicated. For each off-diagonal panel, the x- and y-axes indicate parameter values, and color indicates probability density. Darker shades indicate higher density. Diagonal panels represent marginal posteriors, which are also shown in Panel B. Each marginal posterior is presented as a histogram and as a smooth curve, which was found via kernel density estimation (KDE) as explained in Supplementary Methods. Each two-dimensional marginalization is presented as a contour plot, which was found via KDE.

**SUPPLEMENTARY FIGURE S20 – PANEL A**

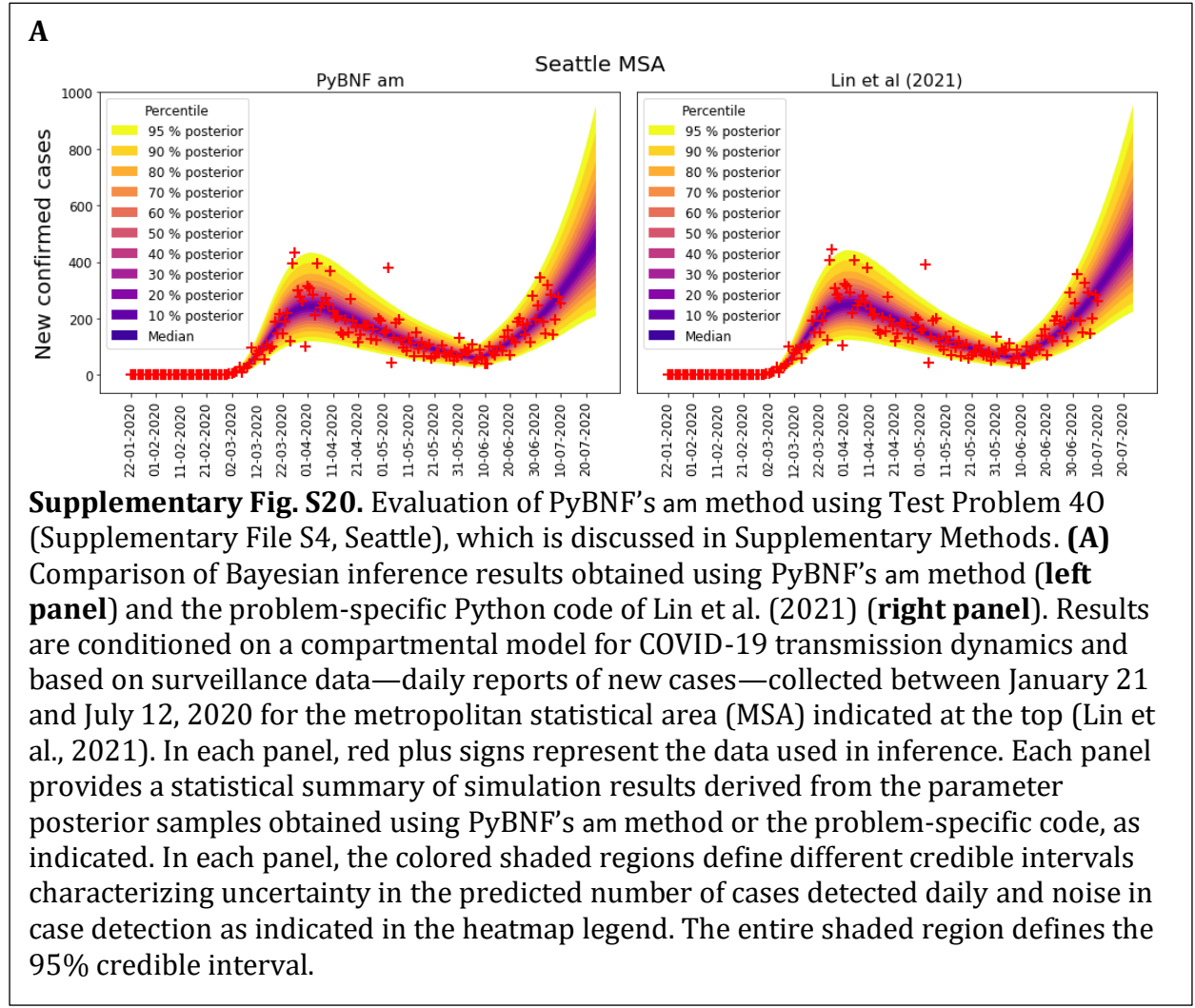

**Supplementary Fig. S20.** Evaluation of PyBNF's ᴀᴍ method using Test Problem 4O (Supplementary File S4, Seattle), which is discussed in Supplementary Methods. **(A)** Comparison of Bayesian inference results obtained using PyBNF's ᴀᴍ method (**left panel**) and the problem-specific Python code of Lin et al. (2021) (**right panel**). Results are conditioned on a compartmental model for COVID-19 transmission dynamics and based on surveillance data—daily reports of new cases—collected between January 21 and July 12, 2020 for the metropolitan statistical area (MSA) indicated at the top (Lin et al., 2021). In each panel, red plus signs represent the data used in inference. Each panel provides a statistical summary of simulation results derived from the parameter posterior samples obtained using PyBNF's ᴀᴍ method or the problem-specific code, as indicated. In each panel, the colored shaded regions define different credible intervals characterizing uncertainty in the predicted number of cases detected daily and noise in case detection as indicated in the heatmap legend. The entire shaded region defines the 95% credible interval.

**SUPPLEMENTARY FIGURE S20 – PANEL B**

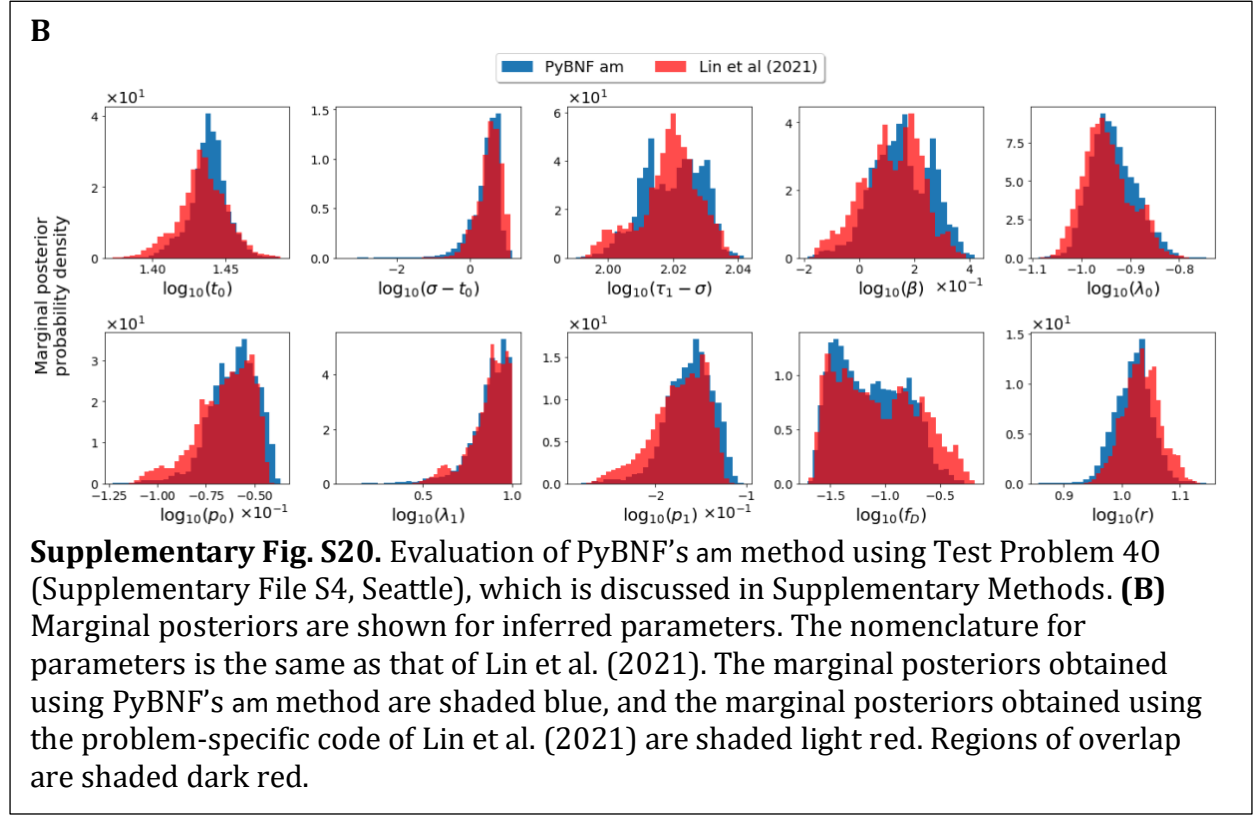

**Supplementary Fig. S20.** Evaluation of PyBNF's am method using Test Problem 4O (Supplementary File S4, Seattle), which is discussed in Supplementary Methods. **(B)** Marginal posteriors are shown for inferred parameters. The nomenclature for parameters is the same as that of Lin et al. (2021). The marginal posteriors obtained using PyBNF's am method are shaded blue, and the marginal posteriors obtained using the problem-specific code of Lin et al. (2021) are shaded light red. Regions of overlap are shaded dark red.

**SUPPLEMENTARY FIGURE S20 – PANEL C**

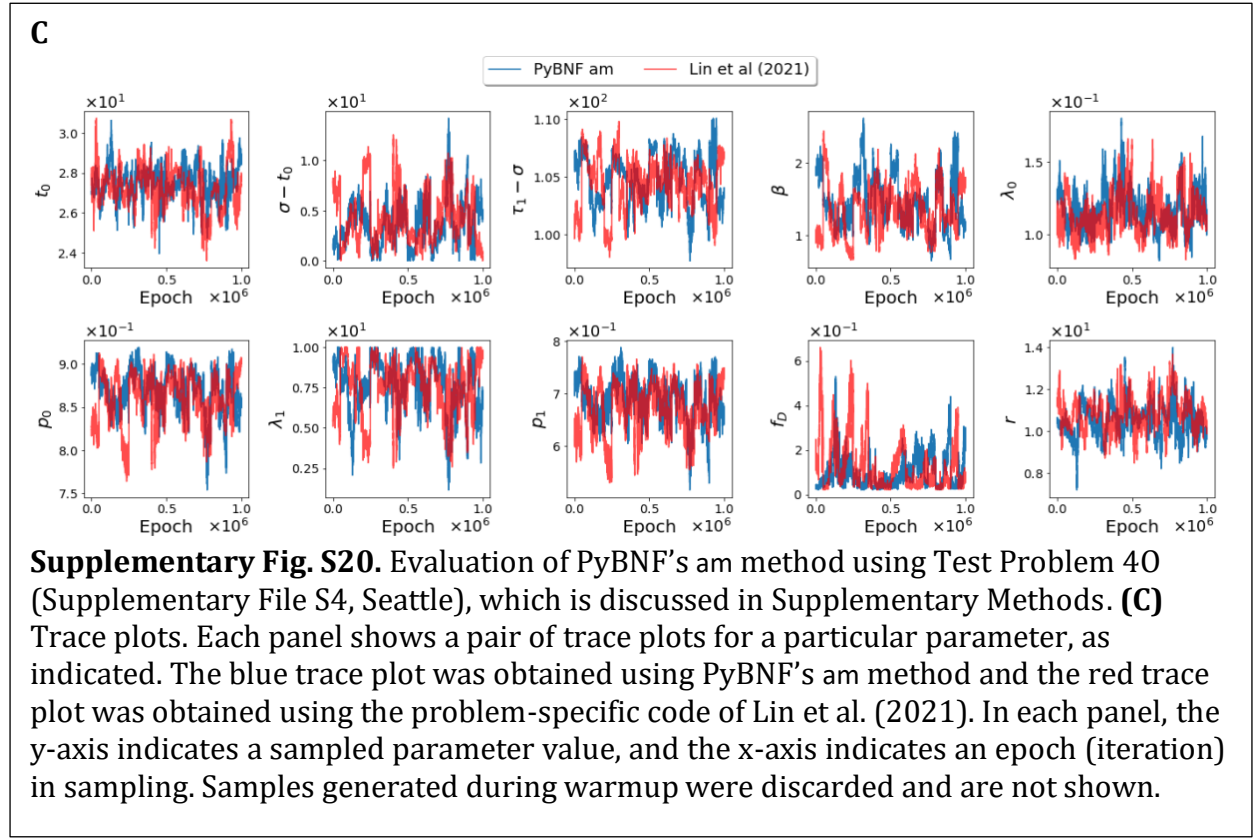

**Supplementary Fig. S20.** Evaluation of PyBNF's am method using Test Problem 4O (Supplementary File S4, Seattle), which is discussed in Supplementary Methods. **(C)** Trace plots. Each panel shows a pair of trace plots for a particular parameter, as indicated. The blue trace plot was obtained using PyBNF's am method and the red trace plot was obtained using the problem-specific code of Lin et al. (2021). In each panel, the y-axis indicates a sampled parameter value, and the x-axis indicates an epoch (iteration) in sampling. Samples generated during warmup were discarded and are not shown.

**SUPPLEMENTARY FIGURE S20 – PANEL D**

**D**

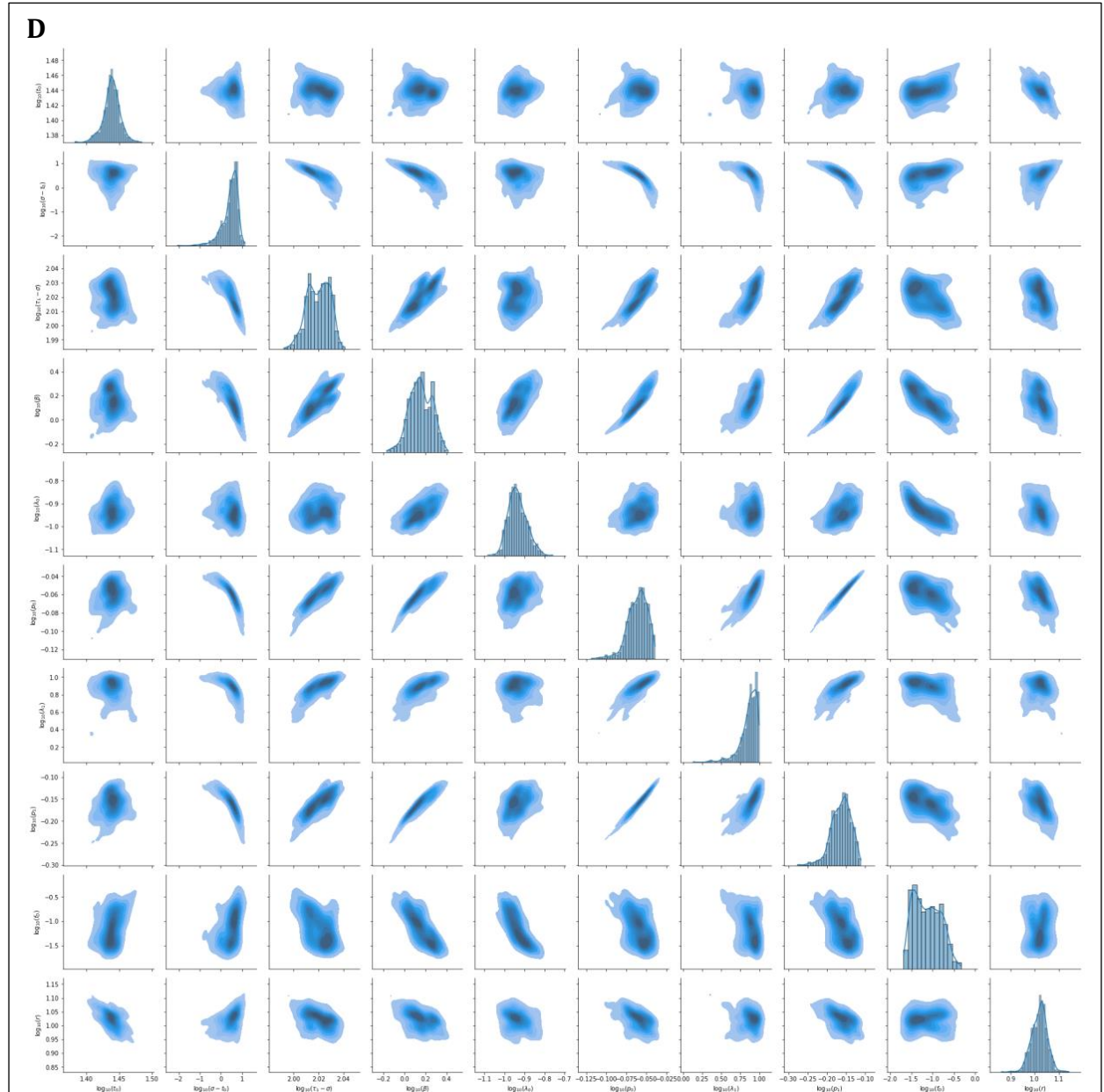

**Supplementary Fig. S20.** Evaluation of PyBNF's am method using Test Problem 4O (Supplementary File S4, Seattle), which is discussed in Supplementary Methods. **(D)** Pairs plot for samples generated using PyBNF's am method (cf. Panel E). Off-diagonal panels represent two-dimensional marginalizations of the multivariate parameter posterior and illustrate relationships between pairs of parameters, as indicated. For each off-diagonal panel, the x- and y-axes indicate parameter values, and color indicates probability density. Darker shades indicate higher density. Diagonal panels represent marginal posteriors, which are also shown in Panel B. Each marginal posterior is presented as a histogram and as a smooth curve, which was found via kernel density estimation (KDE) as explained in Supplementary Methods. Each two-dimensional marginalization is presented as a contour plot, which was found via KDE.

**SUPPLEMENTARY FIGURE S20 – PANEL E**

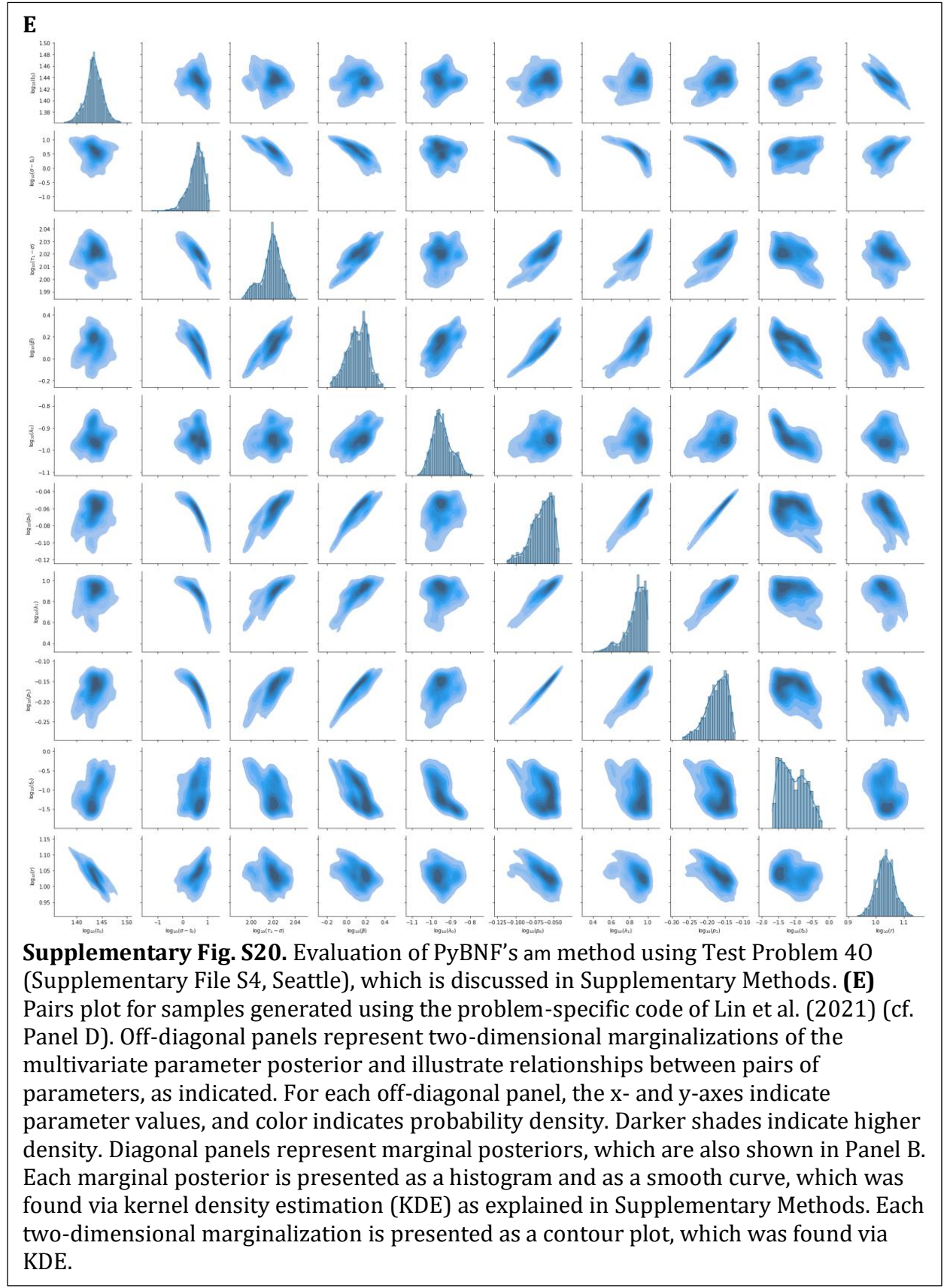

**Supplementary Fig. S20.** Evaluation of PyBNF's am method using Test Problem 4O (Supplementary File S4, Seattle), which is discussed in Supplementary Methods. **(E)** Pairs plot for samples generated using the problem-specific code of Lin et al. (2021) (cf. Panel D). Off-diagonal panels represent two-dimensional marginalizations of the multivariate parameter posterior and illustrate relationships between pairs of parameters, as indicated. For each off-diagonal panel, the x- and y-axes indicate parameter values, and color indicates probability density. Darker shades indicate higher density. Diagonal panels represent marginal posteriors, which are also shown in Panel B. Each marginal posterior is presented as a histogram and as a smooth curve, which was found via kernel density estimation (KDE) as explained in Supplementary Methods. Each two-dimensional marginalization is presented as a contour plot, which was found via KDE.